%% file: ms.tex
\pdfoutput=1
\documentclass[titlepage]{tcibook}
\pdfoutput=1
\usepackage{cite} 
\usepackage{fancyhea}
\usepackage{work}
\usepackage{bm}       %    enables bold math symbols  e.g.  \bm{\gamma}
\usepackage{graphicx}
\usepackage{hyperref}      % hypertext links %%ARXIV
\usepackage[usenames,dvipsnames]{color} %used for font color
\usepackage{amssymb}
\usepackage{amsmath}
\usepackage{chngcntr}
\usepackage{lscape}
\counterwithout{figure}{chapter}
\usepackage{titlesec}
\usepackage{epstopdf} % added to be able to compile with eps figures
\usepackage{subfig} % make it possible to include more than one captioned figure/table in a single float
\usepackage{multicol}

\hypersetup{
    colorlinks,
    citecolor=blue,
    filecolor=black,
    linkcolor=black,
    urlcolor=blue
}

\input workshopsymbols.tex      %   standard macros for common HEP terms

\graphicspath{{Introduction/}{telescopes/}{detector_rf/}{broadband_optics/}{readout/}{readout/TDMfigs/}{readout/FDMfigs/}}

\definecolor{orange}{rgb}{1,0.3,0}

\DeclareUrlCommand\email{\urlstyle{rm}}

\newcommand{\Pb}{{\sc Polarbear}}
\newcommand{\bicepArray}{{\sc Bicep} Array}
\newcommand{\bicepI}{{\sc Bicep}}
\newcommand{\bicepII}{{\sc Bicep}2}
\newcommand{\bicepIII}{{\sc Bicep}3}
\newcommand{\spider}{{\sc Spider}}

\newcommand{\umux}{{\sc $\mu$mux}}
\begin{document}

%\title{CMB-S4 Science Book \\ Working Draft}
%\author{CMB-S4 Collaboration}
%\maketitle

\def\bibname{References}

\bibliographystyle{utphys}  %%%% MODIFIED FOR CF %%%%

\raggedbottom

\pagenumbering{roman}

\parindent=0pt
\parskip=8pt
\setlength{\evensidemargin}{0pt}
\setlength{\oddsidemargin}{0pt}
\setlength{\marginparsep}{0.0in}
\setlength{\marginparwidth}{0.0in}
\marginparpush=0pt

% The content begins here

\renewcommand{\chapname}{chap:intro_}
\renewcommand{\chapterdir}{.}
\renewcommand{\arraystretch}{1.25}
\addtolength{\arraycolsep}{-3pt}

\numberwithin{equation}{section}

\newcommand{\comred}[1] {\textcolor{red}{#1}}

%    
% Abbreviations for ADS bibtex entries
%
\newcommand\mnras{{\it Mon. Not. Roy. Astron. Soc.}}
\newcommand\apj{{\it ApJ\ }} %{{\it Ap.\ J.\ }}
\newcommand\apjl{{\it ApJ\ Lett.}} %{{\it Ap.\ J.\ Lett.}}
\newcommand\araa{{\it Ann. Rev. Astron. Astroph.}}
\newcommand\physrep{{\it Phys. Rept.}}
\newcommand\apjs{{\it ApJ\ Suppl.\ }}
\newcommand\jcap{{\it JCAP }}
\newcommand\prd{{\it Phys.\ Rev.\ }{\bf D}\ }
\newcommand\aap{{\it A \& A\ }}
\newcommand{\procspie}{Proc. SPIE}
\newcommand{\ao}{Ap. Opt.}

%%%%%%%%%%%%%%%%%%%%%%%%%%%%%%%%%%%%%%%%%%%%%%%%%%%
\pagenumbering{roman} 
\chapter*{CMB-S4 Technology Book\\ First Edition}
\vskip -9.5pt
\hbox to\headwidth{%
       \leaders\hrule height1.5pt\hfil}
\vskip-6.5pt
\hbox to\headwidth{%
       \leaders\hrule height3.5pt\hfil}

{\Large\bf
  \begin{center}
   CMB-S4 Collaboration\\
  %\\
   \bigskip
   \today
%  March 21, 2016
 \end{center}
}
\eject

\setcounter{page}{1}

%%%%%%%%%%%%%%%%%%%%%%%%%%%%%%%%%%%%%%%%%%%%%%%%%%%

%\input ExecSum/exec-sum.tex

%%%%%%%%%%%%%%%%%%%%%%%%%%%%%%%%%%%%%%%%%%%%%%%%%%%
%\clearpage
%\begin{center}
%  {\Large \bf Acknowledgements}
%\end{center}
%%%%%%%%%%%%%%%%%%%%%%%%%%%%%%%%%%%%%%%%%%%%%%%%%%%  
%\clearpage
%\begin{center}
%  {\Large \bf Executive summary}
%\end{center}
%
%\input{executive_intro}
%
%\section*{Telescope design}
%\input{telescopes/executive_summary}
%
%\section*{Receiver optics}
%\input{broadband_optics/Executivesummary}
%
%\section*{Focal-plane optical coupling}
%\input{detector_rf/executive_summary_brief}
%
%\section*{Focal-plane sensors and readout}
%\input{readout/exec_summary}
%
%\section*{List of acronyms}
%\input{acronym_execsumm}

%%%%%%%%%%%%%%%%%%%%%%%%%%%%%%%%%%%%%%%%%%%%%%%%%%%  
\clearpage
\setcounter{tocdepth}{2} %Do not show subsubsection in TOC
\tableofcontents
\input{endorsers/endorsers.tex}

%%%%%%%%%%%%%%%%%%%%%%%%%%%%%%%%%%%%%%%%%%%%%%%%%%%
%%%% Author comment macros here

\def\as#1{[{\bf AS:} {\it #1}] }

%%%%%%%%%%%%%%%%%%%%%%%%%%%%%%%%%%%%%%%%%%%%%%%%%%%
%%%%%%%%%%%%%%%%%%%%%%%%%%%%%%%%%%%%%%%%%%%%%%%%%%%
%%%     All of your files should be in a subdirectory.  Here the
%%%     subdirectory is called CF0. The title of your
%%%     report should be wgreportCF0.tex in that subdirectory.  Input
%%%     that file here
%%%%%%%%%%%%%%%%%%%%%%%%%%%%%%%%%%%%%%%%%%%%%%%%%%%%
%%%%%%%%%%%%%%%%%%%%%%%%%%%%%%%%%%%%%%%%%%%%%%%%%%%

\eject
\pagenumbering{arabic}
\setcounter{page}{1}
\input{Introduction/intro_cmbs4.tex}
\input{telescopes/telescope_cmbs4.tex}
\input{broadband_optics/optics_cmbs4.tex}

\input{detector_rf/detectorrf_cmbs4.tex}
\input{readout/detectorreadout_cmbs4.tex} 
\input{conclusion_cmbs4.tex}

\input{acronym_total.tex}

\clearpage

\bibliography{executive_intro,telescopes/telescopes,broadband_optics/bibtex/broadband_in_use,detector_rf/bibtex/detector_rf_in_use,readout/bibtex/readout_in_use}

%\input conclusions_cmbs4.tex

%%%%%%%%%%%%%%%%%%%%%%%%%%%%%%%%%%%%%%%%%%%%%%%%%%
%%%%%%%%%%%%%%%%%%%%%%%%%%%%%%%%%%%%%%%%%%%%%%%%%%
%%%   Your subdirectory (here CF0) should include
%%%    the files:
%%%           wgreportCF0.tex
%%%           authorlistCF0.tex
%%%           bibCF0.bib
%%%         and all needed figures in pdf format
%%%%%%%%%%%%%%%%%%%%%%%%%%%%%%%%%%%%%%%%%%%%%%%%%%%%
%%%%%%%%%%%%%%%%%%%%%%%%%%%%%%%%%%%%%%%%%%%%%%%%%%%%

\end{document}

%% file: workshopsymbols.tex
\newcommand{\nc}{\newcommand}  

\def\eg{{\it e.g.}}

\def\beq{\begin{equation}}
\def\eeq#1{\label{#1}\end{equation}}
\def\eeqn{\end{equation}}

\newenvironment{Eqnarray}%
   {\arraycolsep 0.14em\begin{eqnarray}}{\end{eqnarray}}
\def\beqa{\begin{Eqnarray}}
\def\eeqa#1{\label{#1}\end{Eqnarray}}
\def\eeqan{\end{Eqnarray}}

\nc{\ra}{\rightarrow}  
\nc{\slsh}{\slash\hspace*{-0.22cm}}
\def\Re{{\cal R \mskip-4mu \lower.1ex \hbox{\it e}\,}}
\def\Im{{\cal I \mskip-5mu \lower.1ex \hbox{\it m}\,}}

\nc{\vev}[1]{ \left\langle {#1} \right\rangle }
\nc{\bra}[1]{ \langle {#1} | }
\nc{\ket}[1]{ | {#1} \rangle }
\nc{\fb}{\,{\rm fb}^{-1}}
\nc{\ev}{{\rm eV}}
\nc{\kev}{{\rm keV}}
\nc{\Mev}{{\rm MeV}}
\nc{\gev}{{\rm GeV}}
\nc{\tev}{{\rm TeV}}
\nc{\mev}{{\rm MeV}}

\def\del{\partial}
\def\Dslash{\not{\hbox{\kern-4pt $D$}}}
\def\dslash{\not{\hbox{\kern-2pt $\del$}}}
\def\pslash{\not{\hbox{\kern-2pt $p$}}}
\def\ETmiss{ \not{\hbox{\kern-4pt $E$}}_T }

\def\One{\ensuremath{\bf 1}\xspace}

\def\msb{{\bar{\ssstyle M \kern -1pt S}}}

%% file: endorsers/endorsers.tex
\clearpage
\begin{center}
{\LARGE \textbf{List of Contributors and Endorsers}}
\vspace*{\baselineskip} 
%\vspace{1cm} 
\end{center}
The following people have endorsed survey of instrumentation for the CMB as presented here.  Those contributed to the writing of this document are marked with asterisk:\\

\noindent 
Kevork	N.	Abazajian$^{	38	}$,
Maximilian	H.	Abitbol$^{*	8	}$,
Zeeshan		Ahmed$^{*	30	}$,
David		Alonso$^{	49	}$,
Adam	J.	Anderson$^{	12	}$,
Kam		Arnold$^{	39	}$,
Jason	E.	Austermann$^{	26	}$,
Darcy		Barron$^{*	36	}$,
Ritoban		Basu Thakur$^{*	40	}$,
Andrew		Bazarko$^{	27	}$
% ...

\clearpage
\begin{multicols}{2}{
% ...
$^{	8	}$	Columbia University	\\
% ...
$^{	12	}$	Fermi National Accelerator Laboratory	\\
% ...
$^{	26	}$	NIST	\\
$^{	27	}$	Princeton University	\\
% ...
$^{	30	}$	SLAC National Accelerator Laboratory	\\
% ...
$^{	36	}$	University of California, Berkeley	\\
% ...
$^{	38	}$	University of California, Irvine	\\
$^{	39	}$	University of California, San Diego	\\
$^{	40	}$	University of Chicago	\\
% ...
$^{	49	}$	University of Oxford	\\
% ...

}
\end{multicols}

\clearpage

%% file: Introduction/intro_cmbs4.tex
\chapter{Introduction and overview} %to CMB-S4 technology book} 
\label{ch:introduction}
\vspace*{\baselineskip} % title is multiline, and without this no space between title & text
\vspace{1cm}

\section{Introduction to the CMB-S4 Technology Book}
CMB-S4 is a proposed experiment to map the polarization of the Cosmic Microwave Background (CMB) to nearly the cosmic variance limit for the angular scales that are accessible from the ground.  The science goals and capabilities of CMB-S4 in illuminating cosmic inflation, measuring the sum of neutrino masses, searching for relativistic relics in the early universe, characterizing dark energy and dark matter, and mapping the matter distribution in the universe have been described in the CMB-S4 Science Book~\cite{cmbs4_sciencebook}.   For CMB-S4 to be able to achieve the ambitious goals laid out in the Science Book will require a major step forward in experimental capability from the ongoing Stage-III experiments that are now starting. This CMB-S4 Technology Book is a companion volume to the Science Book and describes the status of the technology for CMB-S4. 

There is range of existing technologies that are promising for meeting the challenging requirements set by CMB-S4;  the optimum set of technologies has not yet been determined. While the focus of this document is on individual technologies, it is exciting to contemplate collaboration across groups, the potential for hybrid approaches that could lead to new breakthroughs, all leading towards developing the best approach to achieve the exciting science goals of CMB-S4. The diversity of approaches under development highlights the vitality of the experimental CMB community.

To guide in the compilation of the Technology Book, the community therefore agreed upon the following charge for this document:

{\it Summarize the current state of the technology and identify R\&D efforts necessary to advance it for possible use in CMB-S4. CMB-S4 will likely require a scale-up in number of elements,
frequency coverage, and bandwidth relative to current instruments. Because it is searching for
lower magnitude signals, it will also require stronger control of systematic uncertainties.}

%This document has four chapters, each discussing the current status and future needs of technologies that could be used in CMB-S4. 

Proceeding from the sky, through the instrument, all the way to the detector data being stored to disk, we have grouped the relevant technologies into the following areas: Telescope Design; Receiver Optics; Focal-Plane Optical Coupling; and Focal-Plane Sensors and Readout. A chapter of the book is dedicated to each of these four areas.  The technology choices for CMB-S4 will be inter-dependent, however. For example, the telescope will need to be designed together with the receiver optics, which in turn will need to be designed jointly with the detector arrays.

To aid in the understanding of the maturity of the technologies considered, we evaluated their current {\it technical readiness} with a 5-level Technology Status Level (TSL) and their {\it manufacturing readiness} with a 5-level Production Status Level (PSL). The criteria for these levels are described in Table~\ref{tab:tslpsl}.  We also evaluate the effort needed to mature the technology if it is to be a viable candidate for CMB-S4. %Future work will improve estimation of systematic errors that are generated or mitigated by each technology, however an experiment-wide perspective is often needed, and therefore results from Stage-III experiments will be crucial.

\begin{table} [!h]
\begin{center}
\begin{tabular} {cl}
\hline
\textbf{TSL} & \textbf{Description} \\
\hline
1 & Lab test of technology to show principle\\
2 & Lab test of technology but with full feature set and performance suitable for ground test\\
3 & Experiment capable version built and tested in the lab\\
4 & Deployed in a CMB experiment and data taken\\
5 & Data fully analyzed, systematic errors understood\\
\hline
\hline
\textbf{PSL} & \textbf{Description} \\
\hline
1 & Fabrication of a TS1/TS2 prototype demonstrated\\
2 & Fabrication of a one or more experimental capable units\\
3 & Conceptual plan of methods for production at scale\\
4 & Demonstrated the critical steps for production at scale\\
5 & Capability for production at scale exists and is demonstrated\\
\hline
\end{tabular}
\caption{Technology Status Level (TSL) and Production Status Level (PSL) definitions.}
\label{tab:tslpsl}
\end{center}
\end{table}

%
%The science case for an ambitious ``Stage-IV" ground-based CMB experiment has been made in the CMB-S4 Science Book. CMB-S4 promises to illuminate the physics of the very beginning of the universe $10^{-36}$ seconds after the Big Bang at an energy scale of $10^{16}$ GeV.  CMB-S4 will also provide a census of relativistic particles present before recombination, measure or limit the sum of the neutrino masses, probe the early time behavior of dark energy, and constrain weakly interacting massive particles (WIMP) as 
%candidates for the dark matter.  

%To achieve these science goals a leap in experimental capability from Stage-III experiments is needed. The basic technologies required to make this leap already largely exist, but research and development is needed to bring them to maturity and develop the manufacturing capability for producing the quantities necessary for CMB-S4. Judicious choices among the available technologies can reduce the cost, complexity, risk, and development time for CMB-S4. This CMB-S4 ``Technology Book" is the companion to the Science Book~\cite{cmbs4_sciencebook} and describes candidate technologies for the CMB-S4 experiment.  

\begin{figure}[h!]
\centering
\includegraphics[trim = 0.4in 0.7in 0.8in 0.7in, clip, width=1.0\textwidth]{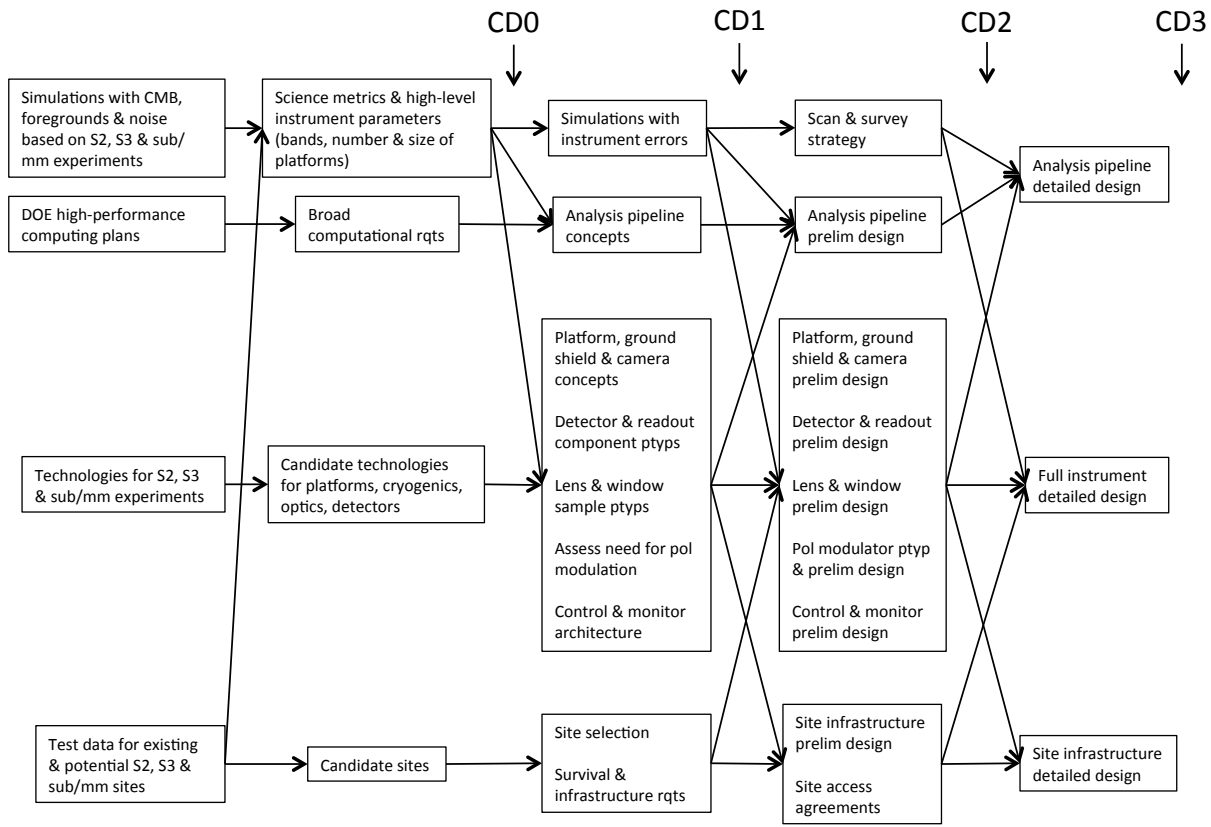}
\caption{CMB-S4 Roadmap. Here S2 and S3 stand for Stage-II and Stage-III, respectively.}
\label{fig:roadmap}
\end{figure}

To understand the timing of the technology development within the overall CMB-S4 schedule, a roadmap of the Critical Decision (CD) development path for the project is shown in Figure~\ref{fig:roadmap}. Overall system design will drive all technology selections. For example, systematic error considerations can drive telescope and optics geometry which in turn impacts decisions on detector design. The strong connections between the many sub-systems in the experiment will require an iterative optimization.  

In the next section we give brief overviews of the requirements for CMB-S4 and status of the technologies.The subsequent chapters of this book provide detailed descriptions of these technologies, their status (including TSLs and PSLs), and the next development steps, as follows: Chapter 2 covers Telescope Design; Chapter 3 covers Receiver Optics; Chapter 4 covers Focal Plane Optical Coupling; and Chapter 5 covers Focal Plane Sensors and Readout.  Lastly, brief concluding remarks are given in Chapter 6.

\section{Overview and Status of  CMB-S4 Technology Areas}
\subsection*{Telescope Design}
\input{telescopes/executive_summary}

\subsection*{Receiver Optics}
\input{broadband_optics/Executivesummary}

\subsection*{Focal-plane Optical Coupling}
\input{detector_rf/executive_summary_brief}

\subsection*{Focal-plane Sensors and Readout}
\input{readout/exec_summary}

%% file: telescopes/executive_summary.tex
To achieve the ambitious science goals as set out in the Science Book, CMB-S4 will require of order $\sim$500,000 effectively background-limited detectors. Since this exceeds the number of detectors that could fit in any current single telescope design, CMB-S4 will be an array of multiple telescopes. Chapter 2 reviews current telescope designs and presents promising future designs that would help meet the throughput challenges presented by CMB-S4.

At least a subset of the telescopes for CMB-S4 must have an optical beam size in the range of 1--4 arcminutes to meet many of the science requirements including those that exploit gravitational lensing, measurements of the damping tail, and galaxy cluster measurements. These large-aperture telescopes will have primary apertures in the 2--10 meter diameter range.
%To date, the most sensitive measurements of the degree scale inflationary recombination peak have been made with purpose built small-aperture telescopes in the 0.3--0.5 meter diameter range. 
To date, the most sensitive constraints on the degree scale inflationary recombination peak have been obtained with purpose built small-aperture telescopes in the 0.3--0.5 meter diameter range. 
Although large-aperture telescopes are, in principle, capable of high-fidelity measurements at these large scales, a conservative design of CMB-S4 would include both small and large telescopes. %Even if the large telescopes become capable of measuring the degree scale signals, 
The large and small telescopes have different and complementary sets of systematic errors, and including both types will likely lead to the most robust results on inflation. Finally, the optics of both telescopes must be designed to minimize beam systematic uncertainties, sidelobes, and spurious instrumental polarization.  
\begin{itemize}

\item \textbf{Small-aperture telescopes: } The small-aperture telescopes for CMB-S4 can be built with entirely cryogenic optics, reducing detector noise due to optical loading from the telescope, and suppressing the coupling of stray light onto the detectors. The optical design can be refractor-based as with \bicepI/\spider/Keck Array or reflector-based as with ABS. {\bicepIII} and the {\bicepArray} are moving forward with high-throughput fully cryogenic optics. CLASS has a mix of cold refractive optics and ambient-temperature reflective optics. Performance from these designs will inform the design of CMB-S4. %if small-aperture telescopes are used.

\item \textbf{Large-aperture telescopes: } Existing designs of fielded large-aperture telescopes are reviewed, as well as  two attractive new large-aperture optical designs. One is a crossed Dragone design, which provides
excellent image quality across a wide field of view. %Another would be a larger number of less expensive Gregorian telescopes, each with smaller field of view. We have made a first attempt at this trade off based on idealized optical designs, but 
The final configuration will require a detailed comparison of realistic estimates of systematic errors as well as consideration of 
mapping speed vs.\ cost. % for designs that include realistic estimates of systematic errors.

\item \textbf{Control of telescope systematic uncertainties:} Beam sidelobes that pick up spurious signals (such as those from the ground, sun and moon), beam asymmetries, and instrumental polarization can all be induced by the telescope if it is not carefully designed. Several small-aperture telescopes have demonstrated stringent control of sidelobes
using a co-moving cylindrical absorptive shield. Existing large-aperture telescopes
do not have the same comprehensive shielding as existing small-aperture
telescopes, but it is plausible that such stringent shielding could be
achieved provided that the 
requirement is built in at the beginning of the design process. CMB-S4 designs must also limit far-field scattering from optical
surfaces including reflector-panel gaps and optical stops. 
To design the telescope to avoid these effects, ray tracing, as well as full beam diffraction effects can be simulated with commercial software. To further mitigate systematic uncertainties, there are also optimal mapmaker and other data analysis techniques. It will also be critical to verify with simulation and measurements that ground shields intended to limit sidelobes
are not in fact adding new sidelobes themselves. Before finalizing the optical design of CMB-S4, end-to-end simulations of the telescope performance, through to simulated science data, would enable us to evaluate the impact of telescope design choices on the CMB science results.

\end{itemize}

%% file: broadband_optics/Executivesummary.tex
Advances in optical technologies have been crucial in moving the community towards realizing the full science potential of the CMB.  Chapter 3 discusses the optical elements contained in CMB receivers which include lenses, filters, polarization modulators, and vacuum windows. The implementation and performance of these technologies are rapidly evolving and many new approaches are being successfully fielded on Stage-III experiments. These innovations draw on advances in materials, processing techniques, and developments in electrical engineering including metamaterial research.  Examples of all of these components with relatively large diameters and broad operating bandwidths have been already successfully deployed to the field, but continued improvements on both diameter and bandwidth would be useful for CMB-S4. In Chapter 3 we summarize the current state-of-the-art and identifies development efforts that would be needed to ready each technology for CMB-S4.
%We now summarize the status of each technology area.

\begin{itemize}
%\vspace{-.15 in}
%\paragraph{Windows:} 
\item \textbf{Windows:} Vacuum windows have been built with closed-cell form sheets (Zotefoam) and from AR-coated polyethylene sheets. Work is needed to realize larger windows for the high-throughput CMB-S4 receiver designs that maintain low loss, scattering, and reflectivity.  

%\vspace{-.15 in}
%\paragraph{Filters:} 
\item \textbf{Filters:} Most experiments currently use hot-pressed metal-mesh filters to help define the spectral band and to block out-of-band radiation coupling onto the detectors.  Mesh filters are also used, along with infrared (IR) absorptive filters of polytetrafluoroethylene (PTFE), Nylon, and alumina ceramic, to reduce thermal loading from incident IR radiation onto the cryogenic stages. Emerging technologies for reducing the IR thermal loading include layers of laser-ablated free-standing metal mesh filters, IR-absorbing foam, and a composite of absorbing crystals and mesh filters on silicon. Further work is needed to realize large-diameter filtering schemes while minimizing in-band loss, scattering, and spurious polarization effects.

%\vspace{-.15 in}
%\paragraph{Anti-reflection coatings (ARC):} 
\item \textbf{Anti-Reflection Coatings:} In the last five years, new approaches to multilayer coatings have emerged including: dielectric metamaterials cut by dicing saws, lasers, or etching; dielectric coatings made by casting, pressing, and thermal spray methods; and artificial dielectrics. Work is needed to realize appropriately wide-bandwidth and large-diameter optical elements with low loss and at a practical cost for CMB-S4.

%\vspace{-.15 in}
%\paragraph{Polarization modulators:} 
\item \textbf{Polarization Modulators:} Polarization modulators are an important tool for realizing precision measurements of polarization as they mitigate systematic errors and $1/f$ noise. Emerging approaches include anti-reflection coated half-wave plates (HWPs) fabricated from sapphire; metamaterial grooved silicon; or metamaterials realized as metal meshes; and variable polarization modulators realized by wire grids. Work is needed to realize  broad-bandwidth, large-diameter, high-uniformity, and  practically manufacturable modulators for CMB-S4.

%\vspace{-.15 in}
%\paragraph{Dielectric materials:} 
\item \textbf{Materials:} Low-loss dielectric materials are common to lenses, half-wave plates, IR filters, and windows.  Work is needed to develop both high index of refraction alumina and silicon and relatively low index of refraction polyethylene in the large diameters required for CMB-S4.   
%Also, work is needed to extend the direct machining of these materials to create metamaterials.

%\vspace{-.15 in}
%\paragraph{Characterization:} 
\item \textbf{Characterization:} Current approaches to optical characterization include transmission and reflection measurements using either coherent or broad-band sources of warm and cold optical elements and direct metrology. Precise characterization of the quality of the materials and performance of completed optical elements at their operating temperature is crucial to achieving the goals of CMB-S4.   

\end{itemize}

%% file: detector_rf/executive_summary_brief.tex
%This CMB-S4 technical paper
After the light from the sky passes through the telescope and is brought to a focus by the receiver optics, it must be coupled to the detectors. Chapter 4 discusses focal-plane coupling elements used in CMB arrays. The detector sensors themselves will be discussed in the next section.  Rather than coupling the CMB signals to the sensors with simple absorbers in the traditional bolometric method, most CMB arrays use antennas.
% Note:  we haven't explained about atmospheric bands yet -- an absorber coupled detector can in general be enormously broad band otherwise (thus the word "bolometric"), so the previous description was jarring.
An antenna-coupled pixel has (i) an antenna that converts the free space wave to a guided wave, (ii) superconducting transmission lines, and (iii) one or several filters that define the passband(s) before the light continues on to the detector sensor(s). This approach allows simultaneous coverage of several distinct
frequency bands in each spatial filter.  

%After the light from the sky passes through the telescope and is brought to a focus by the receiver optics, focal-plane coupling elements are needed to couple the light onto the detectors. This chapter discusses focal-plane coupling elements used in CMB arrays. The detectors themselves will be discussed in the next chapter. An absorber-coupled detector directly couples to the telescope using a simple absorptive element. While sensitive and more straightforward to built, they are necessarily limited to a single band. An antenna-coupled pixel has (i) an antenna that converts the free space wave to a guided wave, (ii) superconducting transmission lines, and (iii) one or several filters that define the passband before the light continues on to the detector. This approach can allow each spatial pixel to be built to simultaneously cover several different frequency bands.

\begin{itemize}
%The antennas don't generically determine the bandwidth -- feedhorns need lowpass filters somewhere -- they just become modally impure at high frequencies.  
\item \textbf{Antennas: } The antenna determines the polarization performance and beam shape of the detector.  Different types of antennas permit different total bandwidths with good optical properties. Controlling the polarization and beam shape is crucial to mitigating systematic errors in the CMB measurements, while the total bandwidth impacts the total sensitivity per unit focal plane area.  
%The antenna naturally determines the total bandwidth, polarization performance, and beam shape of the detector. The total bandwidth drives the total sensitivity of the pixel, and controlling the polarization properties and beam shape of the antenna are crucial to mitigating overall systematic error in the instrument. 
The three antenna types in current CMB experiments are feed horns coupled to planar orthomode transducers (OMTs),  lenslet-coupled planar antennas, and planar phased-array antennas. 
% I checked with Bill Jones who agreed that now that Planck is done, there are no experiments using a feed horn coupled to an absorber for CMB

\begin{itemize}
\item \textit{Feed horn} arrays have been manufactured using gold-plated stacks of etched silicon wafer ``platelet arrays," reducing the manufacturing cost and complexity compared to traditional electroformed horns. Direct drilling of arrays of smooth-walled feed horns has also been demonstrated.  A planar OMT in the circular waveguide at the base of each feedhorn, backed by a quarter-wave short, couples the light onto planar transmission lines leading to the detector sensors.  Some earlier CMB instruments used direct absorbers in the feedhorns instead, typically operating with a single band.  % GSFC did develop FSBs though they were probably never used for CMB - but depending on your definition of  for single-band operation. 
% antennas are being manufactured into arrays using a stack of etched silicon wafer ``platelet arrays", reducing the manufacturing cost and complexity compared to traditional electroformed horns. Direct drilling of arrays of smooth-walled feed horns has also been demonstrated. In all of these systems, the transmission line is a circular waveguide, often with probes to couple the light onto the planar transmission line in the detector chip. Alternately, for single-band detectors, a direct absorber has been successfully used in the waveguide.

\item \textit{Lenslet} arrays using sprayed anti-reflection coatings, as well as stacked-wafer gradient index lenses, are being developed to simplify manufacturing. For these antennas, a planar superconducting antenna is on the chip under each lenslet to couple the light onto the planar transmission line.

\item \textit{Phased array} antennas use grids of small antennas, each coupled to a millimeter-wave transmission line. Each of these transmission lines travels back to the detector through a summing tree architecture, with the length of each transmission line matched such that the waves combine in phase at the detector. This lets the antenna array combine to form a main beam.
\end{itemize}

% is this a reference to multi-mode horns? or multimode direct absorbers i.e. PIXIE? I think it's the latter...I'll have to look later to find which
% Architectures that detect multiple modes using direct absorber coupling have been demonstrated with single pixel prototypes but large-scale arrays require demonstration. 

\item \textbf{Multichroic Architectures: } A single-band antenna-coupled pixel has a single band-defining transmission line filter between the antenna and the lossy detector termination for each polarization mode.   Multichroic pixels have filter banks (channelizers) that divide the total bandwidth of the antenna into multiple simultaneous bands each of which terminates at its own detector. The mm-wave circuitry of the pixel, including the transmission lines, crossovers, filters, and terminations need to be manufactured with stringent control of the circuit parameters and high uniformity across the array. Multichroic operation has been demonstrated for feed horn and lenslet-coupled antennas and is in the prototype stage for phased-array antenna pixels. Given that multiple frequency bands are required for foreground separation, a pixel that can measure more than one band simultaneously can provide more efficient use of the focal-plane area. Multichroic pixels require corresponding broadband receiver and telescope optics. Since the bands for each pixel are coupled through a single aperture, the pixel aperture size has to be chosen to optimize the sensitivity per band, and the improvement does not scale linearly with the number of bands.  
%.  effective aperture for each pixel The need to choose a single effective aperture for each pixel across a broad frequency band 
 %physical pixel spacing for a broad range of wavelengths requires optimization, and limits the potential improvement. This and other considerations means that 
 The number of bands per pixel must be determined by an experiment-wide optimization that includes the performance of all the optical elements, and the  cost of the telescopes, detectors, readout electronics, and receiver optics.

\item \textbf{Fabrication and Testing: } CMB-S4 will require of order 1,000 science-grade silicon detector wafers, and therefore mass manufacturing capability has to be developed.  Among the DOE labs, ANL, LBNL, and SLAC are developing wafer fabrication throughput and consistency, as well as exploring hybrid fabrication using a combination of on-site and commercial foundries. Detector characterization is an essential part of the detector manufacturing process.
%Increasing throughput of detector testing is crucial to achieving the goals of CMB-S4.  

\end{itemize}

%% file: readout/exec_summary.tex
Chapter 5 provides a survey of the state of low-noise sensors and signal readout suitable for CMB-S4, focusing on promising scalable technologies. The order of magnitude leap in detector count from Stage-III experiments to CMB-S4 puts an emphasis on choosing a sensor and multiplexing combination that is straightforward to read out, integrate with the experiment, and manufacture. 

\begin{itemize}
\item \textbf{Sensors: }We have identified Transition-Edge Sensors (TES) and Microwave Kinetic Inductance Detectors (MKIDs) as the leading candidates for the signal transducers in CMB-S4. TESs have a long record of well-characterized performance and CMB science results. They are a natural choice for CMB-S4 and would benefit from production scaling R\&D. MKIDs are an attractive option to combine highly-multiplexed readout with signal transduction. They have demonstrated promising noise performance in laboratory measurements and millimeter-wave astronomy instruments, but are at an earlier stage of technological maturity. Areas for future MKID R\&D include antenna-coupling, sensors for 100\,GHz and below, and on-sky CMB measurements to verify that the detectors continue to have adequate white and 1/f noise in CMB applications.    

\item \textbf{Multiplexed readout: } As with Stage-III experiments, multiplexed readout will be crucial for the large detector arrays needed for CMB-S4.

\begin{itemize}
\item \textit{TES} sensors will be multiplexed using one or more of three candidate technologies: Time-division multiplexing (TDM) using
  Superconducting Quantum Interference Devices (SQUIDs) as switches,
  frequency-division multiplexing (FDM) using in-series MHz resonators, 
  or frequency-division multiplexing using GHz-excitation techniques (\umux{}).
  
\item \textit{MKID} sensors employ frequency-division multiplexing using GHz-excitation techniques, but do not require a cold multiplexer separately from the sensor. 
\end{itemize}
 
 For CMB-S4, every TES multiplexing technology
  considered will benefit from an increased multiplexing
  factor, defined as the maximum number of sensors read out per
  readout channel.  With some development effort, the TDM and FDM
  techniques for TES readout can be scaled to multiplexing factors of $\sim\,$200.
  %Further scalability requires modest to extensive R\&D, but is recommended..
  The \umux\ TES readout technique could potentially produce MKID-like high multiplexing
  factors for TESs ($\sim$1000). All multiplexing techniques will also benefit from improved designs that reduce assembly complexity and provide cost and schedule savings given the large number of detectors necessary for CMB-S4.

\item \textbf{Room temperature electronics: } In all of the technologies under consideration, the warm readout electronics appear scalable with development. Frequency-division techniques for both TESs and MKIDs use similar room-temperature biasing and readout electronics, enabling common development. It will be important to ensure that good individual detector performance (linearity, stability, sensitivity, and other performance parameters) continues even as multiplexing factors increase. 

\end{itemize}

%% file: telescopes/telescope_cmbs4.tex
\chapter{Telescope design}\label{chp:telescopes}
\vspace*{\baselineskip} % title is multiline, and without this no space between title & text
\vspace{1cm}

\input{telescopes/introduction2}

\input{telescopes/current_designs} % Nils
\input{telescopes/engineering_systematics} % Steve
\input{telescopes/future}

%\appendix
\input{telescopes/projects}

\input{telescopes/conclusion}

%% file: telescopes/introduction2.tex
\section{Introduction}

The design of the telescope has a central role in determining both the sensitivity and the level of systematic errors in a cosmic microwave background (CMB) experiment.  The sensitivity of modern CMB detectors is determined by photon arrival statistics, and therefore the total sensitivity is determined by the number of detectors.  CMB-S4 will require of order 500,000 detectors.   Stage-III telescope designs typically contain $2,000-10,000$ detectors, so the development of higher throughput (larger field of view) telescope designs would reduce the cost of CMB-S4. Systematic errors can be introduced by the telescope in several ways.  First, the level of the telescope's sidelobe response generated from diffraction and scattering is critical since the sidelobes will scan the 300K ground and relatively bright sky sources such as the sun, moon, and galaxy.  Second, distortions of the beam (e.g., ellipticity) that depend on polarization angle, and polarization leakage from cross polarization and instrumental polarization will create false signals.  Finally, pointing errors that could arise from thermal distortion of the telescope will create false signals by shifting structures on the sky. 

Currently, ``small-aperture" telescopes such as \bicepI/Keck Array, ABS, \spider, PIPER, and CLASS, which typically have a primary aperture $\lesssim 1$~m, have the best demonstrated noise and systematic error performance at the angular scales of inflationary signals ($\ell < 200$).  The smaller aperture size makes it practical to implement comprehensive co-moving shields, boresight rotation of the entire telescope, and polarization modulators as the first optical element.  All of these features can be advantageous for control of systematic errors at the angular scales of the inflationary signals.  

``Large-aperture" telescopes such as ACT, EBEX, \Pb, QUIET, and SPT have demonstrated high-fidelity mapping of faint CMB structure at small angular scales, such as those from gravitational lensing, the CMB damping tail, and galaxy clusters.  Given the relatively high cost of large-aperture telescopes, it is highly desirable to increase the size of the field of view (FOV) and studies have yielded Stage-IV designs that give up to a factor of 10 improvement in detector count compared to current Stage-III designs.  

Given their complementary performance attributes, a plausible choice for CMB-S4 is to employ a hybrid of small- and large-aperture telescopes.  However, if large-aperture telescopes could be demonstrated to give high fidelity measurements at inflationary angular scales, the use of a homogeneous array of large-aperture telescopes would reduce the cost of CMB-S4 by reducing the total number of detectors, readout, and cryogenic systems including their energy use at remote sites.  

This chapter is organized as follows. In Section~\ref{sec:telescope_designs} we provide a general review of current telescope designs, including both small ($\lesssim$ 1\,m) and large ($\gtrsim$ 1\,m) apertures. In Section~\ref{sec:CD_design} we discuss a concept to achieve high throughput from a telescope. In Section~\ref{sec:engineering_systematics} we provide a non-exhaustive list of methods that can mitigate instrumental systematic errors. In Section~\ref{sec:futurestudies} we propose future studies and development areas that could expand optical designs that are viable for CMB-S4. In Section~\ref{sec:projects} we further review currently fielded telescope designs. We provide our final conclusions in Section~\ref{sec:telescope_conclusion}.

%% file: telescopes/current_designs.tex
\section{Current CMB telescope designs and maturity}
\label{sec:telescope_designs}

% Overview
% - Number of experiments
% - variety of apertures and designs
% - design considerations
%    - Mitigating scattering and far sidelobes
%    - Instrumental polarization
%    - Polarization modulation
%    - Maximizing throughput for mapping speed
%       - wide field optical design
%       - multichroic detectors
%    - Window size/material
%    - AR coatings
%    - lens material

The current generation of CMB telescope designs incorporate lessons
learned from decades of %measurement 
experience \cite{HananyNiemackPage2013}. The ten
current-generation experiments and two previous-generation telescope
designs presented in Table~\ref{tab:telescopes} implement a wide variety of optical design approaches, including
both refractive and reflective primary apertures, Gregorian and
crossed-Dragone mirror configurations, single- and multi-camera
systems, and the use or absence of polarization modulation 
mechanisms such as rotating HWPs, reflective variable-delay polarization
modulators (VPMs), telescope boresight rotation, and sky rotation. 
%or no polarization modulation.

Small- and large-class telescope designs are discussed separately in
the sections below. A table summarizing some relevant optical parameters
is given in Table~\ref{tab:telescopes}.  A detailed description of several current
telescope optical designs is given in Section~\ref{sec:projects}.

\subsection{Current small-aperture telescope designs}

Small mm-wavelength CMB telescopes, with apertures $\lesssim 1$\,m, are
designed to probe the inflationary gravitational wave $B$-mode signal at
large angular scales, $\ell < 200$, but not the lensing signal at
$\ell > 200$. Small telescopes reviewed here include ABS (0.25\,m
physical aperture), \bicepIII\ (0.53\,m), CLASS (0.6\,m), Keck Array/\spider\
(0.25\,m), and PIPER (0.39\,m). \bicepIII\ and Keck Array/\spider\ use all
refractive elements, whereas the other experiments use dual off-axis
reflector designs, and (with the exception of ABS) cold refractive
reimaging optics.

\bicepIII\ (and the future \bicepArray) is an evolution of the
\bicepI/Keck Array optical design. Both are a simple two-lens
objective/field lens design with a stop just behind the
objective. The lenses and stop are cooled to
4\,K. The \bicepIII\ telescope and each Keck Array/\spider\ telescope are
single-frequency, which simplifies the anti-reflection (AR) coating
implementation. Multi-frequency coverage is accomplished via the
deployment of multiple telescopes. Lenses in \spider\ and the Keck Array are
fabricated from HDPE plastic, whereas those in the larger, 520\,mm
aperture, \bicepIII\ telescope are alumina. A comoving absorptive conical baffle
is placed in front of the telescope cryostat, and fixed reflective
ground shields surround the telescopes. The telescope mount allows for full
boresight rotation of the entire telescope to rotate the polarization angle sensitivity of each detector and
check for polarization systematic errors.  \spider's design is very similar,
but uses a 4\,K HWP rotated twice per sidereal day to modulate polarization, 
and reflective rather than absorbing external baffles to avoid the 
associated optical loading to take advantage of lower optical loading from atmosphere.

The CLASS ground-based experiment and PIPER balloon-borne experiment
have similar optical layouts consisting of a VPM as the first optical element at the
entrance pupil, two elliptical mirrors, cold refractive reimaging
optics, and a cold stop. In CLASS, the VPM and mirrors are at ambient
temperature, whereas in PIPER all optics are cooled to 1.4\,K by superfluid helium
boiloff. CLASS uses HDPE lenses and an UHMWPE window, whereas PIPER
uses silicon lenses. Both telescopes have co-moving reflective baffles surrounding
the mirrors and in front of the VPM. Similar to \bicepIII\ and Keck
Array/\spider, multi-frequency coverage is accomplished for CLASS using
multiple telescopes (40, 90, and 150/220\,GHz telescopes) and for PIPER
using multiple flights, which simplifies AR coating implementation.

ABS used a compact crossed-Dragone dual mirror design and no tertiary
re-imaging optics. The mirrors were cooled to 4\,K, with
a cold stop preceding the primary mirror. An ambient temperature
continuously-rotating HWP and co-moving reflective baffle were
placed just above the AR-coated UHMWPE vacuum window. 

%\begin{figure}
%	\centering
%	%\includegraphics[height=8in]{figures/Stage-III_Telescope_Overview}
%	\includegraphics[height=8in]{figures/Stage_III_Telescope_Overview_20170323_v2}
%	\caption{Table of telescope and instrument parameters for current projects sorted by primary aperture. Descriptions of each of these projects are presented in Section \ref{sec:projects}.}
%	\label{tab:telescopes}
%\end{figure}

\begin{table}
\centering
	\begin{center}
	\includegraphics[trim = 0.7in 1.0in 0.8in 1.0in,clip,height=7in]{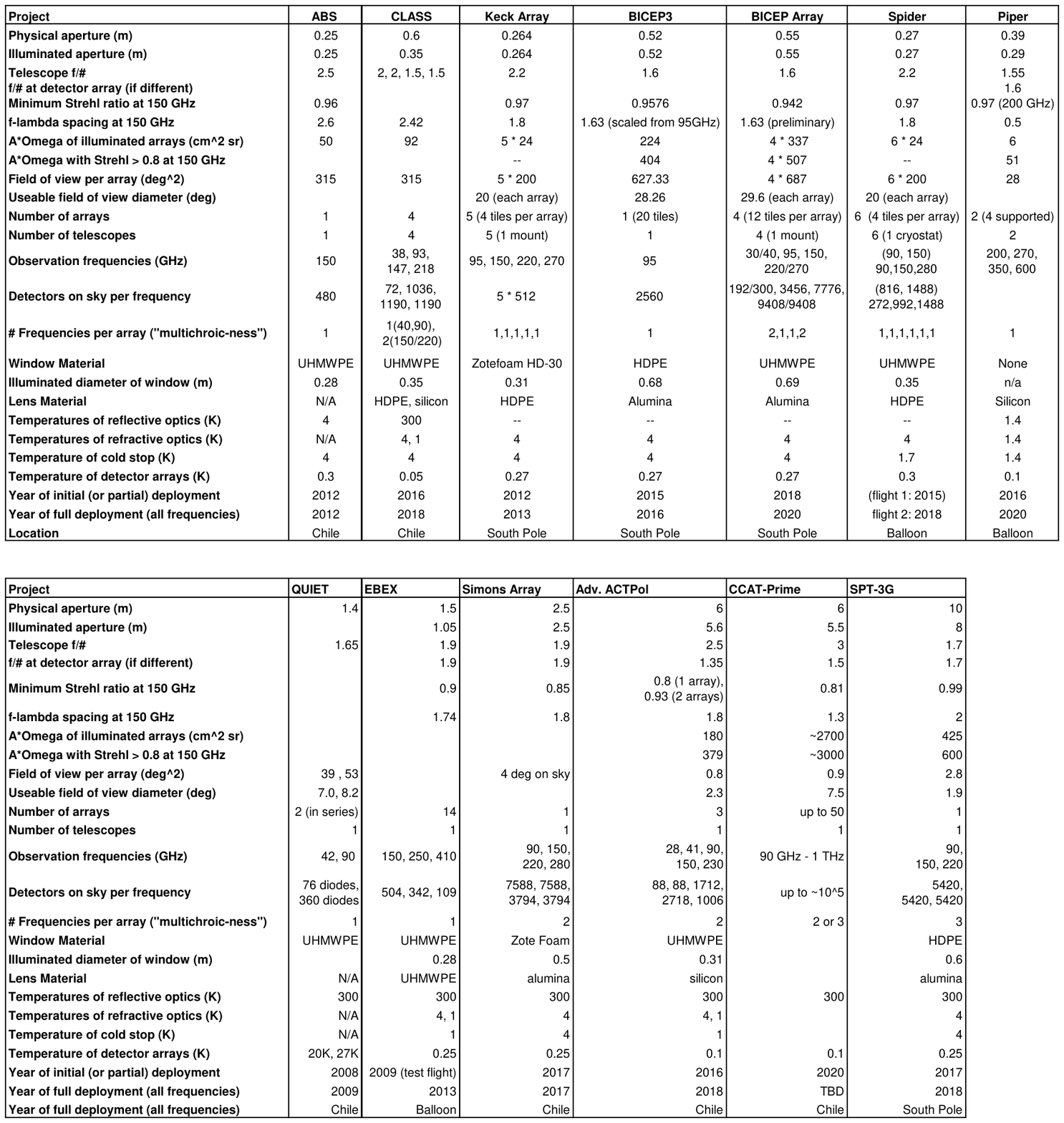}
	\end{center}
	\caption{Table of telescope and instrument parameters for current and recent projects sorted by primary aperture size. Descriptions of each of these projects are presented in Section \ref{sec:projects}.}
	\label{tab:telescopes}
\end{table}

\subsection{Current large-aperture telescope designs}

Large-aperture mm-wavelength CMB telescope designs, $\gtrsim 1$\,m
diameter, can be used to measure the CMB lensing signal at $\ell >
200$, and $> 5$\,m designs can also measure arcminute-scale secondary
anisotropies such as the Sunyaev-Zeldovich effect from galaxies and galaxy clusters, emission from dusty galaxies and active galactic nuclei. Large aperture designs reviewed here include AdvACT
 (6\,m physical aperture), EBEX (1.05\,m), Simons Array (2.5\,m),
SPT-3G (10\,m), and QUIET (1.4\,m). All large telescope designs with the
exception of QUIET use ambient temperature dual-mirror off-axis
Gregorian configurations and cold refractive reimaging optics to form a
cold stop and flat (telecentric) image plane. In contrast to the
small-aperture designs, the large-aperture telescopes are all designed
to conduct simultaneous observations in multiple frequency bands.

The AdvACT receiver design consists of three
independent optics tubes for each of three frequency pairs (28/41,
90/150, and 150/230\,GHz).  Each optics tube uses a UHMWPE vacuum
window and has three cold AR coated silicon lenses and a 1\,K cold stop
to control illumination of the primary. AdvACT will use ambient
temperature continuously rotating half-wave plates (HWPs) just outside each optics tube's
vacuum window. The ACT telescope has a co-moving reflective shield and a stationary reflective ground shield.  
The large primary aperture is comprised of small panels separated by gaps.

The Simons Array (three telescopes) and SPT-3G share a similar optical
design consisting of a single receiver (per telescope) with three cold
alumina lenses and a 4\,K cold stop. SPT-3G uses a HDPE window whereas
the Simons Array receivers (called the \Pb-2 receivers) each use a
10-inch thick laminated Zotefoam window. The first \Pb-2 receiver will
have an ambient temperature continuously-rotating HWP just outside the
receiver window, whereas the second and third receivers will have 50\,K
HWPs inside the cryostat window. SPT-3G does not have plans to use a
HWP. Both the Simons Array telescopes and SPT-3G use a prime focus
baffle and reflective co-moving shields, but no fixed ground
shields. The Simons Array primary mirrors are monolithic, while the
SPT primary mirror is comprised of small ($\sim$0.7\,m) panels
separated by small gaps.

EBEX operated at 150, 250, and 410\,GHz using a receiver containing
five plastic (UHMWPE) lenses, a co-located cold stop, and a continuously rotating
HWP at 1\,K. Polarization sensitivity was achieved by using a
polarizing grid inside the receiver cryostat and two focal planes of
non-polarization sensitive detectors. Each focal plane consisted of
seven detector wafers, with each wafer sensitive to a single frequency
band (150, 250, or 410\,GHz), defined by reflective filters above the
feedhorns and the cylindrical waveguides between the feedhorns and
detectors.

QUIET was a crossed-Dragone telescope operating at 42 and 90\,GHz. The
two mirrors were at ambient temperature, and no tertiary optics were
used in front of the cryogenic focal plane of feedhorns. Unlike ABS,
it did not use a stop above the primary mirror. Instead, the feeds
were sized to under-illuminate the mirrors to minimize
spillover. Sidelobes were also mitigated by using an absorbing baffle
in front of the entrance aperture and surrounding the telescope.

%\subsection{Current refractive and reflective optics designs}

\section{Concept for high throughput large-aperture telescope design}
\label{sec:CD_design}

As described in the CMB-S4 science book, several of the science goals require arcminute-scale resolution, which roughly translates to telescope apertures between a few and ten meters at 150\,GHz. 
This requirement has motivated designs with lower levels of systematic error (e.g., cross polarization) and larger throughput than existing telescopes with the potential for illuminating a much larger number of detectors than current telescopes. 

%As with the development of optical telescopes, the number of detectors for CMB telescopes has been limited by how many detectors could practically be manufactured.  
For CMB-S4, the number of detectors required is nearly two orders of magnitude larger than the number in a typical Stage-III telescope.  It is critical, therefore, to develop telescope designs with much higher throughput then Stage-III designs.  Fortunately, there are a number of candidate designs including the crossed-Dragone, the three-mirror anistigmat, and commonly-used offset Gregorian but implemented with a secondary mirror of similar size to the primary mirror.  Of these, the crossed-Dragone has been studied the most and will be described in detail below as an example case.

%There is a fairly large parameter space of telescope designs that could give increased throughput compared to current designs.

%One optics design concept has been shown to achieve substantially larger optical throughput than existing off-axis Gregorian telescope designs, combined with reduced systematics \cite{niemack2016,tran/etal:2008}. 
The example concept is based on a crossed-Dragone design with higher $f/\#$ than had been studied previously\cite{niemack2016,tran/etal:2008}. Specifically, previous studies of crossed-Dragone CMB telescope designs focused on systems with focal ratios closer to $f/1.5$ and with the detector arrays at the telescope focus \cite{2009AIPC.1185..494E, imbriale2011, taylor/etal:2004, tran/etal:2010, tran/etal:2008}. For ground-based telescopes, controlling spillover with this approach generally requires either severely under-illuminating the primary mirror \cite{imbriale2011} or cooling the entire telescope to cryogenic temperatures \cite{2009AIPC.1185..494E}, which is not practical for a large telescope.

\begin{figure}[h]
	\centering
	\includegraphics[height=7cm]{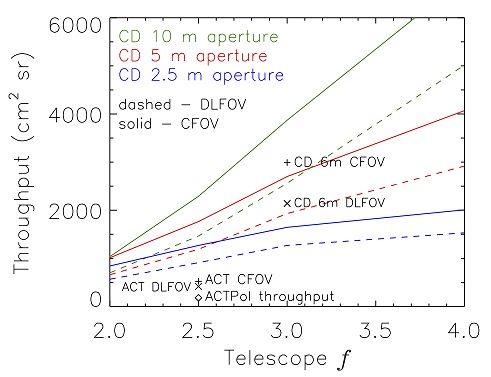}
	\caption{Optical throughput comparison for large-aperture crossed-Dragone telescopes with different $f/\#$ and apertures compared to the off-axis Gregorian ACT design \cite{niemack2016}. Dashed lines are diffraction limited field of view (DLFOV) and solid lines are correctable field of view (CFOV)}
	\label{fig:CD_telescopes}
\end{figure}

The new crossed-Dragone concept instead controls spillover past the mirrors using cryogenic refractive optics similar to those used in the existing large-aperture telescopes ACT \cite{henderson/etal:2016}, Simons Array \cite{Stebor}, and SPT \cite{benson2014}. Refractive optics naturally couple to telescopes with larger focal ratios, which also increases the available optical throughput as shown in Figure~\ref{fig:CD_telescopes}. This crossed-Dragone design has recently been adopted for the CCAT-prime submillimeter astronomy project to study cosmic origins of stars, planets and galaxies.  Preliminary designs and potential systematic error advantages of this design are shown in Figure~\ref{fig:CCAT-prime} and discussed in Section~\ref{sec:ccatp}. When combined with closely-packed optics tubes, a 6\,m telescope based on this design is capable of providing a diffraction-limited field of view for more than $10^5$ detectors, which is roughly ten times more detectors than will be deployed on upcoming Stage-III telescopes \cite{niemack2016}.

As mentioned above, there are several telescope designs that may be capable of providing a similar scale of diffraction-limited field of view. For example, adding an optimized tertiary similar in size to the primary for a more traditional off-axis Gregorian design results in a relatively large field of view, and this design should be studied in further detail. As described in Section~\ref{sec:futurestudies}, such alternate designs represent a high priority area of study for CMB-S4.

%\subsection{Table of telescope/instrument designs} 
%Possible items for table: aperture, f/\#, min Strehl ratio, f-lambda
%at 150 GHz, A*Omega, ?

%\pagebreak
%\href{https://docs.google.com/spreadsheets/d/1u3vzA8KZmLOFIndMwLHxpqJiUk6PAy3qHjipRxy_q_Y/edit#gid=0}{See Google Docs page linked here for current table status.}

%% file: telescopes/engineering_systematics.tex
\section{Telescope engineering to control systematic errors}
\label{sec:engineering_systematics}

CMB-S4 will require exquisite control of systematic errors, and therefore the
telescopes must be designed to have low sidelobe pickup, stable optics, and have the capability to  scan fast enough to minimize atmospheric brightness
fluctuations without introducing vibrations that cause microphonics pickup or temperature instability
in the cryogenics.  It may also be necessary to measure systematic errors
that are fixed relative to the instrument, e.g., by rotating the camera
or the entire telescope about boresight. CMB-S4 will build on experience
with existing telescope designs, but the scale of CMB-S4 may allow
approaches that were deemed impractical for current experiments. Some
of these approaches are described below; all will require design and
manufacturing studies to assess their viability for CMB-S4.

The level of systematic error induced by sidelobe pickup depends on a number of optical aspects that will have to be studied extensively for CMB-S4. To give a sense of the rough order of magnitude involved, the
EPIC, a proposed CMB polarization satellite mission, 1.4\,m aperture design was expected to have about $-80$\,dB sidelobes ($-20$\,dBi),
which corresponds to approximately 0.1\,nK~rms polarized pickup from
the galaxy at 150\,GHz~\cite{EPIC}.
%(\textquotedblleft{}Study of the
%Experimental Probe of Inflationary Cosmology - Intermediate Mission
%for NASA\textquoteright{}s Einstein Inflation Probe,\textquotedblright{}
%NASA, 4 June 2009). 
There are two approaches for controlling pickup:
(i) reduce scattering, which means using off-axis optics with enough clearance to avoid sidelobes due to clipping the beam, and smooth optical surfaces to avoid scattering from gaps
between mirror segments, and selecting low-scatter windows and filters; and (ii) control what does get scattered, which requires reflecting shields and/or absorbing baffles to eliminate
pickup in far sidelobes.

The pointing requirements for CMB-S4 will be stringent: we estimate that pointing reconstruction will need to be accurate at the level of 1\% of beamwidth or about 1.5$''$ for the large-aperture telescopes~\cite{hhz03},
%(W. Hu, M. M. Hedman and M. Zaldarriaga,
%\textquotedblleft{}Benchmark parameters for CMB polarization experiments,\textquotedblright{}
%Phs. Rev. D 67, 043004 (2003))
so the telescope structures must be
stiff.  Limiting spurious signals from flexing of the optics also require stiff structures and schemes to
keep the optical surfaces free of water, snow, and ice.

\subsection{Monolithic mirrors}
\label{sec:monolithic}

The use of a monolithic primary mirror for large-aperture telescopes would
avoid the structured sidelobe response that is seen with all telescopes
with segmented primary mirrors.   The \Pb/Simons Array telescopes have 2.5 meter diameter
monolithic mirrors made from machined cast aluminum.  Fabrication of a monolithic, millimeter-wavelength mirror larger than
a few meters in diameter is challenging however, and therefore 6-10 meter diameter CMB
telescopes (i.e., ACT and SPT) have mirrors made of $\sim$1\,m
segments with about 1\,mm gaps between segments.  Scattering from the gaps generates sidelobes, which account for approximately 1\% of the telescope response \cite{fluxa16}. It is difficult to make the gaps smaller because
some clearance is needed for assembly and manufacturing tolerances. 
Various gap cover/filler schemes have been attempted, but a robust solution has not yet been demonstrated.

%Monolithic mirrors were not practical for the large, Stage-III, CMB
%telescopes, but CMB-S4 will involve multiple telescopes, so the cost
%of developing a fabrication approach for large monolithic mirrors
%may be reasonable. There is obviously no point pursuing monolithic
%mirrors unless the rest of the telescope design is consistent with
%small sidelobes.

\begin{figure}
\begin{center}
\includegraphics[height=6cm]{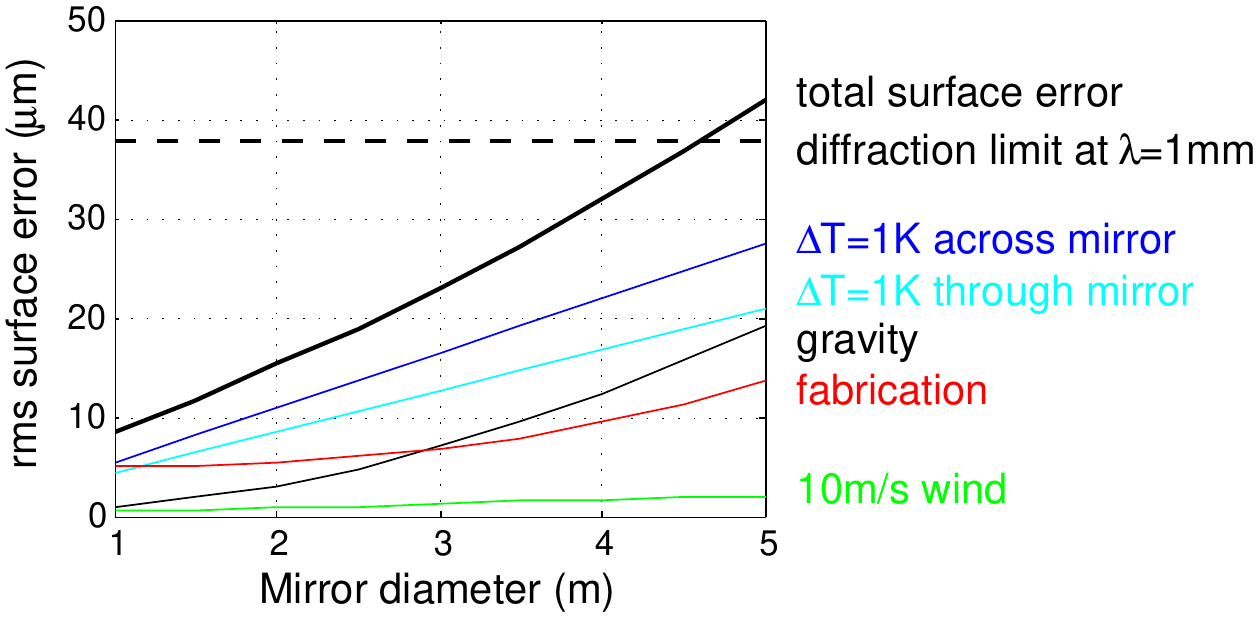}
\caption{\label{fig:MirrorSurfaceError}Surface error vs.\ diameter for a monolithic
aluminum mirror with $\textrm{thickness}=\textrm{diameter}/4$. Temperature
gradients across the mirror change the thickness, while a temperature
gradient through the mirror causes cupping. The gravitational deformation
model represents the deflection of a simply supported plate, and wind-induced
deformation is the gravitational deformation scaled by the ratio of
wind pressure to mirror weight per unit area. The fabrication error
model has 50~$\mu$m rms for a 10\,m
mirror, with error scaling as the square of diameter, combined in
quadrature with a setup error of 5~$\mu$m rms.
The model is based on the OVRO 10.4\,m primary mirrors,
which were machined as a single piece, and the 1\,m segments
for SPT. The horizontal dashed line corresponds to 80\% Strehl ratio
($\lambda/27$ rms surface error) at 300\,GHz.}
\end{center}
\end{figure}

The key issues for monolithic mirrors are: (i) fabrication errors;
and (ii) thermal deformation.  Figure \ref{fig:MirrorSurfaceError} shows surface error contributions
for a monolithic, aluminum mirror, which is an obvious choice for
low cost. A 5\,m diameter, $\lambda=1$\,mm mirror
seems possible if thermal gradients through and across the mirror
can be kept within 1\,K, which is what the $\sim$1\,m diameter and 50\,mm thick
Caltech Submillimeter Observatory primary mirror segments achieve at night. Keeping thermal gradients
below 1\,K in a large aluminum mirror will require insulation
on the back of the mirror, a reflective front coating for daytime
operation, and maybe active control (e.g., cooling the back of the
mirror even at night). A carbon fiber reinforced polymer (CFRP) mirror would have an order of magnitude better
thermal performance, but it is technically challenging to fabricate a
large monolithic CFRP mirror with the required surface accuracy.

\subsection{Boresight rotation}
\label{sec:boresight}

A few experiments (e.g., DASI, CBI, QUAD, QUIET, \bicepI, Keck Array)
have included boresight rotation to measure and potentially cancel systematic errors that
are fixed with respect to the instrument (e.g., instrumental polarization)
and vary slowly (on timescales of tens of seconds). All these
experiments have or had small telescopes, or arrays of small telescopes;
the largest boresight rotator was the 6\,m diameter platform
of the CBI.  Boresight rotation on a off-axis 6m telescope would be technically challenging, but it could play an important role in achieving the systematic error control required to reach CMB-S4 sensitivity levels.

The key issues for boresight rotation are: (i) balancing the telescope
structure while also providing adequate range of motion; and (ii)
protecting the drive mechanisms from the weather.

A mount that supports boresight rotation can wrap around the outside
of the telescope, which allows full range of motion with a naturally
balanced structure (no counterweight), but results in a massive, expensive
mount with large mechanisms that are difficult to protect. Alternatively,
a compact, inexpensive, enclosed mount can be placed behind the telescope,
but this requires a counterweight which results in limited range of
motion because the counterweight interferes with the mount. Figure
\ref{fig:3AxisMountLayout} shows a concept for a compact mount with
boresight rotation. The design provides an optical bench that can
support a single, large, off-axis telescope, or an array of smaller
telescopes inside a deep baffle. The compact drive mechanisms can
be enclosed and are accessible from below, which is appropriate for a site
that has severe snow storms or very low temperatures.

Another approach under study that offers partial boresight rotation is having the telescope elevation axis aligned with the chief ray between the secondary and the tertiary (or between the secondary and the instrument if a tertiary is not used). This approach offers other potential advantages of not tilting the instruments in elevation and enabling instrument rotation independent of the telescope, and it is being pursued by the CCAT-prime project (see Section \ref{sec:ccatp} for details).

\subsection{Shields and baffles}
\label{sec:baffles}

\begin{figure}
\begin{center}
\includegraphics[height=6cm]{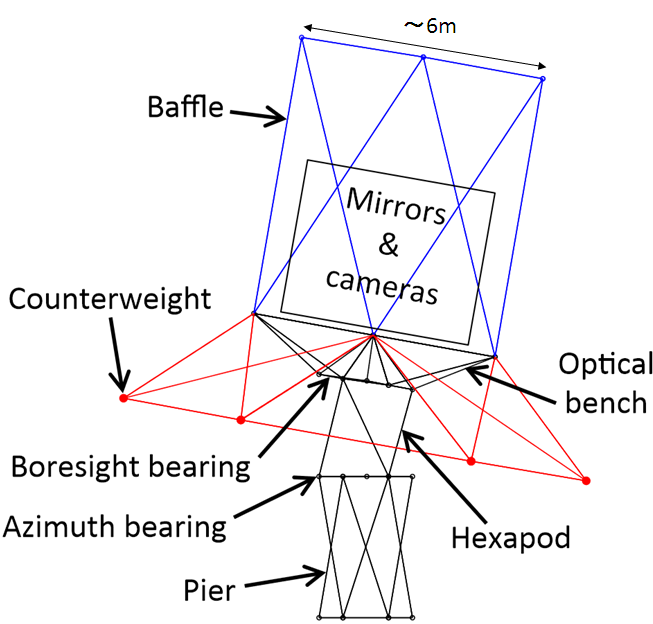}

\caption{\label{fig:3AxisMountLayout}Concept for a telescope mount with boresight
rotation. The mount provides a large, flat optical bench that can
accommodate various arrangements of cameras and off-axis mirrors.
Standard, slewing-ring bearings allow fast scanning in azimuth and
rotation about boresight to modulate/measure polarization. Zenith
angle motion is controlled by a hexapod that provides a stiff connection
between the azimuth and boresight slewing rings. The blue structure
is a lightweight CFRP spaceframe that supports an absorbing baffle
to reduce pickup. Dimensions in the diagram are based on a 6\,m
diameter optical bench and 2\,m diameter slewing rings.}
\end{center}
\end{figure}

Co-moving reflective shields and/or absorbing baffles that fully shield the optics will be
needed to control pickup in the far sidelobes.   Stringent shielding is a key
ingredient in the success of small CMB telescopes making measurements
at low $\ell$, but a full shield or baffle may also be practical for a
large telescope.  For example, the mount in Figure
\ref{fig:3AxisMountLayout} can accommodate a 5\,m telescope inside a
deep, cylindrical baffle that is supported by a light, CFRP
spaceframe.

The key issues for shields and baffles are: (i) maintaining adequate
mechanical stability to avoid time-varying pickup, e.g., due to wind
buffeting; (ii) keeping surfaces clear of water, snow, and ice, which
change the optical loading; (iii) limiting baffling temperature variations,
which cause variations in optical loading; and (iv) ensuring the
survival of absorbing coatings.

Mechanical stability is more challenging for a reflective shield because
any part of the surface that sees scattered light must be stable. 
For an absorbing baffle, the rim must be stable, but the rest of the
baffle can move relative to the telescope beam as long as the baffle
is truly black. There is no practical experience with large absorbing
baffles, so the effect of temperature variations needs consideration.
Some work must also be done to identify or develop a light, robust,
weather-resistant absorber. 

%% file: telescopes/future.tex
\section{Potential future studies and development areas}
\label{sec:futurestudies}

%Section ubsection{Control of systematic uncertainties}

%Describe various systematic design considerations (polarization modulation, boresight rotations, ... )

%Section ubsection{Studies to inform telescope designs}

%Section ubsection{Cost trends? Designing cheaper telescopes?}

The CMB-S4 effort would profit from a variety of studies with 
respect to telescope and receiver design. A comprehensive list is beyond the scope 
of this document, but we list a few potential studies and areas of research that could significantly help in
%categories and items to be considered in 
the course of designing the CMB-S4 telescopes.

\paragraph{Comprehensive optical design study for large-aperture telescopes }

It has been shown that crossed-Dragone designs can provide substantially larger throughput than current large-aperture telescopes (Section \ref{sec:CD_design}). However, current telescopes were not designed with the goal of maximizing throughput, and therefore it is important to perform a comprehensive study to determine if modifications to more traditional off-axis Gregorian or Cassegrain designs, such as implementing a larger secondary mirror, could achieve similar performance. %For example, adding an optimized tertiary mirror to an off-axis Gregorian might significantly increase throughput. 
As mentioned, the three-mirror anistigmat (TMA) is a well known design that can have very high throughput that should be investigated.  Introducing higher-order correction terms to the mirrors of a conventional crossed-Dragone also has the potential to increase diffraction-limited throughput.

\paragraph{Optimization study for number of aperture sizes for CMB-S4}

Given the strong relationship between cost and telescope diameter, an optimization study should be performed to determine the number of aperture sizes to include in CMB-S4.  By including two or more telescope sizes, resolution could be matched to some degree over the roughly factor of ten range in frequency to be measured by CMB-S4.  However, the savings in telescope construction cost must be balanced against additional design and logistics costs of a larger number of telescope types.  

%The optimal configuration might differ for ``large" versus ``small'' aperture telescopes; for example, it may make sense to adjust the apertures of telescopes $<1\,$m by frequency, but to only build one or two designs for the more costly larger aperture telescopes, or vice versa.

\paragraph{Study on measuring inflationary angular scales with large-aperture telescopes}

This study is critical since significant cost savings are possible if large-aperture telescopes can cover at least the degree angular scales of the Inflationary recombination peak.   ACT (Section \ref{sec:ACT}), the Simons Array (Section \ref{sec:SA}), and SPT (Section \ref{sec:SPT}) are currently making measurements aimed at the recombination peak with different combinations of HWPs and scan strategies, but none have published high-fidelity results.  
Some new large-aperture telescope designs have a FOV approaching that of small-aperture telescopes (e.g., $\sim$8$^\circ$ for CCAT-prime, Section \ref{sec:ccatp}), which could be beneficial for measuring degree-scale polarization, although results may not be available before a CMB-S4 telescope is designed.

\paragraph{Study on the feasibility of reionization peak measurements by CMB-S4} 

Measurements of both $E$-mode and $B$-mode polarization of the reionization peak would be highly valuable for increasing the confidence in the detection of an inflation signal and to improve the error on the sum of neutrino masses by improving constraints on the optical depth to reionization. CLASS (Section \ref{sec:CLASS}),  PIPER (a NASA balloon, Section \ref{sec:Piper}), and GROUNDBIRD are the only current sub-orbital projects targeting this largest angular scale range. CMB-S4 science can be pursued independently of these measurements, but it may be worth expanding the CMB-S4 science case if these projects produce compelling results down to $\ell\approx5$.

%\paragraph{Impact, feasibility, and cost study for large-diameter monolithic primary mirrors}
\paragraph{Tradeoff study between segmented and monolithic primary mirrors}

As mentioned earlier, the gaps between panels in the segmented primary and secondary mirrors generate sidelobe structure in large-aperture telescopes. A study should be made of the impact of these sidelobes for CMB-S4 science.   Additional simulations and calculations are needed to better understand the impact of the sidelobes and the potential benefit of monolithic primary mirrors. Also, this study would address the cost and feasibility of large-diameter ($\geq 5$ m) monolithic mirrors (Section \ref{sec:monolithic}).

\paragraph{Study of baffle design for CMB-S4 telescopes}

There is currently diversity in the design of optical baffling for CMB telescopes.  Both absorptive and reflective baffles are common.   It is important to study the optimal design of baffles, especially in the case of co-moving baffles on large-aperture telescopes (Section \ref{sec:baffles}). It is particularly important to understand how to control systematic errors and to minimize spillover onto absorptive baffles. Spill over onto absoptive baffles gives additional optical loading and the consequent increase in photon noise. 

\paragraph{Study of systematic error mitigation with boresight rotation}

Boresight rotation of the entire telescope is very effective in reducing systematic errors from beam shape imperfections.  Boresight rotation has been implemented only on $\leq 1.5$m-aperture telescopes. Additional simulations including a variety of sources of systematic error could help address whether the additional cost of implementing full boresight rotations or partial boresight rotations (Section \ref{sec:boresight}) on large-aperture telescopes is worthwhile.

\paragraph{Study of optimal receiver envelope} 

As the physical aperture of receiver windows increases, more cooling power is needed to remove the additional radiative loading. 
%For example, the receiver design in \cite{niemack2016} (Figure~\ref{fig:CCAT-prime}) includes 50 optics tubes, and would require a larger number of pulse tube refrigerators than existing receivers. 
Improved cryogenic modeling would help to assess the trade-offs in the amount of optical throughput per receiver, and in particular the trade-off in single versus multiple receivers for large-aperture telescopes.  The study would include the effect of cooldown time on testing efficiency and explore methods to reduce the cooldown time of large-envelope receivers such as the implementation of an integrated liquid nitrogen pre-cool system.

\paragraph{Optimization study on the number and diameter of optics tubes per receiver}

There are a set of trade-offs on the number and diameter of optics tubes per receiver for 
large-aperture telescopes.   A choice of larger diameter optics tubes reduces the number of tubes 
required arguably giving a reduction in system complexity.   A single optics tube is currently 
used by Simons Array (Section \ref{sec:SA}) and SPT-3G (Section \ref{sec:SPT}).  However, the optical bandwidth of those systems is limited 
to that achievable with practical anti-reflection coatings, and the systems have additional loss due 
to having relatively thick refractive optics.  

Having many optics 
tubes enables each one to be optimized for a different wavelength range, and relaxes space 
constraints around each detector array (e.g., Section \ref{sec:ACT} and Section \ref{sec:ccatp}); 
however, it can lead to increased cryogenic complexity and requires fabrication 
of more refractive optical elements. Use of many optics tubes also appears to 
increase the usable field of view \cite{niemack2016}.   It is important to understand the optical 
design and implementation trade-offs between these options and, relatedly, the optimal diameter of each tube in the multiple-tube design.

%% file: telescopes/projects.tex
\section{Optics designs for current projects}
\label{sec:projects}

\subsection{Advanced ACTPol} %Mike
\label{sec:ACT}

\begin{figure}[h]
	\centering
	\includegraphics[height=6cm]{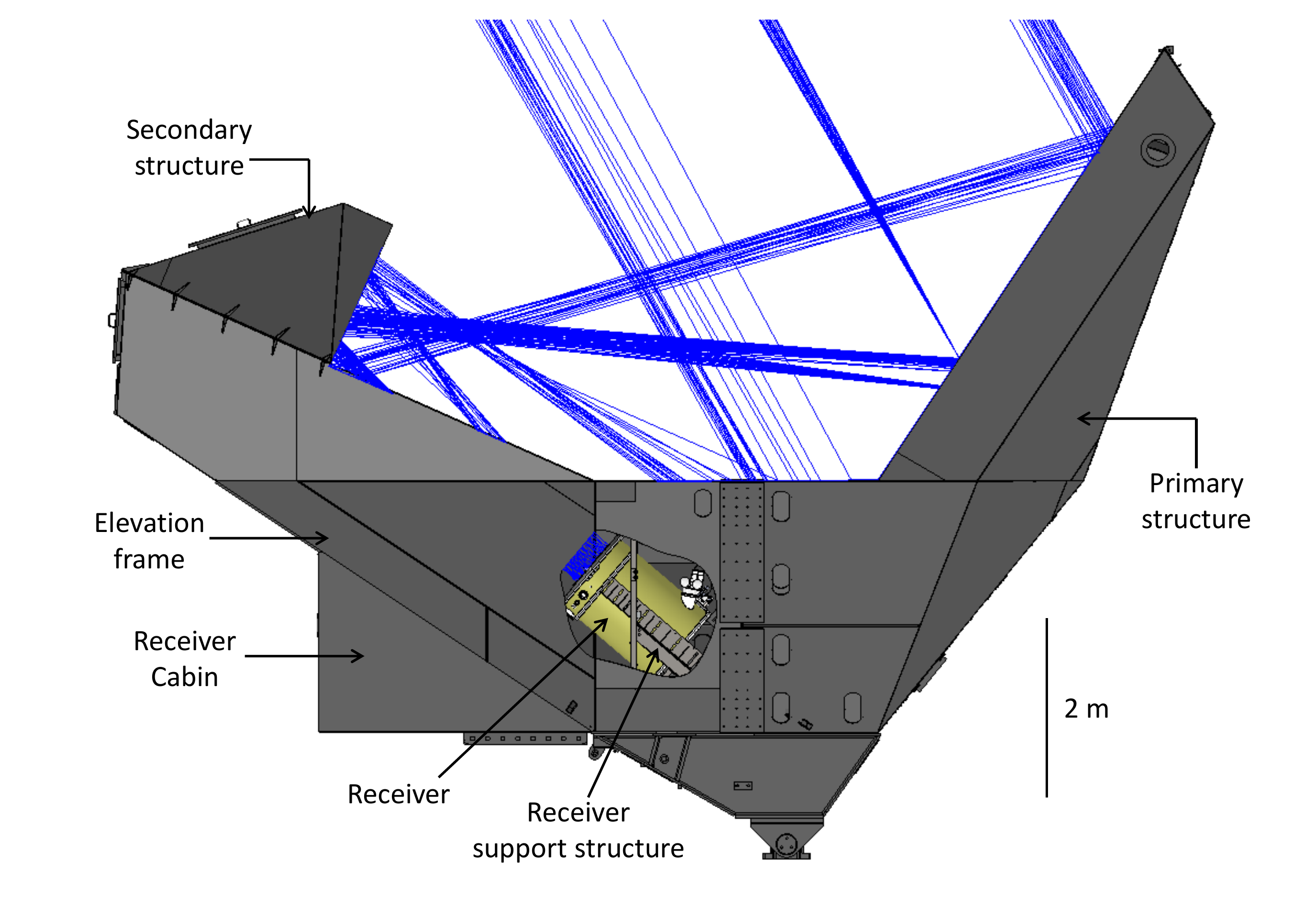}
	\includegraphics[height=6cm]{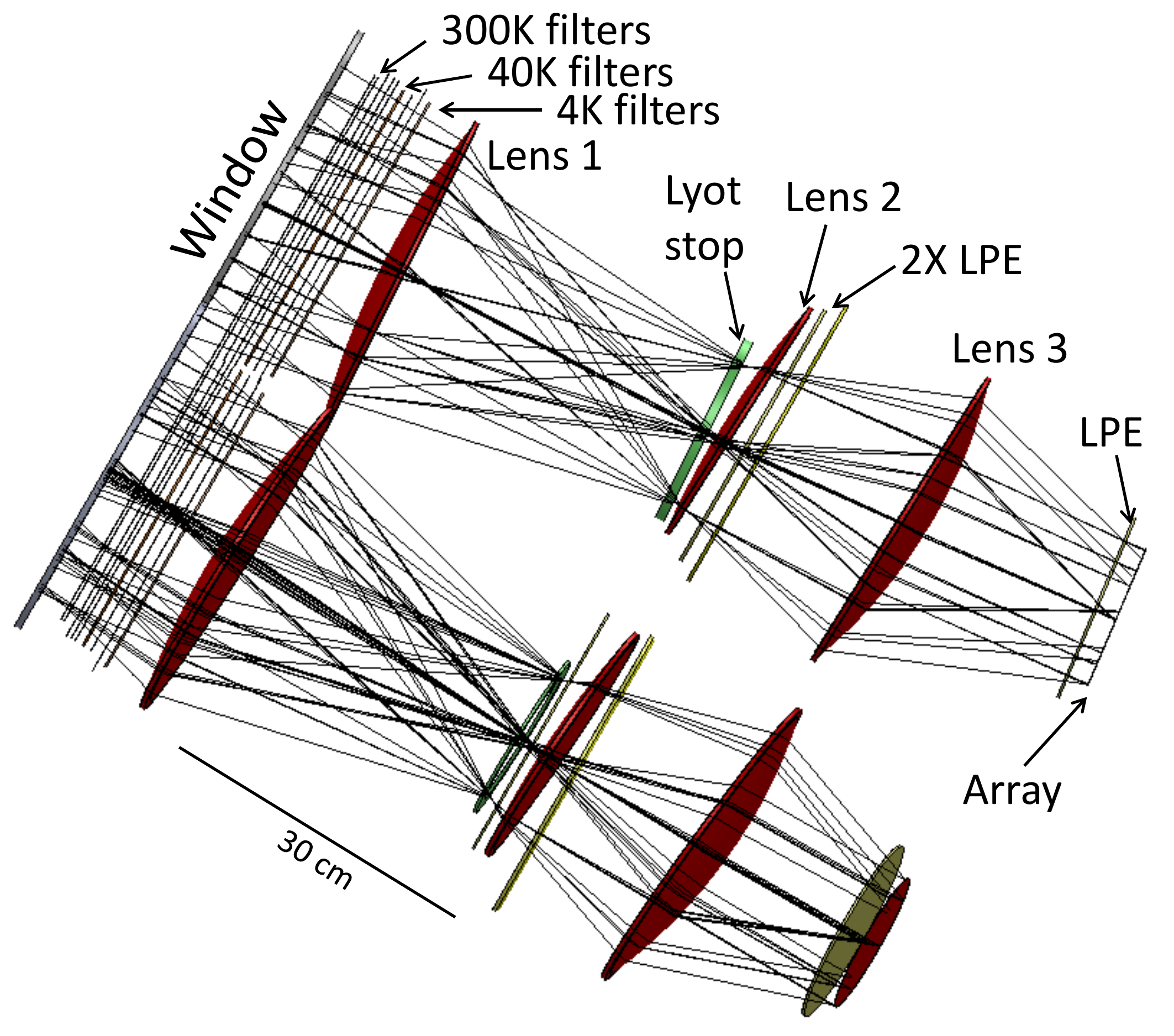}
	\caption{Left: ACT telescope optics and mechanical structure. Right: Ray trace of Advanced ACTPol receiver optics, which includes three optics tubes: one on the top and two symmetric tubes on the bottom~\cite{thornton/etal:2016}.}
	\label{fig:AdvACT}
\end{figure}

AdvACT is the third instrument upgrade for the 6\,m ACT. The 6\,m primary and 2\,m secondary are arranged in a compact off-axis Gregorian configuration to give an unobstructed image of the sky. The details of the telescope optics design are presented in~\cite{fowler/etal:2007}, while the ACTPol and AdvACT receiver optics designs are presented in~\cite{niemack/etal:2010, thornton/etal:2016}. Figure~\ref{fig:AdvACT} shows a ray trace through the ACT mechanical structure as well as through the AdvACT receiver optics. Illumination of the primary mirror is controlled using a 1\,K Lyot stop. To minimize ground pickup during scanning, the telescope has two ground screens. A large, stationary outer ground screen surrounds the telescope and a second, inner ground screen connects the open sides of the primary mirror to the secondary mirror and moves with the telescope during scanning.

ACTPol and AdvACT use the same receiver with three independent optics tubes. Both use large silicon lenses with two and three layer metamaterial anti-reflection (AR) coatings~\cite{datta/etal:2013}. These coatings offer the advantages of negligible dielectric losses ($< 0.1$\%), sub-percent reflections, polarization symmetry equivalent to isotropic dielectric layers, and a perfect match of the coefficient of thermal expansion between coating and lens. Each optics tube focuses light onto a two-frequency multichroic detector array at one of the following frequency pairs: 28/41\,GHz, 90/150\,GHz, or 150/230\,GHz~\cite{henderson/etal:2016}. The AdvACT reimaging optics have $f$/1.35 at the array focus. A pixel-to-pixel spacing of 4.75\,mm in the recently deployed 150\&230 GHz array leads to approximately 1.8\&2.5 $f$-$\lambda$ spacing. A UHMWPE vacuum window is used combined with metal mesh filters to control out of band radiation.

\subsection{BICEP3} %- Keith, Zeesh 

\begin{figure}[h]
	\centering
	\includegraphics[width=5in]{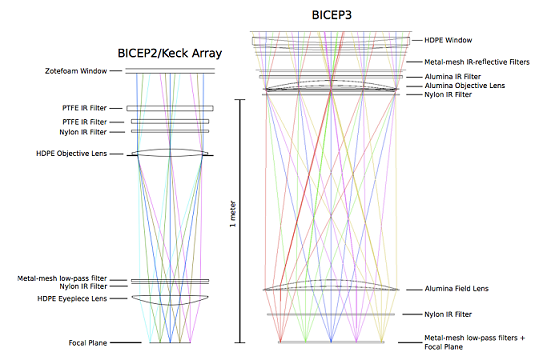}
	\caption{Left: \bicepII/Keck Array ray trace.  Right: \bicepIII\ ray trace.}
	\label{fig:keckbicep}
\end{figure}

\bicepIII\ is a cryogenic refractor of aperture 0.52\,m, with two alumina
lenses~\cite{ahmed2014,grayson2016} in an $f$/1.6 system.
The main cryostat volume is 29~inches in diameter and 95~inches high.
It operates at 95\,GHz on a three-axis mount at the South Pole.
The field used is 14.1$^\circ$ half-opening angle, nearly the unvignetted
FOV. At 2\,mm wavelength the design gives Strehl $> 0.96$ over
the full unvignetted field (top-hat illumination assumed).

The lenses are 99.6\% pure alumina.  The lenses and stop are at
4\,K. The window is made using HDPE.  The optical filters consist of metal mesh filters at
(nominal) ambient temperature, an alumina filter at 50\,K, another mesh
filter below that, two Nylon filters at 4\,K, and Ade edge filters at
250\,mK over each detector module. The window, alumina components, and
Nylon filters are single-layer AR coated; the alumina AR is built from an epoxy
mix. A co-moving absorptive forebaffle and a reflective groundshield
mitigate ground source contamination.

Starting in the 2016 season, \bicepIII\ has 2400 light
detectors.  The pixels are phased slot antenna arrays with tapered weighting to approximate
Gaussian beams~\cite{obrient2012spie}.
The 1/$f$ noise knee after atmospheric common-mode rejection from
detector pair differencing is well below the degree-scale science
band~\cite{ade2014,ogburn2010}.
Beam systematic errors are averaged down by boresight rotation and
residual temperature to polarization beam leakage is removed
by deprojection~\cite{sheehy2013}.
Thus, a (fast) polarization modulator is not used in \bicepIII\
(as is also true with \bicepII\ and Keck Array).

The mount (originally built for \bicepI) provides elevation down
to $\sim 50^\circ$, full azimuth and boresight
rotation of $255^\circ$. The latter provides 
%for two $45^\circ$ offset pairs of 180$^\circ$ complement boresight angles (four angles total) for
full $Q$/$U$ discrimination and
cancellation of several beam related systematic errors.
Mapping is performed with a sequence of constant elevation scans
at $2.8^\circ$/s in azimuth.

The \bicepArray\ receivers will be substantially the same as \bicepIII, with small improvements planned to windows, filters, and optical throughput.

\subsection{CLASS} %- Tom?
\label{sec:CLASS}

\begin{figure}[h]
	\centering
	\includegraphics[height=6cm]{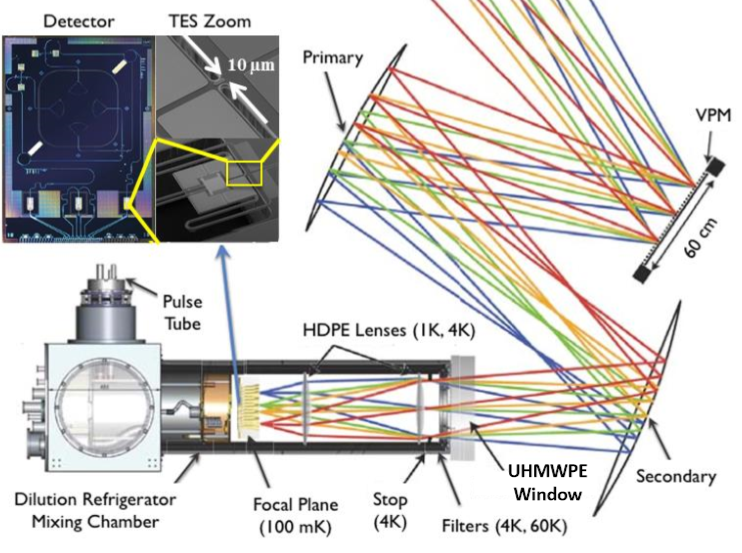}
	\includegraphics[height=6cm]{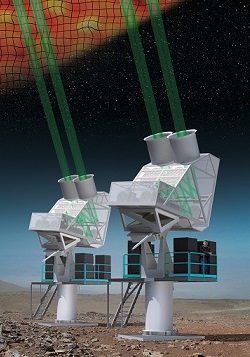}
	\caption{Left: CLASS system overview. Right: CLASS site rendering, showing the two mounts with four telescopes.}
	\label{fig:class_site_rendering}
\end{figure}

%The Cosmology Large Angular Scale Surveyor (CLASS) 
The CLASS experiment consists of four telescopes sharing similar optical layouts~\cite{2012SPIE.8452E..20E}. One telescope operates at 40\,GHz, two at 90\,GHz, and the final telescope is a dichroic 150/220\,GHz, hereafter the high-frequency (HF) telescope. A 60~cm-diameter VPM is the first element in the optical chain, providing approximately 10\,Hz front-end polarization modulation~\cite{2012ApOpt..51..197C}. Ambient-temperature, off-axis, elliptical, 1-meter primary and secondary reflectors reimage the cold stop of the receiver at 4\,K onto the VPM. Cryogenic reimaging lenses, one at 4\,K and one at 1\,K, focus light onto the focal plane of feedhorn-coupled TES bolometers. The CLASS design emphasizes per-detector efficiency and sensitivity with 10\,dB edge-taper illumination of the cold stop. The CLASS telescopes provide diffraction-limited performance over a large, $20^{\circ}$ FOV with resolutions ranging from $90^{\prime}$ at 40\,GHz to $18^{\prime}$ at 220\,GHz. Three-axis mounts give azimuth, elevation, and boresight rotations, with two telescopes on each of two mounts (See Figure~\ref{fig:class_site_rendering}). Co-moving ground shields and baffles reduce ground pickup.

The lenses for the 40 and 90\,GHz telescopes are made of HDPE, while the HF telescope employs silicon lenses. All of the lenses are AR coated with simulated dielectrics cut directly into the lens material. The receivers have vacuum windows approximately 50\,cm in diameter made of UHMWPE. A combination of capacitive-grid metal-mesh filters, absorptive polytetrafluoroethylene (PTFE) filters, and Nylon filters reject infrared radiation. 

\subsection{EBEX} %- Shaul

\begin{figure}[h]
  \includegraphics[width=\textwidth]{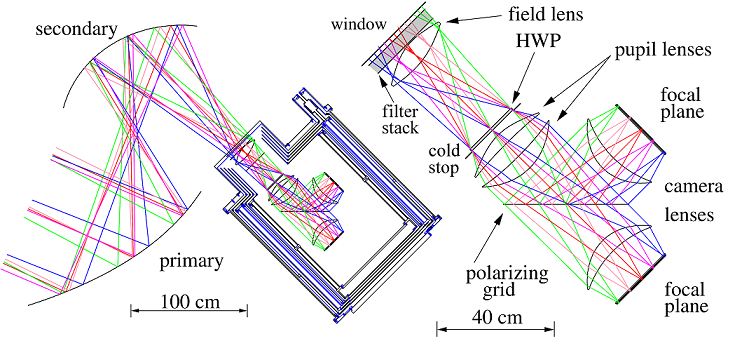}
  \caption{EBEX optical design ray trace schematic consisting of two
    ambient temperature reflectors in an off-axis Gregorian
    configuration and a cryogenic receiver (left). Inside the receiver
    (right), cryogenically cooled polyethylene lenses formed a cold
    stop and provided diffraction limited performance over a flat,
    telecentric, $6.6^\circ$ FOV. A continuously-rotating achromatic
    HWP placed near the aperture stop and a polarizing grid provided
    the polarimetry capabilities.  }
  \label{fig:raydiagram}
\end{figure}

The EBEX telescope was a balloon-borne CMB polarimeter observing at 150, 250, and 410\,GHz.  
The EBEX design achieved flat telecentric focal planes, a large diffraction limited FOV defined as Strehl ratio $>0.9$, a cold stop to control sidelobe response, as well as a continuously-rotating achromatic HWP~\cite{klein2011} and polarizing grid to provide polarimetry, all while remaining sufficiently compact to fit on a balloon payload~\cite{Milligan_thesis}.  

To achieve this, the EBEX optical system consisted of a 
1.05\,m, $f$/1.9, ambient temperature, Gregorian Mizuguchi-Dragone~\cite{dragone82,mizuguchi78} 
reflecting telescope and a cryogenic receiver containing five UHMWPE 
re-imaging lenses  (see Figure~\ref{fig:raydiagram}). The mirrors were 
oversized to suppress sidelobe pickup; the illuminated aperture is 1.05\,m while the 
physical aperture is 1.5\,m. The reimaging lenses preserved the $f$/\# of the system while
forming a 1\,K cold stop, the location of the continuously-rotating achromatic HWP, 
enlarging the diffraction limited FOV to $6.6^\circ$, and forming two flat, 
telecentric focal planes~\cite{Milligan_thesis, EBEXPaper1}. On the focal planes conical 
feedhorns coupled the detectors, TES bolometers, 
to free space. Each focal plane consisted of seven wafers, four at 150\,GHz, two at 250\,GHz, and one at 410\,GHz. 
%Each wafer contained 128 usable detectors; the system was readout limited~\cite{EBEXPaper2}.
%Observing bands were defined by reflective filters above the feedhorns and cylindrical waveguides 
%between the feedhorns and bolometers. 
%Reflective IR filters and one absorptive Teflon filter were used to reduce load on the cryostat 
%\cite{Zilic_thesis}. 

%\textcolor{red}{How should we cite or refer to the upcoming EBEX papers?}

%\textcolor{red}{The main thing I don't have here that has numbers in the table is throughput.  I would need to calculate the full throughput of EBEX.  Unless you (Shaul) have that number?}

\subsection{Keck Array/SPIDER} %- Keith, Zeesh 

The Keck Array and \spider\ are close relatives of \bicepI\ and \bicepII.
Both consist of multiple cryogenic refractors with an approximately 
250\,mm aperture and $f$/2.2 of essentially the same optical
design as \bicepII~\cite{aikin2010}.
Both use JPL dual-polarization slot antenna array coupled TES
bolometers~\cite{obrient2012spie}.

Keck consists of five telescopes co-aligned in their ground-based
mount at the South Pole, each in its own independent vacuum jacket. Individual 
telescopes have been assigned each observing season to different
frequency bands from 95 to 270\,GHz~\cite{ade2015}. Apertures are
264\,mm and FOVs are 15$^\circ$~\cite{karkare2016}.

The Keck telescopes have 120\,mm thick Zotefoam
windows, 50\,K PTFE and Nylon filters, 4\,K HDPE lenses and a Nylon
filter, and Ade edge filters~\cite{ade2015,karkare2016}.
The lenses and filters (except
the edge filter) are single-layer AR coated, matched to the
frequency band of the detector in use. The stop is at 4\,K,
on the bottom of the first lens. Absorptive co-moving forebaffles
surround each telescope aperture, and along with a reflective
groundscreen minimize ground pickup.

The Keck Array is on a three-axis mount (built for DASI). Mapping
is performed by a sequence of constant EL scans at each of eight
boresight rotation angles, four pairs of 180$^\circ$ complements for
complete $Q$/$U$ discrimination and mitigation of beam systematic errors.
The azimuth scan speed is 2.8$^\circ$/s.
The 1/$f$ noise knee after atmospheric common-mode rejection from detector
pair differencing is well below the degree-scale science band
\cite{ade2014,ogburn2010}.
Beam systematic errors are averaged down by boresight rotation and residual
temperature to polarization beam leakage is removed by deprojection
\cite{sheehy2013}.
Thus, a (fast) polarization modulator is not used in Keck (as
with \bicepII\ and \bicepIII).

\spider\ is a balloon experiment with six co-aligned
telescopes in one large LHe cryostat~\cite{rahlin2014,Gudmundsson2015}.
It has some optical differences from the Keck Array
(and \bicepII) to take advantage of the lower sky loading at float altitude (35\,m).
Specifically, the filter stack is predominantly comprised of 
reflective metal mesh (vs.\ absorptive) filters at 250\,K, 130\,K and 35\,K, 
with multi-layer mesh lowpass edge filters at 4\,K and 2\,K, and an absorptive Nylon 
filter at 4\,K. The optical sleeve baffles are cooled to 1.6\,K. This configuration
led to less than 0.35\,pW optical loading on the detectors. 
%The {\sc Spider} telescope have apertures of 250mm and 20$^\circ$
%fields of view.  
%The window is 1/8in UHMWPE.  Metal mesh filters
%at ambient, 120K and 30K are above a 4K sapphire half wave plate,
%Nylon filter, and HDPE lenses (\cite{gudmundsson2015}).
%Between the lenses is a 1.5K cooled
%sleeve with optical baffles.  
%Between the 2nd lens and the 300mK
%focal plane is a low pass filter at 1.5K.  
%The dielectrics are single-layer AR-coated.

The detectors (for the first circum-Polar flight in January,
2015) are slot antenna array-coupled TES bolometers, with 95
and 150\,GHz bandpasses.
The detector 1/$f$ noise knee is low~\cite{rahlin2014}, the
science goals for $10 < \ell < 300$ are accommodated with available
scan rates (see below),
and fast polarization modulation is not needed, as with
the Keck Array and the \bicepII\ and 3 instruments
(see also~\cite{mactavish2008}). The second circum-Polar flight is planned to include
feedhorn arrays and orthomode transducer (OMT)-coupled detectors~\cite{hubmayr2016}.

The gondola provides for AZ and EL scanning. It does not
have boresight rotation, but uses cryogenic waveplates
(rotated 22.5$^\circ$ every 12 sidereal hours)
for $Q$/$U$ discrimination and polarization modulation, with a
contribution as well from sky rotation.  

% (JR:  I'm not sure how any of this is relevant here;  the biggest difference
% is sky coverage, which is partially due to platform (avoiding sun), and 
% partly due to choice (large sky coverage).)
%The
%scan strategy (\cite{rahlin2014,shariff2014,shariff2015}) is
%quite different from Keck and the \bicepI\
%telescopes.
%due to gondola inertial limitations; a sindusoidal
%scan was adopted in both AZ and EL with variable amplitudes.
%The waveplates are stepped 22.5$^\circ$ every 12 sidereal hours.
%This generates good mapping coverage, Q/U descrimination, and
%cross-linking.

\subsection{PIPER} %- Eric
\label{sec:Piper}

\begin{figure}[h]
\begin{center}
\includegraphics[height=6cm]{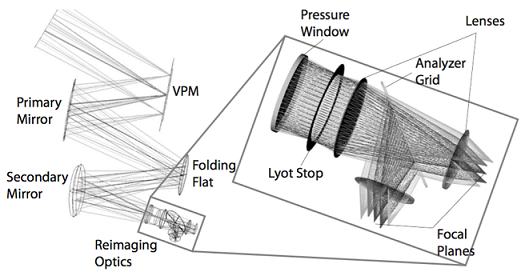}
\includegraphics[height=6cm]{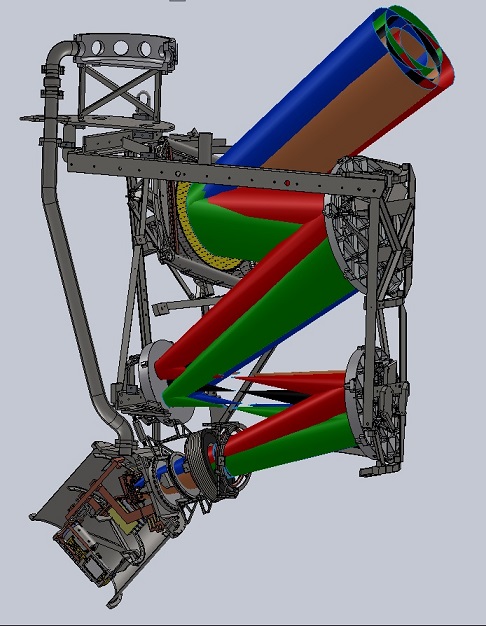}
\caption{\label{fig:PIPER} PIPER. Left: Ray trace of original PIPER optics design. Right: Current PIPER implementation.}
\end{center}
\end{figure}

PIPER is a balloon-borne instrument to observe CMB polarization at 200, 270, 350 and 600\,GHz~\cite{2016arXiv160706172G}. Twin co-pointed telescopes survey Stokes $Q$ and $U$. Like CLASS, the first optical element of each telescope is a VPM. The VPM separates sky signal from instrument drifts by modulating the incoming polarized signal at 3\,Hz, aiding reconstruction of the polarized CMB sky on the largest angular scales. The VPM efficiently mitigates instrument polarization systematic errors by being the first optical element. Each of the PIPER VPMs have a 40\,cm clear aperture with $36\,\mu{\rm m}$ wires at $115\,\mu{\rm m}$ pitch. PIPER uses the bucket dewar from ARCADE, which carries 3000\,L of liquid helium. Helium boiloff allows operation without emissive windows. Superfluid fountain effect pumps draw LHe to cool all optics to $1.4$\,K. Cold optics and the lack of windows reduce photon noise and allow PIPER to take full advantage of the float conditions (especially at high frequencies) and to conduct logistically simpler, conventional flights from Palestine/Ft. Sumner and Alice Springs. Each flight is optimized for one band, and flights from the northern and southern hemispheres cover $85\%$ of the sky.

Two aluminum mirrors image a 12\,cm diameter cold aperture stop ($1.4$\,K) onto the central region of the front-end VPM. The entrance pupil is 29\,cm in diameter and is undersized to limit edge illumination of the VPM (33\,dB edge taper). The stop is a corrugated stack of Eccosorb. The 1.4\,K environment of the bucket dewar mitigates stray light and acts as a co-moving ground screen. The reflective fore-optics feed silicon re-imaging optics that use metamaterial anti-reflection layers~\cite{datta/etal:2013}. The off-axis nature of the fore-optics creates aberrations that can be corrected by de-centering the reimaging lenses. The reimaging lenses remain planar to the stop and are oversized to retain cylindrical symmetry for diamond turning. The final lens focuses light onto a $32 \times 40$ free-space backshort-under-grid detector array at $f$/1.6. The resolution at 200\,GHz is 21$'$ and its Airy disk spans approximately six bolometers.The minimum Strehl ratio within the $6 \times 4.7^\circ$ FOV is $0.97$~\cite{2010SPIE.7733E..3BE}. PIPER uses a common detector array for all frequencies. Between flights, the VPM throw, band-defining filters, and (when necessary) lenses are swapped. This strategy is facilitated by a backshort that is optimized for $200$\,GHz and is less efficient at high frequencies where the atmosphere and dust emission are brighter. A narrower passband toward higher frequency also limits loading.

\subsection{POLARBEAR-2/Simons Array} %- Toki
\label{sec:SA}

\begin{figure}[h]
\centering
\includegraphics[width=0.60\textwidth]{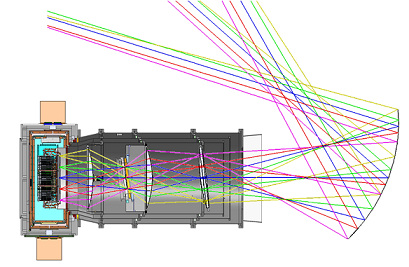}
\caption{Ray trace of the \Pb-2 and the Simons Array optics. Secondary mirror and cryogenic receiver are shown. The length of the cryogenic receiver is 2\,m. The diameter of the three cryogenic lenses are 500\,mm.}
\label{fig:LensSinuous_proj}
\end{figure}

The three telescopes that comprise the Simons Array are identical off-axis Gregorian designs that utilize a 2.5\,m monolithic primary mirror~\cite{Stebor}. 
The telescope and receiver optics are designed to provide a flat, telecentric focal plane over a wide diffraction-limited FOV. 
The angular resolution of the telescope is 5.2$'$, 3.5$'$, and 2.7$'$ at 95, 150 and 220\,GHz, respectively. 
Relative positions of the primary mirror and the secondary mirror obey the Mizuguchi-Dragone condition to minimize instrumental cross-polarization~\cite{Tran:08}. 
Each telescope has a co-moving shield to prevent sidelobe pickup from ground emission and an optical baffle around prime focus to block stray light from reaching the window and scattering into the receiver. 
The first telescope comprising the Simons Array --- the Huan Tran Telescope --- was installed in Chile in 2011 and has been operating nearly continuously with the \Pb-1 experiment since. 
The second and third telescopes were installed in early 2016.

The receivers have windows made out of laminated 10-inch thick
Zotefoam.  Radio-Transmissive Multi-Layer Insulation and a
2-mm thick AR coated alumina plate are anchored to the 50-K stage
as infrared filters~\cite{Choi, Inoue:14}.  The first \Pb-2 receiver,
\Pb-2a, will deploy with ambient temperature continuously-rotating
HWPs~\cite{Hill}.  The second and third \Pb-2 receivers, \Pb-2b and
\Pb-2c, will have cryogenically cooled HWPs at the 50-K stage.
All three receivers have three 500\,mm diameter alumina re-imaging
lenses cooled to 4-K.  The high index of refraction of alumina
allowed for an optics design with lenses that have moderate radii of
curvature. The first lens is a double convex lens, whereas the second
and third lenses are plano-convex.  The optics design has a cold stop
between the second (aperture) lens and the third (collimator) lens.
Metamaterial infrared blocking filters and a Lyot stop are mounted at
the stop.  The final $f$/\# of the focal plane is 1.9.  The optics
design provides diffraction limited illumination that extends over
the 365\,mm diameter of the focal planes.  The Strehl ratio at the edge of
the focal plane is 0.95 for 95\,GHz and 0.85 for 150\,GHz.

\subsection{SPT-3G} %- Nils
\label{sec:SPT}

SPT-3G is a third generation wide-field trichroic (95, 150, 220\,GHz)
pixel camera for the South Pole Telescope (SPT), a 10-m off-axis
Gregorian telescope first fielded in
2007~\cite{benson2014,carlstrom2011}. The new optical design features a
4\,K Lyot stop with a 7.5\,m primary illumination, Strehl ratios $>
0.97$ at 220\,GHz across the 1.9$^\circ$ linear FOV, and
angular resolutions of $1.4'$, $1.0'$, and $0.8'$ at 95, 150, and 220\,GHz,
respectively. The telescope is fitted with a reflective primary guard
ring, side shields, and a prime focus baffle to mitigate far sidelobe
pickup.

\begin{figure}[h]
	\centering
	\includegraphics[width=0.9\textwidth]{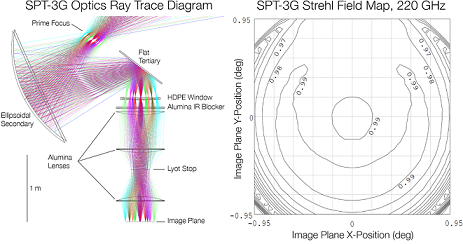}
	\caption{Left: SPT-3G optics ray trace diagram. The new optics
	include a warm secondary and flat tertiary mirror, and three cold
	~4\,K alumina lenses which serve to form a 4\,K Lyot stop and a
	flat telecentric $f$/1.7 focal plane. Right: Strehl ratio field
	map of the 1.9$^\circ$ diameter image plane at 220\,GHz. The Strehl
	ratio is $> 0.97$ at 220~GHz across the usable 430-mm
	image plane diameter.}
\label{fig:SPT3G}
\end{figure}

SPT-3G employs a new optical design consisting of a warm 1.8-m
off-axis ellipsoidal secondary mirror positioned at the
Mizuguchi-Dragone angle~\cite{dragone82,mizuguchi78}, a tertiary folding
flat mirror, and a single receiver with three 720-mm diameter 4\,K
plano-convex alumina lenses which serve to form a 4\,K Lyot stop and a
flat telecentric $f$/1.7 focal plane (see Figure~\ref{fig:SPT3G}). The usable
FOV is limited by vignetting from the lens apertures, not optical
aberrations. The 6.8-mm pixel pitch translates into a pixel spacing of
$1.3 f \lambda$, $2.0 f \lambda$ and $2.9 f \lambda$ at 95, 150,
220\,GHz, respectively, and was chosen to optimize mapping speed given
readout constraints.

The vacuum window is 600-mm inner diameter HDPE with a triangular-grooved AR coating. IR
filters consist of multiple layers of closed cell polyethylene foam
behind the window, an alumina IR blocker and metal mesh IR shaders at
50\,K, and low-pass metal mesh IR filters at 4\,K and 300\,mK.  The
lenses are three-layer AR coated using alumina
plasma spray~\cite{jeong16} and laminated expanded PTFE.

%\section{Projects using crossed-Dragone telescopes}
%\label{sec:CDprojects}

\subsection{ABS} %Tom?

\begin{figure}[h]
\centering
\includegraphics[width=2.5in, clip=true, trim=0in 0in 0in 0in, angle=270, origin=c]{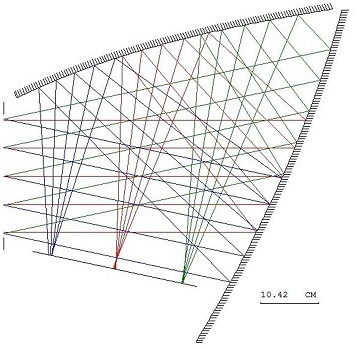}
\includegraphics[width=2.5in, clip=true, trim=0in 0in 0in 0in]{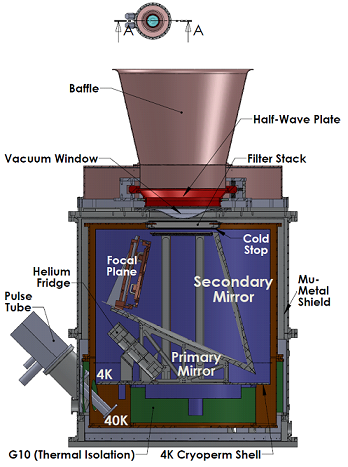}
\caption{Left: Ray trace of ABS optics. Right: Overview of the ABS Receiver.}
\label{fig:receiver_overview}
\end{figure}

%The Atacama B-Mode Search (ABS) 
The ABS telescope consists of 60-cm, cryogenic
primary and secondary reflectors in a crossed-Dragone configuration
held at the 4\,K stage of the receiver (see
Figure~\ref{fig:receiver_overview})~\cite{2009AIPC.1185..494E}. This
optics design was chosen for its compactness for a given focal-plane
area and low cross-polarization. The reflectors were machined out of
single pieces of aluminum. A 25-cm stop at 4\,K limits illumination of
warm elements. The reflectors couple the 25-cm-diameter array of 240
feedhorn-coupled, polarization-sensitive, TES bolometer pairs (480
detectors) operating at 145\,GHz to the sky with $33^{\prime}$ FWHM
beams over a 20$^{\circ}$ FOV. The telescope is an $f$/2.5 system. The
polarization directions of the detectors within groups of ten adjacent
detectors were oriented to minimize cross-polarization and each group
was tilted to minimize truncation on the cold stop. Although neither
the orientations of the ten elements within each group nor the
orientations of different groups are parallel, the detectors are
largely sensitive to polarizations $\pm 45 ^{\circ}$ to the symmetry
plane of the optics.

An ambient-temperature, 33-cm-diameter continuously-rotating HWP is placed at the entrance aperture of the receiver~\cite{2016arXiv160105901E}. The HWP is made of 3.15-mm-thick $\alpha$-cut sapphire AR coated with 305 $\mu$m of Rogers RT-Duroid 6002, a fluoropolymer composite. An air-bearing system provided smooth rotation of the HWP at 2.55\,Hz and polarization modulation in the detector timestreams at 10.20\,Hz. Infrared blocking is provided by capacitive-grid metal-mesh filters patterned on 6~$\mu$m Mylar with grid spacings of 150 and 260$\,\mu$m, along with absorptive 2.5-cm PTFE filters AR coated with porous PTFE at 4\,K and 60\,K. A 0.95\,cm Nylon filter AR coated with porous PTFE at 4\,K provides additional filtering below 1\,THz. The receiver has a 3-mm thick UHMWPE vacuum window AR coated with porous PTFE. A reflective baffle, shown in Figure~\ref{fig:receiver_overview}, and a co-moving ground shield reduce ground pickup.

\subsection{QUIET}

\begin{figure}[h]
\begin{center}
\includegraphics[height=6cm]{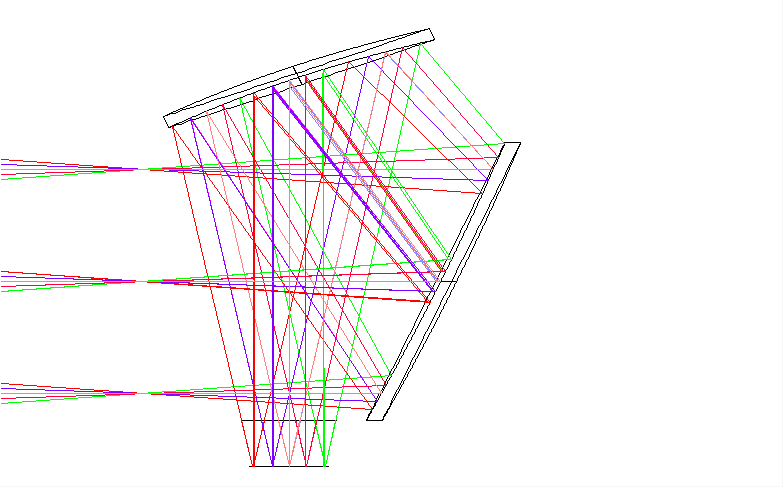}
\caption{\label{fig:QUIET} QUIET ray trace.}
\end{center}
\end{figure}

QUIET was a crossed-Dragone telescope and receiver sited in Chile 
with 1.4\,m diameter mirrors~\cite{imbriale2011,quiet2013}.  
It operated with 42 and 90\,GHz 
receivers using corrugated feedhorns (19 and 91 feeds, respectively) 
and no tertiary optics. It did not have a stop above the 
primary mirror; the absorptive entrance aperture was 
large enough to miss any ray-traced beam from the 
receiver and only intercepted scattered or strongly 
diffracted radiation. Above the entrance aperture an 
absorptive fore-baffle caught several known sidelobes. 

At a wavelength of 2.0\,mm, the design would give a Strehl $> 0.8$ across a
field size (assuming uniform illumination at $f$/1.65) of about
$6.6^\circ$ half-angle. When considering realistic detector beams the
Strehl-limited field size was considerably larger.

The receivers had slow feeds to minimize spillover through 
the telescope, with FWHM of 7--8.6$^\circ$ (for 
the two edges of the 90\,GHz band, similar for the 42\,GHz 
band). The resulting beam sizes on the sky were $0.5^\circ$ for 
the 42\,GHz band and $0.22^\circ$ for the 90 GHz\,band.  
The unvignetted $f$-ratio for the 90\,GHz receiver's 
feed locations was 1.65 (full angle 33.7$^\circ$), resulting 
in less than 0.25\% spillover for any feed in the 91 pixel 
95\,GHz receiver (modeled, not measured). The telescope 
was surrounded by a box of 
Eccosorb (HR-10 on sheet aluminum, protected by Volara foam), 
so that all spillover was intercepted at ambient temperature except 
the small percentage that made it through the entrance 
baffle onto the sky or back into the receiver itself. The cross-polar response of both the telescope and the feed horns 
was also exceptional~\cite{monsalve2010}.  
A larger receiver with 397 identical feeds in the same 
hex pattern on an unmodified QUIET telescope would 
reach about 2.2\% spillover for the edge feeds. 
However, redistributing the feed pattern and widening 
the mirrors out of the plane of symmetry would reduce 
that number. The design is not Strehl-limited.  

The QUIET telescope was operated on the CBI mount, with three axes 
including boresight rotation.

\subsection{CCAT-prime}
\label{sec:ccatp}

\begin{figure}[h]
	\centering
	\includegraphics[height=4cm]{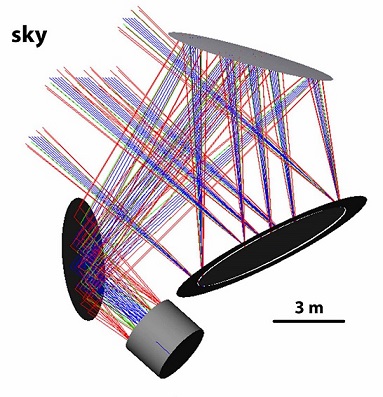}
	\includegraphics[height=4cm]{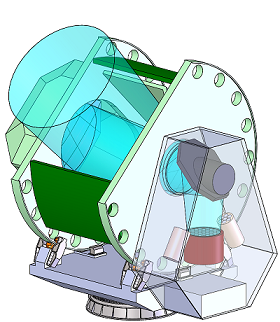}
	\includegraphics[height=4cm]{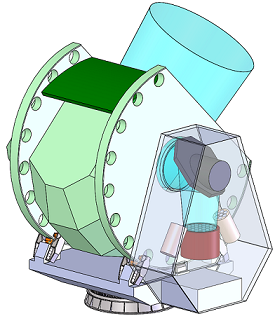}
	\includegraphics[height=4cm]{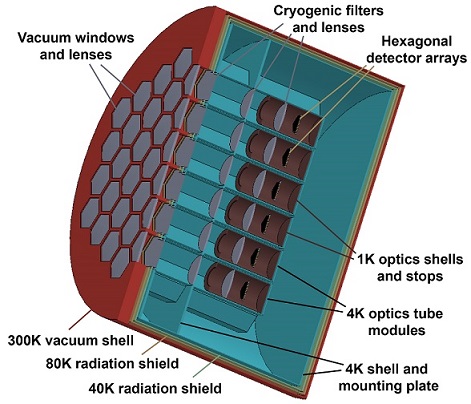}
	\caption{Left: Preliminary high-throughput crossed-Dragone telescope design~\cite{niemack2016}. The CCAT-prime telescope design evolved from this. Middle pair: A preliminary design for the CCAT-prime telescope provides a 180$^\circ$ elevation range without tilting the cryogenic instruments. This is accomplished by rotating the telescope in elevation about the optical axis between the secondary and flat tertiary; the telescope is shown at 45$^\circ$ and 135$^\circ$ elevation. The extended elevation range enables the equivalent of one telescope boresight rotation at each observing elevation. These two telescope boresight positions could be combined with an instrument rotator that provides arbitrary boresight rotations for the cryogenic instrument. Right: Concept for a CMB-S4 receiver with 50 optics tubes that could each illuminate between 2,000--3,000 detectors, providing more than $10^5$ detectors on a single telescope.}
	\label{fig:CCAT-prime}
\end{figure}

The crossed-Dragone telescope design presented in~\cite{niemack2016} and described in Section~\ref{sec:CD_design} has recently been adopted by the CCAT-prime project 
%(\href{http://www.ccatobservatory.org}{www.ccatobservatory.org}) 
and is being studied in greater detail as a candidate telescope design for CMB-S4. A preliminary CCAT-prime engineering study is shown in Figure~\ref{fig:CCAT-prime}. This design offers a large increase in diffraction-limited field of view compared to existing large aperture telescopes, with a $\sim$8$^\circ$ diameter usable field of view. The compact nature of this design enables an unusual optical layout in which the elevation axis is aligned with the optical axis between the secondary and tertiary (or between the secondary and the instrument for designs without the flat tertiary), providing a single Nasmyth-like position for instruments. A fixed, flat tertiary can be used to fold the focal plane down, keeping the overall size compact while improving stability for heavy instruments by shifting them down, closer to the azimuth platform. A design without the flat tertiary is also being considered to reduce loading from the optics. The instruments only rotate with the azimuth structure; they are not required to tip in elevation. This simplifies instrument design and allows for implementation of an instrument rotator to help control systematic errors. Due to the symmetric nature of the telescope mount, the bore sight can be flipped on the sky by rotating in elevation beyond zenith ($>90^\circ$ elevation, see Figure~\ref{fig:CCAT-prime}) and coming back around 180$^\circ$ in azimuth. Some baffling is inherent in the structure, as the optics are mounted inside it, and more baffling or a co-moving ground screen would be straightforward to add.  %Careful analysis is required to determine mechanical keep-out zones, and more baffling could be added, or the plate structure partially replaced with a truss to open up clearances. A co-moving ground screen would also be straightforward to add. 
A clear advantage of having the optics ``buried" inside the mount is lower wind loading on the mirrors as they are effectively inside an enclosure, and elimination of the typical secondary support structure.  
The design also lends itself to having an integrated shutter to provide protection from weather during poor observing conditions.

%% file: telescopes/conclusion.tex
\section{Conclusion}
\label{sec:telescope_conclusion}

The telescope design(s) for CMB-S4 can likely be drawn from the set of existing and new design concepts already available.  For each CMB-S4 science goal, telescopes have been demonstrated with good control of systematic errors, although further study and measurements will be required to prove the designs at the sensitivity level of CMB-S4.   Large-aperture designs with up to a factor 10 larger optical throughput than Stage-III designs exist, and these designs have the potential to greatly reduce the telescope cost for CMB-S4.  We have identified a set of studies to improve our understanding of the trade-offs for choosing the final configuration of telescopes for CMB-S4.   These studies will feed into a global systems engineering process that will include the relationship of the telescope design with other design choices in areas such as cryogenics, detectors, and readout electronics.

%% file: broadband_optics/optics_cmbs4.tex
\chapter{Receiver optics}\label{chp:optics}
\vspace*{\baselineskip} % title is multiline, and without this no space between title & text
\vspace{1cm}

\section{Introduction}\label{sec:opt_intro}
\input{broadband_optics/Introduction}

%%%%%%%%%%%%%%%%%%%%%%%%%%%%%%%%%%%%%
\clearpage
\section{Windows}\label{sec:window}
\input{broadband_optics/Window_intro}
\input{broadband_optics/Window_UHMWPE} %\label{sec:winuhmwpe}
\input{broadband_optics/Window_Zotefoam} %\label{sec:winzotefoam}
%\subsection{HDPE} TBW

%%%%%%%%%%%%%%%%%%%%%%%%%%%%%%%%%%%%%
\clearpage
\section{Filters}\label{sec:filter}
\input{broadband_optics/Filter_Intro}
\input{broadband_optics/Filter_MetalMesh} %\label{sec:filtermmf}
\input{broadband_optics/Filter_LaserIRShader} %\label{sec:filterlaser}
\input{broadband_optics/Filter_Plastic} %\label{sec:filterplastic}
\input{broadband_optics/Filter_Alumina} %\label{sec:filteralumina}
\input{broadband_optics/Filter_SiHybrid} %\label{sec:filtersi}
\input{broadband_optics/Filter_RTMLI} %\label{sec:filterrtmli}

%%%%%%%%%%%%%%%%%%%%%%%%%%%%%%%%%%%%%
\clearpage
\section{Lens material}\label{sec:lens}
\input{broadband_optics/Lens_Intro}

\input{broadband_optics/Lens_Silicon} %\label{sec:lenssi}
\input{broadband_optics/Lens_Alumina} %\label{sec:lensalumina}
\input{broadband_optics/Lens_UHMWPE} %\label{sec:lensuhmwpe}
\input{broadband_optics/Lens_MetaMaterial} %\label{sec:lensmm}

%%%%%%%%%%%%%%%%%%%%%%%%%%%%%%%%%%%%%
\clearpage
\section{Anti-reflection coatings}\label{sec:ar}
\input{broadband_optics/AR_Intro}
\input{broadband_optics/AR_Plastic} %\label{sec:ARplastic}
\input{broadband_optics/AR_ThermalSpray} %\label{sec:ARthermalspray}
\input{broadband_optics/AR_Epoxy} %\label{sec:ARepoxy}
\input{broadband_optics/AR_Silicon} %\label{sec:ARdicedsi}
\input{broadband_optics/AR_Silicon_DRIE} %\label{sec:ARdriedsi}
\input{broadband_optics/AR_Laser} %\label{sec:ARlaser}
\input{broadband_optics/AR_PlasticHole} %\label{sec:ARplastichole}
\input{broadband_optics/AR_MetalMesh} %\label{sec:ARmmarc}

%%%%%%%%%%%%%%%%%%%%%%%%%%%%%%%%%%%%%
\clearpage
\section{Polarization modulators}\label{sec:polmod}
\input{broadband_optics/PolMod_Intro}
\input{broadband_optics/PolMod_AHWP}

\input{broadband_optics/PolMod_Sapphire} %\label{sec:sapphire}
\input{broadband_optics/PolMod_SiliconHWP} %\label{sec:materialSI}
\input{broadband_optics/PolMod_MetalMeshHWP} %\label{sec:mmpolmod}
\input{broadband_optics/PolMod_Rotator} %\label{sec:polmodrotator}
\input{broadband_optics/PolMod_VPM} %\label{sec:polmodvpm}

%%%%%%%%%%%%%%%%%%%%%%%%%%%%%%%%%%%%%
\clearpage
\section{Characterization}\label{sec:char}

\input{broadband_optics/Material_Testing20160911}

%%%%%%%%%%%%%%%%%%%%%%%%%%%%%%%%%%%%%
\clearpage
\section{Conclusion}\label{sec:opt_conclusion}
Optical technologies for observations of the CMB are rapidly evolving with many exciting approaches reaching full maturity through deployments on Stage-III experiments. Work is needed to select and optimize the optical materials needed for CMB-S4 as was discussed in each of the technology sections.   

While this document has focused on presenting the current state of the art, there are opportunities for new ideas to develop during the CMB-S4 process including synergistic combinations of approaches. For example, combinations of metamaterials with  low loss dielectrics could result in wave plates, lenses, and filters with superior optical properties and simplified manufacturing. This document represents the first step in the initiation of a community-wide conversation about how to implement the optical system for CMB-S4. We look forward to the exciting technological developments that will come out of this process.
%
%%%%%%%%%%%%%%%%%%%%%%%%%%%%
% Sean Bryan, testing commit ability

\clearpage
\section{Summary of optics technologies}\label{sec:optsum}
%\input{broadband_optics/OptSum}
\input{broadband_optics/OptSumTable}

%% file: broadband_optics/Introduction.tex
\begin{table}[b!]
\begin{center}
\begin{tabular}{ l  c  c c c c}
\hline
\hline
 			& \textbf{AdvACT} & \textbf{BICEP3} & \textbf{CLASS}  & \textbf{PB/SA} & \textbf{SPT-3G}\\
\hline
\hline
\textbf{Window} & HDPE 	& HDPE 	& UHMWPE 	& Zotefoam 	& HDPE \\
\hline
\textbf{Pol Mod} & Silicon HWP	&  	Instrument	& VPM 		& Sapphire HWP &  \\
 & +Sky Rot	&  	Boresight Rot	& + Sky Rot 		& +Sky Rot &  \\
  & 	&  		& + Boresight Rot 		&  &  \\
\hline
\textbf{IR Filter} &  Silicon	& LAIS 	& MMF 		& RT-MLI 		& Zotefoam  \\
 			& MMF 	& Alumina 	& 		& Alumina 		& LAIS  \\
  			&  		& Nylon 	&			& MMF 		& MMF  \\
\hline
\textbf{Lens} 	& Silicon	& Alumina & HDPE / Silicon	& Alumina 		& Alumina \\
\hline
\textbf{AR Coating} & Meta-Material & Epoxy & Meta-Material		& Thermal Spray & Thermal Spray \\
 			& 		&  		&  		& Epoxy		& Plastic \\
\hline
\hline
\end{tabular}
\end{center}
\caption{Summary of optical elements for ground based Stage-III experiments.}
\label{tab:expsummary}
\end{table}

The science requirements for CMB observations are driving rapid
progress in millimeter wave optical technology. Unambiguous detection
of the B-mode signal from inflationary gravitational waves (IGW)
requires broad frequency coverage to handle foregrounds and tight
control over beam systematic errors. The need for sensitivity drives the
push for high efficiency and wide bandwidth optics to compliment
multiband detector technologies. Similarly, polarization modulators
can be used to mitigate multiple detector systematic errors that are 
associated with differences between two orthogonally polarized detectors, 
and it can also be used to eliminate atmospheric $1/f$ noise that currently
degrades sensitivity on large angular scales where the IGW B-mode
signal peaks. The optical and infrared blocking requirements for the
system can be met with a combination of technologies including vacuum
windows, polarization modulators, IR filters, high index of refraction
lenses, and advanced optical coatings.

Stage-II and Stage-III CMB experiments have developed numerous
innovative solutions to address these instrumental
challenges. Table~\ref{tab:expsummary} summarizes these receiver
optics choices.  Advanced ACTPol uses the transparency and high
dielectric constant of Silicon to implement receiver optics with
broadband lenses and polarization modulators.  \bicepIII\ achieved low
optical loading from the cryostat by using simpler single band optics
and only instrument rotation as the polarization modulator.
\Pb/Simons Array and SPT-3G designed large aperture broadband optics
with high purity alumina.  \Pb{} will deploy with a polarization
modulator whereas SPT-3G has no polarization modulator and no sky
rotation.  CLASS uses plastic lenses with proven performance from past
CMB experiments, as well as Variable Polarization Modulators (VPMs), a
polarization modulation scheme currently unique to their system. The
fact that most of these technologies did not exist five years ago
illustrates the vitality of the field. The fact that no two
experiments have chosen to use identical technologies illustrates the
complexities of optical design. Developing and optimizing these
optical design approaches further is a critical goal of the CMB-S4
effort.

In this write up, we survey state-of-the-art optical technologies for
CMB polarimetry experiments. This includes windows
(Section~\ref{sec:window}), IR filters (Section~\ref{sec:filter}),
lenses (Section~\ref{sec:lens}), coatings (Section~\ref{sec:ar}), and
polarization modulators (Section~\ref{sec:polmod}). For this survey,
research groups prepared notes on their technologies which give a
basic introduction to each technology, descriptions of existing
implementations, and suggestions for necessary research and
development to achieve the technological readiness required to meet
CMB-S4's scientific goals. This survey seeks to present the current
technological landscape in order to aid in the development of
optimized optical system designs for CMB-S4.

%% file: broadband_optics/Window_intro.tex
Receiver windows maintain the cryostat vacuum and maximize the transmission of mm-wave radiation. 
The first condition requires a robust material and the second requires materials with low dielectric losses and small reflections in the band of interest. The required clear diameter (i.e. the diameter through which light can pass, necessarily smaller than the diameter of any mounting rings or hardware) ranges from tens of centimeters to a meter or more depending on the optical design.  In this section we review the state of the art in mm-wave window technology.

%% file: broadband_optics/Window_UHMWPE.tex
\subsection{Polyethylene windows}\label{sec:winuhmwpe}

%%%%%%%%%%%%%%%%%%%%%%%%%%%%%%%%%%%%%%%%%%%%%%%%%%%%%%%%%%%%
\paragraph{Description of the technology}
Polyethylene windows maintain vacuum over large ($>$ 400 mm) clear 
apertures and can be made thin to minimize dielectric losses.  

Polyethylene is manufactured in multiple grades, broadly divided
into 3 categories: low density (LDPE), high density (HDPE), and
ultra high molecular weight (UHMWPE).  These have similar and somewhat
overlapping physical and optical properties.

\begin{figure} [h]
\centering
\includegraphics[width=6in]{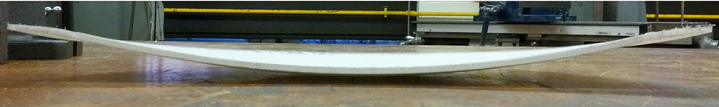}
\includegraphics[width=6in]{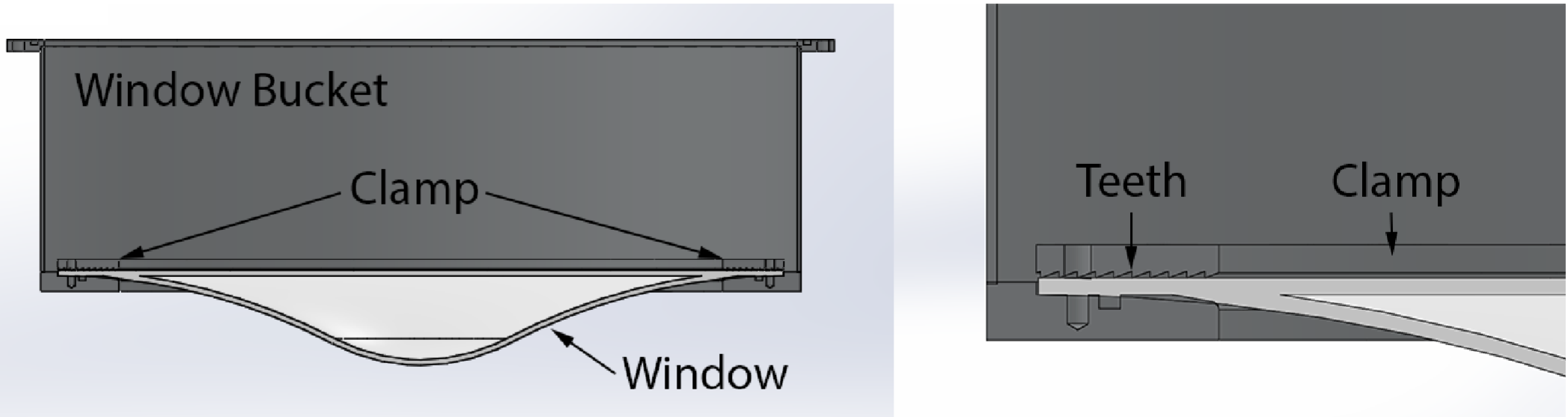}
\caption{(Top) A window sliced in cross section after 3 months of
  continuous use.  (Bottom) On the left is a cross section of the
  window bucket which holds the UHMWPE window. The window is held down
  using the clamp ring which is screwed into threaded holes at the
  bottom of the window bucket. On the right is a zoom in on the
  interface between the clamp and the window. Note the teeth milled
  into the clamp ring. AR coating not drawn.}
\label{fig:UHMWPEWindow}
\end{figure}

%%%%%%%%%%%%%%%%%%%%%%%%%%%%%%%%%%%%%%%%%%%%%%%%%%%%%%%%%%%%
\paragraph{Demonstrated performance}
%\subparagraph{UHMWPE}
UHMWPE has high impact strength allowing for the use of very thin
sheets as windows which minimizes absorption losses without
sacrificing strength.  
Absorption loss is often quoted as loss-tangent 
$\tan\delta = \epsilon_i/\epsilon_r$ which is the tangent of the 
angle between the real and imaginary components of the dielectric function.
UHMWPE has a loss tangent, $\tan\delta <
3\times10^{-4}$ at 150 GHz~\cite{Dumitrescu}, and lab measurements of
the windows built for \spider\ show that the mm-wave photon scattering
is less than 1\% for those windows.  Additionally, UHMWPE has a
relatively low refractive index, $n = 1.525$, allowing simple
anti-reflection coatings with expanded Teflon glued with LDPE to be
effective over wide observing bands \cite{quiet2013, Lamb}.  Because
the windows are thin, they plastically deform when holding a vacuum.
This deformation occurs the first time vacuum is pulled on the
windows.  For the windows used on \spider\ there was no measurable
creep after 48 hours under vacuum, even after repeated pressure
cycles.  Figure~\ref{fig:UHMWPEWindow} shows a cross-sectional view of
a window that successfully held vacuum for three months.  It was then
cut in half for visual inspection, but no signs of damage were
found. For lab testing and deployment, the windows were mounted on the
cryostat in a recessed structure, and a cross sectional view of a
window assembly is shown in Figure~\ref{fig:UHMWPEWindow}.  UHMWPE has
a low coefficient of friction, making it difficult to hold the windows
in place with clamping force alone.  To provide additional gripping
force, concentric teeth were milled into the clamp which push into the
plastic material, as shown on the right of
Figure~\ref{fig:UHMWPEWindow}.

%\begin{figure} [h] \centering
%\includegraphics[width=6in]{figure/UHMWPEWindow1.png}
%\includegraphics[width=6in]{figure/UHMWPEWindow2.png} \caption{Top: A
%window sliced in cross section after three months of use.  Bottom: on
%the left is a cross section of the window bucket which holds the
%UHMWPE window. The window is held down using the clamp ring which is
%screwed into threaded holes at the bottom of the window bucket. On
%the right is a zoom in on the interface between the clamp and the
%window. Note the teeth milled into the clamp
%ring.}  \label{fig:UHMWPEWindow} \end{figure}

The QUIET Q-band and W-band vacuum windows were composed
of Teflon-coated UHMWPE \cite{quiet2013}.  The cryostat windows were 22 inches in diameter.  Multiple tests proved that 2 mm
thick UHMWPE could withstand multiple cycles with 75 mm of bowing.  The
Teflon coating was adhered by melting a layer of LDPE between the
Teflon and the UHMWPE in a large vacuum chamber and pressing the
materials together. The UHMWPE thickness was chosen from
commercially-available stock (6.35 mm for W-band, 9.53 mm for Q-band)
to be close to an integer wavelength in the material. The QUIET window
was anti-reflection coated with expanded Teflon (Zitex) because it has
a well-matched index of refraction ($\sim$1.2) for an anti-reflection
coating for polyethylene, and the required thickness of $\lambda/4$.
All windows and coatings maintained physical integrity during receiver
testing in the laboratory and 1-2 years of deployment in Chile.

The calculated reflection coefficients gave transmission minima of
84\% for uncoated windows, while the Teflon AR-coated window has
minimum transmission of 95\% for the W-band window and 98\% for the
Q-band window.  The absorptive losses increase the noise temperature
by 4~K for the W-band window and 3~K for the Q-band window.  The
reflection values for the coated windows were confirmed in VNA tests
of small samples, and the noise temperature value was confirmed in
laboratory measurements by placing a second window in front of the
receiver.  Finally, GRASP\footnote{Commercially available software for
  electromagnetic simulations,
  http://www.ticra.com/products/software/grasp.} simulations were
performed to estimate the level of induced polarization from the
curvature of the window.  For the QUIET window, the central feedhorn
had negligible instrumental polarization while the off-center pixel
had instrumental-polarization induced by the window curvature of
0.01\%, occurring only at the edge of the bandpass.

HDPE is also used as vacuum window by CMB experiments.
The \bicepIII\ team considered both UHMWPE and HDPE for the receiver
window.  The window was required to be thick enough not to bow down
into the volume occupied by the nearby infrared thermal blocking
filters.  Testing showed that HDPE and UHMWPE had similar performance
(the former's slightly higher room temperature absorption was offset
by its slightly higher stiffness and strength), so HDPE was chosen as
it had been used within the collaboration for lenses (\bicepI,
\bicepII, Keck Array).  The window is a conservative 31.75~mm thick,
with a span of 684.53~mm.  It bows under vacuum by $\sim$35~mm.  The
AR treatment is the same as described above, although a different
source for the ePTFE was chosen since Porex (described in
Section~\ref{sec:ARplastic}) is not wide enough to cover the clear
aperture in one sheet-width.

The \bicepIII\ window was designed to be conservatively thick to be
robust and reliable, but it therefore contributes a measured $\sim$6~K
of in-band loading.  Plans are in place to replace it with a thinner
window, following on the successful \spider\ and QUIET experience.

%%%%%%%%%%%%%%%%%%%%%%%%%%%%%%%%%%%%%%%%%%%%%%%%%%%%%%%%%%%%
\paragraph{Prospects and R\&D path for CMB-S4}

A ``catenary'' style window like \spider\ and QUIET could in principle
be linearly scaled to larger aperture diameters. The window material
is under nearly pure tensile stress, and if the thickness increases
along with the diameter (and circumference), the same (scaled) bowing
should be seen and the same maximum material stresses should
result. This natural scaling means that in principle a new material is
not required to increase to a larger diameter. As with the currently
deployed windows, careful design of the supporting ring and gripping
mechanism will be necessary.

Other polyethylene variant materials may prove useful, given that the
plastics industry continually develops new materials.  Polyethylenes
with embedded high-strength fibers may also prove interesting if the
index of refraction of the fibers is sufficiently close to that of the
polyethylene to minimize scattering.  Investigating these options
would require a modest dedicated effort to procure samples and test
them both optically and mechanically.  Further investigations that
could be carried out are measuring the polarization qualities of
highly stressed polyethylene to determine the expected instrumental
polarization contribution, and understanding any potential changes in
optical properties over time and/or with ambient and temperature.

The technology status level of the solid plastic window is 5. 
QUIET and the series of BICEP experiments has analyzed data taken with plastic windows.
Many Stage-III experiments deployed with solid plastic windows. 

The production status level of the solid plastic window is also 5.
As mentioned above, design can be challenging. 
But once design is complete it can be manufactured in large quantity.
Plastic is readily available commercially, and industrial machining can be utilized to fabricate machined part quickly.

%%%%%%%%%%%%%%%%%%%%%%%%%%%%%%%%%%%%%%%%%%%%%%%%%%%%%%%%%%%%

%% file: broadband_optics/Window_Zotefoam.tex
\subsection{Zotefoam windows}\label{sec:winzotefoam}
\paragraph{Description of the technology}
Closed-cell foam has been considered as a window material for microwave and radio receivers since as early as 1992~\cite{Kerr92}. 
% (Kerr, Baily, \& Boyd, 1992, MMA Memorandum \#90, NRAO).
The first foam CMB receiver windows used Zotefoam PPA-30
(Zotefoams PLC, Croyden, UK), a polypropylene based foam, expanded with 
nitrogen gas (N$_2$).
Zotefoam has been used on CMB receivers ACBAR, \bicepI, \bicepII, Keck Array,
\Pb-1, SPT-SZ, and SPTpol.
In recent years, since the supply of PPA-30 dwindled given a halt 
in manufacture, 
HD-30 has been used, which is based on high density polyethylene.
%\comred{should check which Rx'es used HD-30 vs. PPA-30; KLT only knows for 
%sure about BICEPs and Keck}

\paragraph{Demonstrated performance}
PPA-30 and HD-30 have substantial appeal given (i) their very high 
transparency in the mm band, and (ii) their near-unity indices of 
refraction, eliminating the need for anti-reflection treatment.
The practical diameter limit is of order 500 mm, approached by the 
\Pb-2 receiver, due to the low modulus of elasticity and strength.
For example,  the \Pb-2 receiver uses $\sim$200 mm thick laminated 
HD-30 Zotefoam.
The low thermal conductivity of Zotefoam also helps cryogenic 
receiver performance, with the
cold side of a foam window cooling to of order 200$\,$K or less.

\paragraph{Prospects and R\&D path for CMB-S4}
At higher frequencies PPA-30 (and presumably HD-30) becomes
lossy~\cite{Fixsen01}, suggesting more limited application for
broadband or high frequency instruments.  The scattering of the
material is significant enough that SPT-3G and \bicepIII\ (2017
season) will both use HD-30 as a low-pass FIR filter (i.e. the RT-MLI
filters described in Section~\ref{sec:filter}).  Smaller cell foam
could reduce scattering losses, but its mechanical strength is not as
strong as commonly used HD-30.  With current commercially-available
foam thicknesses, multiple laminated layers of foam are required to
hold vacuum on a window of $\sim$200~mm diameter or greater.  These
laminations provide a mechanical advantage, but are known to increase
loss at higher frequencies.  The ideal foam material for a receiver
window would be a thick closed-cell foam without lamination.  Both the
use of smaller cell size and the elimination of lamination would lead
to lower scattering losses.

The technology status level of the Zotefoam window is 5. 
Many Stage-II experiments used the Zotefoam windows. 

The production status level of the Zotefoam window is also 5.
As long as foam thickness is less than 10 inch, it can be purchased commercially. 
Cutting foam into desirable shape can be done easily. 
Making window from the Zotefoam requires epoxying the foam to metal frame.
It is a simple process that can be parallelized to fabricate large quantities of windows simultaneously.

%Another approach is to support a thin layer of Zotefoam with 
% a cryogenically cooled mechanically strong material.
%An unlaminated 1-inch thick HD-30 layer could provide the vacuum 
%barrier, and an alumina plate could be used for mechanical support.  
%While this is no better than using the alumina directly as the 
%window, a performance gain would be achieved if that alumina 
%is anchored 
%to a cryogenically cooled stage (and isolated thermally from the 
%vacuum jacket).  Cooling, of course, reduces the intrinsic 
%thermally emitted power from the alumina, but additional gains 
%come from an actual reduction in its microwave-band emissivity at 
%lower temperatures.  
%Multiple stacked unlaminated pieces of zotefoam could be placed 
%between the outer vacuum-barrier layer of foam and the supporting 
%plate to provide additional thermal isolation.
%The alumina will then work as an effective IR filter.
%The IR filtering characteristics of alumina filters are 
%discussed in Section~\ref{sec:filter}.

%% file: broadband_optics/Filter_Intro.tex
CMB detectors are subject not only to in-band photon loading, but also
to radiative IR loading from warm optical elements such as the window,
lenses, and telescope and receiver structures. The in-band
contribution increases photon noise on the detectors, reducing mapping
speed. Additionally, there may be loading from the cold stages which
could compromise the bath temperature seen by the detectors, or reduce
the efficiency or duty cycle of the sub-Kelvin cooler that maintains
the bath temperature. Various reflective and absorptive optical filter
technologies are employed in current CMB experiments to mitigate this
infrared radiation from the 300$\,$K, 50$\,$K, 4$\,$K and sub-Kelvin
stages. In the case of wideband CMB detectors, the photons of interest
might also be selected using optical bandpass filters. Below we review
the performance of these filters and the prospects for their use in
CMB-S4.

%% file: broadband_optics/Filter_MetalMesh.tex
\subsection{Metal mesh filters}\label{sec:filtermmf}

%%%%%%%%%%%%%%%%%%%%%%%%%
\paragraph{Description of the technology}
For many years mm and sub-mm experiments have employed multilayer
metal-meshes embedded in polymeric dielectrics to define detector
passbands, reject unwanted optical/near IR (NIR) radiation, and
control the thermal environment in cryogenic instruments.  By
employing several such filters at sequential temperature stages it is
possible to reject optical/NIR radiation while maintaining the
transmission performance of the filter stack to $>80\,$\%.  These
filters are composed of patterned metal-mesh layers which essentially
act as reflectors of the high-frequency radiation, thus rejecting
thermal power that would otherwise be absorbed, and therefore reducing
the thermal loading in the instrument.

\begin{figure}[h!]
\centering
\includegraphics[width=0.6\linewidth]{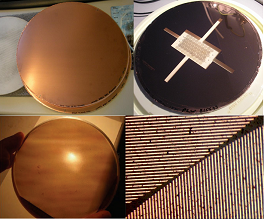}
\caption{Top left: two hot-pressed low-pass filters; top right:
  photolithographed polarizers used in BLASTPol; bottom right: macro
  detail of BLASTPol polarizers; bottom left: metal-mesh
  HWP.}\label{fig:filters_fig1}
\end{figure}

\begin{figure}[h!]
\centering
\includegraphics[width=0.5\linewidth]{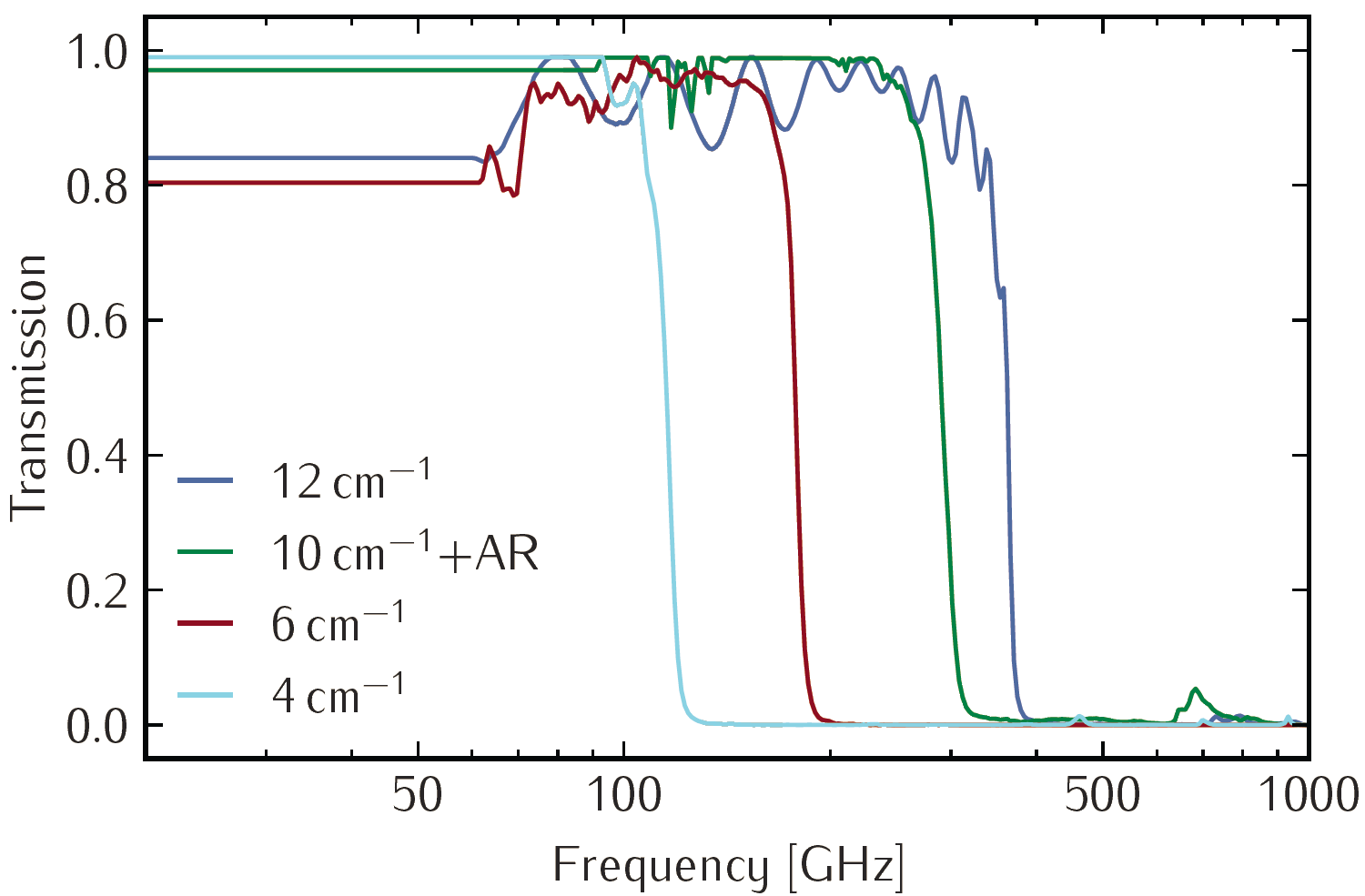}
\caption{Typical measured transmission coefficients at 300\,K for a
  series of hot-pressed low-pass filters. For clarity, the data has
  been extrapolated with a horizontal line below the minimum
  measurement frequency.}
\label{fig:filters_fig2}
\end{figure}

\paragraph{Demonstrated performance}
%\subsection{Low-pass, high-pass and band-pass mesh filters} 
By using multiple layers of inductive, capacitive, or resonant metal
mesh patterns and combinations thereof it is possible to achieve
high-pass, low-pass, and band-pass optical filtering, respectively
\cite{ade:meshfilters,marcuvitz1951waveguide,Ulrich1967}. Heritage
from lab measurements, and modeling with commercial software (e.g.,
HFSS and CST), enables precise filter design.  The technology for all
such filters relies upon highly accurate and reproducible
photolithography performed on Cu layers on a polypropylene substrate,
with standard patterns composed of elements with feature sizes ranging
from 1 to 1000\,$\mu$m or greater.  The basis of a composite filter is
the accurate embedding of many such metal-mesh layers (typically 6 to
12) within a solid polypropylene disc, through a hot-pressing
technique.  The same technology has been used to produce filters for
operation from 30\,GHz to 25\,THz.  Good band-pass filters and
dichroic beam splitter performance can be achieved over an octave,
whereas other devices (low- and high-pass filters) can perform
excellently in transmission or reflection over much broader ranges.
Photos of several metal-mesh devices are shown in
Fig.\,\ref{fig:filters_fig1}, and lab spectral measurements are shown
in Fig.~\ref{fig:filters_fig2}.  The devices are stable and robust and
have been cryogenically-qualified for wide used in space and
suborbital missions.  Finally, high-pass filters can be installed just
above the detectors at the focal plane to mitigate radio-frequency
interference originating outside of the receiver.

%\subsection{IR Blocking filters (thermal filters)}
Thermal filtering of optical/NIR radiation is needed for cryogenic
instruments to minimize the cryogenic cooling requirements. Here it is
vital to reject as much excess heat from the cryogenic chain as early
as possible, since the re-emission of heated dielectric components in
the optics would lead to excess thermal loading onto the detectors. A
combination of very thin scattering and single-layer metal-mesh
devices have been designed, built, and deployed at various temperature
stages to reject optical and NIR radiation. The porous material
scatters most of the incident optical/NIR radiation for wavelengths
equivalent to the pore size of the material, while the metal mesh
reflects longer NIR wavelengths.  As a result the composite device
maintains high transmission ($>$ 95\%) for mm-waves. The in-band
millimeter wave absorption loss in these devices is typically $\ll
1$\% \cite{Tucker2007}.

\paragraph{Prospects and R\&D path for CMB-S4}
CMB receivers are using larger detector arrays, necessitating larger-diameter optics and filters. Metal mesh filters can currently be manufactured with excellent uniformity and reproducibility with an optically-active diameter up to 300\,mm. Expansion of fabrication capability up to 530\,mm diameter is under way for Stage-III experiments \cite{Pisano2016a}. CMB-S4 is likely to require receiver optics and filters of diameter $\sim$500--1000\,mm. To increase metal-mesh filter diameter further, R\&D is necessary and should include establishment and verification of high-fidelity photolithography and uniform thermal pressing of multilayer metal-mesh structures up to 1000\,mm diameter. No major technical challenges are foreseen for this R\&D effort.

The technology status level of the metal mesh filter is 5. 
Every Stage-II and Stage-III experiments has been deployed with some metal mesh filters.

The production status level of the metal mesh filter is 3.
Many metal mesh filters were produced for Stage-II and Stage-III experiments, but request for metal mesh filter production the production needs have been spread out over time.
Depending on receiver design for CMB-S4, the project may require a large number of metal mesh filters in short a amount of time.

%% file: broadband_optics/Filter_LaserIRShader.tex
\subsection{Laser-ablated infrared blocking filters}\label{sec:filterlaser}
\begin{figure} [h]
\centering    
\includegraphics[width=80mm]{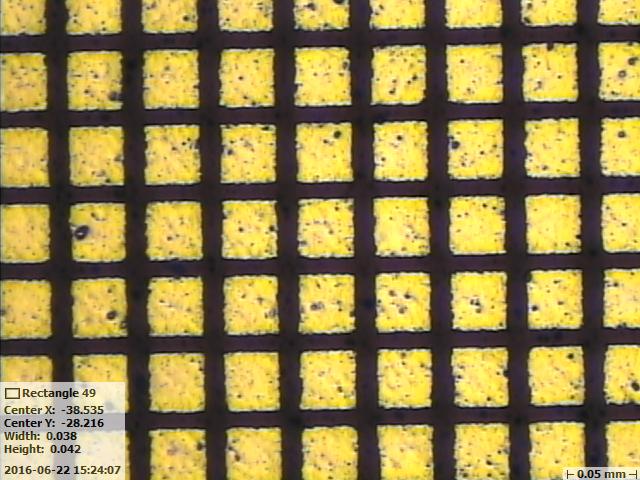}
\caption{Microphotograph of laser-ablated metal mesh features on
  infrared shaders. Squares are 400~nm-thick aluminum $40$~$\mu$m on a
  side with $15$~$\mu$m pitch.}
\label{fig:lasermesh}
\end{figure}

\paragraph{Description of the technology}
An alternative technique to photolithography for metal mesh filter
fabrication is laser ablation, which is a standard industrial process.
In this approach, a single- or double-sided metal-coated dielectric
such as aluminized Mylar or copperized BOPP film is mounted on a
precision XY linear translation stage. A 355~nm NdYAG laser with a
beam waist of ${\cal O}(10)$~$\mu$m is focused on the metal and pulsed
at ${\cal O}(100)$~Hz. The translation stage controller moves the substrate while
the laser pulses to ablate away lines of metal in the two
perpendicular axes, leaving behind metal squares on the dielectric
substrate~\cite{ahmed_large-area_2014}. A typical pattern of 40~$\mu$m
squares separated by 15~$\mu$m spacing achieved using this technique
is shown in Figure~\ref{fig:lasermesh} and has a reflection resonance
around 1~THz.

\paragraph{Demonstrated performance}
Laser-ablated FIR shaders with optically-active diameters of
$\sim600$~mm were successfully fabricated and used in
\bicepIII~\cite{ahmed2014}. Installing a stack of these filters at
ambient temperature leads to an $\sim85$$\%$ reduction of IR power
absorbed by the second-stage absorptive alumina filter. The filter
stack consisted of ten $3.5$~$\mu$m-thick Mylar filters with
400~nm-thick aluminum squares of side $40$~$\mu$m on a $15$~$\mu$m
pitch. In-band mm-wave transmission of $\sim\,98\%$ and reflection of
$\sim1.5\%$ was measured at 95~GHz.

\paragraph{Prospects and R\&D path for CMB-S4}
800~mm optically-active diameter shaders are currently in fabrication
for SPT3G~\cite{benson2014}, and diameters of up to 1~m are possible
without significant technical challenges. Laser ablation provides a
solution for fabrication of large-area metal mesh filters where
feature sizes smaller than ${\cal O}(10)$~$\mu$m are not
necessary. This technique has not been demonstrated with sufficient
feature tolerance for use in band-defining filters where ${\cal
  O}(1)$~$\mu$m tolerance is necessary. Beam-waist shaping of the
laser could improve filter tolerances. Additionally, a program of IR
and mm-wave measurements of transmission, reflection, and loss will be
necessary to validate the choice of technology for CMB-S4.

The technology status level of the laser ablated filter is 5. 
The laser ablated aluminum filter was deployed for a single color receiver in BICEP-3.

The production status level of the laser ablated filter is 3.
Aluminized mylar and laser ablation are done commercially, and in principle the technology is scalable.
No explicit steps were taken to demonstrate mass production, thus PSL of 3 is assigned.

%% file: broadband_optics/Filter_Plastic.tex
\subsection{Nylon and Teflon filters}\label{sec:filterplastic}

%%%%%%%%%%%%%%%%%%%%%%%%%%%%%%%%%%%%%%%%%%%%%%%%%%%%%%%%%%%%
\paragraph{Description of the technology}
%Nylon and Teflon (DuPont, trade names) have strong absorption in the THz and IR region, yet acceptable transparency to mm-waves, making them useful for absorptive IR filtering~\cite{aikin2010}.
%Teflon has a higher frequency cutoff and also higher mm-wave transparency, thus has often been used as a first stage filter, held at 77$\,$K or some other intermediate temperature. Nylon has a lower frequency cutoff but also higher in-band losses, so it is generally used on colder stages.

Nylon and Teflon (DuPont, trade names; generically, a class of
polyamide, and polytetrafluoroethylene or PTFE, respectively) have
strong absorption in the THz and IR region, yet acceptable
transparency to mm-waves, making them useful for absorptive IR
filtering~\cite{aikin2010}. Teflon has a higher frequency cutoff and
also higher mm-wave transparency and thus has often been used as a
first stage filter, held at 77~K or some other intermediate
temperature.  Nylon has a lower frequency cutoff but also higher
in-band losses, so it is generally used on colder
stages~\cite{halpern1986}.

%%%%%%%%%%%%%%%%%%%%%%%%%%%%%%%%%%%%%%%%%%%%%%%%%%%%%%%%%%%%
\paragraph{Demonstrated performance}

Several CMB receivers have used one or both of these filter materials
in lab and on the sky. \bicepI, \bicepII, and Keck Array, for example
used thick PTFE filters at the intermediate stage and thin Nylon
filters at the 4~K stage~\cite{BICEP2Keck_optical}.  The thickness of
the PTFE filters was driven by the need for thermal conductance to
carry the heat out, and not by the optical opacity at high frequencies.
\bicepIII\ exchanged the PTFE filters for an alumina filter given its
higher thermal conductivity, but still had two Nylon
filters~\cite{Wu2016}.  \Pb, on the other hand, used a porous PTFE
filter in combination with mesh filters, its waveplate, and its
Zotefoam window for thermal IR blocking~\cite{pb}.

\spider\ uses a 2.8 mm thick Nylon 6/6 filter mounted at 4~K and
located just skyward of the primary lens.  
Nylon 6/6, one of commercially available type nylon, was selected forits excellent balance of strength, ductility and availability.
This filter has a clear
diameter of 285 mm and is AR coated with Porex ePTFE (expanded Teflon)
using the LDPE vacuum bonding technique described in Section
\ref{sec:winuhmwpe}.  The induced polarization was constrained to be
less than 1$\%$ by warm 90 GHz measurements in samples of Quadrant
extruded Nylon 6/6, Quadrant cast Nylon 6, and Tecamid extruded Nylon
6/6.  Although a second Nylon filter was considered for \spider's 30~K
stage, it was not used for flight because of the potential for high
emission due to poor heat sinking and the redundancy of the filter
stack~\cite{Sasha_rahlin_thesis}.

%%%%%%%%%%%%%%%%%%%%%%%%%%%%%%%%%%%%%%%%%%%%%%%%%%%%%%%%%%%%
\paragraph{Prospects and R\&D path for CMB-S4}
The limited thermal conductivity of plastic means that the center of a
large-diameter plastic filter can be heated by its absorption of the
incident IR radiation, causing the filter itself to re-radiate. This
makes it difficult for these filters to manage IR power for larger
receivers. This absorbed power also loads the cryogenic cooling
system. In contrast, reflective and/or scattering filters reject the
incident power back to the outside of the cryostat, meaning these
filters have a cryogenic advantage for large filter diameters. To
design for and control these effects, accurate measurements of the
frequency-dependent absorption and thermal conductivity of these
materials are needed.

The technology status level of the plastic filter is 5. 
Many Stage-II experiments used plastic filters. For example, series of BICEP receivers used both teflon and nylon filters to reduce infrared loading successfully.

The production status level of the plastic filter is also 5.
Large quantity of plastic can be easily purchased from vendors.
It simply needs to be cut to necessary diameter. Also its low dielectric constant makes application of AR coating simple.

%\bibitem[Aikin et al.(2010)]{aikin2010} Aikin, R.~W.,  Ade, P.~A., Benton, S., et al.\ 2010, \procspie, 7741, 77410V

%% file: broadband_optics/Filter_Alumina.tex
\subsection{Alumina IR filters}\label{sec:filteralumina}
\paragraph{Description of the technology}
Alumina IR filters are attractive for CMB experiments due to the high
mm-wave transmission, high IR absorption, and high thermal
conductivity of alumina at cryogenic temperatures. In particular, the
high conductivity prevents the center of the filter from heating when
loaded with incident radiation.  Standard industrial manufacturing
processes can be used to make filters that are larger than 1 meter in
diameter.  The optical and thermal properties, as well as the
commercial availability of large diameter plates make alumina a
promising IR blocking filter candidate for future CMB experiments.

\paragraph{Demonstrated performance}
The current state-of-the-art alumina manufacturing process can produce
alumina with a loss tangent less below $10^{-4}$ at
100~K. Measurements in \cite{Inoue:14} show $>$95\% transmission at 95
and 150~GHz in bands with 30\% fractional bandwidth, limited by the AR
coating and not the loss tangent.  Multiple methods for AR coating
alumina are discussed in Section~\,\ref{sec:ar}.  Alumina has high
IR absorption with a rapid cutoff frequency near 1~THz.  The high
thermal conductivity of the material ($\sim\,$100 W/m$\cdot$K at 50~K,
roughly three orders of magnitude greater than PTFE) allows for a
smaller temperature gradient across the filter, even when placed near
the cryostat window. For example, a 50~K alumina IR filter design of
2~mm thickness and 500~mm diameter for \Pb-2 has a measured temperature
gradient of $<$6~K between the center and edge \cite{Inoue:14}.
%, in contrast to 75 K of PTFE of the same dimensions.~\ref{fig:b}
%Lastly, unlike reflective IR filters such as the metal mesh filter,
Lastly, the technology should be scalable in size, since companies
have been identified that can make alumina plates larger than 1 meter
in diameter.
%there is no difficulty is fabricating large diameter ($>$700 mm)
%alumina plates.

\begin{figure} [h]
\centering    
\includegraphics[width=80mm]{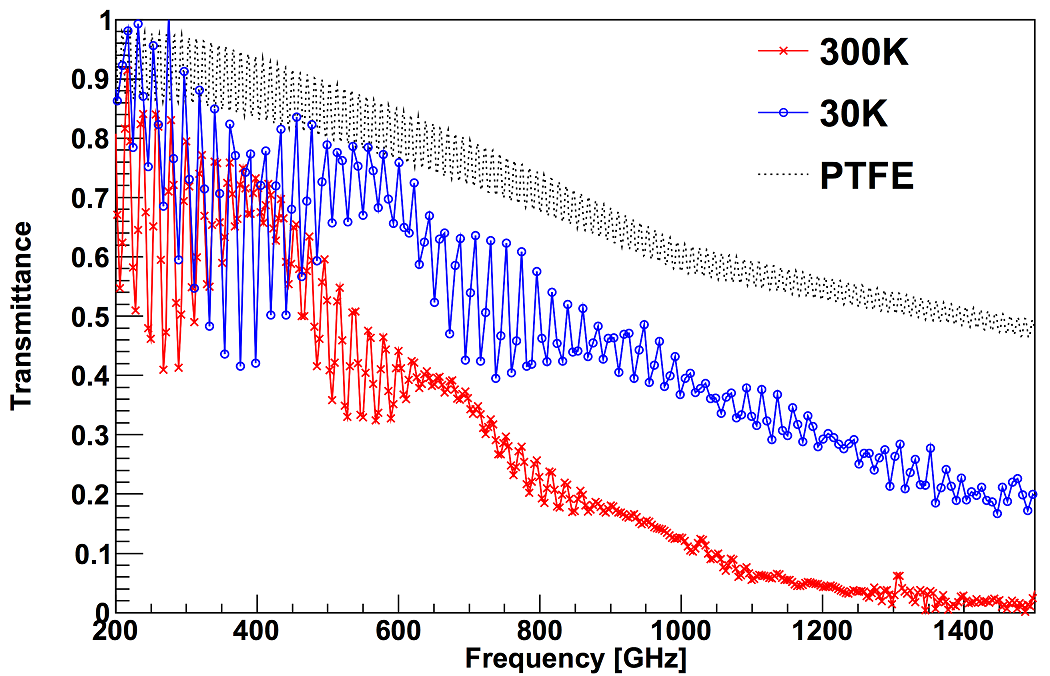}
\caption{Measured spectra of a 2~mm thick IR filter with a single-layer AR coating at 300~K (red) and 30~K (blue). For comparison, a calculated spectrum for a 20~mm thick PTFE filter is also shown~\cite{Inoue:14}.}\label{fig:aluminair}
\end{figure}

\bicepIII, \Pb-2, and SPT-3G use alumina absorptive IR filters at 50~K~\cite{ahmed2014,suzuki15,benson2014}. 
These ex\-pe\-ri\-ments apply the same AR coating between alumina reimaging lenses and alumina IR filters to maximize mm-wave transmission while efficiently absorbing IR photons. 
Figure~\ref{fig:aluminair} shows the measured absorptive performance of a one-layer AR-coated, 2~mm thick alumina IR filter, with a 3~dB cutoff frequency of 450 and 700~GHz at 300 and 30~K respectively.

\paragraph{Prospects and R\&D path for CMB-S4}
In-band transmission and out-of-band absorption have both been shown to vary with different alumina powders. This means that continuing to explore various powder compositions may further improve the performance of alumina filters. 

The technology status level of the alumina filter is 5. 
The alumina filter was deployed for single color receiver in BICEP-3, and it was deployed for multichroic operation in SPT-3G. Alumina filter is also being prepared for POLARBEAR-2 for dichroic operation.

The production status level of the alumina filter is 3.
Alumina plates are available from commercial companies with high throughput production rate.
Application of broadband anti-reflection coating may become throughput limiting aspect of the technology.
There are R\&Ds for rapid application of broadband AR coating as discussed in Section~\ref{sec:ar}.

%% file: broadband_optics/Filter_SiHybrid.tex
%\documentclass[10pt,onecolumn,oneside]{article}
%
%\usepackage{cite}
%\usepackage{natbib}
%\usepackage{graphicx}
%
%\bibliographystyle{abbrv}
%
%\begin{document} 
%
%\title{Silicon Substrate Filters}
%\author{C. D. Munson, Jeff McMahon, Kevin Coughlin}
%\maketitle
%
%

\subsection{Silicon substrate filters}\label{sec:filtersi}

%%%%%%%%%%%%%%%%%%%%%%%%%%%%%%%%%%%%%%%%%%%%%
\paragraph{Description of the technology}
\begin{figure}
	\centering
	\includegraphics[width=0.8\linewidth]{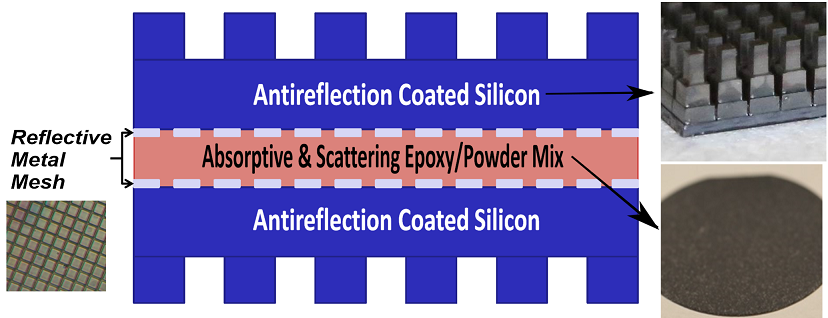}
	\caption{Photographs and an overview drawing of a
          silicon-substrate composite filter. In the filter pass-band
          the metamaterial AR coated silicon couples light into and
          out of the filter stack from free space. In the stop-band, a
          set of lithographically patterned reflective metal features
          reflect a significant portion of the incident light, and an
          absorptive and scattering layer of optical epoxy loaded with
          powdered Reststrahlen materials blocks much of the remaining
          light.}
\label{fig:filter-schematic}
\end{figure}

A silicon substrate infrared filter is a hybrid filter based on
reflective frequency-selective structures patterned on silicon
substrates, scattering/absorptive layers consisting of crystal powders
embedded in an epoxy binder, and metamaterial AR coatings to reduce
in-band reflections from the vacuum-silicon interfaces.

Figure \ref{fig:filter-schematic} shows a conceptual drawing and
photographs of a composite absorptive/reflective IR-blocking filter.
Proceeding from top to bottom, the first surface is a groove
metamaterial AR coating cut into a silicon wafer. A
lithographically-defined frequency-selective surface is patterned on
the bottom of this wafer. Below that is a $\sim$25~$\mu m$ layer of
an absorptive mixture of epoxy and Reststrahlen (frequency selective reflective) powders. Finally,
below that is another frequency-selective surface, patterned on
another silicon wafer, with another groove metamaterial AR coating cut
into the very bottom of the entire stack. At IR wavelengths, light is
reflected off the front silicon wafer and the frequency selective
surface. The front metamaterial surface specularly and diffusely
scatters light at frequencies above the single-moded limit of the
metamaterial structure. IR light not reflected by the metal mesh is
subject to both scattering and absorption by the powder-epoxy
composite layer. The second metal-mesh layer reflects most of the
remaining light back into the epoxy-powder layer, boosting absorption
and (to a lesser extent) reflection. This approach reduces the load on
the cryogenic stage by reflecting a significant portion of the IR
power, and also uses an absorbing layer to further attenuate IR power
passing the first reflective layer even at non-normal incidence
angles.

At mm- and sub-mm wavelengths the frequency-selective surfaces have
high transmission, and since the absorbing layer is much thinner than
a wavelength it has low in-band absorption \cite{datta/etal:2013}.

%%%%%%%%%%%%%%%%%%%%%%%%%%%%%%%%%%%%%%%%%%%%%
\paragraph{Demonstrated performance}
\begin{figure}
\begin{center}
\includegraphics[width=0.8\textwidth, keepaspectratio]{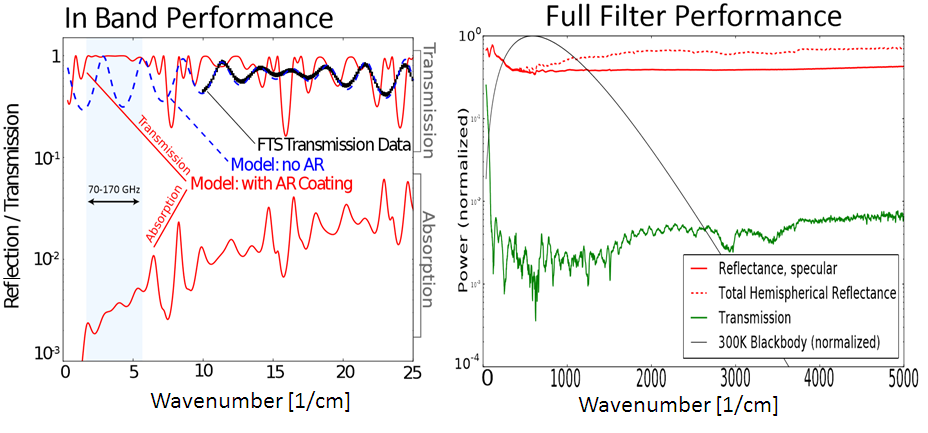}
\end{center}
\caption{Measured and calculated performance of composite filter
  parts. Left: The measured (black dots), and simulated (blue dashed)
  low frequency performance for a single 75~$\mu$m layer of
  Reststrahlen powder mix, in epoxy, on a silicon wafer. Additionally
  shown are the best fit simulated performance (transmission and
  absorption) for a stack consisting of AR coated silicon on either
  side of a 75~$\mu$m powder mix layer. The target transmission band
  for the AR coating is 70-170~GHz and is marked with the blue band
  on the plot.  Right: The IR blocking performance of a full composite
  filter is shown, with a 300~K blackbody overlaid. 
  %Right: A
  %drawing of the integrated test cryostat, wherein the filter (shown
  %in white, and held at 20~K) is used to block power from a 300K
  %blackbody falling on a 5~K disk bolometer (black). A total blocking
  %efficiency of \textgreater 98\% was demonstrated using this setup,
  %in keeping with the prediction from the FTS measurements. This was
  %determined by measuring the heating of the bolometer when exposed to
  %an aperture open to 300~K, and determining the incident power
  %relative to the same environment without a blocking filter. 
}
\label{LF}
\end{figure}

The performance of composite filters was evaluated using Fourier
transform spectrometer (FTS) measurements, as well as integrated
measurements made with a disk bolometer in a cryostat receiving
radiation from a 300~K blackbody.  The IR blocking performance of
these filters was measured on an FTS up to 150~THz, giving a full
characterization of the transmission across the spectrum of a 300~K
blackbody.  In these measurements, the composite filter specularly
reflected \textgreater 40\% of the light incident from the 300~K
blackbody (indicating reflection off the front silicon surface and
metal mesh features), and diffusely reflected another $\sim\,$10\%,
indicative of backscattering off the powder layer.  It transmitted
\textless 1\% of the 300~K blackbody in FTS tests, and \textless 2\%
in an integrated cryostat test, confirming excellent IR blocking
performance.

%\paragraph{Low frequency performance:}
The low frequency performance of a 75~$\mu$m layer of the powder
filter component was measured down to 300 GHz using an FTS.  These
data were then fit with a simple transmission line model.  This model
was then used to extrapolate down to the mm-wave band and to simulate
the effect of adding a three-layer antireflection coating.  This model
shows that the filter introduces minimal loss (dominated by the epoxy
carrier) in a signal band from 70-170~GHz, and that an instrument-band
transmission of \textgreater 99\% should be achievable for a filter
using this technology with the total transmission limited by the AR
performance.

%\paragraph{Cold performance:}
%The Cold Performance was characterized to ensure the proper functioning of these filters at cryogenic temperatures. 
It is a known phenomenon that some Reststrahlen materials have
absorption that significantly decreases when the material is cooled
down.  In particular, alumina (${\rm Al_2 O_3}$) is known to have a section
of its absorption band (between 30 and 300 microns) that decreases at
temperatures of tens of Kelvin \cite{Dobrov:sapphire,Hadni:65}.  A
powder filter consisting of a mixture of calcium carbonate (${\rm CaCO_3}$)
and magnesium oxide (${\rm MgO}$) was measured in an FTS at a range of
temperatures between 4~K and 300~K to confirm that there is not
a significant performance change upon cooling.
%\paragraph{Thermal performance and cryostat testing:}
An integrated test of the composite filter performance was carried out
in a cryostat, to measure the total blocking efficiency of a 150~mm
diameter prototype.  For this %150~mm diameter composite 
filter, there was no measurable heating 
%of the center of the filter 
when the filter was cooled to 20~K and used to block the power from a
70~mm diameter window open to 300~K.  In this configuration, the
power deposited on a carbon disk bolometer at $\sim\,$5~K was
measured, and this measurement yielded the lower limit that at least
98\% of the ambient 300~K blackbody radiation was blocked. This
limit is in agreement with the FTS measurements of the full composite
filter.
\paragraph{Prospects and R\&D path for CMB-S4}
In addition to forming effective free-space IR blocking filters, this
filtering approach offers several novel possibilities for
silicon-substrate optical elements.  Lower frequency-selective metal
elements can be incorporated into these filters to aid in defining the
instrument signal band.  These filters can also be easily and
inexpensively integrated into other optical components, such as
silicon lenses.

To scale this technology for CMB-S4, demonstrations of a full scale
prototype, of merging this approach with a silicon lens, and of band
defining filters are required. Also diffraction and scattering from
these filters should be explored for wider frequency bands.
%All of these prototypes should be tested cryogenically, and optically.  
The first generation prototype filter has smaller than expected
reflection of IR power (40\% vs 90\% predicted). This could be
attributed to a near-field coupling of IR power into the absorbing
layer. Further study would be helpful to reveal origin of this
discrepancy.

The technology status level of the silicon substrate filters filter is 2 and the production status level is 1.
Laboratory testings were performed to demonstrate the IR blocking properties of the filter.
A filter suitable for deployment in a CMB experiment should be fabricated and demonstrated.

%A second prototype that spaces out the first reflector from the absorbing layer is an important next step. The optical measurements must test for diffraction and scattering to wide angles at a wide range of frequencies.

%% Bibliography
%\bibliography{FilterPaper}
%
%\end{document}
%
%
%
%
%
%\pagebreak
%
%%\begin{thebibliography}{99}
%\bibliographystyle{plain}
%\bibliography{FilterPaper}

%\end{thebibliography}

%\end{document}

%% file: broadband_optics/Filter_RTMLI.tex
%\documentclass[a4paper,12pt]{article}
%\usepackage{setspace}
%\usepackage{graphicx}
%\usepackage{fullpage}
%\usepackage{parskip}
%\usepackage{multicol}
%\usepackage{subfigure}
%\usepackage{multirow}
%\usepackage{tabularx}
%
%\usepackage{float}
%\floatstyle{boxed}
%
%\begin{document}
%\noindent {\bf \Large Radio-Transparent Multi-Layer Insulation}
%
%\noindent{Osamu Tajima}

%\subsection{Radio-transparent multi-layer insulation}\label{sec:filterrtmli}
\subsection{Foam based infrared blocking filters}\label{sec:filterrtmli}
% - Technology intro, how it works
% One paragraph summary

\begin{figure}
	\centering
	\includegraphics[height=2in]{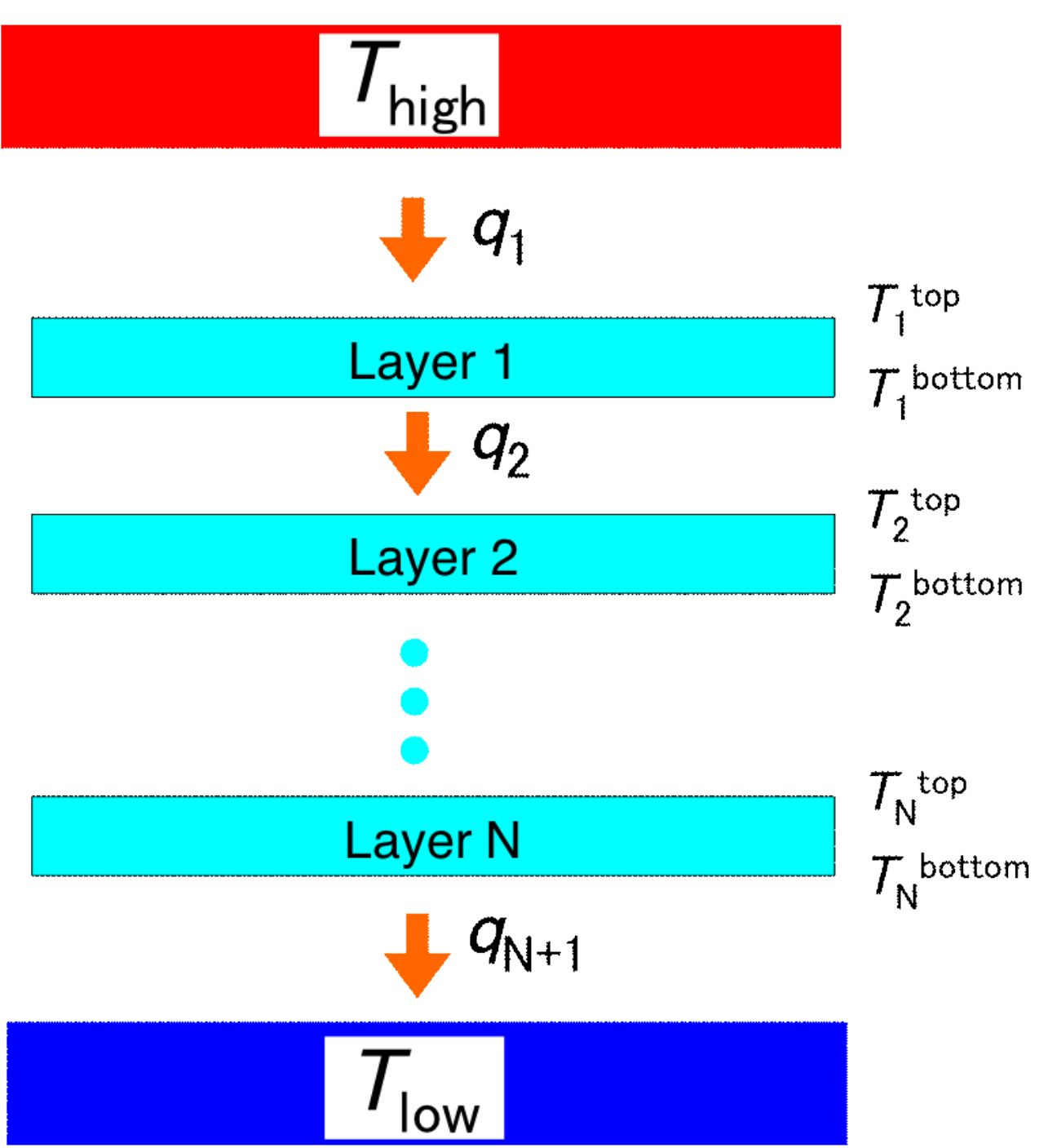}
	\includegraphics[height=2in]{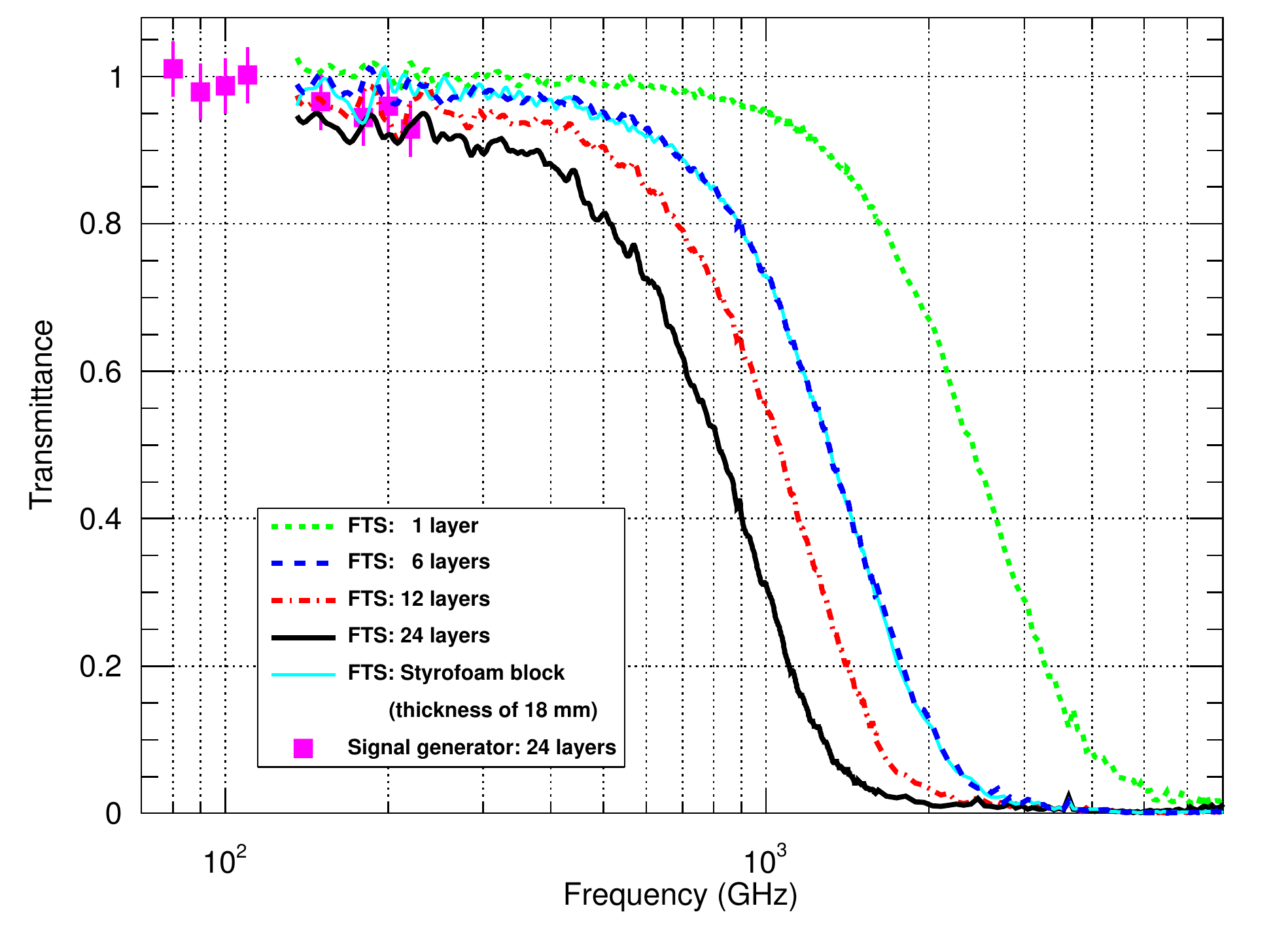}
	\caption{
	Left: Conceptual drawing showing the principle behind RT-MLI. The exchange of thermal radiation between each intermediate layer is balanced, and thermal gradients develop within each individual layer. These two effects permit each layer to have a colder bottom side, and to float to successively lower temperatures, reducing the heat flow from $T_{high}$ to $T_{low}$ by at least a factor of $N+1$.
	Right: Measured transmission of RT-MLI at room temperature. Using an FTS, we measured four different configurations: the number of layers in each configuration was 1, 6, 12, and 24. We also measured the transmittance of a styrofoam block for comparison. Below the 220$\,$GHz region, the transmission through a 24-layer sample was also measured using signal generators. 
	}\label{fig:RTMLI}
\end{figure}

\paragraph{Description of the technology}
Foam based infrared blocking filters, such as stacks of Zotefoams and radio multi-layer insulation (RT-MLI)-based filters are
constructed using commercially-available thermally-insulating
foam~\cite{bib:RT-MLI}.  The insulator is transparent to mm-waves but
absorptive to IR radiation as shown in Figure~\ref{fig:RTMLI}.  The
working principle is similar to conventional aluminized-Mylar
multi-layer insulation (MLI) used for cryostat thermal isolation. Heat
sinking to a cryogenic stage is not required for this filter to work
effectively, simplifying the installation process.

% principle
Figure~\ref{fig:RTMLI} shows the working principle of the RT-MLI
filter. A stack of $N$ isolated layers of the material is assembled,
and each intermediate layer is allowed to reach radiative
equilibrium. Just as with conventional aluminized-Mylar MLI, to first
order this has the effect of reducing the thermal radiation by a
factor of $N+1$. When the simplified formula is compared to measured
load on the RT-MLI, the measured radiation tends to be even smaller
than predicted by this formula because of thermal gradients inside
each layer. A more accurate model \cite{bib:RT-MLI} can be built by
simultaneously solving $N+1$ coupled equations for radiative
equilibrium between the layers, and accounting for the thermal gradients
inside each layer.

Since RT-MLI filters do not rely on heat sinking to a temperature
stage, or on thermal conduction radially across filter layers, the
filters should work as well for large-aperture systems as for
small-aperture systems. Therefore, RT-MLI is naturally suitable for
large-aperture systems, a major advantage for future CMB
receivers. Because this technology relies on passive cooling, maximum
performance is obtained by attaching the RT-MLI on the cold side of
the space between two surfaces.

% - Current state, which experiment uses them?
\paragraph{Demonstrated performance}
In general, any material that is transparent to radio frequency but opaque to infrared radiation can be used as a filter material.
The low index of refraction of foamed polystyrene makes an anti-reflection coating unnecessary. Figure~\ref{fig:RTMLI} shows measurements of 2 mm layers of this material, verifying its performance as an effective IR blocking filter. The transmission through an $N$-layer stack of this material is approximately $0.997^N$ at 200$\,$GHz, i.e., 97\% in the case of 10-layer RT-MLI stack. So far, GroundBIRD~\cite{bib:GB}, and the \Pb-2a receiver for the Simons Array~\cite{suzuki15} employ this technique. GroundBIRD uses RT-MLI with a metal mesh filter in a 300$\,$mm aperture in the space between the 300$\,$K window and the 50$\,$K filter, as well as in the space between 50$\,$K and 4$\,$K. \Pb-2a uses RT-MLI behind the vacuum window (500$\,$mm diameter). RT-MLI is also used at SRON, RIKEN, and NAOJ, in laboratory test cryostats.  In addition, KUMODeS uses RT-MLI for a cryostat that houses a cold calibration source~\cite{bib:KUMODeS}.

% - What is current limitation? Direction for R&D for CMB-S4
% Usage 
\paragraph{Prospects and R\&D path for CMB-S4}
RT-MLI is expected to work equally well for large apertures, but building larger diameter systems to test their cryogenic performance will be informative.  Evaluations of the performance of different RT-MLI materials, studies to determine the optimal number of foam layers, and further development of stack assembly techniques are some additional developments that are required to develop this technology further for CMB-S4.

The technology status level of the foam filter is 3. 
Foam filter deployed in SPT-3G receiver. It is also implemented in POLARBEAR-2 and GroundBIRD receivers. 

The production status level of the foam filter is 5.
Foam used for foam filters is available commercially.
No AR coating is required because of its low dielectric constant.
Also cutting foam into required diameter is very easy and fast.

%
%\bibliographystyle{unsrt}
%\bibliography{cmbs4_RTMLI.bib} 
%
%\end{document}

%% file: broadband_optics/Lens_Intro.tex
Multiple design studies have found that high index of refraction lenses ($n \gtrsim 3$) are required for refracting re-imaging optics to realize large fields of view on $>3$ m telescopes %(what can we reference, ask Niemack, Halverson, and others for ideas) 
and to maximize the number of detectors per telescope in fully-refracting telescope systems.
This does not imply that plastic lenses (in particular polyethylene)
are not suitable for some of the optical components in CMB-S4, but it does mean that the high index materials currently being
used in Stage-III telescopes will likely play a role in CMB-S4.

Silicon, alumina, and sapphire are naturally occurring materials which have high index of refraction and low dielectric losses, and for which AR coatings are currently in use and under further development. 
These materials have trade-offs that drive their use for different applications.   
Sapphire has extremely low dielectric loss (loss tangent $\lesssim 10^{-4}$) and is available in single crystal pieces up to 510$\,$mm in diameter, but it is birefringent. 
Its birefringence makes it useful for waveplates, but unsuitable for lenses. 
Silicon has a high index of refraction ($n = 3.4$) and extremely low dielectric loss (loss tangent below $7 \times 10^{-5}$), but is only available in pieces up to 460$\,$mm in diameter.  
Alumina also has a high index of refraction ($n \sim 3$), reasonably  low dielectric loss (loss tangent below 1 $\times 10^{-4}$), and has the advantage that it can be fabricated as a single piece for parts up to $\sim0.78\,$m in diameter. 
Alumina's material property could vary depending impurities and sintering method. 
Using alumina as optical element requires accurate characterization of actual alumina being used.
These dif\-fe\-ren\-ces in performance and availability drive the application of these materials in different optical systems. 
For example, ACTPol requires lenses only up to 330$\,$mm diameter and uses silicon to take advantage of its machinability and low loss, while \bicepIII, SPT-3G and \Pb{} use alumina since they require larger diameter lenses.

%% file: broadband_optics/Lens_Silicon.tex
\subsection{Silicon}\label{sec:lenssi}

\paragraph{Description of the technology}
\label{sec:materialSI}
Silicon is an excellent material for the fabrication of millimeter wave optics. It has good strength, a high index of refraction ($n$ = 3.4), high thermal conductivity, and low dielectric loss tangents. Its low hardness (6 on Mohs scale) permits machining with diamond tooling. For the highest purity silicon crystals, grown with the float zone process, the loss tangent at 300 K can be $\sim10^{-5}$, and for Czochralski (CZ) grown crystals it is typically $\sim10^{-4}$. When cooled most samples realize good loss tangents in the $10^{-5}$ range where the measurements are limited by the difficulties of cryogenic mm-wave testing.

Currently CZ silicon can be fabricated in boules up to 460 mm in diameter, while float zone silicon is restricted to 200$\,$mm by the limits of surface tension in the zone refining process. The CZ process introduces a number of oxygen defects into the lattice that could increase absorption loss. However with thermal donor annihilation the impact of these defects can be mitigated to the point where the material approaches the performance of float zone silicon.

\paragraph{Demonstrated performance}
300$\,$mm diameter silicon lenses were deployed for ACTPol and Advanced ACTPol receivers.

\paragraph{Prospects and R\&D path for CMB-S4} 
The primary challenges for CMB-S4 are acquiring larger diameter samples and further reducing the dielectric losses. It may be possible to contract with a company to develop the capability to grow boules larger than 460 mm. A lower cost alternative is to develop the ability to glue multiple single crystal wafers together to form a larger diameter composite piece. The glue process that has been developed for the silicon metamaterial HWPs as described above could be adapted for this purpose and direct wafer bonding may offer a higher performance alternative which could be pursued. If even lower loss material is required in large diameters, requiring the CZ growth process, there are processing techniques such as neutron doping which could be applied to fabricate that material.

The technology status level of the silicon lens is 5.
Silicon lenses were deployed in ACTpol and Adv-ACT receivers for single color and multichroic operation.
Data from the experiment was analyzed in detail.

The production status level of the silicon lens is 3.
Commercial vendors are available to provide silicon ingots and to machine them.
Large production throughput that can meet CMB-S4's demand is yet to be demonstrated.

%% file: broadband_optics/Lens_Alumina.tex
%\documentclass[11pt]{article}
%\usepackage[pdftex]{graphicx}
%\usepackage{url}
%\setlength{\topmargin}{0in}
%\setlength{\headheight}{0in}
%\setlength{\headsep}{0in}
%\setlength{\textheight}{7.7in}
%\setlength{\textwidth}{6.5in}
%\setlength{\oddsidemargin}{0in}
%\setlength{\evensidemargin}{0in}
%\setlength{\parindent}{0.15in}
%\setlength{\parskip}{0.10in}
%\usepackage{dsfont}
%\usepackage{amssymb}
%\usepackage{setspace}
%\usepackage{subfigure}
%
%\date{}
%\begin{document}
%%\maketitle

\subsection{Alumina}\label{sec:lensalumina}

\paragraph{Description of the technology}
Alumina is a suitable lens material for CMB polarimetry applications due to its commercial availability, high index of refraction, low loss tangent, and high thermal conductivity at cryogenic temperatures. Its high index of refraction ($\sim$ 3) allows for higher optical power with lower curvature lens surfaces, resulting high-throughput lens systems that can fit in a compact cryostat. Its low loss tangent ($\leq 10^{-4}$ at 4$\,$K) results in low absorption of in-band photons for high optical efficiency. Current state-of-the-art alumina reduces the loss tangent even further below 10$^{-4}$. Its high thermal conductivity facilitates easier cooling and lowers the operating temperature of the optics, which in turn lowers the loss tangent and thermal emission of the optics. Alumina has a thermal conductivity of $\sim$ 100 W/m$\cdot$K at 50$\,$K, which is three orders of magnitude better than typical plastics used for CMB optics. Alumina also has high strength with a fracture toughness of 4$\,$MPa$\cdot$m$^{1/2}$ so that the alumina refractive elements are mechanically robust. In addition to its excellent physical properties, alumina lens fabrication is extremely precise (25 $\mu$m accuracy on aspheric surface) and relatively inexpensive (\$10,000 per lens). 

\paragraph{Demonstrated performance}
Experiments such as \bicepIII~\cite{ahmed2014}, \Pb-2~\cite{suzuki15}, and SPT-3G~\cite{benson2014} currently use or plan to use alumina with high purity ($\ge$ 99.5\%) for their refractive elements. The material has also been used to make cryogenic filters~\cite{Inoue:14}. Unlike silicon, alumina lenses are not limited in diameter or thickness, with the largest existing lens being one of the SPT-3G reimaging lenses which is 720$\,$mm in diameter and 65$\,$mm thick. 

\paragraph{Prospects and R\&D path for CMB-S4}
Alumina is an attractive lens material since it is the material that is currently enabling the largest lenses.  Manufacturing throughput and the consistency of material properties should be further studied for CMB-S4.  In order to support accurate optical design, precision setups for measuring the index and loss-tangent of alumina at operating temperatures should be developed. 
%For ARC, plasma spray AR~\cite{jeong16} and epoxy-based dielectric AR~\cite{Rosen:13} are already existing technologies that can be applied to CMB-S4 alumina optics.

The technology status level of the alumina lens is 5. 
BICEP-3 has taken data with receiver with alumina lens. SPT-3G also deployed receiver with three alumina lenses. POLARBEAR-2 is preparing alumina lens for 2017 deployment.

The production status level of the alumina lens is 3.
Alumina lens is fired and machined at a commercial company. 
The company has many production lines to work on lens in parallel. Large production is possible in principle, but it has not been demonstrated.

%% this information is already in the filter section...
%Alumina IR filter is effective for CMB experiments due to its high in-band microwave transmission, high IR absorption, and high thermal conductivity at cryogenic temperature which prevents the filter from heating due to high incident radiative power. \bicepIII, \Pb-2, and SPT-3G use alumina absorptive IR filters at 50$\,$K. Figure~\ref{fig:b} shows the excellent absorptive performance of a one-layer AR coated, 2$\,$mm thick alumina IR filter, with 3$\,$dB cutoff frequency at 450 and 700$\,$GHz at 300 and 30$\,$K respectively, compared to that of a PTFE IR filter.
%Alkali impurities have shown to great affect the absorptive properties of alumina. Current state-of-the-art alumina reduces loss tangent to $<$ 10$^{-4}$ at 100$\,$K with tighter control on the impurities, which are dominantly alkali metal oxides. Investigate colder pressing processes to produce alumina which is more robust to localized pressure and direct machining, while retaining its in-band transmissive properties.
%\comred{[Toki Comment: Edit last few sentences]}

\begin{figure}
\centering    
\includegraphics[width=73mm]{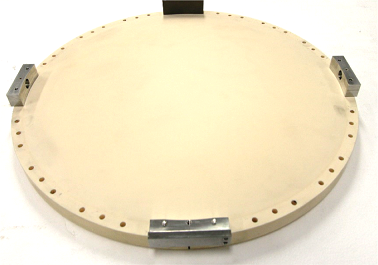}\label{fig:b}
\caption{500 mm diameter \Pb-2 Alumina reimaging lens.}
\end{figure}

%\bibliographystyle{unsrt}
%\bibliography{CMBS4_optics}
%\nocite{*}
%
%\end{document}

%% file: broadband_optics/Lens_UHMWPE.tex
\subsection{Polyethylene lenses}\label{sec:lensuhmwpe}

%%%%%%%%%%%%%%%%%%%%%%%%%%%%%%%%%%%%%%%%%%%%%%%%%%%%%%%%%%%%
\paragraph{Description of the technology}
Polyethylene has been successfully used as a lens material for multiple CMB experiments.
Polyethylene is manufactured in multiple grades, broadly divided
into 3 categories: low density (LDPE), high density (HDPE), and
ultra high molecular weight (UHMWPE).  
These have similar physical and optical properties.

The materials have moderate indices of refraction $n \sim 1.53$, which makes AR coating relatively easy.
Large ($>$ 500 mm, 100 mm or more thick) plates of all grades are easily available, and have good machinability.
Since LDPE has a slightly lower melting point, it may be used in a hot-press process to adhere ePTFE AR coatings to HDPE and UHMWPE lenses. 
Alternately, antireflection grooves, pyramids, or posts can be
machined directly onto lens surfaces.

The low thermal conductivity of polyethylene makes the cryogenic
design challenging for the larger diameter optics.  Also, the moderate
index of refraction puts practical limits on the refraction power of any
particular lens, since higher power (thick) lenses may have too high
an absorption, too steep a curvature, and/or too long a thermal time
constant for some applications.

Due to the moderate index of polyethelene, a single thick HDPE or
UHMWPE lens with front and back surfaces can have individual aspheric
surfaces to reduce aberrations, whereas an alumina or silicon lens
would be in the ``thin lens'' limit.  Also, for polyethelene, a single
AR layer may be adequate over a very broad band, while for the higher
index dielectrics a multi-layer AR coating would be needed.  Finally,
the cost of polyethylene and its fabrication is considerably lower
than the high index materials.

%{\it We need here a discussion of annealing.  Reference might be 
%Chin Lin Wong's dissertation.  One or two sentence paragraph.  }

%%%%%%%%%%%%%%%%%%%%%%%%%%%%%%%%%%%%%%%%%%%%%%%%%%%%%%%%%%%%
\paragraph{Demonstrated performance}
APEX-SZ, QUaD, \bicepI, \bicepII, Keck Array, and SPTpol
successfully deployed cryogenic receivers with multiple HDPE lenses.
\Pb-1 and EBEX deployed with UHMWPE lenses.  
These lenses are up to 350~mm in diameter and 
100~mm thick.  
APEX-SZ deployed lenses with anti-reflection grooves on their surfaces, while the others
deployed with bonded expanded Teflon anti-reflection
coatings. %({\it CHECK THIS, esp. EBEX (true for EBEX -Karl Young Feb. 2017)})

%% refs and other info
%% APEX-SZ  HDPE x2, AR grooves    Schwan, etal 2012
%% QUaD, \bicepI, \bicepII, Keck HDPE x2, ePTFE
%% \Pb-1, UHMWPE x3, ePTFE    Arnold, et al. 2010
%% SPTpol, HDPE x1, ePTFE, Austermann, et al 2012
%% EBEX UHMWPE x4(5) (4th lens duplicated for the 2 focal planes),
%%         unknown AR in Reichborn-Kjennerud, et al. ,2010
%%         AR details (ePTFE) in EBEX Paper 1.  in prep. 2017.
%% also note EBEX had UHMWPE window

There have been some cryogenic challenges with UHMWPE lenses.
For example, the lenses were the last optical elements to get cold during a
cool down for the \Pb-1 receiver.
Also in the \Pb-2 receiver optics design, it was not possible
to design large FOV optics using UHMWPE due to its low index of refraction.
\bicepIII\ experienced similar cooling problems with a set of HDPE back-up lenses 
used only in the lab, 
although the HDPE optical design had nearly the same performance as the deployed 
alumina lens design excepting its (uncertain) absorptive losses.

%%%%%%%%%%%%%%%%%%%%%%%%%%%%%%%%%%%%%%%%%%%%%%%%%%%%%%%%%%%%
\paragraph{Prospects and R\&D path for CMB-S4}
The CMB community has extensive experience with polyethylene
(HDPE and UHMWPE) lenses.
Polyethylene continues to be a good candidate material to
consider for CMB-S4 lenses, especially for
small aperture optics where thermal conductivity might not be as important
and total lens thickness is smaller.
Good machinability, low cost, and transparency to mm-waves
also makes it a good lens material candidate for lab test setups,
particularly for laboratory measurements where AR coatings 
are not required.
%especially since AR coatings are optional for some lab measurements.

An advancement that would benefit CMB-S4 would be to make
comprehensive low temperature transmission and index measurements of
the commonly used grades and brands of polyethylene.  Although some
exist in the literature (e.g. \cite{Lamb}), from grade to grade and
manufacturer to manufacturer we expect some variation.  Removing that
uncertainty would make the best polyethylene materials more appealing
to the optical designer.

The technology status level of the UHMWPE lens is 5.
UHMWPE lenses were used by nearly every Stage-II experiments that successfully detected B-mode polarization.

The production status level of the UHMWPE lens is also 5.
Bulk UHMWPE is readily available from commercial companies. Also it is easy to machine to desired shapes.

%%%%%%%%%%%%%%%%%%%%%%%%%%%%%%%%%%%%%%%%%%%%%%%%%%%%%%%%%%%%
%\begin{figure} [h]
%\centering    
%\includegraphics[height=2in]{figure/UHMWPElens.png}
%\caption{WriteCaption}
%\label{fig:UHMWPELens}
%\end{figure}

%% file: broadband_optics/Lens_MetaMaterial.tex
\subsection{Metal mesh lenses}\label{sec:lensmm}
\paragraph{Description of the technology}
%Mesh technology has been recently used to realize devices with focusing properties. 
%This can be achieved either by manipulating the effective refractive index of the medium or manipulating the phase across the surface of the lens. Graded Index (GrIn) lenses require changing refractive indices across the lens surface following speciffic rules. 
%Mesh-embedded artificail dielectrics with variable geometries have been used to develop this type of thin lenses \cite{Savini12}.%(Savini et al., 2012).
%
%A device able to modify a planar wavefront into a converging wavefront has the focusing effect of a lens. 
%A mesh-lens can be imagined as a simple planar transmissive device that locally modifies  the phase of the radiation across its surface to re-create the effect of a classical lens. 
%The mesh-lens is discretized into pixels that are optimized to provide high transmission and a phase-shift, relative to a central point, with the required frequency dependence. 
%Each pixel is a column of aligned capacitive unit-cells designed like a normal mesh-filter. 

Metal mesh technology has been recently used to realize flat devices with focusing properties similar to conventional curved dielectric lenses. This can be achieved either by manipulating the effective refractive index of the medium \cite{Savini2012} or the phase across the surface of the lens. %\subsection{Graded index metamaterial lenses} Graded Index (GrIn) lenses require changing refractive indices across the lens surface following specific rules. Mesh-embedded artificial dielectrics with variable geometries have been used to develop this type of thin and flat lenses \citep{Savini2012}. %\subsection{Mesh lenses}
In the latter case, a mesh-lens can be imagined as a simple planar transmissive device that changes a planar wavefront into a converging wavefront by locally modifying the phase of the radiation across its surface. The mesh-lens is spatially discretized into pixels optimized to provide a phase-shift relative to the lens center with the required frequency dependence. Each pixel is a column of aligned high-transmission capacitive unit-cells designed like a normal mesh-filter.

\paragraph{Demonstrated performance}
 A 54 mm diameter, $\sim 2.3$ mm thick mesh lens for the 75 -- 110 GHz band has been designed, built, and tested \cite{Pisano2013}. The beam measurements show very good agreement down to the fourth sidelobe. This device did not require anti-reflection coatings and the overall modeled transmission was above 97\%. The diameter of these lenses (currently 10-50\,mm) can be in principle increased by adopting a Fresnel-lens like approach. This should only modestly increase the lens thickness, of order one wavelength. Although the operational frequency ranges are in principle the same as those of successful existing mesh filters ($\sim$ 30 -- 300\,GHz), misalignments and grid non-idealities can affect the performance at the higher frequencies of interest.

\paragraph{Prospects and R\&D path for CMB-S4}
%Future development of metal mesh lenses is closely linked with the future development path of metal mesh filters. This means that the development already planned to increase the diameter and fabrication precision for metal mesh filters will benefit metal mesh lenses as well. Building and testing a large-diameter prototype lens, and integrating a multi-lens system will be helpful in further maturing this technology.

Metal mesh lenses are in an early stage of development although results are promising. While demonstrations have occurred on small scales, large-area fabrication requires R\&D similar to that of metal mesh filters (Section \ref{sec:filtermmf}). CMB-S4 is likely to require receiver optics and filters of diameter $\sim$500--1000\,mm. To increase metal-mesh lens diameter further, R\&D is necessary and should include establishment and verification of high-fidelity photolithography and uniform thermal pressing of multilayer metal-mesh structures up to 1000\,mm diameter. 

The technology status level of the metal mesh lens is 1. 
Laboratory demonstration shows good agreement with simulation. 
On-sky demonstration would further advance technology status level.

The production status level of the metal mesh lens is also 1.
Demonstration of large diameter lens and fast production rate would further advance production status level.

%% existing text is a good status report, but does not describe future developments
%There is an increasing need for pixel integration at focal plane level in millimeter-wave telescopes. 
%Hundreds to thousands of detectors need to be coupled to the telescope optics necessarily avoiding massive and expensive horn antennas. 
%Different solutions can be adopted such as lens-let and phased-array antennas, although their manufacturing processes might not be straightforward. 
%An alternative solution consists in realizing arrays of miniaturized mesh-lenses. 
%Arrays of small diameter mesh lenses can be designed to mimic the behaviour of the well-known and more complex lenslet arrays \cite{Pisano2016b}. A whole mesh-lens array consists of a single flat device manufactured using exactly the same processes required for a single mesh-filter. In this case the alignment of the grids must be accurate enough to guarantee the correct local phase-shifts.

%% file: broadband_optics/AR_Intro.tex
Broad-band AR coated lenses are required for nearly all of the currently proposed CMB-S4 optical designs, including small-aperture refracting telescopes and high angular resolution telescopes using reimaging lenses.      
Similar optical coatings can also be used to realize efficient HWP polarization modulators which could dramatically improve the ability of CMB-S4 to measure polarization on large angular scales. They can also be used to coat vacuum windows.
Various approaches for AR coating are explored such as adding stack of sheets of dielectric layers, modifying the dielectric constant of surfaces by modifying material density, and stacks of metal mesh layers.
Broad-band detector designs have evolved such that 2:1 ratio bandwidth detectors have been deployed, 3:1 ratio bandwidth detectors will soon be deployed, and even broader bandwidths are envisioned. 
In this note we review the requirements for these coatings, discuss the state of the art, and outline the next steps required to prepare these technologies for the CMB-S4 project.

%% file: broadband_optics/AR_Plastic.tex
\subsection{Plastic sheet coatings}
\label{sec:ARplastic}
\paragraph{Description of the technology}
Plastic sheets with a wide range of dielectric constants are commercially available, making them an attractive AR coating option. 
The dielectric constant of a plastic sheet can be tuned by changing its porosity or by adding high dielectric constant filler material. 
CMB experiments have explored PTFE, expanded PTFE (Teflon), loaded PTFE, polyimide (Kapton) and polyetherimide (Ultem) as AR coating materials. 
A list of plastic AR coating materials is given in Table~\ref{tab:plasticAR}.

\begin{table}[th]
\begin{center}
\begin{tabular}{l c c l}
\hline
\hline
Material & $\epsilon_r$ & $\tan\delta [\times10^{-3}]$ & Description\\
\hline
ePTFE & 1.1 - 2.0 &  & expanded PTFE\\
PTFE & 2.1 &  & Teflon \cite{Lamb}\\
RO3035 & 3.5 & 1.7 & PTFE loaded with high dielectric ceramics \cite{ro15}\\
RO3006 & 6.5 & 1.6 & PTFE loaded with high dielectric ceramics \cite{ro15}\\
RT/Duroid & 2.9 & & Glass-reinforced, ceramic-loaded PTFE \cite{ro15}\\
TMM & 3.3 - 12.9 & 2.0 & Loaded plastics. Thermoset laminates \cite{ro15}\\
Cirlex & 3.4 & 0.8 & Pressure-formed laminate of polyimide \cite{Lau}\\
Skybond & 2.1 & 2.5 & Skybond foam is an expanded polyimide \cite{Inoue:2016kyq}\\
Polyetherimide & 3.15 & & Ultem \cite{Quealy}\\
\hline
\hline
\end{tabular}
\end{center}
\caption{Summary of plastic AR coating materials. The loss tangent values shown here are room temperature values that may decrease upon cooling to cryogenic temperatures.}
\label{tab:plasticAR}
\end{table}

\begin{figure}[h]
\centering    
\includegraphics[height=2in]{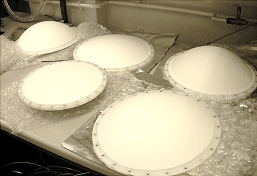}
\includegraphics[height=2in]{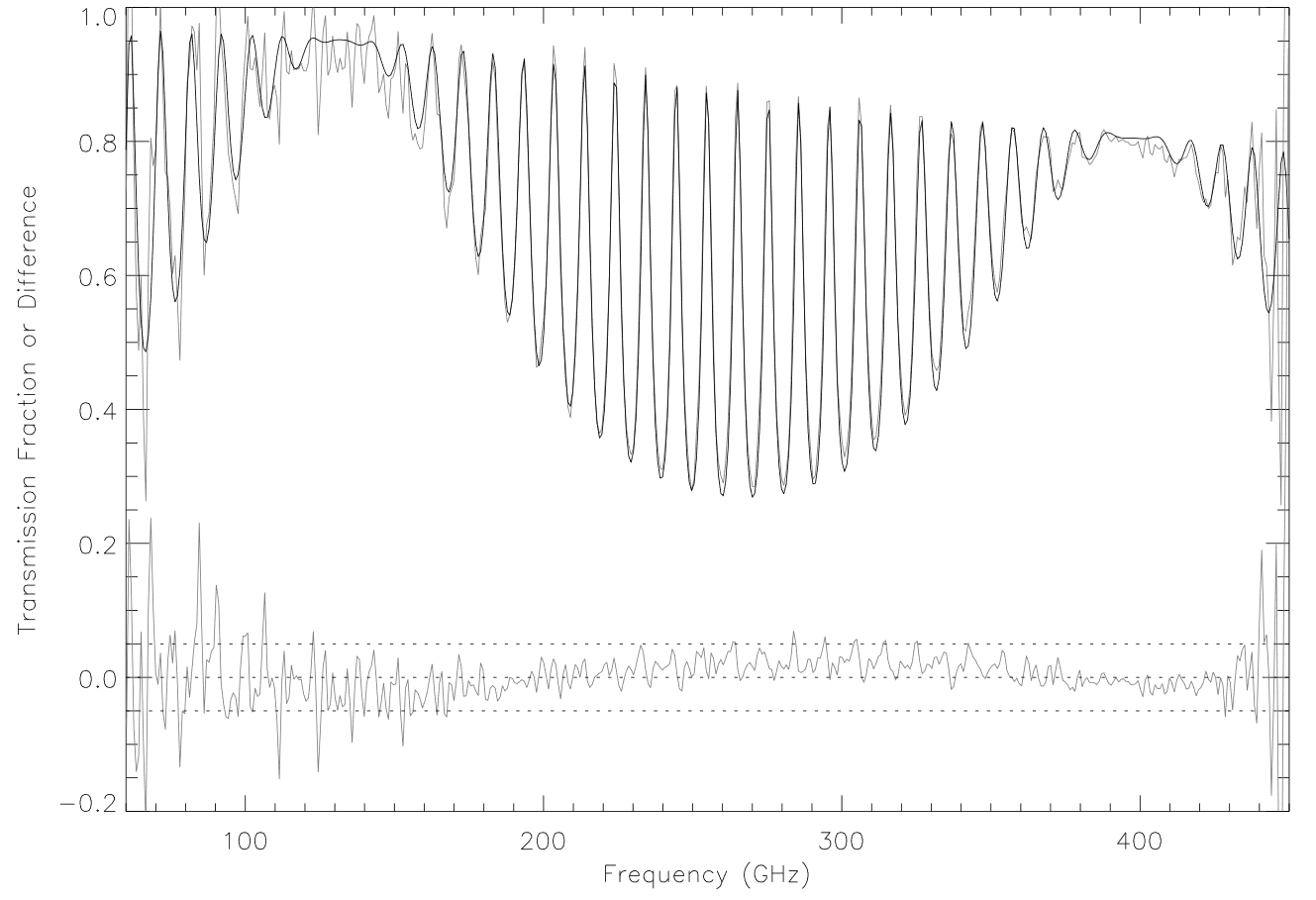}
\caption{Examples of plastic AR coated optical devices. The left panel shows lenses made for the EBEX experiment with the AR-coatings applied. The right panel shows a measured spectrum of the AR coating used on an ACT lens.}
\label{fig:plasticAR2}
\end{figure}

There are several methods to bond multiple plastic sheets onto a lens.
Since the melting temperature of LDPE is below that of commonly used plastics for AR coating, a thin layer of LDPE can be melted between plastic AR coat layer(s) and the lens material. Multiple groups have also used Stycast 1266 epoxy to adhere a plastic layer to a lens.
%The epoxy we use throughout the fabrication process is Stycast 1266, available from Emerson \& Cuming.
%The epoxy is clear, low-viscosity, and has dielectric constant $\epsilon_{r}=3$, measured at \SI{60}{\hertz}; the loss tangent of the epoxy is tan$_{\delta}=0.02$, measured at \SI{60}{\hertz} \citep{ec03}.
It is possible to bond the various PTFE layers directly into a single, monolithic sheet (self-bonding) through a controlled heating and cooling cycle as shown in Figure~\ref{fig:plasticAR2}.
This technique eliminates any intermediate bonding layers which could cause additional loss and unexpected coating performance. 

Application of plastic sheets to a lens requires a uniform adhesive layer across the lens, and the sheet must be applied without any wrinkles.
This is especially challenging on a curved surface.
If the plastic is soft, such as expanded PTFE, it can simply be stretched over the curved surface.
Some plastics, such as polyetherimide, can be thermoformed before coating \cite{Quealy}.
Other plastics, such as PTFE, are not suited to thermoforming processes.
Consequently, a spring loaded press was designed to mold PTFE; the SPT-3G experiment is using this technique to coat lenslet arrays.
Cirlex has been machined to the correct shape before being adhered to a silicon lens \cite{Lau}.
Otherwise, for low radius of curvature shapes, vacuum-bagging produces a uniformly adhered, smooth coating.

Differential thermal contraction between plastic and other common lens materials such as silicon, alumina, and sapphire is problematic for cryogenic operation.
Adhesion promoter, such as Lord AP-134, helps to increase the epoxy bond strength to make a bond strong enough to withstand thermal contraction.
Dicing stress relief grooves into plastic sheets also helps to mitigate the thermal contraction issue as shown in Figure~\ref{fig:materialsetup3}.
Multilayer plastic coatings of sizes up to 300 mm in diameter are robust to violent thermal cycling without dicing.
The mechanical modulus of expanded PTFE lowers as the material becomes more porous. 
A lower modulus helps to mitigate the delamination problem that occurs as a result of differential thermal contraction.

%
%\begin{figure}[h]
%\centering    
%\includegraphics[height=3in]{figure/3glenslet.png}
%\caption{Caption}
%\label{fig:plasticAR1}
%\end{figure}

\paragraph{Demonstrated performance}
Many deployed instruments have successfully used this technology to AR coat a range of optical devices. Expanded PTFE sheets were glued with LDPE to an UHMWPE cryostat window and reimaging lenses for the \spider, \Pb{} and EBEX experiments as shown in Figure~\ref{fig:plasticAR2} \cite{SpiderFilippini, Hargrave,pb}.
The HWP for the ABS experiment was coated with Rogers RT/Duroid using rubber cement as a glue layer \cite{abs}.
The 150 GHz HWPs for the \spider\ experiment were coated with hot-pressed Cirlex using HDPE as an adhesion layer, and this is baselined for the 285 GHz \spider\ HWPs as well \cite{SBryan}.
Machined Cirlex was adhered to a silicon lens with Stycast 1266 and Lord AP-134 adhesion promoter for the ACT experiment \cite{Lau}.
The measured performance from the coating is shown in Figure~\ref{fig:plasticAR2}.
%Thermoformed polyetherimide was glued to silicon lenslets with Stycast 1266 for the \Pb-1 
A sheet of Skybond foam was attached to an alumina IR filter that was coated with a thermal sprayed mullite ceramic layer to cover the 90\,GHz and 150\,GHz bands simultaneously \cite{Inoue:2016kyq}. 
%Stacks of ePTFE, RO3035, and RO3006 were used to anti-reflection coat alumina lenslet array for the South Pole Telescope Third Generation experiment. 
%Three plastics were chosen to form broadband anti-reflection coating that covers 90 GHz, 150 GHz and 220 GHz bands simultaneously. 
%Three layers were bonded with thermal cycling method. 
%30W CO$_{2}$ laser was used to dice around each lenslet, minimizing the total contractile area of any one PTFE element as shown in Figure~\ref{fig:plasticAR1}.
%The laser ablation is fast, accurate, repeatable, and---unlike traditional dicing and cutting methods---exerts no tool pressure on the fragile silicon substrate.
Multiple sheets of Roger's corporation's TMM laminates and expanded PTFE were used to coat a broadband HWP of the EBEX experiment which covers 150\,GHz, 220\,GHz and 410\,GHz simultaneously \cite{Reichborn2010}.

\paragraph{Prospects and R\&D path for CMB-S4}
The performance of plastic sheet AR coatings is primarily limited by the commercial availability of suitable materials.
It is possible to custom order loaded PTFE sheets with different types of dopant to optimize the dielectric property of the material.
Such an idea was not practical for smaller scale Stage-III experiments, but with larger commercial orders, it may become feasible for CMB-S4.

The technology status level of the plastic coating is 5. 
Plastic AR coating was used for multiple Stage-II experiments on UHMWPE lens. 
Plastic AR coating has not been used for high dielectric lens such as silicon lens and alumina lens. 
For multi-chroic application, 5 layer plastic coating was applied on sapphire for a cryogenic half-wave plate.

The production status level of the plastic coating is 5.
Stage-II experiments demonstrated that coating is relatively easy to apply once proper jig is setup.
%However, actual demonstration of plastic AR coating lens at large quantity has not been done. 

%% file: broadband_optics/AR_ThermalSpray.tex
\subsection{Thermal spray coatings}
\label{sec:ARthermalspray}
\begin{figure}[h]
\centering    
\includegraphics[height=2in]{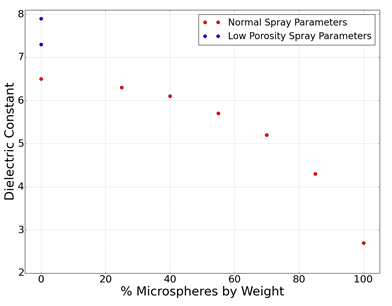}
\includegraphics[height=2in]{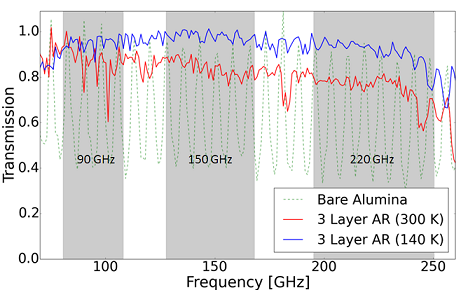}
\caption{Left: Tunability of dielectric constant (i.e. the square of the index of refraction) for plasma spray AR technologies. The dielectric constant of an alumina-based coating is controlled by mixing hollow microspheres (red) and/or varying plasma energy with different spray parameters (blue), such as the flow rate of plasma gas \cite{jeong16}. Right: Measured transmission of a three layer AR coating on two sides of a 6~mm thick slab of alumina.  The bottom two AR layers were thermal spray coatings. The top AR layer was expanded Teflon adhered to the alumina with LDPE.}
\label{fig:thermspray1}
\end{figure}

\paragraph{Description of the technology} 
Plasma spray AR is a process by which a base material of alumina and silica is melted with a plasma jet and sprayed onto a lens surface, cooling immediately upon impact to form a strongly adhered coating without the need for any glue or adhesion promoters. 
The ability to tune the dielectric constant by varying porosity within the coatings, as shown in Figure~\ref{fig:thermspray1}, and the low loss-tangent (below $10^{-3}$ at 140~K) of plasma sprayed coatings allow for AR coatings with the range of dielectric constants suitable for broadband multi-layer applications \cite{jeong16}.
The spraying process is technically straightforward. 
The alumina-silica coating material has a matching coefficient of thermal expansion to the alumina optics, meaning it is robust against cryogenic delamination. 
Additionally, the robotic arm that controls the spraying allows for a fast and accurate programmable spraying pattern to uniformly coat a variety of surface profiles, whether they are flat filters or waveplates, curved lenses ($\sim$ 700 mm diameter), or a large array of small hemispherical lenslets ($\sim$ 6.35 mm diameter). 

\paragraph{Demonstrated performance}
The \Pb-2 receiver will be deploying alumina infrared filters coated with this approach. 
The SPT-3G receiver will be deploying with a 720~mm diameter infrared filter and lenses, with multi-layer sprayed AR coatings that simultaneously cover the 90, 150, and 220~GHz bands \cite{benson2014}. 
For a broadband AR coating, it is desirable to have some layers with an index of refraction as low as 1.25. The lowest index currently achieved by the plasma spray technique is 1.6. SPT-3G and POLARBEAR-2 experiments combined plasma sprayed layers with a porous plastic sheet (index of 1.25) to create the broadband AR coating shown in Figure~\ref{fig:thermspray1}.

\paragraph{Prospects and R\&D path for CMB-S4}
Having dielectric layer with an index below 1.4 
facilitates design with low level of reflection for a broadband AR coating. 
Currently, the lowest demonstrated index for
alumina ceramic thermal spray is 1.6.  The spray process has several
variables that could be tuned to further lower the index, such as the
powder feed rate, spray distance and flame temperature.  Other
dielectric powders compatible with the spray process should also be
explored, to explore if the index can be lowered using this existing
process but a different material.

The technology status level of the thermal spray AR coating is 3
Multilayer anti-reflection coating with thermal spray coating and expanded kapton sheet was applied on POLARBEAR-2 lenses.
POLARBEAR-2 lenses are on going through integration test for 2017 deployment.
SPT-3G used thermal spray coating for bottom two layers of three layer anti-reflection with plastic layer as top layer.
Second and third receivers for Simons Array and SPT-3G is planning upgrade of alumina lens anti-reflection coating with fully thermal sprayed lenses.

The production status level of the thermal sprayed AR coating is 3.
Production rate was demonstrated on POLARBEAR-2 lens, and development is on going for thermal spray coating to cover required range of dielectric constat.

%% file: broadband_optics/AR_Epoxy.tex
\subsection{Epoxy coatings}
\label{sec:ARepoxy}
\begin{figure}[h]
\centering    
\includegraphics[height=2in]{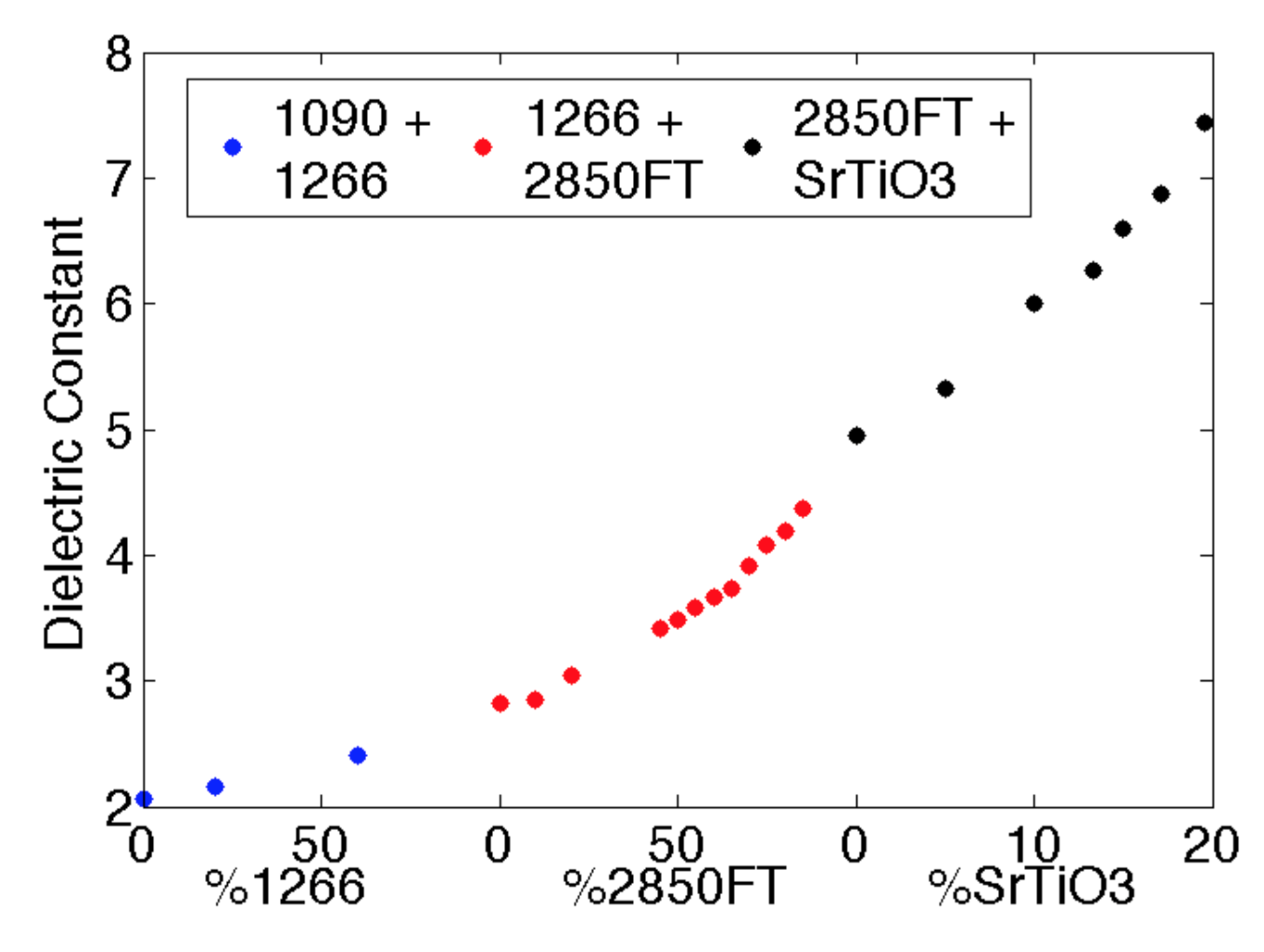}
\includegraphics[height=2in]{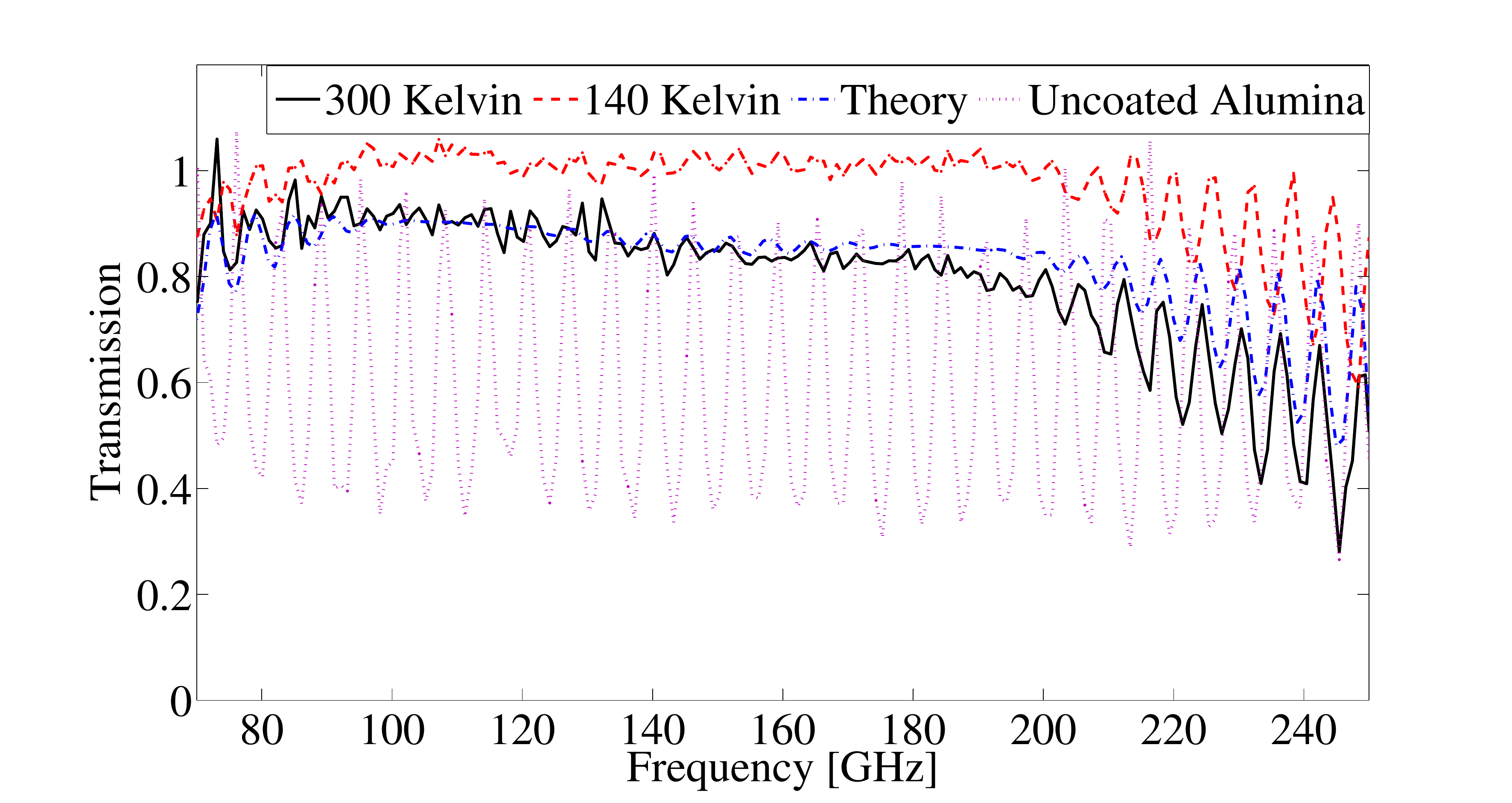}
\caption{Tunability of the dielectric constant for epoxy-based AR coatings. The dielectric constant is controlled by mixing different Stycasts and SrTiO$_{3}$\cite{Rosen:13}. The spectra in the right panel show that the coatings have low loss when cooled to 140~K or lower.}
\label{fig:epoxyspray1}
\end{figure}

\paragraph{Description of the technology}
Epoxy can be used as an AR coating material \cite{Rosen:13}. 
Different types of epoxies have different indices of refraction, and the index can be tuned by mixing two epoxies in various ratios.
For example, Stycast 1090 and Stycast 2850 FT have indices of refraction of 1.43 and 2.23, respectively. 
Even higher indices have been obtained by mixing other powdered material, such as strontium titanate, into the epoxy.
The tunability of the index and the low loss of epoxy-based coatings at cryogenic temperatures are shown in Figure~\ref{fig:epoxyspray1}.
 An epoxy-based AR coating can be applied on a lens with a negative mold to coat the lens surface, and then after the epoxy hardens the surface can be precisely machined to the correct thickness with a computer numerical controlled (CNC) milling machine. 
Laser machined stress relief grooves have been shown to relieve mechanical stress between the epoxy AR coating and lens due to the thermal contraction mismatch.
The groove width can be made smaller than 25 microns to prevent scattering.

\paragraph{Demonstrated performance}
An epoxy-based dielectric single-layer AR coating was applied to the
600\,mm diameter IR filter and lenses of the 95\,GHz
\bicepIII\ receiver \cite{ahmed2014}. A multi-layer AR coat on 500\,mm
diameter lenses will be deploying on the \Pb-2a receiver covering the
95 and 150\,GHz bands \cite{suzuki15}.  The thicknesses of each
coating were assessed by measuring the profile of the lens before and
after coating with a coordinate measuring machine, showing that the
thickness of the coatings were machined to 10 to 20 micron accuracy.
The loss tangent of epoxy and epoxy-filler mixture increases at
frequencies above the CMB passbands, which means that the AR coating
itself provides additional IR filtering when used with alumina
infrared filters or lenses \cite{Inoue:14}.
 
\paragraph{Prospects and R\&D path for CMB-S4} 
For single-layer and two-layer AR coatings of alumina, a combination of Stycast 1090 and Stycast 2850-FT provides the necessary range of dielectric constants with low levels of absorption loss.
For three- or more layer AR coatings, an index above 2.23 is required for some of the layers, which has been achieved by adding the strontium titanate powder to the mix. While it has performed well in a three layer coating, unfortunately the loss tangent of a mixture of Stycast 2850-FT and strontium titanate mixture is relatively high. 
It would therefore be useful to develop a high index filler powder with a lower loss tangent. Silicon and alumina should be explored to see if they will reduce the total loss tangent of the mix to below 5$\times$10$^{-4}$.

The fabrication process for epoxy AR coatings requires a somewhat laborious process of moulding and machining, which may limit its applicability for high volume fabrication. Furthermore, epoxy-based AR coating requires a large CNC to machine large diameter lens coatings. A dedicated machining center, or potentially an industry contract, could solve these scalability challenges.

The technology status level of the epoxy AR coating is 5. Epoxy coated alumina lens has been deployed on BICEP-3 receiver. Multichroic version of epoxy coated alumina lens is prepared for POLARBEAR-2 deployment, and it is in final stage of integration and test. 
The production status level for the epoxy AR coating is 2. Although it was successfully deployed for Stage-III experiment, epoxy coating requires laborious coating and machining process. 

%% file: broadband_optics/AR_Silicon.tex
%\documentclass[11pt, oneside]{article}   	% use "amsart" instead of "article" for AMSLaTeX format
%\usepackage{geometry}                		% See geometry.pdf to learn the layout options. There are lots.
%\geometry{letterpaper}                   		% ... or a4paper or a5paper or ... 
%%\geometry{landscape}                		% Activate for rotated page geometry
%%\usepackage[parfill]{parskip}    		% Activate to begin paragraphs with an empty line rather than an indent
%\usepackage{graphicx}				% Use pdf, png, jpg, or epsÂ§ with pdflatex; use eps in DVI mode
%								% TeX will automatically convert eps --> pdf in pdflatex		
%\usepackage{amssymb}
%\usepackage{sidecap}
%%SetFonts
%
%%SetFonts
%
%
%
%\begin{document}
%%\maketitle
%%\section{}
%%\subsection{}
%
%\noindent {\bf \large Metamaterial Silicon AR coatings} \\
%{Kevin Coughlin, Charles Munson, Rahul Datta, Jeff M$^c$Mahon}
%%\date{}							% Activate to display a given date or no date

\subsection{Diced silicon}
\label{sec:ARdicedsi}
\begin{figure}
\centering    
\includegraphics[height = 2 in]{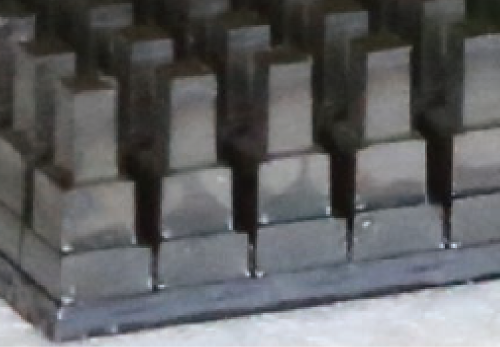}
\includegraphics[height = 2.5 in]{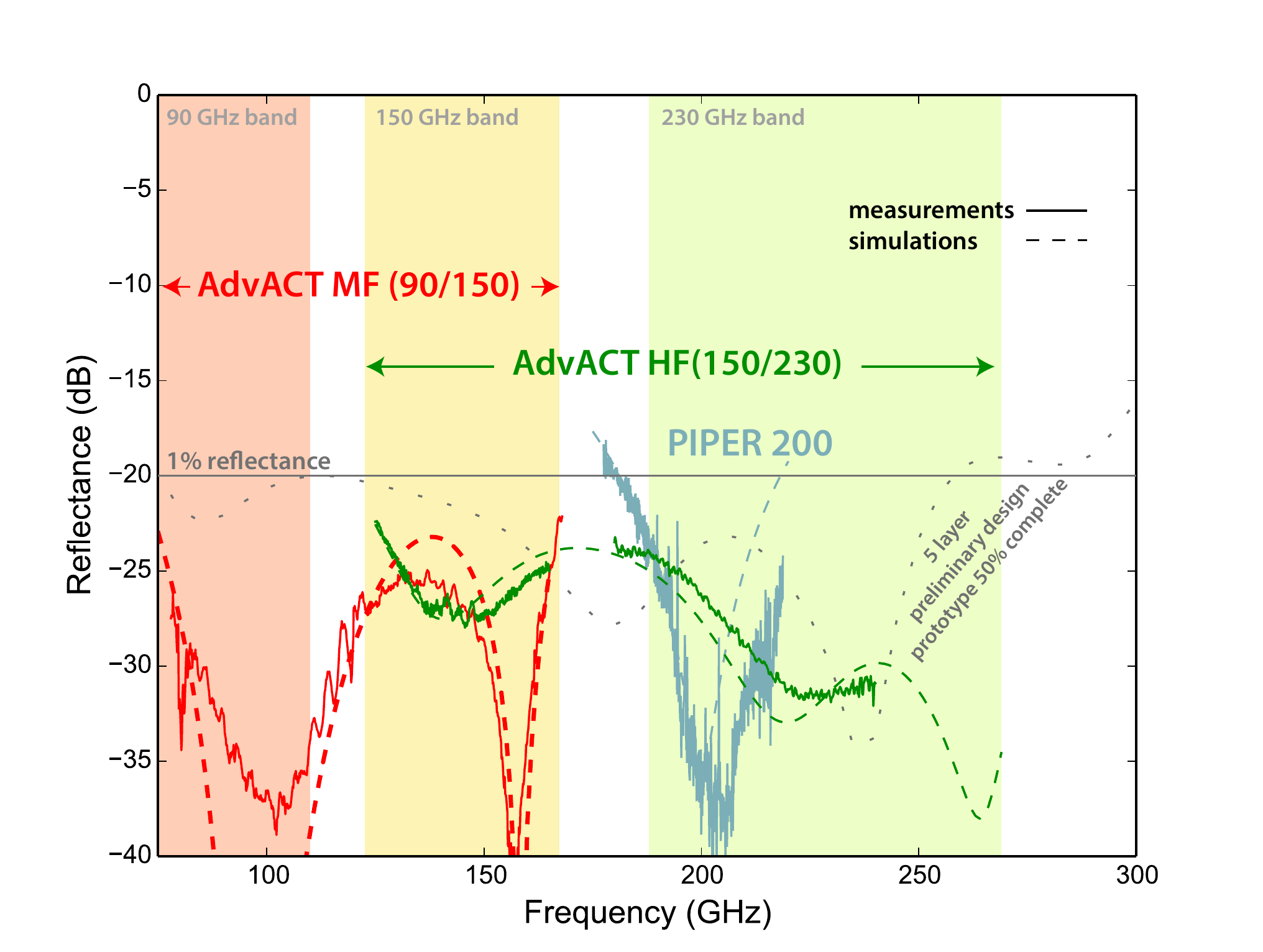}
\caption{Left: Photograph of a mechanical prototype of a three-layer coating for 75 -- 165\,GHz. The top cut is 250\,$\mu\mathrm{m}$ wide, and 500\,$\mu\mathrm{m}$ deep. The middle layer is 110\,$\mu\mathrm{m}$ wide, 310\,$\mu\mathrm{m}$ deep. The last layer is 25\,$\mu\mathrm{m}$ wide, 257\,$\mu\mathrm{m}$ deep. The pitch between the sets of cuts is 450\,$\mu\mathrm{m}$. Right: Performance of metamaterial silicon AR coatings fabricated on lenses that have been or will soon be fielded including: the medium frequency (MF) (90/150 band) lenses for ACTPol, the HF (150/220 band) lenses for AdvACT, and the soon-to-be-deployed PIPER 200\,GHz lenses. Simulations are shown as dashed lines and measurements are shown as solid lines. The MF and HF lenses use three layer AR coatings while the PIPER lenses use a single layer coating. A preliminary design for a five layer coating is shown by the gray dashed line. A prototype of this five layer coating has been fabricated on one side of a 250 mm diameter test wafer. \label{fig:HWP}}
\end{figure}

\begin{figure}
\centering    
\includegraphics[height = 2 in]{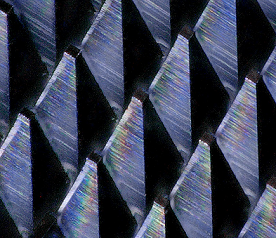}
\caption{Photograph of smooth sided pyramids cut on silicon using a bevelled dicing saw.  The pyramids are 
1000\,$\mu\mathrm{m}$ tall, have flat tops 50\,$\mu\mathrm{m}$ wide, and a pitch of 350\,$\mu\mathrm{m}$.
\label{fig:pyramids}}
\end{figure}

\paragraph{Description of the technology}
Diced silicon metamaterial AR coatings are fabricated by cutting sub-wavelength features into surfaces of refractive optical elements. Since AR coatings are machined directly into the surface of the cryogenic optical element, there are no concerns about thermal contraction mismatch or delamination during cryogenic cycling.
The fill factor tunes the dielectric constant of the machined layer.
Multi-layer coatings can be produced by cutting progressively deeper and thinner square grooves centered on the wider first layer, creating a stepped pyramid as shown in Figure~\ref{fig:HWP}.~\cite{datta/etal:2013} Alternatively, a single pass using a bevelled dicing saw blade produces smooth sided pyramids, Figure~\ref{fig:pyramids}, which provide a continous gradient in index.~\cite{young_2017_arxiv}

%It is possible to fabricate these coatings on silicon optics using a custom three-axis silicon dicing saw. \cite{datta/etal:2013}
%This system allows production of micron-accurate arrays of square-based stepped pyramids, as shown in Figure~\ref{fig:HWP}.  
%The fill factor tunes the dielectric constant of each layer. 
%Multi-layer coatings can be realized by cutting progressively thinner grooves at greater depths centered on the wider first layers. 
%We have created a number of full scale optics with near zero defect rate, and the quality of our process is continually being refined.

\paragraph{Demonstrated performance}
Stepped coatings, Figure~\ref{fig:HWP}, with micron level accuracy have been produced using a three-axis dicing saw.~\cite{datta/etal:2013}
Twelve lenses were deployed on the ACTPol and the AdvACT experiments and lenses have been delivered to the PIPER experiment. 
The largest lens produced is 330 mm in diameter, but lenses up to the maximum diameter available for single crystal silicon, 460 mm, can be fabricated. 
Coatings on both concave and convex surfaces have been demonstrated.
%The measured bandwidth of these coatings is in excess of an octave for three-layer coatings. 
%In addition, we have the ability to achieve wider bandwidth with the proven three layer coatings by trading reflection performance for bandwidth.  (this is true for any coating)
Current stepped coatings achieve $\sim 0.1\%$ average reflections in bands centered on 90 GHz and 150\,GHz.
Coatings using smooth sided pyramids, Figure~\ref{fig:pyramids}, have been demonstrated with 97\% fractional bandwidth.~\cite{young_2017_arxiv}

%This proof of principle shows that 4:1 bandwidth is possible with these coatings. 
%We will complete the second side of this five layer prototype in the coming months to confirm its performance. 

\paragraph{Prospects and R\&D path for CMB-S4} 
The primary challenge for applying this technology to CMB-S4 is reducing the time it takes to AR coat a single lens.  
Currently, fabricating the stepped coatings takes roughly two weeks per lens.   
Automating some of the most time intensive setup tasks, it could be possible to build a machine that reduces that time to 1-2 days.  Alternatively, the number of cuts required to produce a coating could be reduced by using bevelled blades.
%The key features of this new system would be: (i) a rotation stage to
%change the orientation of the the lens, (ii) automated metrology to
%acquire lens positioning after mounting and rotations, and (iii) multiple
%independent spindles set up with different dicing blades to minimize
%time intensive blade changes. This system would make it practical to
%fabricate the large number of AR-coated lenses required for CMB-S4.

It is possible to design silicon pyramids for wider bandwidth.  Using
pyramids, a three-layer coating designed for $1\%$ reflection could
achieve 3:1 ratio bandwidth.  Wider bandwidth (5:1) may also be
achieved by exploring more pyramid profile options.  A prototype
five-layer AR coating covering 75-300\,GHz has been
fabricated, and optical performance measurements will be made soon.

The technology status level of the diced stepped silicon metamaterial AR coatings is 5. 
Diced silicon metamaterial AR coatings on silicon lenses have been deployed on the ACT-pol and Adv-ACT receivers. 
The technology status level of the coatings using smooth sided pyramids is 1. These smooth pyramid coatings cut using 
bevelled saws have been demonstrated on flat samples.

The production status level of the diced stepped silicon metamaterial AR coatings is 3. 
Current dicing method takes non-negligible time to complete a lens, but an upgrade for the dicing setup is planned to speed up production rate.
The production status level of smooth sided pyramid coatings is 1; prototypes have been fabricated.

%[Toki Comment: This is silicon issue that should be covered in silicon section]
%The next potential challenge is in fabricating lenses larger than 46 cm. 
%The issue here has to do with the availability of large silicon rather than limits of the machine.  
%Therefore we discuss this challenge in the note about silicon properties.
%The most important R\&D issue is to build a new machine and demonstrate a speed up of the fabrication of these optics. 
% An additional issue is fabricating larger diameter silicon lenses.   This requires bonding multiple pieces of silicon together as discussed in the single crystal silicon white paper.

%
%
%
%\end{document}  

%% file: broadband_optics/AR_Silicon_DRIE.tex
% Primary contributor M. D. Niemack

\subsection{Deep reactive ion etched silicon}
\label{sec:ARdriedsi}
\paragraph{Description of the technology}
Deep reactive ion etching (DRIE) has recently been developed \cite{wheeler/etal:2014,gallardo/etal:2016} as an alternative to dicing saw machining to produce diced silicon AR coatings as described in previous section \cite{datta/etal:2013}. This approach is significantly less mature than the dicing saw approach, and has only thus far been fully demonstrated on flat 100 mm diameter silicon wafers for a wavelength of 350\,$\mu$m.

DRIE offers the potential for significant advantages for optics as large as 300 mm diameter. Specifically, the DRIE process acts on the entire wafer simultaneously, unlike the dicing saw and laser ablation approaches which cut individual grooves or holes. Simultaneous processing of entire 300 mm wafers should make this approach easier to scale to fabrication of hundreds of lenses.

\paragraph{Demonstrated performance}
The DRIE approach has been demonstrated for single layer AR coatings on flat 100 mm diameter silicon wafers for a wavelength of 350\,$\mu$m. Preliminary results of this work are presented in \cite{wheeler/etal:2014}. Results of high-efficiency double-sided coatings including silicon bonding of two samples and a prototype two-layer coating are presented in \cite{gallardo/etal:2016}.

\paragraph{Prospects and R\&D path for CMB-S4} 
The primary challenges for this technology include: (i) bonding silicon wafers with DRIE coatings onto curved lenses and (ii) demonstrating this approach at the 300 mm scale. A final requirement for this technology is to demonstrate performance at longer wavelengths; however, this is expected to be straightforward for wavelengths up to $\sim$3\,mm. The largest diameter wafers that could be coated using this approach are limited to 300 mm by the size of available DRIE machines. This may be sufficient for optics designs that use modular optics tubes (e.g., \cite{niemack2016}).

The technology status level of the DRIE etched silicon metamaterial AR coatings is 1. 
Laboratory demonstration has been done on flat silicon wafer. 

The production status level of the diced silicon metamaterial AR coatings is 1.
Laboratory demonstration has been done, but deployable model has not been demonstrated.

%% file: broadband_optics/AR_Laser.tex
%\documentclass[12pt]{article}
%\usepackage{graphicx}
%\usepackage{amsmath}
%\usepackage{color}
%
%\begin{document}

\subsection{Laser ablated surface}
\label{sec:ARlaser} 
\paragraph{Description of the technology}
Laser ablation has been used to fabricate metamaterial AR coatings similar to the dicing saw and DRIE techniques described in previous sections.
Since hard materials are difficult to process with mechanical machining, laser ablation is advantageous when working with materials such as alumina or sapphire. 
Producing structures on any material with features smaller than $\sim$25 microns is challenging with mechanical machining, but laser ablation can readily make features down to nearly 1 micron, approximately the scale of the laser spot diameter.

\begin{figure}[htbp]
   \centering
   \includegraphics[width=\linewidth]{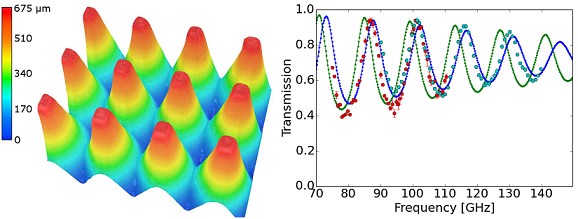}
   \caption{Left: Image of a laser-ablated sub-wavelength AR coating fabricated on sapphire. The patterned 
            pitch is $320\,\mu$m and the height is $700\,\mu$m. Right: Measured transmission for 
            this structure between 75 and 110\,GHz (red) and 90 and 140\,GHz (cyan) matches predictions 
            that are based on the tallest (blue) and shortest (green) structure measured on this sample. 
            The data is normalized to the blue curve at 87\,GHz.
            }
   \label{fig:laser_arc}
\end{figure}

\paragraph{Demonstrated performance}
Matsumura et al. \cite{matsumura_2016} recently demonstrated laser ablated pyramidal structures on sapphire and alumina; 
see Figure~\ref{fig:laser_arc}. The structures were produced by repeated raster scanning of the sample using a 
pico-second laser operating at green wavelengths (515\,nm) with total power of 30\,W. There is reasonably good agreement between 
the designed and as-built structures, and between the measured and predicted transmission; see Figure~\ref{fig:laser_arc}. 
Young et al. \cite{young_2017_arxiv} demonstrated similar structures on silicon using the same laser system at 28\,W.

\paragraph{Prospects and R\&D path for CMB-S4} 
For CMB-S4, several areas can benefit from further development. Plastics can be explored to see if laser ablation can be used 
on those materials. Investigating smaller features for higher frequencies, and higher aspect ratio structures for wider 
bandwidth, are both useful areas for future study.  Finally, it would be advantageous to work to increase the ablation rate of the process 
to decrease the cost and increase the speed of fabrication.

The technology status level of the laser ablation on sapphire, alumina, and silicon is 1. 
Development is still in the early stages. 
Laser ablated AR coatings have been demonstrated on flat samples of all three materials in the laboratory.

The production status level of the laser ablation on sapphire, alumina, and silicon is also 1. 
Although this technology is an attractive way to achieve low loss and wide bandwidth AR coatings, ablating over large surface areas still takes non-negligible time.

%
%
%\bibliographystyle{unsrt}
%\bibliography{arc}
%
%
%\end{document}
%
%
%

%% file: broadband_optics/AR_PlasticHole.tex
%\documentclass{article}
%\usepackage[utf8]{inputenc}
%\usepackage[letterpaper]{geometry}
%\usepackage{graphicx}
%%\usepackage{sidecap}
%
%
%\begin{document}

\subsection{Machined plastic} 
\label{sec:ARablatedplastic}

\begin{figure}
\centering
\includegraphics[width=0.5\textwidth]{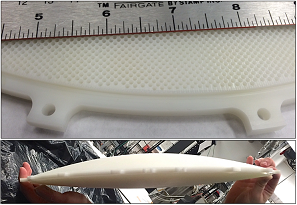}
\caption{ \label{fig:classq_lens} A 400\,mm diameter CLASS 40\,GHz lens that is AR coated using a grid of sub-wavelength holes (top) to create a simulated dielectric layer of lower mean density and tuned index of refraction.}
\end{figure}

\textbf{Description of the technology} 
Metamaterial AR coatings, as shown in Figure \ref{fig:classq_lens}, have been created using a grid of sub-wavelength holes cut into the surface of plastic cryogenic optics. The hole diameter and grid spacing are tuned for the frequency band and the required index of refraction. These AR coatings are machined on a conventional CNC milling machine with minimal alterations and standard tooling.
\\

\noindent \textbf{Demonstrated performance} Simulated dielectrics were used to AR coat sixteen 140\,mm HDPE lenses for CAPMAP \cite{mcmahon2006}. More recently, simulated dielectrics were applied to two 400\,mm diameter HDPE lenses, three flat Teflon filters, and one Nylon filter for the CLASS 40\,GHz telescope \cite{harrington2016}. The lenses and filters for the two CLASS 90\,GHz telescopes are also AR coated with simulated dielectrics and are currently in production. 
AR coating for HDPE window for SPT-3G is CNC machined pyramidal grooves on its surface. Two orthogonal groove patterns were machined on opposite sides of HDPE window. Pyramidal grooves were designed for broadband AR coating application that covers 90 GHz, 150 GHz, and 220 GHz. 
\\

\noindent \textbf{Prospects and R\&D path for CMB-S4} The size of the optics AR coated in this fashion is only limited by the travel of the CNC mill used in fabrication and standard commercial machines have sufficient range for 600\,mm diameter pieces. 
Demonstration of broadband coating over curved surface would make this method more versatile. 

The technology status level of the machined plastic coating is 4. 
Broadband AR coated plastic window has been deployed on SPT-3G. 

The production status level of the machined plastic coating is 5.
Machining of SPT-3G window demonstrated that broadband AR coating on plastic lens can be applied easily with CNC mill.

%\bibliography{sim_dielectric_bib}
%\bibliographystyle{spiebib}   %>>>> makes bibtex use spiebib.bst
%
%\end{document}

%% file: broadband_optics/AR_MetalMesh.tex
\subsection{Metal mesh}
\label{sec:ARmmarc}

\paragraph{Description of the technology}
There are cases where the refractive indices required for AR coatings are hard to obtain.
%This is the case for broadband multilayer coatings or high refractive indices materials to be AR-coated like sapphire, quartz or silicon. 
In these cases quarter-wavelength layers made of artificial dielectrics can be synthesized and used for very broadband applications, more than 100\% in bandwidth. 
Artificial dielectrics have been realized by loading dielectric materials with stacked metal-mesh grids. 
In addition to the requirement of having sub-wavelength structures as in mesh-filter-type applications, the periodic grids need to be stacked within their near field distances. 
The stacked grids will look like a uniform medium to the electromagnetic radiation passing through them. 
The equivalent refractive index will depend on the number of grids and their spacing. The effective index can be tuned by modeling with commercial software such as HFSS, compared with previous experience in metal mesh filter design, and measured in the lab after fabrication.

\begin{figure} [h]
\centering    
\includegraphics[height=1.8in]{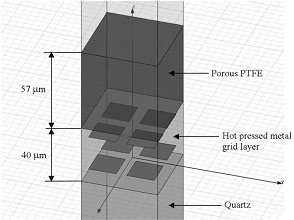}
\includegraphics[height=1.8in]{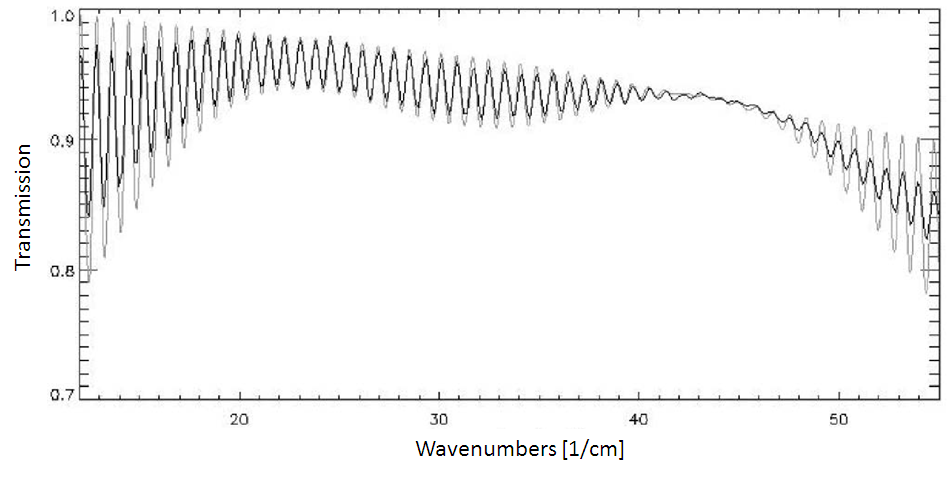}
\caption{Left: Full model of an AR coating on both sides of a quartz substrate. Right: Measured transmission spectrum of the stack (PPTFE-ADMquartz-ADM-PPTFE). The black line is the best fit of a scattering matrix model with varying optical constants to fit the data as a function of frequency \cite{Zhang09}.} %(Zhang et al., 2009)
\label{fig:mmfar}
\end{figure}

\paragraph{Demonstrated performance}
%\comred{Significant research and rewriting needed here... -Sean}
Refractive indices ranging from 1.2 to 4 can be easily achieved with negligible losses \cite{Zhang09}.
%\comred{missing Zhang09 ref}%(Zhang et al., 2009). 
Measurements of a two-layer AR coating with artificial dielectrics and porous PTFE (PPTFE) on both sides of a quartz substrate are shown in Figure\,\ref{fig:mmfar}.

\paragraph{Prospects and R\&D path for CMB-S4}
%Similar to other metal mesh technologies, development for larger size will widen possibility for this technology to be implemented for CMB-S4 receiver.

Broadband metal-mesh AR coatings are in an early stage of
development. While demonstrations exist on small scales, large-area
fabrication requires R\&D similar to that for metal mesh filters
(Section \ref{sec:filtermmf}). CMB-S4 is likely to require receiver
optics and filters of diameter $\sim$500--1000\,mm. To increase the
diameter of metal-mesh AR coatings further, R\&D is necessary and
should include establishment and verification of high-fidelity
photolithography and uniform thermal pressing of multilayer metal-mesh
structures up to 1000\,mm in diameter.

The technology status level of the metal mesh AR coating is 1.
Demonstration of dichroic AR coating was demonstrated, but it has not deployed on CMB experiment yet. 

The production status level of the metal mesh AR coating is 1.
Metal mesh infrared filters have been used by multiple CMB experiments successfully. 
Basic production concept is similar, however, production of metal mesh AR coating has not been demonstrated beyond laboratory level.

%% file: broadband_optics/PolMod_Intro.tex
\label{sec:pol_mod_intro}

Polarization modulation is a generic term for all techniques that are used to change the orientation of instrument's polarization sensitivity, relative to fixed sky coordinates. This is a powerful technique that enables consistency checks to reveal and mitigate several kinds of systematic errors in the instrument. Effects that can be mitigated with a suitably-implemented modulator include beam systematic errors, performance differences between each detector in a polarized pixel, and spurious temperature-to-polarization and polarization-polarization coupling in the instrument. Moving towards the specific implementation, it is useful to consider two different regimes of modulation speed: one that is much slower (i.e. stepped rotation), and the other much faster (i.e. continuous rotation), than the characteristic 1/$f$ knee of the detector+atmospheric noise spectrum. The modulator can be installed either at ambient temperature or inside the cryostat. In both cases, there are several different places in the optical chain that modulators have been used. Conventional motors dissipate a fraction of a watt of heat, meaning that installing the drive mechanism inside the receiver cryostat has required special consideration. Several mechanical design solutions and their associated instrumental challenges are reviewed in in Section~\ref{sec:polmodrotator}. 

At mid-latitude and equatorial observing sites, a significant degree of polarization modulation happens naturally through the daily rotation of Earth. Rotating the entire instrument is also possible, and has recently been used by the \bicepI/Keck Array teams. In this section, we will discuss polarization modulation by motion of an optical 
element in the light path. The rotation of a half-wave plate (HWP), realized either 
through (a) crystalline plate(s) or metamaterial, and the translation 
of a mirror in front of a polarizer (VPM) are among the standard 
techniques to achieve modulation with an optical element in the light path. Polarization modulation using 
these approaches have been implemented by a number of CMB 
experiments~\cite{SpiderFilippini, Hill, klein2011, maxipol, abs, bicepFRM, henderson/etal:2016,nikalekid, pilot1,qubic1}.

Continuous rotation modulates the polarization signal up to higher audio frequencies in the detector, which reduces the impact of atmospheric fluctuation noise that can be polarized by instrumental effects. ABS and other instruments have shown that atmospheric noise can be reduced this way. So far, this has been done with the instrument scanning slowly, such that there is more than one complete modulator rotation in the time it takes to scan a single beamwidth across the sky. This slow scan meant that atmospheric fluctuations added excess noise to the CMB temperature data. Since temperature maps are still important to achieve lensing and other science goals, it has been proposed (and testing is underway with ACT) to scan the instrument much faster than this to enable simultaneous temperature and polarization mapping with a rapid modulator.

Stepped modulation changes the sensitivity angle of the instrument periodically. The instrument then scans rapidly to make a map in this configuration. Data taken at several modulator states are combined in software to yield polarized maps, as well as allowing for assessment and removal of systematic errors. The \spider\ balloon mission used a stepped half wave plate, and the \bicepI/Keck Array instruments used stepped instrument rotation. In this approach, the temperature and polarization data appear at the same audio frequencies. Detector differencing was used in \bicepI/Keck Array to remove atmospheric fluctuation noise, instead of using rapid modulation to remove atmospheric fluctuation noise.

The rotation of the HWP, at a mechanical frequency $f$, completely rotates the polarization vector of each pixel of the observed sky at twice that rate. This means that after passing through a polarized detector, polarized signals from both $Q$ and $U$ appear in the detector timestreams at a frequency of 4$f$. Spurious signals from HWP emission and non-uniformity appear at 2$f$ and other frequencies~\cite{abs, Salatino10}.  Instruments using stepped rotation also take advantage of this separation between polarization signal and systematic errors~\cite{Bryan2010}. The HWP can be located at room temperature at the entrance aperture of the entire telescope~\cite{abs,Hill} or nearby the Lyot stop at cryogenic temperatures~\cite{Reichborn2010}. In the first case, since it is the first optical element seen by the incoming light, it completely separates the instrumental polarization from the sky polarization. In the second case the thermal emission of the HWP is reduced, but it will unfortunately be unable to fully separate instrumental and sky polarization. Several experiments have demonstrated data analysis techniques to separate the two in software.

In the following sections, we review the technical issues and outline the challenges to implementing these polarization modulation techniques for CMB-S4, focusing on HWPs and VPMs. Subsections~\ref{sec:AHWP}--\ref{sec:polmodrotator} describe elements of HWP systems.  Subsection~\ref{sec:AHWP} describes the principles and properties of AHWP, which achieves wide bandwidth required for CMB-S4 and can generically be applied to both of two types of HWP materials discussed later: sapphire (Section~\ref{sec:sapphire}) and metamaterial silicon (Section~\ref{sec:materialSI}). Subsection~\ref{sec:mmpolmod} describes a metal-mesh HWP, which also achieves wide-band polarization modulation. All of these HWPs require rotation mechanisms; subsection~\ref{sec:polmodrotator} discusses them. Subsection~\ref{sec:polmodvpm} describes the VPM, for both working principle and mechanical implementation.

The Section concentrates on technical issues in implementing specific technologies, not on sources of systematic errors. It also ignores technical solutions that require rotation of the entire instrument.

%CMB experiments are susceptible to low-frequency noise induced by unpolarized atmospheric fluctuations leaking into polarization via instrumental effects. A continuous polarization modulator suppresses this noise by rapidly modulating the sky signal and hence allowing the instrument to achieve white-noise sensitivity on long timescales in polarization. 

%Additionally, CMB experiments are susceptible to instrumental effects that cause mismatch between orthogonal detectors. A half-wave plate mitigates these effects by rotating the input polarization signal, allowing a single polarimeter to measure polarized stokes parameters and hence eliminating systematic errors associated with differencing of orthogonal detectors. 

%Polarization modulation has already been demonstrated on several CMB experiments~\cite{maxipolHWP, abs, klein, bicepFRM, spiderHWP, charlieSPIE, actHWP}. In this Section we review the technical issues and outline the challenges to implementing polarization modulation for CMB-S4. 

%% file: broadband_optics/PolMod_AHWP.tex
\subsection{Achromatic half-wave plates} \label{sec:AHWP}

\paragraph{Description of the technology} 
A crystalline, single-plate HWP is a narrow-band, birefringent optical element 
whose thickness is tuned to give a precise half-wave difference in the phase of the electric field traversing the two 
orthogonal optical axes. The half-wave difference only occurs at a single frequency. A Pancharatnam AHWP is a stack of several single plates, each oriented at an angle relative to the next to give a half-wave difference over a broad range of frequencies. This makes AHWPs suitable 
for multi-frequency CMB polarization experiments \cite{ahwp}.

%A half wave plate (HWP) is naturally a narrow-band instrument, having a thickness tuned to preserve linear %polarization at a single frequency. Deviation of the input signal from the HWP design frequency results in %linear-to-circular polarization leakage. However, a Pancharatnam achromatic HWP (AHWP) can preserve %linear polarization over a broad range of frequencies, hence making AHWPs practical polarization %modulators for multi-frequency CMB polarization experiments \cite{ahwp}.

An AHWP consists of an odd number of identical ``single HWPs'' stacked in an optimized orientation \cite{savini}. A useful approximation for considering the performance of an AHWP is to treat it as rotating incoming light with linear polarization fraction $P_{\mathrm{in}}$ by twice the angle $\theta_{\mathrm{in}}$ with respect to the principle axes of the AHWP, plus a frequency-dependent phase $\phi(\nu)$, and with a frequency-dependent modulation efficiency $\epsilon(\nu)$ \cite{tomoFormalism}.

\begin{equation}
	\Delta \theta = 2 \Big[ \theta_{\mathrm{in}} + \phi(\nu) \Big] \; \; ; \; \; \epsilon(\nu) = \frac{\sqrt{Q_{out}^{2} + U_{out}^{2}}}{\sqrt{Q_{in}^{2} + U_{in}^{2}}}
\label{eq:rot}
\end{equation}

\begin{figure}
\centering
\includegraphics[trim={1.0cm, 0.0cm, 1.5cm, 2.5cm}, clip, width=0.4 \linewidth]{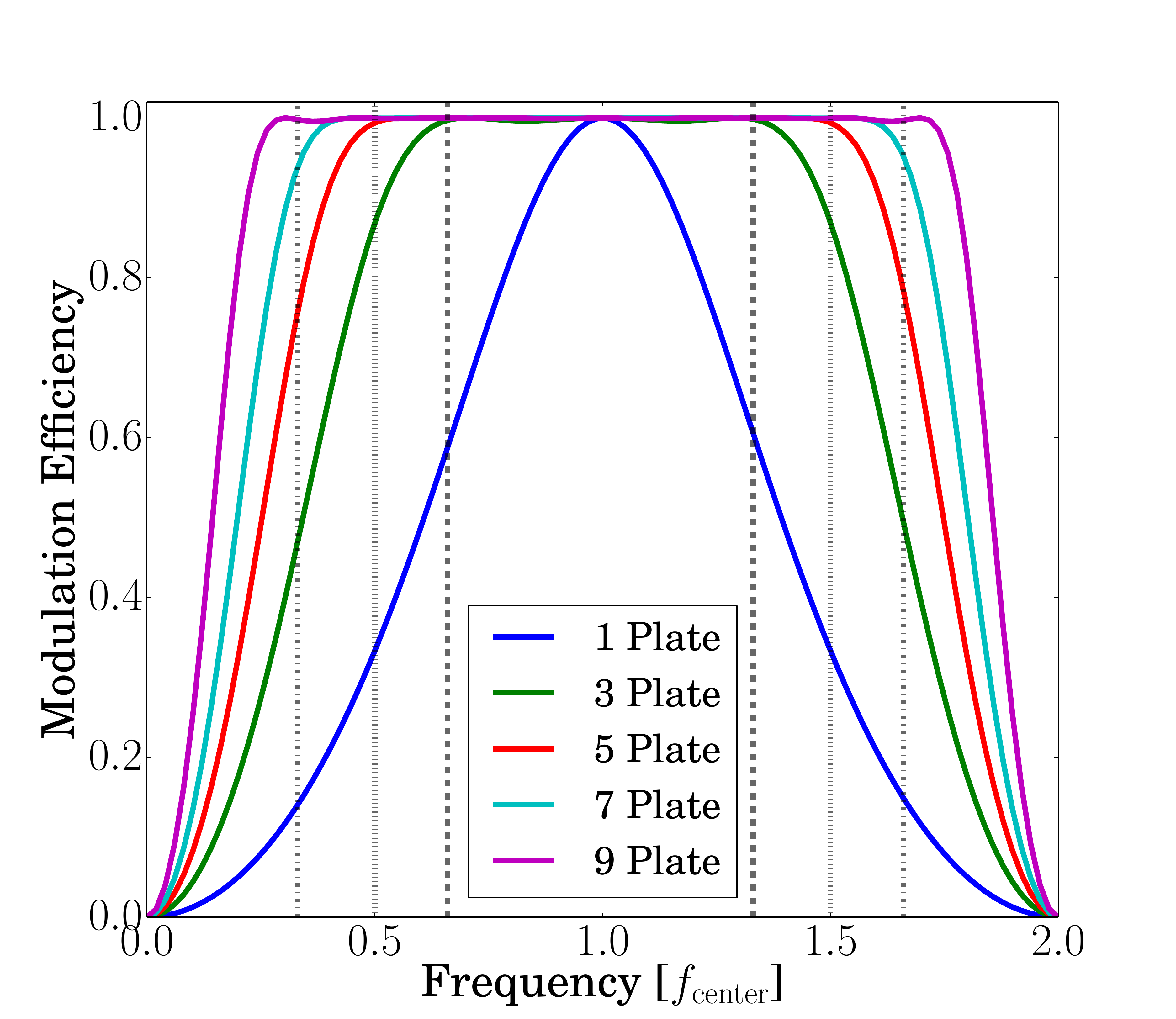}
\includegraphics[trim={1.0cm, 0.0cm, 1.5cm, 2.5cm}, clip, width=0.4 \linewidth]{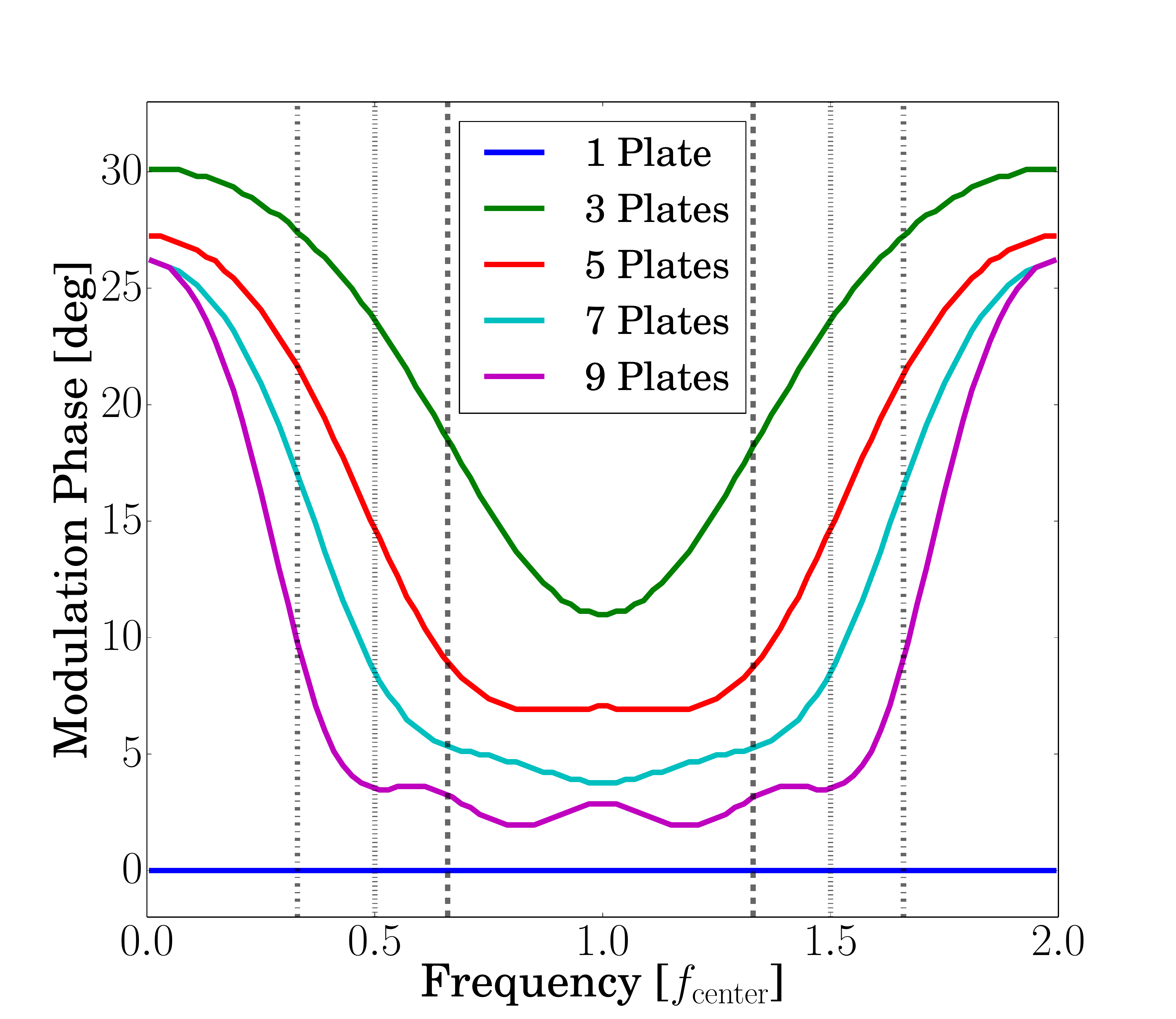}
\caption{The calculated modulation efficiency and phase for various AHWP stacks, referenced to the modulator's central frequency. Increasing the number of plates increases the polarization efficiency and decreases the phase variation across an increasing bandwidth. Various percent bandwidths are shown for reference: 2:1 (dash), 3:1 (dot), and 5:1 (dash-dot). \label{fig:AHWP}} 
\end{figure}

The calculated modulation efficiency and phase for various AHWP stacks is shown in Figure \ref{fig:AHWP}.
%, referenced to the central frequency of the modulator, which is set by the thickness of the identical ``single HWPs.'' (leave this to the caption.)
A greater number of plates gives increased polarization efficiency and decreased phase variation across a larger bandwidth. 
However, the larger the number of plates leads to a thicker device, and therefore increased absorption loss and thermal emission. 
This could be reduced by operating the waveplate at cryogenic temperatures, however this has thermal and mechanical challenges.

\paragraph{Demonstrated performance} 
The implementation of the AHWP technology is relatively new to CMB polarization experiments 
but has shown early promise. During an 11-day balloon observation of 150, 250, and 410\,GHz in 2012/2013, 
EBEX flew a cryogenically cooled five-stack sapphire AHWP that completed a half-million revolutions at 4\,K \cite{Reichborn2010}. 
Advanced ACTPol (AdvACT) has deployed a three-stack, ambient-temperature silicon metamaterial AHWP on their 90/150\,GHz receiver and is planning to utilize this technology on future AdvACT receivers \cite{henderson/etal:2016}. Simons Array will deploy an ambient-temperature three-stack sapphire AHWP on \Pb-2a to observe at 90 and 150\,GHz starting in 2017 \cite{Hill}. 

\paragraph{Prospects and R\&D path for CMB-S4}
The primary technological advances required for implementing AHWPs for CMB-S4 include the availability 
of large-diameter birefringent plates, the development of broadband AR coatings, mitigation of increased 
absorption and thermal emission due to multi-plate stacks, and control of frequency-dependent effects such as 
modulation efficiency and phase \cite{Hill}. Each of these topics is being actively addressed within the CMB community. 

Various birefringent materials --- including sapphire, metamaterial silicon, and metal-mesh substrates --- have been suggested for large-diameter AHWP design \cite{shaul, pisanoAHWP, henderson/etal:2016}. Additionally, various anti-reflection techniques --- such as laser-ablated sub-wavelength structures, thermal-sprayed ceramic, and epoxy --- have been suggested for large-bandwidth AHWP construction \cite{laserAblation, jeong16, Rosen:13}. Cryogenic rotation stages have been developed to facilitate AHWP cooling and suppress thermal emission \cite{klein2011, spiderHWP, lbHWP, Hill}. Hardware and analysis techniques have been proposed to control AHWP frequency-dependent effects \cite{bao, matsumura}. Lessons from EBEX and AdvACT HWP data analysis, from in situ characterization of the \Pb-2a AHWP, and from HWP R\&D associated with other broadband CMB experiments such as LiteBIRD will help define the role and construction of AHWP polarization modulators for CMB-S4 \cite{lb}.

%% file: broadband_optics/PolMod_Sapphire.tex
\subsection{Sapphire}
\label{sec:sapphire}
\begin{figure}
	\centering
	\includegraphics[height=1.7in]{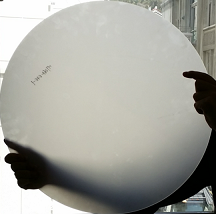}
	\includegraphics[height=1.7in]{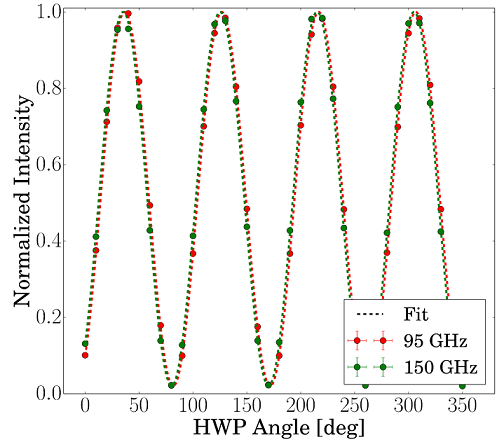}
	\includegraphics[height=1.7in]{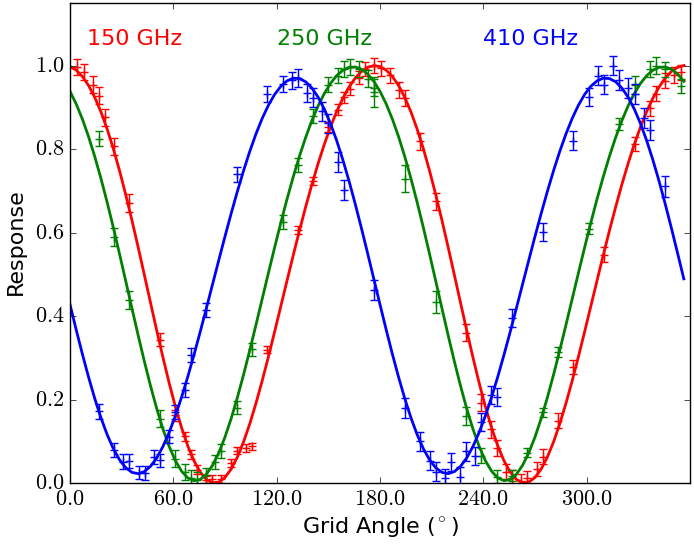}
	\caption{(Left) 512$\,$mm-diameter sapphire plate cut from a 200$\,$kg ingot of HEM sapphire grown at Tuizhou Haotian Optoelectronics Technology in China. (Center) Polarization modulation performance of a three-layer sapphire AHWP for the \Pb-2a receiver. Polarization modulation efficiencies are 99\% and 98\% for 90\,GHz and 150\,GHz band, respectively. (Right) Polarization modulation performance of a five layer sapphire AHWP for the EBEX experiment. Polarization modulation efficiencies are 98\%, 98\%, and 92\% for 150\,GHz, 250\,GHz, and 410\,GHz band respectively. \label{fig:lrgSapp}}
\end{figure}

\paragraph{Description of the technology} 
Sapphire is an appealing HWP candidate due to its low loss tangent (tan$\delta \sim 10^{-4}$ at 300\,K, tan$\delta < 10^{-6}$ at 50 K) and large differential index ($n_{o} \sim 3.1$, $n_{e} \sim 3.4$) at millimeter waves~\cite{parshin,Bryan2010SPIE}. Additionally, sapphire has already been successfully demonstrated on several CMB polarization modulators~\cite{maxipol, pb, abs, Reichborn2010, Fraisse2013}. 
For example, the polarization modulation efficiencies measured for the AHWPs of the \Pb-2a and EBEX experiments are shown 
in Figure~\ref{fig:lrgSapp}~\cite{Hill,Zilic_thesis,EBEXPaper1}. 
%with three sapphire plates covers 90 GHz band and 150 GHz band. The HWP has polarization modulation efficiency of 99\% for 90 GHz band and 98\% for 150 GHz band as shown in Figure~\ref{fig:lrgSapp}~\cite{charlieSPIE}. 
%Achromatic HWPs with multiple sapphire plates have shown expected performance. For example, achromatic HWP fabricated for the \Pb-2a experiment with three sapphire plates covers 90 GHz band and 150 GHz band. The HWP has polarization modulation efficiency of 99\% for 90 GHz band and 98\% for 150 GHz band as shown in Figure~\ref{fig:lrgSapp}~\cite{charlieSPIE}. 

%The challenge with Sapphire is the availability at large diameters and appropriate purities. 
%However, sapphire is difficult to manufacture at large diameters and high purities. Fortunately, industry techniques are evolving such that sapphire HWPs may be practical for S4 experiments.

\paragraph{Demonstrated performance} 
The Heat Exchanger Method (HEM) is the standard growth technique for large sapphire boules~\cite{sapphireGrowth}. As shown in Figure~\ref{fig:lrgSapp}, GHTOT (in China) is now reaching $>$ 500$\,$mm diameters while achieving low levels of impurities and crystal defects via their Advanced HEM method~\cite{ghtot}. Arc-Energy has developed a Controlled Heat Extraction System (CHES) furnace which controls seed orientation during HEM growth to push beyond 500$\,$mm~\cite{arcenergy}. Despite its successes producing HWPs for Stage-III CMB experiments~\cite{Hill}, the HEM process is limited by its need for large chambers that are difficult to clean and inherent thermal gradients that tend to cause crystal defects.

\paragraph{Prospects and R\&D path for CMB-S4} 
In reaction to demand for larger plates, industry is developing alternative sapphire growth techniques. The edge-defined, film-fed growth (EFG) method aims to create plates during growth rather than via post-process machining by drawing the crystal through shaping aids~\cite{sapphireGrowth}. The clear, large aperture sapphire sheets line of EFG products at Saint-Gobain crystals reach 300$\,$mm. Kyocera (in Japan) can go up to 200$\,$mm and is pushing towards larger diameters~\cite{saintgobain, keyocera}. 

In the event that single-crystal growth does not meet its diameter and purity requirements, CMB-S4 can turn to other sapphire solutions, including composite plates. For example, sapphire bonding is a common technique that can be pushed to large diameters for low-stress applications~\cite{bond}. Combining the power of precision dicing and novel bonding techniques may further accommodate large fields of view in CMB-S4 optical systems.

The technology status level of the sapphire plate for polarization modulator is 4. 
Sapphire was used as birefringent material for EBEX, ABS and POLARBEAR-1 polarization modulator.
ABS and POLARBEAR-1 used single plate for single color operation. EBEX used multiple sapphire plates for broadband operation.

The production status level of the sapphire plate for polarization modulator is 5.
Sapphire ingot is commercially available. It can be purchased at desired thickness.

%% file: broadband_optics/PolMod_SiliconHWP.tex
%\documentclass[11pt, oneside]{article}   	% use "amsart" instead of "article" for AMSLaTeX format
%\usepackage{geometry}                		% See geometry.pdf to learn the layout options. There are lots.
%\geometry{letterpaper}                   		% ... or a4paper or a5paper or ... 
%%\geometry{landscape}                		% Activate for rotated page geometry
%%\usepackage[parfill]{parskip}    		% Activate to begin paragraphs with an empty line rather than an indent
%\usepackage{graphicx}				% Use pdf, png, jpg, or epsÂ§ with pdflatex; use eps in DVI mode
%								% TeX will automatically convert eps --> pdf in pdflatex		
%\usepackage{amssymb}
%\usepackage{sidecap}
%%SetFonts
%
%%SetFonts
%
%
%
%\begin{document}
%%\maketitle
%%\section{}
%%\subsection{}
%
%\noindent {\bf \large Metamaterial Silicon Broadband Half-Wave Plates} \\
%{Kevin Coughlin, Charles Munson, Rahul Datta, Jeff M$^c$Mahon}
%%\date{}							% Activate to display a given date or no date

\subsection{Diced silicon broadband half-wave plates}

\paragraph{Description of the technology}
Birefringent metamaterial silicon is fabricated by making asymmetric features in the surface of Silicon plates. There are several advantages of this technology. First, the difference in the index of refraction between the two principal axes can be made large ($\Delta n > 1$). Second, the loss of this material can be extremely low as silicon has a low loss tangent and with the high $\Delta n$ each HWP layer can be made very thin. The warm loss tangent of silicon is typically $10^{-4}$ which drops to $\sim 10^{-5}$ at cryogenic temperatures. Larger achievable $\Delta n$ with diced silicon half-wave plate technique allows thinner substrate, which in turn helps to reduce absorption loss.

\begin{figure}[ht!]
\center
        \includegraphics[width = 0.40\textwidth]{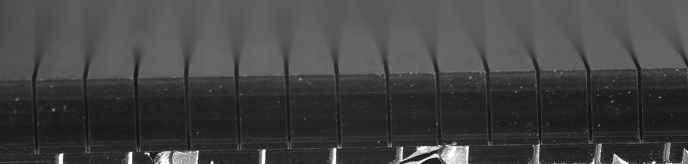}
        \includegraphics[width = 0.40\textwidth]{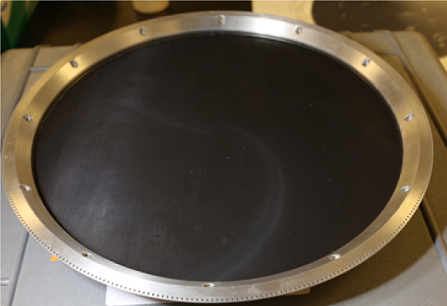}
        \caption{ \footnotesize (Left) Birefringent silicon is cut only in one direction, with evenly spaced cuts. To form the three-layer HWP, one silicon wafer is cut all the way through, leaving only strips which remain in place due to the glue layer between the two wafers. By varying the width and depth of the cuts, the index of refraction can be tuned within a range. (Right) Pictured is the fully fabricated HWP currently deployed on ACT. It is placed in an encoder ring to measure the angle of the HWP as it rotates in front of the telescope.
        \label{fig:BRsi}}
\end{figure}

\paragraph{Demonstrated performance}
A three-layer metamaterial Silicon broadband HWP with a Pancharatnam geometry and three-layer AR coatings on both sides was fabricated as shown in Figure$\,$\ref{fig:BRsi}. 
The HWP layers consist of a set of evenly spaced grooves cut into the silicon.  
The AR layers consist of two orthogonal sets of grooves, leaving rectangular cross-sectioned stepped pyramids. 
The coating is designed to be birefringent, to better match to the birefringent of the HWP itself, which in turn will minimize the polarization dependence of reflections. 
To make a three-layer HWP, the central HWP layer is cut into one side of the thickest Si wafer.  
The patterned side of the thickest silicon and a thinner silicon layer are then glued together, and the outer layers are cut. The HWP layers just touch, leaving no interstitial silicon, so small holes permeate the entire plate. Despite this, the assembly is mechanically robust.

\begin{figure}[ht!]
\center
        \includegraphics[width = 0.90\textwidth]{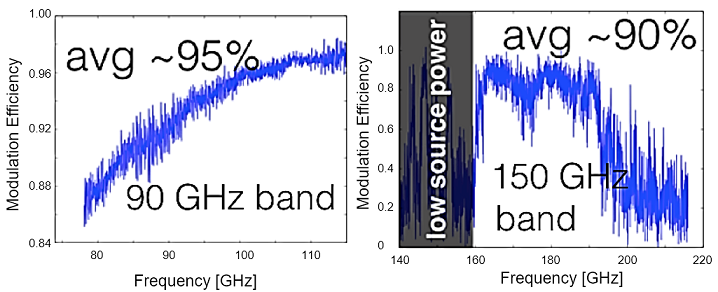}
        \caption{ \footnotesize Measured modulation efficiency of a metamaterial silicon HWP in two bands.  
        %was measured using a vector network analyzer at Goddard Space Flight Center.  This was done in two bands, roughly corresponding to the bands of the telescope.  
        The modulation efficiency was found to be approximately 90\% in the high band (120-185)\,GHz and 95\% in the low band (75-110)\,GHz. These measurements should be interpreted as lower limits due to the presence of a small amount of uncleaned wax in the grooves at the time of the measurement. Since the wax has $n>1$ it reduced the modulation efficiency. The wax was fully cleaned before shipping it to the field. Initial on-sky measurements show promising performance, and measurements and observing is underway. The modulation efficiency of the next AdvACT HWP will be measured in more detail to inform future applications of this technology.
        \label{fig:BR_si}}
\end{figure}

One broadband HWP has been deployed on the Atacama Cosmology Telescope as part of the Advanced ACTPol upgrade. The HWP had a diameter of 340 mm, was optimized for the 75-165\,GHz range, and operates at ambient temperature in front of the cryostat. In lab, it was demonstrated to have a modulation efficiency larger
than 90\% (see Fig.~\ref{fig:BR_si}) and reflections averaging less than 3\%. On the telescope, it was measured to have emission equivalent to approximately 
2 and 4\,K at 90 and 150\,GHz, respectively. Analysis of polarized astrophysical sources confirms that it functions as a polarization modulator.

\paragraph{Prospects and R\&D path for CMB-S4} 
The potential challenges for this approach include: (i) extending beyond 460 mm diameters, (ii) frequency scaling, (iii) bandwidth and (iv) fabrication efficiency.

Diameters beyond 460 mm could be achieved by tiling silicon, but this would need to be developed. Frequency scaling for the CMB-S4 science bands (25-300\,GHz) will soon be demonstrated: the AdvACT HF array covers 120-280\,GHz and its HWP is nearly complete; and the AdvACT low frequency (LF) array which will cover 24-50\,GHz will be fabricated in the coming year. Increasing the bandwidth is possible by adding additional layers to the broad-band stack; however our design process, which relied on full wave simulations with a carefully chosen square superlattice, would need to be expanded to handle these new layers. Given that a five-layer AR coating has already been successfully demonstrated, the broadband coatings needed for a waveplate with more layers should be tractable.

The major challenge to overcome scaling this technology to CMB-S4 level production is the fabrication time and yield. The current system can fabricate one HWP in approximately three weeks. A fully automated system (as described in Section~\ref{sec:ARdicedsi}) could get the fabrication time down to three days. More development on the gluing procedure and associate cleaning procedure would mitigate the risk of delamination of the two silicon plates during the fabrication process which so far has reduced the yield to 50\% for the first two HWPs produced.

The technology status level of the diced silicon for polarization modulator is 4. 
Silicon half-wave plates were deployed for ACTpol and Adv-ACT. Data is currently being analyzed. 

The production status level of the diced silicon for polarization modulator is 3. 
Silicon ingot can be purchased, and it comes sliced at desirable thickness.
Dicing setup is being upgraded to dice larger silicon at faster pace. 

%Finally, it would be advantageous to have a cold HWP rotator.  Work should be invested in developing such a system as it would reduce the requirements on the HWP simplifying design and fabricaiton.
%% (Toki) this will be discussed in mechanical rotation section

%%\paragraph{R\&D path forward}
%The next steps of R\&D is to continue to refine our fabrication technique to increase our yeild.  More development of our gluing procedure can mitigate the risk of delamination of the two silicon plates durring the fabrication process.  
%After working on the gluing process, the next step would be to begin working on further automation of our system by constructing a new saw, including the upgrades listed above.  We should also invest in developing broader band HWPs and a cold rotator.

%\begin{figure}
%        \includegraphics[width = 0.40\textwidth]{figure/HWP_full.png}
%        \caption{ \footnotesize Picured is the fully fabricated HWP currently deployed on ACT.  It is placed in the encoder ring to measure the angle of the HWP as it rotates in front of the telescope.
%        \label{fig:BR_si}}
%\end{figure} 

%
%\end{document}  

%% file: broadband_optics/PolMod_MetalMeshHWP.tex
\subsection{Metal mesh polarization modulators}
\label{sec:mmpolmod}
%
%\paragraph{R\&D path forward}
%[Toki Comment] Size ?

\begin{figure}[h!]
\centering
\includegraphics[height = 3in]{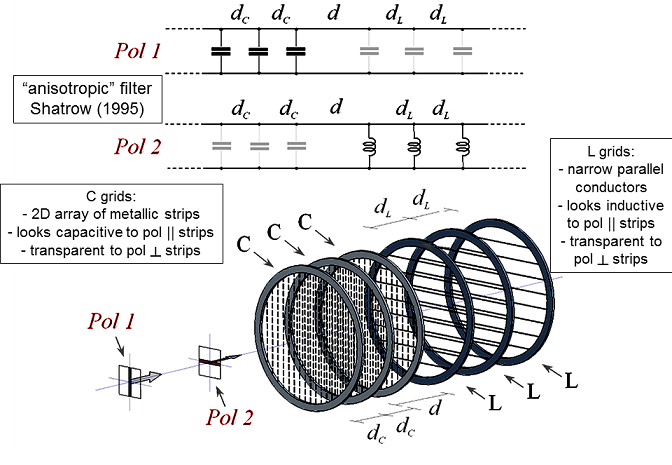}
\caption{Transmission-line model and grid configuration for metal-mesh HWPs \cite{Pisano2008}.}
\label{fig:hwp_fig1}
\end{figure}

\paragraph{Description of the technology}
%The symmetry in the geometrical patterns of normal mesh filters guarantees their polarization independence when used on-axis. 
%However, when the symmetry is broken the grids show anisotropic behaviour. 
Anisotropic patterns of conducting material, similar to the isotropic structures used in metal mesh filters (discussed in Section~\ref{sec:filtermmf}), have been used to create birefringent metamaterials. Parallel continuous lines and parallel dashed-lines are examples of structures with strong 
inductive and capacitive reactance to incident mm-waves of one polarization, yet they are almost transparent to the orthogonal polarization. 
By appropriately stacking capacitive and inductive grids in orthogonal directions, it is possible to create 
an arbitrary relative phase-shift between the two polarizations as shown in Figure~\ref{fig:hwp_fig1}.  
The overall effect is similar to that introduced by the ordinary and the extra-ordinary axes in birefringent 
crystals and so, by using the appropriate number of grids and geometries, it is possible to realize phase retarders. 
These, in turn, can be used to manipulate the polarization state of the light.

\paragraph{Demonstrated performance}
{\textbf Quarter-Wave Plates:} A stack of three capacitive and three inductive grids is enough 
to achieve 90$^{\circ}$ differential phase-shift between two orthogonal axes, which was used to make a metal-mesh quarter wave-plate (QWP) used to convert linear polarization into circular and vice-versa. Mesh QWPs used in combination with polarizers 
have been used to rotate the polarization angle. Bandwidths ranging from 30\% to 90\% can be achieved \cite{Pisano2012a}. 

{\textbf Half-Wave Plates:} Differential phase-shifts of a half-wave can be achieved using capacitive and 
inductive stacks made of four to six grids, depending on the bandwidth required. The challenge of the large bandwidths potentially required by CMB-S4 is to maintain high in-band transmission 
while keeping the differential phase-shift close to 180$^{\circ}$. The first mesh-HWPs had bandwidths of 
the order of $\sim$30\% \cite{Pisano2008,Zhang2011,Pisano2012b}. More 
recent broadband realizations have exceeded 90\% bandwidths.

{\textbf Reflective Half-Wave Plates:} Simple reflective HWPs can be built by locating a polarizer at a 
quarter-wavelength distance from a plane mirror. These are also called  VPMs, and 
a specific application is also discussed in Section~\ref{sec:polmodvpm}. These devices work 
only within periodic narrow bands. However, it is possible to realize dielectrically-embedded reflective HWPs 
with bandwidths larger than 150\% by using polarizers and artificial dielectrics \cite{Pisano2014}.

%\comred{What is the state of the art and what are the challenges in making LARGE aperture modulators? What is the state of the art in terms of central frequencyand what are the challenges in extending to higher frequencies? } 

\paragraph{Prospects and R\&D path for CMB-S4}
%Similar to other metal mesh technologies, development for larger size will widen possibility for this technology to be implemented for CMB-S4 receiver.
Broadband metal-mesh HWPs are under active development. R\&D similar to that of metal mesh filters (Section \ref{sec:filtermmf}) is necessary to increase size of structures as CMB-S4 is likely to require receiver optics and filters of diameter $\sim$500--1000\,mm. To increase metal-mesh HWP diameter further, R\&D is necessary and should include establishment and verification of high-fidelity photolithography and uniform thermal pressing of multilayer metal-mesh structures up to 1000\,mm diameter.

The technology status level of the metal mesh polarization modulator is 3. 
The mtal mesh polarization modular is designed and fabricated for QUBIC experiment.

The production status level of the metal mesh polarization modulator is 2. 
Metal mesh technology was used in almost every CMB experiment, however demonstration of large thorughput production for metal mesh polarization modulator has not  modulation done.

%% file: broadband_optics/PolMod_Rotator.tex
% Maria Salatino, salatino@princeton.edu

%\documentclass[a4paper]{article}
%\usepackage{lscape}
%\usepackage{graphicx}
%\begin{document}

\subsection{Rotation mechanisms}
\label{sec:polmodrotator}

\paragraph{Description of the technology}

Stepped and continuous rotation mechanisms have been deployed for CMB experiments. Mechanical design and engineering challenges are very different between room temperature and cryogenic HWP. As reviewed in the introduction to Section~\ref{sec:pol_mod_intro}, continuous rotation could potentially offer additional benefits than stepped rotation, and a HWP mounted at cryogenic temperatures will have less thermal emission (and therefore contribute less to detector noise and/or spurious signals) than a room temperature one. However from the mechanical engineering and system integration perspective, continuous and/or cryogenic systems are more challenging than stepped and/or room temperature systems. The optimal solution will take both the benefits and challenges into consideration. In this subsection we discuss three different rotation mechanisms: cryogenic step rotation, cryogenic continuous rotation, and room-temperature continuous rotation.

In addition to rotating the modulator, the rotation mechanism must also measure 
%(and therefore control) 
the rotation angle. The absolute accuracy and repeatability requirement for both stepped and continuous rotation mechanisms is stringent, set by the sensitivity of the experiment, and would likely be of order 0.1$^{\circ}$ for CMB-S4. 
%For a continuously rotating HWP, the absolute accuracy requirement is the same, but it may be acceptable to have a higher, possibly $\sim1^{\circ}$, instantaneous noise level in the angle measurement since this noise will average down with the large number of continuous rotations. 
Even though such stringent control was not needed by the current generation of CMB experiments, rotation mechanisms have been successfully deployed that meet all of these requirements, and will be reviewed in this section.

\paragraph{Demonstrated performance}
\begin{figure} [h]
\centering
\includegraphics[height=3in]{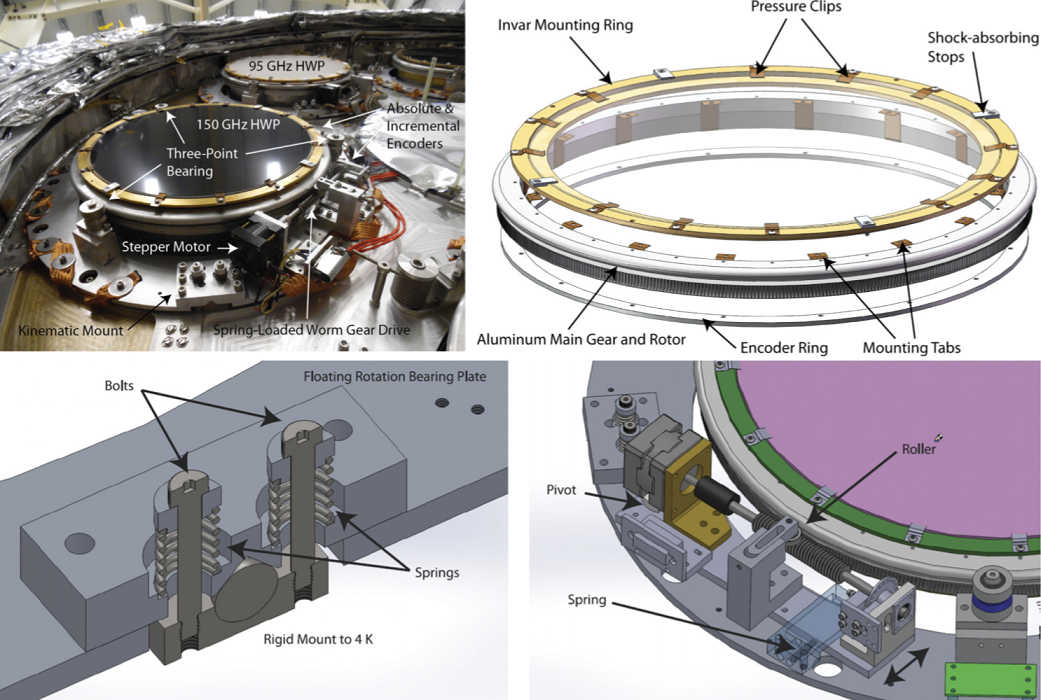}%[height=1.8in]
\caption{The \spider\ HWP rotation mechanism: the rotator in the \spider\ flight cryostat (top left), detailed view of the HWP mount (top right), close view of the mechanical system (bottom)~\cite{spiderHWP}.}
\label{fig:rot_spider}
\end{figure}

\textit{Ambient-temperature continuous rotation} In ABS~\cite{abs}, a 330$\,$mm diameter HWP, at the entrance aperture 
of the telescope, rotated in front of the cryostat window by means of an air bearing system. Compressed air, forced through 
three porous graphite pads around an aluminum rotor, suspended the HWP so it could be rotated at a frequency of 2.5$\,$Hz. 
The angle was monitored by an incremental encoder disk with 2.4' resolution. Based on this success, ACTPol has implemented this strategy for rotating the HWP and Advanced ACTPol is planning to do the same.

%Based on the observational success of the ground experiment 
\Pb-1 observed with a 300$\,$K continuously rotating HWP and \Pb-2a is planning to use the same rotator strategy, a 500$\,$mm diameter HWP rotating at 2$\,$Hz. A mechanical system based on rails, rotational stages, thin-section ball bearings, and an AC servo motor rotates the HWP. The AC servo avoids electrical switching noise present in typical stepper motors. Independent rubber sandwich mounts tangentially and axially oriented to the HWP rotation axis isolate the HWP vibrations from the telescope, while a thin rubber gasket isolates the sapphire from vibrations in the bearing.

\textit{Cryogenic stepped rotation} In January 2015 the \spider\ balloon experiment successfully deployed six cryogenically stepped HWPs rotating at 4$\,$K.~\cite{spiderHWP}  A worm gear driven by a commercially available modified cryogenic stepper motor rotated the HWP, and the rotation angle was monitored with a custom-built optical encoder with an absolute accuracy of 0.1$^\circ$. 
Each HWP was supported by three bearings positioned equidistant around its circumference.
%A three-point bearing was used, meaning that this technology can easily be scaled to larger diameters without needing to modify any of the moving parts. 
%By rotating in steps and completely turning off the motors between the rotations, the total power dissipated onto the cryogenic system was kept very low. 
For each HWP, the rotation mechanism was estimated to boil off 4$\,$ml of helium per 22.5$^\circ$ of motion.~\cite{spiderHWP} 
%The thermal emission of the HWP was significantly reduced by cooling it to cryogenic temperatures. Installing the HWP inside the cryostat also ensured that any reflections from the HWP would terminate on cold surfaces.
%In its first flight in January 2015, \spider\ used six HWPs 330\,mm in diameter, rotated to a different angle twice each day. After a step, the \spider\ field was scanned, then the HWPs were stepped to another angle, then the field scanned again, continuing in this manner for the duration of the flight.The rapid scan speed of the gondola modulated the sky signal well above the detector $1/f$ knee. The HWPs were mounted skyward of each telescope. This means that beam and other systematics were not modulated by the HWP, enabling those systematics to be separated in data analysis from the true sky signal.
\Pb{} observed the sky with a stepped HWP cryogenically cooled down to 80$\,$K~\cite{pb} for first season observation.

\textit{Cryogenic continuous rotation} The balloon-borne EBEX experiment~\cite{EBEXPaper1,Reichborn2010} demonstrated, with a flight in 2013, continuous rotation of the HWP at 4$\,$K using a superconducting magnetic bearing (SMB). A ring-shaped permanent magnet and the HWP constituted the rotor which was levitated 3.2$\,$mm above the stator (a high temperature, 80$\,$K, superconductor YBCO ring)~\cite{EBEXPaper1, klein2011}. The HWP, mounted inside the magnetic ring, rotated continuously at a frequency of 1.235\,Hz. A motor mounted outside the cryostat was connected to a shaft that went through the cryostat wall and turned the rotor with a tensioned kevlar belt. The HWP angle was monitored with an encoder system to better than 0.02$^{\circ}$. The absence of stick-slip friction did not produce vibrations. 
%Given the low coefficient of friction the bearing can rotate for many hours (up to 33) before a reset in their motion become necessary~\cite{Hanany03}.
\Pb-2b and \Pb-2c are planning to use the same cryogenic bearing strategy for a 500$\,$mm diameter HWP rotating at 2$\,$Hz.
Drive mechanism for the rotators are magnetic driven system.

\begin{figure} [h]
\centering
\includegraphics[width=0.8\textwidth]{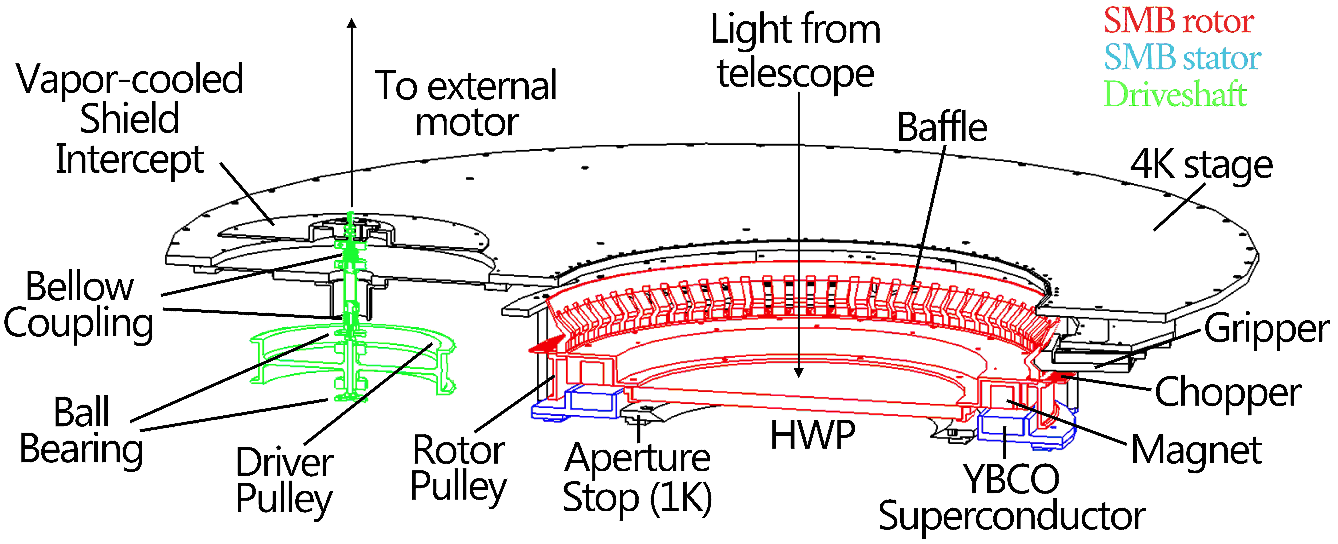}
\caption{Cross section view of the EBEX HWP rotation mechanism which 
        exploits magnetic levitation~\cite{EBEXPaper1}.}
\label{fig:rot_ebex1}
\end{figure}

\begin{figure} [h]
\centering
\includegraphics[height=3in]{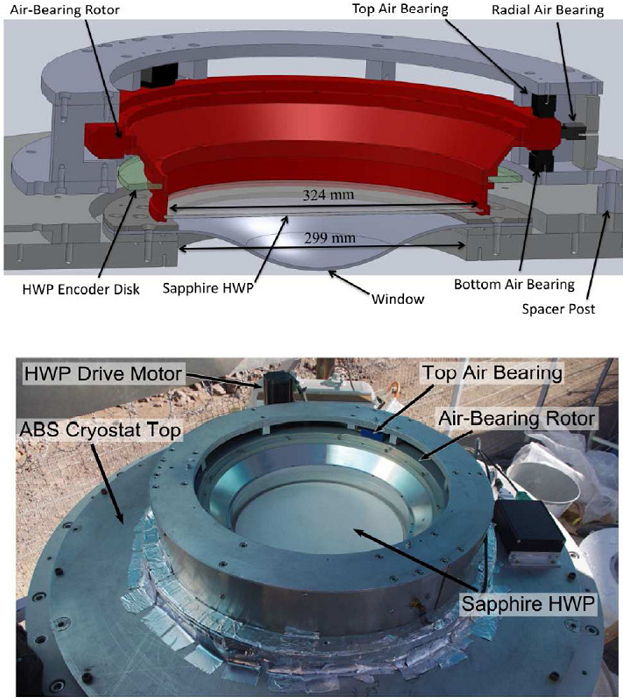}

\caption{The ABS rotation mechanism. Cross-section drawing of the air-bearing system (Top), the rotation mechanism
on the ABS cryostat at the Chilean site HWP
and air-bearing system showing the 3.2$\,$mm thick UHMWPE vacuum window,
sapphire HWP mounted in its rotor, air bearings, encoder
disc, and the overall HWP support. (Bottom) Photograph of
the HWP installed on the ABS cryostat at the Chilean site~\cite{abs}.}
\label{fig:rot_ebex2}
\end{figure}

\Pb{} observed the sky with a stepped HWP cryogenically cooled down to 80$\,$K~\cite{pb}. 
%Based on the observational success of the ground experiment 
\Pb-1 observed with a 300$\,$K continuously rotating HWP and \Pb-2 is planning to use the same rotator strategy, a 500$\,$mm diameter HWP rotating at 2$\,$Hz. A mechanical system based on rails, rotational stages, thin-section ball bearings, and an AC servo motor rotates the HWP. The AC servo avoids electrical switching noise present in typical stepper motors. Independent rubber sandwich mounts tangentially and axially oriented to the HWP rotation axis isolate the HWP vibrations from the telescope, while a thin rubber gasket isolates the sapphire from vibrations in the bearing.

%In all the room temperature working strategies, the HWP emission is reduced choosing low loss materials and suitably characterizing them in laboratory tests.

%The HWP rotation mechanism involves the study of the following topics which will be addressed in a separate document:
%- requirements in the accuracy of the HWP position for each configuration;
%- map making;
%- impact on the detector noise;
%- coupling of each configuration with the scan strategy of the entire experiment;
%- level of the HWP thermal emission;
%- instrumental systematics;
%- impact on beam asymmetry;
%- rotation frequency of the HWP.

\paragraph{Prospects and R\&D path for CMB-S4}
CMB-S4 may require a HWP with a diameter larger than current experiments, and scaling up a rotation mechanism creates challenges. 
Dissipated power from the HWP rotation typically scales linearly with diameter.  
Large cyrogenic bearings may be difficult to buy or produce; the \spider\ mechanism avoids this scaling issue since it has a three point bearing, no moving parts span the entire circumference.
A larger CMB-S4 HWP rotator needs more powerful motor(s) with larger inertia and will be heavier.  
Increased vibrations in a large mechanism could damage the HWP or introduce systematic errors. 
A large HWP will likely have a longer thermalization time as well as increased thermal gradients and thermal fluctuations, controlling these will impact the choice of rotaion mechanism.

The technology status level of the ambient roation mechanism is 4. 
ACTpol, ABS and POLARBEAR-1 have taken data with rotating half-wave plate. Analysis on large angular scale data is on going.

The production status level of the ambient mechanism is 3.
Parts are available commercially. Also once mechanical design is finalized, machining can be done at industrial scale machine shop.
Assembly of the POLARBEAR-1 half-wave plate was done on a short time scale.

The technology status level of the continuous cryogenic rotation mechanism is 4. 
A continuously rotating cryogenic half-wave plate was used successfully on the EBEX balloon flight, and one is currently being assembled for the second receiver of POLARBEAR-2. . %Analysis of the data is actively being pushed.

The production status level of the continuous cryogenic rotation mechanism is 3.
%There is only one company that can make bearing that is large enough. Machining of neodymium, crucial part of cryogenic bearing, is challenging. 
A continuously rotating cryogenic HWP was implemented on EBEX. Superconducting magnetic bearings are commercially available. Lead time could be many month, but it could be planned such that bearing won't be a bottle neck in the process.

%% file: broadband_optics/PolMod_VPM.tex
%\documentclass[10 pt, oneside]{article}   	% use "amsart" instead of "article" for AMSLaTeX format
%\usepackage{geometry}                		% See geometry.pdf to learn the layout options. There are lots.
%\geometry{letterpaper}                   		% ... or a4paper or a5paper or ... 
%%\geometry{landscape}                		% Activate for rotated page geometry
%%\usepackage[parfill]{parskip}    		% Activate to begin paragraphs with an empty line rather than an indent
%\usepackage{graphicx}				% Use pdf, png, jpg, or eps§ with pdflatex; use eps in DVI mode
%								% TeX will automatically convert eps --> pdf in pdflatex		
%\usepackage{amssymb}
%
%%SetFonts
%
%%SetFonts
%
%
%\title{Variable-delay Polarization Modulators}
%\author{D.T. Chuss}
%%\date{}							% Activate to display a given date or no date
%
%\begin{document}
%\maketitle

\subsection{Variable-delay polarization modulators}
\label{sec:polmodvpm}
\begin{figure}[htbp]
   \centering
   \includegraphics[width=4in]{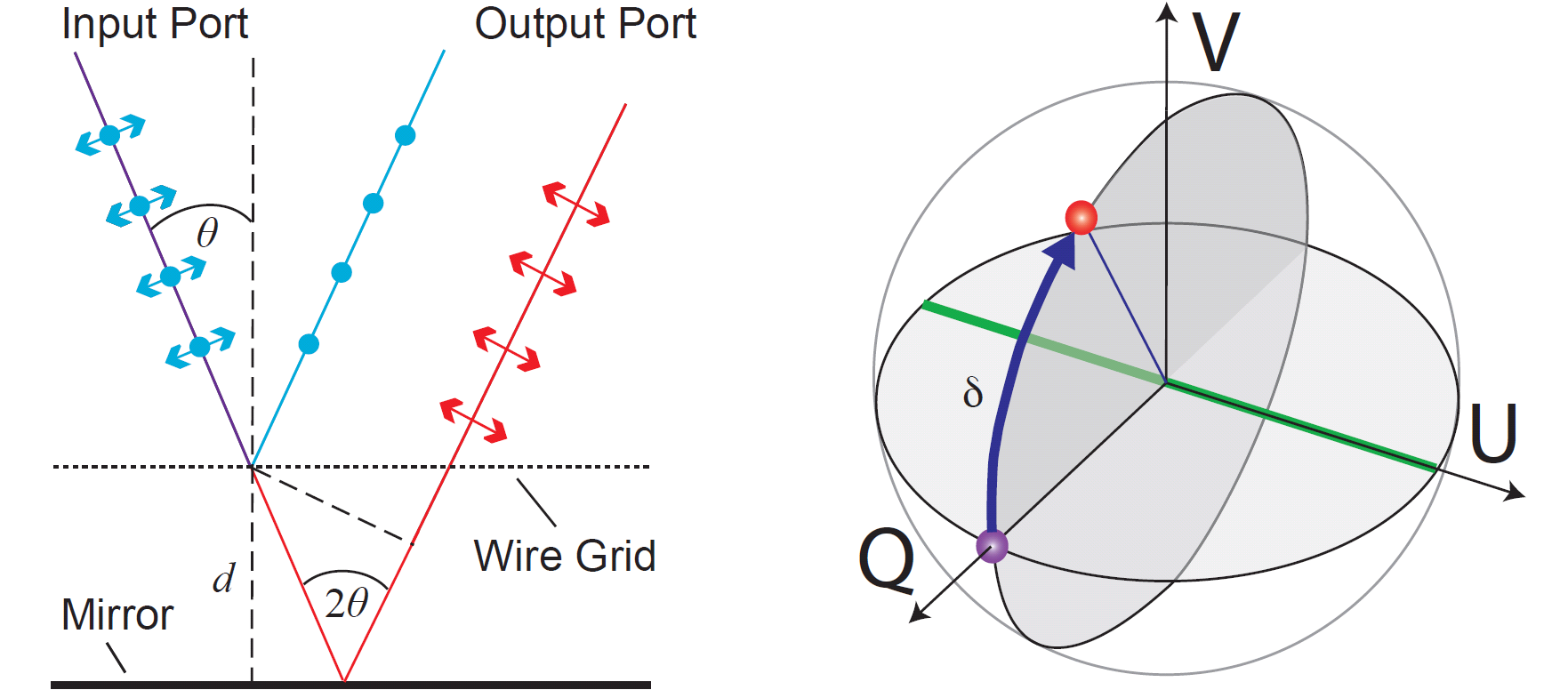} % requires the graphicx package
   \caption{(Left) The VPM introduces a variable phase delay between orthogonal linear polarizations as the distance, $d$, is varied \cite{Chuss12b}. (Right) As the phase delay, $\delta$, changes, the polarization state transitions from $Q\rightarrow V\rightarrow -Q$ with no mixing between $Q$ and $U$.}
   \label{fig:example}
\end{figure}

\paragraph{Description of the technology}
VPMs modulate polarization by introducing a controlled, variable phase delay between linear orthogonal polarizations \cite{houde01,Chuss06}. VPMs have been implemented with a wire grid polarizer and a mirror that is positioned behind and parallel to the polarizer. In this configuration, the polarization component of the incoming light that has its electric field parallel to the grid wires is reflected by the wires without delay; the perpendicular component passes through the wires and is reflected by the mirror, with that extra distance inducing a phase delay. The output polarization state is determined by the incoming state and the delay introduced by the path difference between the grid and the mirror (see Fig.~\ref{fig:example}). This electrical delay can be modulated by varying the separation between the grid and the mirror. Alternately, the delay can be fixed, and the entire device can be rotated.

In the variable-distance mode, as the grid-mirror separation is changed, the VPM will modulate between the linear polarization oriented at an angle of 45$^\circ$ with respect to the grid wires (taken to be defined as Stokes $Q$) and circular polarization (Stokes $V$). In this way, VPMs can be used to switch an instrument's sensitivity between $Q$ and $-Q$. There is no conversion between Stokes $U$ and $Q$ during the modulation cycle, so residuals in the phase delay couple to the $V$ mode, which is expected to be negligible for the CMB. This has the consequence of avoiding $U\rightarrow Q$ leakage due to non-zero cross-polarization across the telescope beam \cite{Chuss10}. This is important because $U\rightarrow Q$ leakage leads to systematic $E\rightarrow B$ leakage. While the motion profile can be selected to reduce the impact of this effect, since the CMB is not expected to be circularly polarized, for CMB polarimetry the inherent $V$ sensitivity of the VPM is a sensitivity disadvantage. Also, the Q to U modulation of other modulators enables simultaneous detection of both linear Stokes parameters, VPM-based instruments require a combination of instrument and sky rotation to fully sample the linear polarization space. 
%Also, the $Q$ to $U$ modulation of other modulators allows full polarimetry with a single detector or pair, whereas a VPM would require data from two detector pairs to be combined (or sky rotation or other modulation) for full reconstruction of both $Q$ and $U$. 
Performance modeling and design of the VPM requires physical optics modeling, similar to dielectric and metal-mesh HWPs \cite{Chuss06}.

An advantage of VPM-based systems includes the capability of potentially building the modulator sufficiently large to be positioned at the primary aperture of a $\sim$meter-scale CMB experiment. As apertures and modulators get larger, it may be easier to implement the small linear motions associated with a VPM than to implement rotational motion required for a wave plate. Scaling the freestanding wire grid to large diameters would then become the limiting engineering challenge. A VPM built with a freestanding wire grid does not require AR coating. Since thermal emission only arises due to the finite conductivity of the wires and metal mirror, low thermal emission is achieved even at room temperature. 
%(If the grid were supported with a substrate to enable larger diameters, naturally that substrate would need AR coating, and both materials would have associated thermal emission.)
In addition, for space applications, VPMs can be implemented without the use of high quality dielectrics that are vulnerable to damage from electrons. 

The modulation scheme of VPM-based systems can be tuned to trade sensitivity to Stokes $V$ for increased sensitivity to linear polarization ($Q$). The limit of this is a square wave motion of the mirror for which polarization sensitivity of the instrument is rapidly switched between the $Q$ and $-Q$ state with little time being spent in the $V$ state. For sinusoidal mirror strokes, a polarization modulation efficiency of $\sim$85\% has been realized for a $\sim$26\% bandwidth \cite{Essinger-Hileman14}, with a decrease in efficiency similar to that for a single-layer HWP for broader bands when used in this mode. Birefringent and metal-mesh HWPs use multiple layers to broaden the modulation bandwidth. Since something similar can not naturally be done with a VPM system, strategies for using VPMs for broader bandwidths and for multichroic focal planes are under development. One strategy includes the optimization of bands to operate at the harmonics of a common VPM modulation function. VPM-based systems could also in principle be used as polarization spectrometers \cite{Chuss06} as their polarization transformation is similar to a Martin-Puplett interferometer. 

\paragraph{Demonstrated performance}
VPMs were prototyped in the submillimeter using the Hertz polarimeter \cite{Krejny08}. These devices utilized kinematic double-bladed flexures \cite{Voellmer06} to maintain parallelism between the mirror and grid. Piezoelectric drives were used to actuate the mirror, and capacitive sensors were implemented to measure the distance and provide feedback to the control system. The construction of large ($>$0.5 m) polarizing grids has been developed \cite{Voellmer08} for the implementation of VPMs as the first optical element of CMB polarimeters. CLASS \cite{Essinger-Hileman14} and PIPER \cite{Gandilo16} are utilizing VPMs in this capacity. PIPER employs 39\,cm diameter VPMs on each of its two telescopes, enabling it to modulate and measure Stokes $Q$ and $U$ simultaneously. The VPMs have been constructed to be cryogenically compatible and will operate at 1.5\,K \cite{Chuss14}. The grid-mirror separation is actuated via a linear voice coil. The parallelism is maintained using a double-blade flexure similar to that used for Hertz, but with a larger operating throw to accommodate the longer wavelengths. 

 The CLASS VPMs are 600 mm in diameter and are operated at ambient temperature. Because of the longer wavelengths (CLASS operates down to 38\,GHz), a four-bar-linkage flexure was used in place of the single-material flexures. A voice coil is used for actuation and an optical encoder is used to measure the distance and close the feedback loop. To fully cover the $Q-U$ space, CLASS employs instrument rotation around the boresight. 
 
The characterization of the Hertz prototype VPM has led to an improved understanding of the transfer function of VPMs \cite{
2012ApOpt..51..197C}. The resulting model enables the characterization of non-ideal properties of the VPM, including its emission properties. These effects have informed simulations of ground-based, VPM-modulated CMB surveys \cite{Miller16}. These forecasts have provided guidance for survey implementation. %in tel bib

\paragraph{Prospects and R\&D path for CMB-S4}
CLASS is currently observing in the Atacama desert, and the first flight of PIPER will be soon. These experiments will inform and refine the data analysis pipeline and systematic error mitigation for VPM-based systems. Beyond CLASS and PIPER, for potential inclusion in CMB-S4 and in a space mission, one of the key aspects of technology maturation would be to scale the VPMs up to larger sizes to accommodate larger focal planes and higher angular resolution. VPMs can likely be developed up to $\sim$1 meter diameters using current grid manufacturing techniques and flexure technologies (perhaps larger with some development). Strategies for operating VPM-based systems over broader bands would need to be explored and developed. 

The technology status level of the VPM is 4. 
VPM has been deployed in CLASS telescope for 40 GHz operation. 

The production status level of the VPM is 2. 
For a wire grid production, once the machinery (CNC or other) is set up, grids can be made fairly rapidly and reliably.
For apertures approaching 2 meters, development efforts to maintain grid flatness and grid-mirror parallelism will need to be undertaken.

%\bibliography{VPM_S4} % bibliography data in report.bib
%\bibliographystyle{spiebib} % makes bibtex use spiebib.bst
%\end{document}  

%% file: broadband_optics/Material_Testing20160911.tex
Accurate characterization of optical elements is crucial for designing 
high performance CMB receivers. 
Mechanical, thermal and optical properties of optical elements need 
to be carefully measured. 
To reflect actual operation conditions, most of the optical elements 
need to be characterized at cryogenic temperatures.
Most material properties vary enough between manufacturers and grades 
that literature values can only be used as a guide.
However, cryogenic measurements are challenging, and often values 
are extrapolated from either room temperature or liquid nitrogen 
temperature, and old property values are adopted in the design 
of new receivers.   
In this section, we review material properties that are important 
for CMB receiver optics, and examples of measurement techniques 
will be presented.

\subsection{Mechanical properties}
\paragraph{Vacuum windows}
A vacuum window needs to support atmospheric pressure while being transmissive to millimeter-wave photons. Because optical loading from room temperature optical material can 
be significant, it is usually desirable to make the window material 
as thin as possible, but this requirements works against making it mechanically robust.
3-D mechanical simulators such as ANSYS and COMSOL have been used to study 
mechanical stress on CMB windows. 
It is straightforward to model if a window is a simple circular 
solid plate of a well known plastic, though some subtle details such 
as the curvature of the inner edge of the supporting ring requires 
some effort to study properly. 
The scenario can become complicated for laminated layers of foam 
or solid plastic with machined features. 

To confirm these models and guide the design, multiple experiments have built simple vacuum chambers to test 
windows for mechanical performance. This allows testing of a fully built window assembly independently of the receiver, and allows observation and study of window failure modes and life testing without putting the receiver itself at risk. It would also be helpful if the fundamental mechanical properties of potential 
window materials were better understood, which will in turn guide the 3D and pen-and-paper simulations.

\paragraph{Material defects}
Stage-III experiments are using silicon, alumina and sapphire as 
lens and half-wave plate material.
These materials have desirable optical and thermal properties, 
but both silicon and alumina are brittle. 
Stage-III experiments that use these material developed flexible 
metal mounting schemes to relieve mechanical stress from differential 
thermal contraction while maintaining optical alignment.
One problem with these materials are that they are very strong
{\it as long as there is no material defect}.
It is hard to find defects and cracks in these materials, although 
a low-tech technique (application of ink followed by solvent 
cleaning) can be useful in cases where the surface is already 
smooth. 
Identifying techniques to produce single crystal silicon, large 
alumina blanks 
and sapphire boules with low defect rates is important.  
However, as CMB-S4 expects to use so much of these materials, 
screening both material 
and finished optical elements for defects may be helpful.
X-ray and ultrasound are used to find such defects, but so far 
no demonstration of such techniques were done on stage-III lens materials.

\begin{figure}[htbp]
   \centering
   \includegraphics[height=2 in]{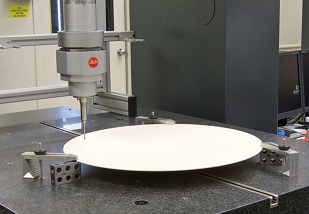}
   \includegraphics[height=2 in]{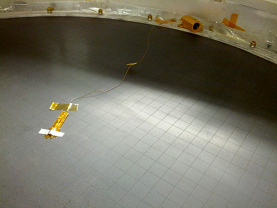}
   \caption{\textbf{Left:} Bridge-type coordinate measuring machine used to measure the profile of a lens before and after applying anti-reflection coating. \textbf{Right:} Photograph of laser-diced, machined-epoxy, anti-reflection coating.}
   \label{fig:materialsetup3}
\end{figure}

\paragraph{Delamination}
Differential thermal contraction is a challenge for anti-reflection 
coatings. 
Some methods, such as silicon dicing and thermal spray of ceramic 
powder, get around this problem by using the same material as the lens.
Epoxy coating and plastic coating relies on bonding strength 
to overcome mechanical stress from thermal contraction as shown 
in Figure~\ref{fig:materialsetup3}.
This problem is mitigated by chosing a plastic that has small 
difference in thermal contraction relative to the optical element material,
and in other cases stress relief grooves have been cut into some 
plastic coatings to relieve mechanical stress. 
Further benefits result from details of the application of the AR 
material, such as surface preparation, use of adhesion promoter, 
and exact conditions of the cure or fusion of the parts.  
Due to lack of knowledge of adhesion properties at cryogenic 
temperatures and mechanical stress from thermal contraction, 
stage-III experiments cryogenically tested anti-reflection coating 
delamination on witness samples.  
To build enough confidence for deployment, it is always  
desirable to test on a full-size optical element.
Such a test is very expensive if failure means the lens or filter is 
no longer usable, and also adds time to the development phase.
It would be useful to build a setup to measure mechanical 
stress and adhesion properties at cyrogenic temperature such that 
mechanical/cryogenic performance can be predicted.

\paragraph{Profile}
Stage-III experiments used highly accurate bridge type coordinate 
measuring machines (CMM) at national labs to measure profiles of 
lenses and thickness of anti-reflection coatings, as shown in 
Figure~\ref{fig:materialsetup3}.
There exist CMM machines at national labs that have a large enough 
throw and accuracy at the micron level, easily able to 
meet CMB-S4's requirements for cryogenic optics.  However, 
CMM operation requires trained technicians to operate, setup can 
take a significant time, and it is expected that they will be 
shared facilities in large labs, so this phase could easily become 
a bottleneck for CMB-S4. 

\subsection{Thermal properties}
Stage-III experiments are using silicon, alumina, plastics and 
copper mesh filters as optical elements.
Understanding thermal conductivity, emissivity, and scattering at 
infrared frequencies are necessary to calculate accurately the final 
temperature of these optical elements. 
For detector sensitivity calculations, emission from filters 
anchored at higher temperature stages can be a significant 
contribution to in-band loading, and for thermal design 
the out of band loading on the cold stages from these ``warm'' 
filters is also critical.  

Thermal conductivity measurement at 4~K and 50~K are 
routinely done with a heater and well-calibrated thermometers.
The community does have a compendium of material property values 
that do inform us in the design phase, but some aspects, like 
the performance of material interfaces between dielectrics and 
metal, are not well-established.  Some additional testing 
will be beneficial to CMB-S4, including filling out thermal 
conductivity and specific heat vs. temperature tables for some 
materials, and determining optimum use of interface materials like 
indium, Apiezon-N grease, and varnish on our various dielectrics.

\subsection{Optical properties}
\begin{figure}[htbp]
   \centering
   \includegraphics[height=2 in]{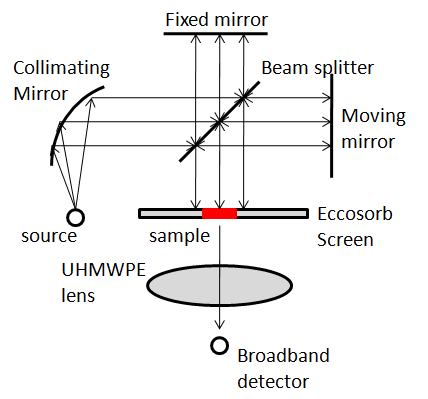}
   \includegraphics[height=2 in]{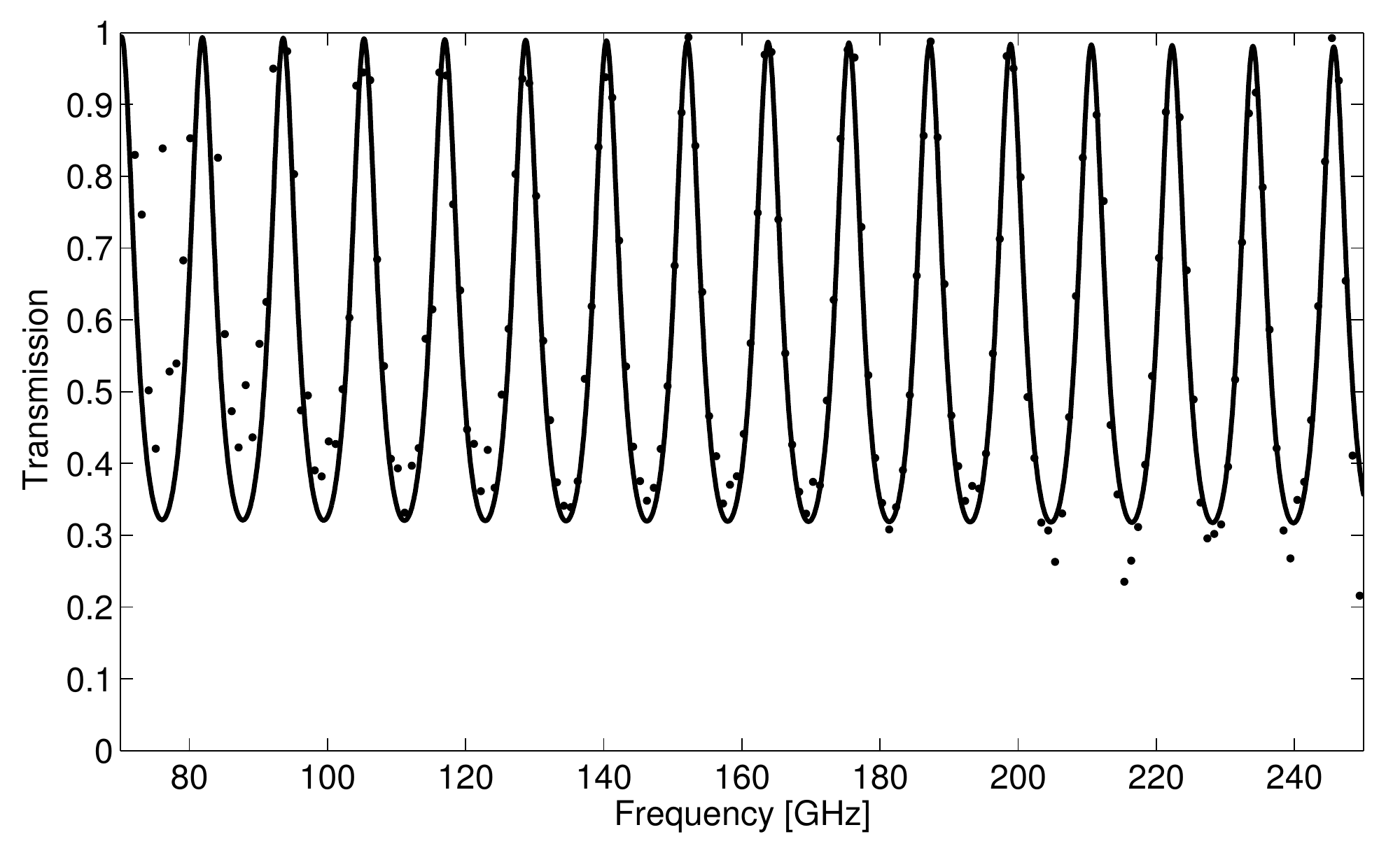}
   \caption{\textbf{Left:} Left: Fourier transform spectrometer example.  
   The sample also may be placed between a collimating mirror and beamsplitter, 
   or (as described below) in one of the arms.  Right: Spectrum of an 
   alumina sample from an FTS scan.  The high frequency oscillations 
   are Faby-Perot fringes from interference associated with surface 
   reflections, and 
   their spacing and amplitude indicate index of refraction when 
   combined with knowledge of the sample thickness.  The difference 
   between unity and the values at the peaks of the fringes represent 
   losses in the material, and one can derive the 
   loss tangent as a function of frequency with an 
   accurate measurement.  }
   \label{fig:materialsetup1}
\end{figure}

\paragraph{Cut-off frequency}
Key parameters for an infrared filter are emissivity at infrared 
frequencies, thermal conductivity, and bandpass parameters.  
The latter consist of in band transmission, cut-off frequency, roll-off speed and out-of-band attenuation and scattering. 
These are essential inputs to calculations 
of sensitivity and cryostat thermal performance. 
Fourier transform spectrometers (FTS) can be used to characterize 
the optical performance of filters.
A schematic drawing of a setup is shown in 
Figure~\ref{fig:materialsetup1}.
A broadband signal from the FTS is transmitted through the sample 
and detected at a detector (often a cryogenically cooled, 
NTD-Ge bolometer with JFET readout). 
Measurements are made with and without a sample in the optical path, 
the latter to normalize the response of the former, 
giving transmission as a function of frequency.
An example of such plot is shown for the RT-MLI section 
in Figure~\ref{fig:RTMLI}
From the plot, it is possible to extract in band transmission, 
cut-off frequency, roll-off speed and out of band attenuation.

\paragraph{Dielectric constant}
Dielectric constant (alternatively, index of refraction) 
is necessary for optics and 
anti-reflection coating design. 
The dielectric constant of a material can be measured accurately 
with an FTS or a frequency tunable coherent source.
The measurement setup with an FTS can be the same as that used for infrared filters 
described above, although the source and detector may be optimized for 
in-band performance.  Fabry-Perot (FP) fringes in frequency space 
are generated by interference between the direct pass-through of 
radiation and the portion of the E-field that reflects band and forth 
on the sample surfaces.  
And example of a measurement of an alumina sample with 
an FTS is shown in Figure~\ref{fig:materialsetup1}.

A measurement setup with a frequency tunable coherent source 
involves the source, a diode detector, and lenses or mirrors 
to collimate the radiation to pass through the sample and then refocus 
for the detector.  
Just like a measurment with an FTS, the measurement with a sample is 
divided by one without the sample to normalize the response.
An example of a measurement of an alumina sample 
with a coherent source setup is shown in Figure~\ref{fig:laser_arc}.
As with the FTS example, the spectral features of FP 
fringes in tranmission data are used to determine the dielectric 
constant.

The index of refraction of a dielectric material can also be 
determined by measuring 
the focal lenth of a lens of known shape, or the angular deviation of 
a prism of known geometry, with lower precision.

\paragraph{Absorption}
\begin{figure}[htbp]
   \centering
   \includegraphics[height=2 in]{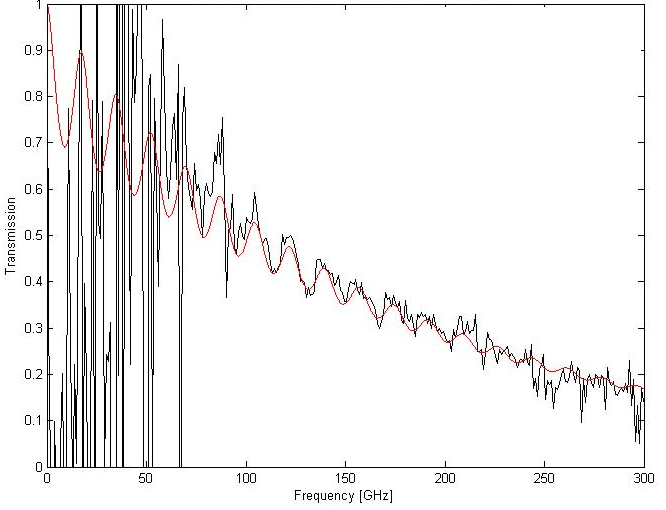}
   \caption{Transmission measurement of epoxy loaded with titanium oxide. Transmission as a function of frequency is fitted with a model to extract the absorption coefficient (and therefore the loss tangent) of a material}
   \label{fig:materialsetup5}
\end{figure}
Absorption loss in the optical elements hurt sensitivity of 
an instrument by decreasing in-band optical efficiency and 
increasing optical loading on the detectors.
%Absorption loss is often quoted as loss-tangent 
%$\tan\delta = \epsilon_i/\epsilon_r$ which is the tangent of the 
%angle between the real and imaginary components of the dielectric function.
Loss-tangent can be calculated from transmission versus frequency 
curve from a FTS measurement or frequency tunable coherent source 
measurement as shown in Figure~\ref{fig:materialsetup5}.
It can also be calculated by measuring transmitted power as a 
function of thickness of a sample at single frequency. 

\paragraph{Dielectric Characterization with Fabry-Perot Resonators} 
An alternative FTS scheme used is to place the sample in
the collimated beam between the beamsplitter and the fixed mirror.
For materials with slowly varying transmission through the band,
this allows simultaneous determination of transmission loss, dielectric
constant, and optical sample thickness.  The key is that the
absolute phase shift as a function of frequency is measured through
the sample as well as the Fabry-Perot fringes, thus adding additional
information to the analysis.  

The technique can also be used to 
measure the index of refraction of near-unity materials like 
Zotefoam when combined with mechanically measured thickness. 
They have such small surface reflections that 
FP fringes are unmeasurable while the net phase shift 
through the material can still be measured, thus allowing 
determination of refractive index.

\paragraph{Reflection}
\begin{figure}[htbp]
   \centering
   \includegraphics[height=2 in]{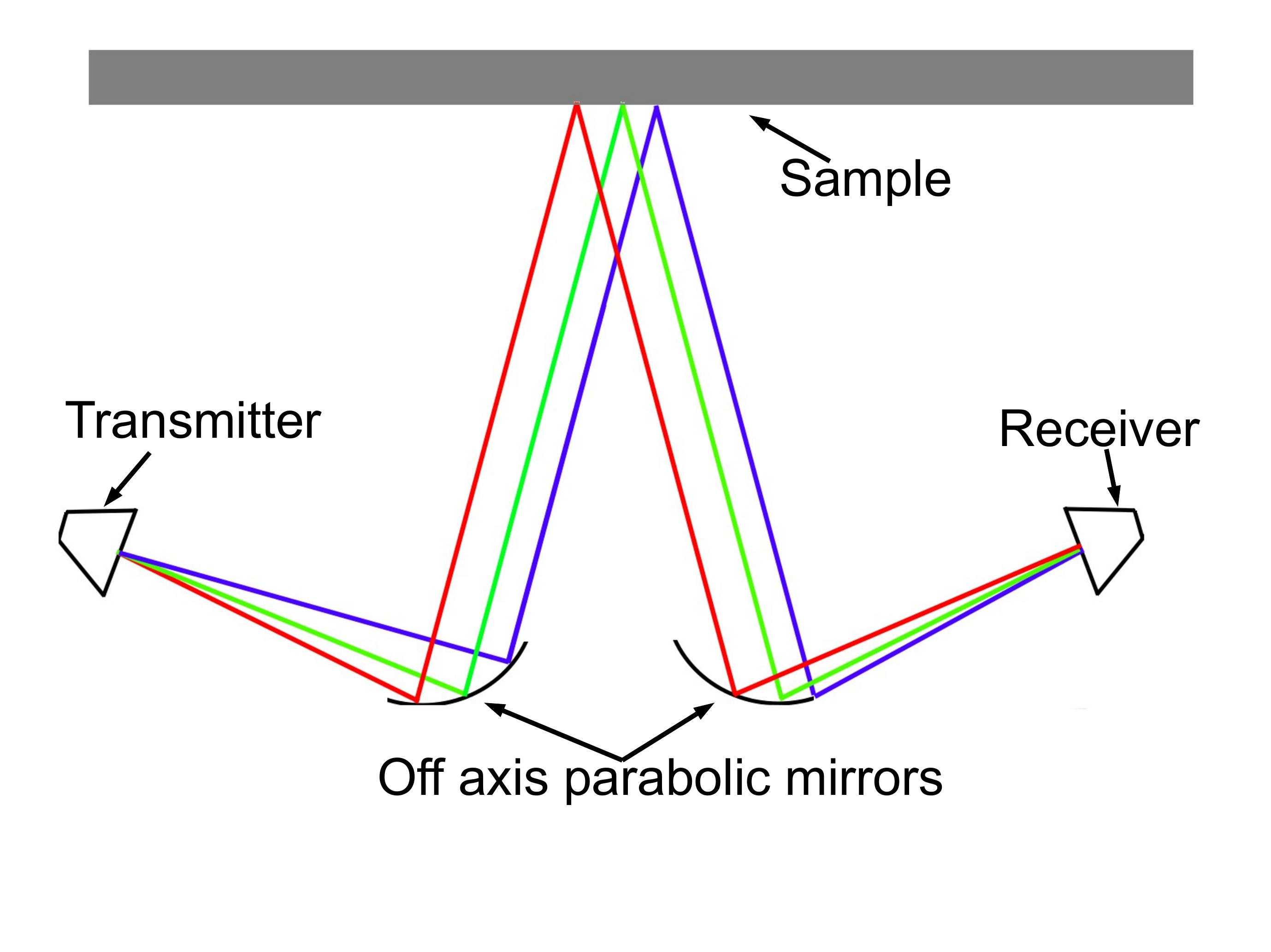}
   \includegraphics[height=2 in]{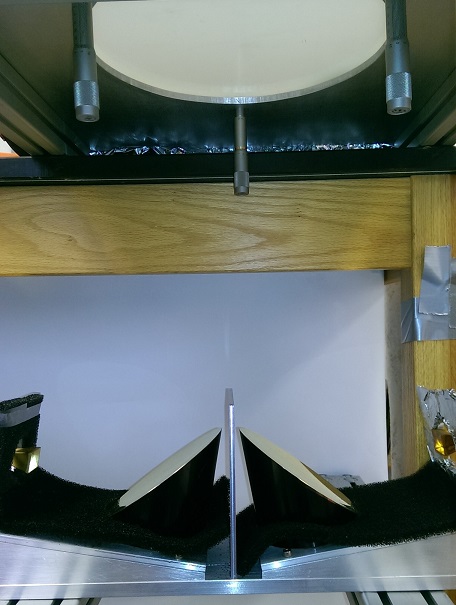}
   \caption{Reflectometer setup at University of Michigan. The reflectometer 
    uses a variable microwave source to measure the reflection off a 
    flat sample. An off-axis parabolic mirror collimates the beam 
    from the source and directs the beam at the sample. Another 
    mirror collects the beam and directs it at a horn-coupled 
    photodiode. Calibrating with known highly reflective sources, 
    this set-up can measure absolute reflection down to 0.1\%, with 
    a relative accuracy of a few percent. }
   \label{fig:materialsetup4}
\end{figure}

Reflection measurements are important for characterization of 
anti-reflection coatings and absorbers.
An example of a reflectometer is shown in Figure~\ref{fig:materialsetup4}.
In the setup, two off-axis parabolic mirrors are used to 
collimate and re-focus the beam from a coherent source.
A goniometer stage, a stage for the precise manipulation angles, is used to align sample to measure 
reflection accurately. 

There is also metrology techniques for reflectometry which rely on phase modulation of the sample \cite{Eimer01,Fixsen01,Murk01}. 
These approaches enable characterization of smaller reflectances if desired.

\paragraph{Scatter}
Scatter of in-band optical signal from porous material could 
increase parasitic optical loading on the detectors.  It is 
a special concern for higher frequency bands, where scattering 
from irregularities in the granularity of various materials 
(such as metal mesh infrared shaders) may cause an increase 
in Rayleigh scattering at higher frequencies, which scales rapidly with increasing frequency as $\sim\nu^4$.
Even a small level of scattering from room temperature optics 
or those near the aperture of a receiver affect 
detector sensitivity strongly.
The expected level of scattering from candidate optical materials is 
usually low, since we have already rejected any materials with 
significant in-band losses.  A candidate infrared filter at 50~K 
with 1\% in-band loss, for example, may be acceptable if it is 
all from absorption (adding 0.5~K loading), but may be intolerable 
if it were from scattering (adding up to 3~K loading if the scattered rays terminate on an ambient temperature surface), and this 
does not even take into account spurious polarization effects, 
for which we are particularly sensitive.  
The small signal that is nonetheless critical makes scattering 
measurements challenging.
A bright, coherent source such as a Gunn diode can be used to illuminate 
a test sample, while a detector is placed off-axis from the line of sight 
to measure the scattered signal.
In this setup, the measured signal is weak since the detector catches 
scattered light within a limited solid angle.
An integrating sphere may be used to improve sensitivity.
For the wide spectral range coverage required for CMB-S4, the ability 
to characterize scattering will be important.

\begin{figure}[htbp]
   \centering
   \includegraphics[height=2 in]{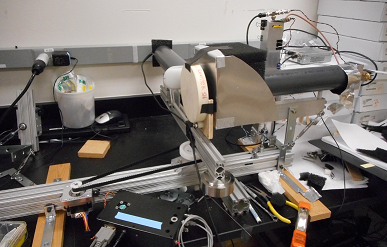}
   \caption{Stanford scattering test prototype.  Gray tubes to the right and 
   on the far side are, respectively, the source collimator and the detector camera.  
   Both are absorber-lined tubes with HDPE lenses.  Installed are a 95\,GHz 
   broadband source and detector, both linearly polarized.  
   A 25\,mm aperture in 
   the aluminum shield just upstream of the sample (not seen in photo)
   defines the beam.  
   As seen, the detector camera is mounted on a motorized swing-arm to sample 
   a 1D cut through the forward-scattering hemisphere.  
   A large absorber (to left, out of the photo) absorbs most of the un-captured,
   un-scattered radiation.  To sample a more complete fraction of the 
   hemisphere, one can in principal rotate source, detector, and sample 
   through various angles and re-scan the 1D arc, but this procedure 
   is cumbersome, thus motivating a more sophisticated setup.  See text.}
   \label{fig:stanfordscatter}
\end{figure}

    Stanford's prototype scattering test setup is shown
in Figure~\ref{fig:stanfordscatter}.  
Only a 1-D sweep of angle is measured, reducing the sensitivity to
total power scattered and introducing uncertainties since 2D beam
shapes must be estimated based on incomplete information.
Also, the beam from the source convolved with the beam seen by the detector
is several degrees across, so small-angle scattering is difficult to
measure.  Nonetheless it can put limits on scattering.  A next-generation
prototype is envisioned making use of larger optics to narrow the
beams and a 2D gimbal hinge to measure a greater fraction of the full sphere.  An alternate setup is considered having a Winston cone 
to capture broader solid angles to increase detection 
sensitivity off-axis.
Enclosure of the entire rig in an absorber-lined box may be needed.

\paragraph{Cryogenic sample testing}
CMB receivers cryogenically cool optical elements to take advantage 
of improved material properties at cryogenic temperatures.
For example, silicon, alumina, and sapphire's absorption losses 
decrease significantly at cryogenic temperature.
Thermal conductivity also changes strongly with temperature, 
and the refractive index of some materials change significantly as 
a function of temperature as well.
Since many optical elements are used at cryogenic temperatures, 
characterizing properties of material at cryogenic temperatures 
is essential for predictable design of a receiver.
There are multiple challenges associated with measurement of a 
sample at cryogenic temperature.

\begin{figure}[htbp]
   \centering
   \includegraphics[height=2 in]{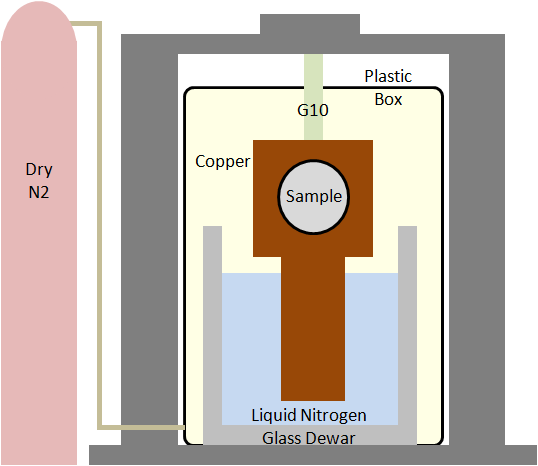}
   \includegraphics[height=2 in]{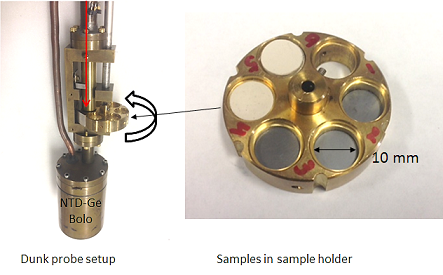}
   \caption{\textbf{Left:} In this setup, a sample is conductively 
    cooled through a copper holder that is immersed into liquid nitrogen.
    Dry nitrogen fills plastic container around the sample to keep 
    the sample try. A Zotefoam window was added to the plastic box to 
    let millimeter wave signal through.  }
   \label{fig:materialsetup2}
\end{figure}

For testing at temperatures above 77~K, LN2-cooled atmospheric-pressure 
sample chambers have been used in several labs.  
Introduction of dry nitrogen (or the LN2 
evaporation) keeps the sample dry, otherwise water vapor, CO2, etc. 
will condense on the sample under test.  
A setup that was used to cool samples to approximately 100~K 
is shown in Figure~\ref{fig:materialsetup2}.
For larger samples and those with lower thermal conductivity such as 
polyethylene, additional infrared blocking becomes necessary or the 
sample will not become sufficiently cold.

Lenses for the CMB receivers are mounted at 4~K, 
and cooling a sample to approximately 4~K in a test chamber 
is more challenging. 
One approach is to immerse samples and detector in liquid helium, 
as shown in Figure~\ref{fig:materialsetup2}. 
An optical signal is transmitted to a sample though a light pipe, 
and mechanical feedthrough allows rotation of sample holder 
immersed in liquid helium. 
To perform measurement with and without samples, mechanical 
motion at 4~K is required.
A cryogenic stepper motor, or a mechanical feedthrough can be 
installed in a dewar to move a sample.
While this LHe direct contact technique is good for absorption loss 
measurements, it is more difficult for dielectric constant measurements 
since the LHe optical properties must be accounted for.  A vacuum 
4~K system, with significant infrared blocking, is important for the 
latter.  

%
%
%\subsection{Technology distribution}
%
%A large part of this category of CMB instrumentation technology is 
%materials selection, sourcing and working with existing commercial vendors.  Materials testing has been performed  
%in many of our labs over the years, and documented internally within individual projects. However, full documentation and distribution of the results 
%to the community has not been critical to existing programs. With the community coming together for CMB-S4, this will now be more of a priority.  
%We will therefore have a more organized effort to document and share 
%new (and old) results, as well as technical techniques and skills, to 
%benefit the entire CMB-S4 project.  
%
%\paragraph{Publications and databases} Naturally we expect publications 
%to fully disseminate results of optical testing and design schemes. 
%In addition (and on shorter timescales), 
%a database of newly measured data should 
%be considered, to augment the existing Lamb and NIST material properties compilations.  
%
%
%
%\paragraph{Labor} Some processes, especially those involving 
%commercial shops, have taken a large dedicated effort to come to fruition.  
%In these cases the additional effort required to transfer the capabilities 
%to another institution needing the new process 
%may not be justified. 
%We should consider the sharing of experienced personnel, for both 
%technology transfer and overseeing design/fabrication directly, 
%as a critical process that is occasionally needed for the success 
%of the CMB-S4 endeavor.  

%% file: broadband_optics/OptSumTable.tex
%TSL and PSL definition.
%
%\begin{table} [!h]
%\begin{center}
%\begin{tabular} {cl}
%\hline
%\textbf{TSL} & \textbf{Description} \\
%\hline
%1 & Lab test of technology to show principle.\\
%2 & Lab test of technology but with full feature set and performance suitable for ground test.\\
%3 & Experiment capable version built and tested in the lab.\\
%4 & Deployed in a CMB experiment and data taken.\\
%5 & Data fully analyzed, systematic errors understood.\\
%\hline
%\hline
%\textbf{PSL} & \textbf{Description} \\
%\hline
%1 & Fabrication of a TS1/TS2 prototype demonstrated.\\
%2 & Fabrication of a one or more experimental capable units.\\
%3 & Conceptual plan of methods for production at scale.\\
%4 & Demonstrated the critical steps for production at scale.\\
%5 & Capability for production at scale exists and is demonstrated.\\
%\hline
%
%\end{tabular}
%\end{center}
%\end{table}

\begin{landscape}
\begin{table}
\begin{center}
\begin{tabular} {|l|c|c|c|c|c|}
\hline  
 & \textbf{Lab Demonstration} &  \textbf{Sky Demonstration} & \textbf{Path to CMB-S4} & Section & T/PSL\\
\hline 
\hline
\textbf{Window}  &  &  &  &  & \\
UHMWPE  & $\surd$ & ABS/ \bicepIII/CLASS/EBEX & large diam, emission & \ref{sec:winuhmwpe} & 5/5 \\
  &  & \spider/SPT-3G, 700\,mm &  &  & \\
Zotefoam  & $\surd$ & ACT\bicepII/PB/SPT, 500\,mm & large diam, scatter & \ref{sec:winzotefoam} & 5/5 \\
Cryowindow  & Foam/alumina & - & full scale lab demo & \ref{sec:winzotefoam} & 1/1 \\
\hline
\textbf{IR Filters}  &  &  &  &  & \\
MMF  & $\surd$ & Mult. exp. $< 300\,$mm & 530\,mm dia, prod rate & \ref{sec:filtermmf} & 5/3 \\
Laser Ablated  & $\surd$ & \bicepIII/SPT-3G & prod rate & \ref{sec:filterlaser} & 5/3 \\
Plastic & $\surd$ & \bicepII, \bicepIII & temp/emission & \ref{sec:filterplastic} & 5/5 \\
Alumina & $\surd$ & \bicepIII/PB2/SPT-3G, 700\,mm & large dia, prod rate &\ref{sec:filteralumina}  & 5/3 \\
Silicon  & 300\,mm & - & Prod rate, diameter & \ref{sec:filtersi} & 2/1 \\
RT-MLI  & $\surd$ & PB2, GroundBIRD, SPT3G & $\surd$ & \ref{sec:filterrtmli} & 3/5 \\
\hline
\textbf{Lens}  &  &  &  &  & \\
Silicon  & $\surd$ & ACTPol $< 480$\,mm & Size, AR prod rate & \ref{sec:lenssi} & 5/3 \\
Alumina  & $< 880$\,mm & \bicepIII/PB2/SPT-3G $< 700$\,mm & loss & \ref{sec:lensalumina} & 5/3 \\
Plastic  & $\surd$ & Stage-II, CLASS $< 300$\,mm & low index & \ref{sec:lensuhmwpe} & 5/5 \\
MML  & Measured beam & - & 530\,mm, prod rate & \ref{sec:lensmm} & 1/1 \\
\hline
\end{tabular}
\end{center}
\end{table}
\end{landscape}

%\textbf{MISSING CAPTIONS. To be inserted in the caption: prod rate stands for production rate, large diam stand for large diameter.}

\begin{landscape}
\begin{table}
\begin{center}
\begin{tabular} {|l|c|c|c|c|c|}
\hline  
 & \textbf{Lab Demonstration} &  \textbf{Sky Demonstration} & \textbf{Path to CMB-S4} & Section & T/PSL\\
\hline 
\hline
\textbf{AR Coating}  &  &  &  &  & \\
Thermal Spray  & PB-2 in prep$\surd$   & SPT-3G (deployed) &  low dielectric outer layer & \ref{sec:ARthermalspray} & 3/3 \\
Epoxy  & 3:1 bandwidth PB-2 in prep $\surd$  & BICEP-3 & CNC fabrication  & \ref{sec:ARepoxy} & 5/2 \\
Diced Silicon  &   4:1 bandwidth: $\surd$  & ACTPol/AdvACT  &  efficient fabrication & \ref{sec:ARdicedsi} & 5/3\\
DRIE Silicon  &  1.5:1 bandwidth: $\surd$  & - & apply to curved surfaces &  \ref{sec:ARdriedsi} & 1/1\\
Laser Ablated  &  3:1 bandwidth $\surd$  & -  & larger areas / lower freq & \ref{sec:ARlaser}  & 1/1 \\
Plastic  & $\surd$  & EBEX (on sapphire) & application on high $\epsilon$ lens & \ref{sec:ARplastic} & 5/5\\
Machined Plastic  & CLASS 90 GHz $\surd$  & SPT-3G window & application for curved lens & \ref{sec:ARablatedplastic} & 4/5\\
MMARC  & $\surd$ & - & on sky demo & \ref{sec:ARmmarc} & 1/1 \\
\hline
\textbf{Pol Mod}  &  &  &  &  & \\
Sapphire  & $\surd$ & PB, ABS, EBEX & broad-band / large diameter & \ref{sec:sapphire} & 4/5\\
Silicon & $\surd$  & ACTPol & efficient fabrication / improve glue & \ref{sec:materialSI} &4/3 \\
MMPolMod  & $\surd$  & - & design / 60\,cm fabrication & \ref{sec:mmpolmod}  & 3/2 \\
AmbRotation  & $\surd$  & PB, ACT/ABS  & reliability / encoder & \ref{sec:polmodrotator} & 4/3\\
CryoRotation  & PB-2 $\surd$  & EBEX & electronic drive electronics & \ref{sec:polmodrotator}  & 4/3\\
VPM  &  $\surd$ & CLASS & large diameter, efficient fabrication & \ref{sec:polmodvpm} & 4/2\\
\hline 
\textbf{Characterization}  &  &  &  &  & \\
reflection & $\surd$ &  &  cover all CMB-S4 bands & & \\
transmission &  &  &  improve accuracy &   & \\
metrology &  $\surd$ &  & extend to larger diameters &  & \\
scattering &  &  &   improve sensitivity & & \\
emission &  &  & measure emission of cold optics  & & \\
\hline
\end{tabular}
\end{center}
\end{table}
\end{landscape}

%% file: detector_rf/detectorrf_cmbs4.tex
\clearpage
\chapter{Focal plane optical coupling}\label{ch:detetcorrf}
\vspace*{\baselineskip} % title is multiline, and without this no space between title & text
\vspace{1cm}
\section{Introduction}\label{sec:detrf_intro}
%\input{detector_rf/Intro_Goal} %\label{sec:goal}
\input{detector_rf/Intro_Introduction} %\label{sec:introduction}

%%%%%%%%%%%%%%%%%%%%%%%%%%%%%%%%%%%%%%%%%%%%
\clearpage
\section{Background}\label{sec:background}
\input{detector_rf/Background_Foreground} %\label{sec:foreground}
\input{detector_rf/Background_Bandwidth} %\label{sec:bandwidthresolution}

%%%%%%%%%%%%%%%%%%%%%%%%%%%%%%%%%%%%%%%%%%%%
\clearpage
\section{Antennas}\label{sec:feed}
\input{detector_rf/Feed_Intro}

\input{detector_rf/Feed_Horn} %\label{sec:horn}
\input{detector_rf/Feed_OMT} %\label{sec:hornomt}
\input{detector_rf/Feed_MKIDHorn2} %\label{sec:mkid_coupling}
\input{detector_rf/Feed_LensletAntenna} %\label{sec:lensletantenna}
\input{detector_rf/Feed_LensletArray} %\label{sec:lensletarray}
\input{detector_rf/Feed_GRIN} %\label{sec:grin}
\input{detector_rf/Feed_AntennaArray} %\label{sec:antennarray}
\input{detector_rf/Feed_Multimode} %\label{sec:multimode}

%%%%%%%%%%%%%%%%%%%%%%%%%%%%%%%%%%%%%%%%%%%%
\clearpage
\section{RF components}\label{sec:rf}
\input{detector_rf/RF_Intro}
\input{detector_rf/RF_Tline} %\label{sec:rf_tline}
\input{detector_rf/RF_Filter} %\label{sec:onchipfilter}
\input{detector_rf/RF_CrossOver} %\label{sec:crossover}
\input{detector_rf/RF_Termination2} %\label{sec:termination}

%%%%%%%%%%%%%%%%%%%%%%%%%%%%%%%%%%%%%%%%%%%%
\clearpage
\section{Array layout, pixel size, and wiring considerations}\label{sec:array}
\input{detector_rf/Array_PixelSize}

%%%%%%%%%%%%%%%%%%%%%%%%%%%%%%%%%%%%%%%%%%%%
\clearpage
\section{Detector characterization}\label{sec:detchar}
\input{detector_rf/DetectorCharacterization}

%%%%%%%%%%%%%%%%%%%%%%%%%%%%%%%%%%%%%%%%%%%%
\clearpage
\section{Conclusion}\label{sec:detrf_conclusion}
\input{detector_rf/ConclusionShort}

\clearpage
\section{Summary of detector-RF technologies}\label{sec:detsum}
%\input{detector_rf/DetSum}
\input{detector_rf/DetSumTable}

%% file: detector_rf/Intro_Introduction.tex
%\subsection{Introduction}
%\label{sec:introduction}

The performance of a CMB experiment depends critically on the design of the focal plane. The focal-plane feed determines the shape and polarization properties of the pixel beams and therefore plays a strong role in controlling systematic errors. The feed design also can determine the total bandwidth and number of photometric bands of each pixel, which is important for the efficient use of a telescope's focal plane area. This chapter discusses the detector system from the focal-plane feed up to the power detection element. Chapter~\ref{chp:readout} discusses the detector itself (TES or KID) and the readout multiplexing system.  

%Detector design (feed and RF architecture) is a critical technology for CMB experiments. The RF and feed design realizes multiples functions for CMB detectors including: beam synthesis, polarization analysis, and total optical bandwidth and throughput. The challenge of CMB-S4 requires continued R\&D into detector RF design to optimize performance across all of these functionalities: good polarized beams (for low systematics), large bandwidth (for foreground removal), and high coupling efficiency (for sensitivity). 

%Detector design drives frequency bandwidth, beam quality, polarization sensitivity and raw sensitivity. Wide bandwidth increases optical throughput of a single pixel. Systematics from beam distortion and polarization sensitivity should be minimized. Also, optical efficiency of a detetor should be high to maximize experiment's sensitivity. 

%In order to characterize the B-mode polarization with increasing precision, rejection of polarized foregrounds is of increasing importance. Because the thermal dust emission and synchrotron radiation follow a different spectrum than the CMB, component separation can be used to remove the foreground contamination, requiring a measurement at multiple frequencies. As a consequence, the next generation of ground-based CMB experiments aims to cover a broad range of frequencies that lie in ``atmospheric windows,'' or spectral regions of high microwave transmission between water and dioxygen emission lines.

There are a number of successful approaches that have been or are being implemented by different experiments. They include using a telescope with a receiver observing at a single frequency band with single-color lenslet-coupled antennas or with corrugated horns (\Pb-1, ABS)~\cite{pb,abs}, using one telescope with multiple receivers each observing at one frequency with corrugated horns (ACTPol)~\cite{actpol}, using multiple telescopes each observing at a single frequency with antenna-array feeds or with horn-coupled antennas (Keck Array, \bicepArray, CLASS 40 \& 90)~\cite{keck,bicepthree,Essinger-Hileman14}, using a multichroic receiver observing on one telescope with single color corrugated horns or smooth wall profiled horns (SPTpol, CLASS 150/220)~\cite{sptpol,Essinger-Hileman14}, or using a multichroic receiver observing on one or more telescopes with multichroic lenslet-coupled detectors (\Pb-2, SPT-3G, Simons Array) or with feedhorns (ACTPol, Advanced ACTPol)~\cite{suzuki15,benson2014,henderson/etal:2016,Stebor}. Experiments with single color detectors first successfully detected B-mode polarization, and multichroic detectors have since been deployed and years of data has been collected.
%SPT3G replaced with benson2014, Simons Array with sa, Advanced ACTPol with Henderson16.
The diversity of detector designs used in these experiments emphasizes the complexity of global experimental optimization. In this chapter, we survey the current state of technologies for antennas and RF circuit architectures developed for CMB polarization experiments. In each section, we give a basic introduction to the technology, describe the current implementation, and identify necessary research and development to bring the technology to a readiness level required for CMB-S4. 

%[Manually merged Toki and Clarence's edit]
%There are a number of successful approaches which have or are being implemented by different experiments. One approach is using multiple small-aperture receiver-telescopes with each receiver observing at a single frequency with a phased-array RF architecture (Keck Array, \bicepArray)~\cite{keck, bicep3}. Another approach is a single large-aperture telescope having multiple receivers with each receiver observing at a single frequency with a feedhorn and OMT architecture (ACTPol)~\cite{actpol}, or similarly a single large aperture with a multi-chroic receiver using a feedhorn coupled architecture (SPTpol)~\cite{sptpol}. Recently, experiments have started using multi-chroic receivers observing on one or more telescopes using multi-chroic pixels using either broadband lenslet-coupled antennas (\Pb-2, SPT3G, Simons Array) or broadband profiled feedhorns (Advanced ACTPol)~\cite{pb2, spt3g, advact, sa}.

%The different approaches to RF design implemented by multiple experiments shows both the complexity and depth of expertise associated with optimizing the detector RF design. In this write up, we survey the current state-of-the-art technologies for various RF feeds and circuitry developed for Cosmic Microwave Background polarimetry experiments. In each write up we give a basic introduction of the technology, current implementation and identification of necessary research and development to bring the technology to a readiness level relevant to CMB-S4. 

%% file: detector_rf/Background_Foreground.tex
% Subsection motivating choice of frequency bands based on astrophysical
% foregrounds.
\subsection{Foreground considerations for frequency band selection}
\label{sec:foreground}

\begin{figure}[h]
\centering
\includegraphics[height=1.8 in]{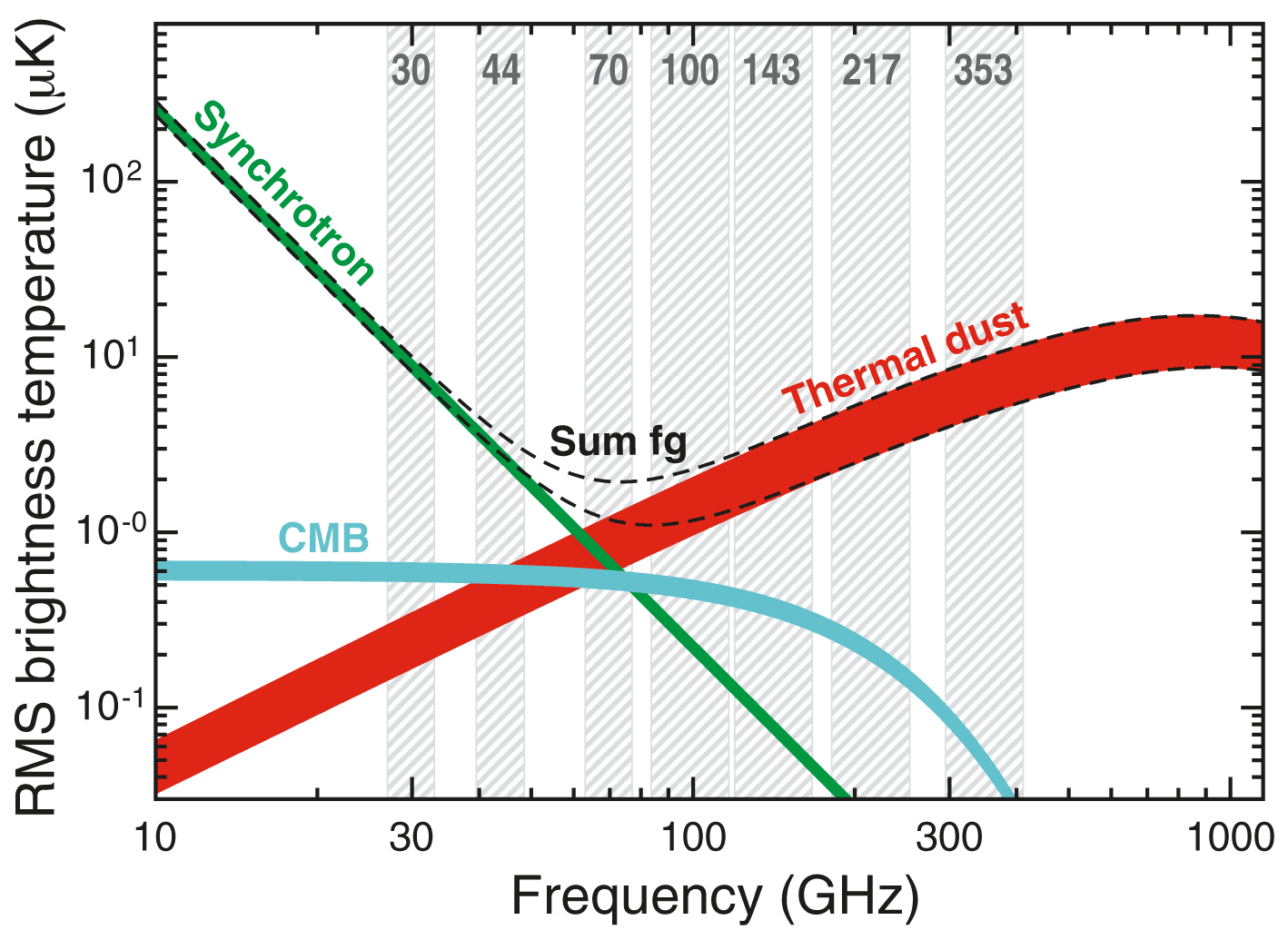}
\includegraphics[height=1.8 in]{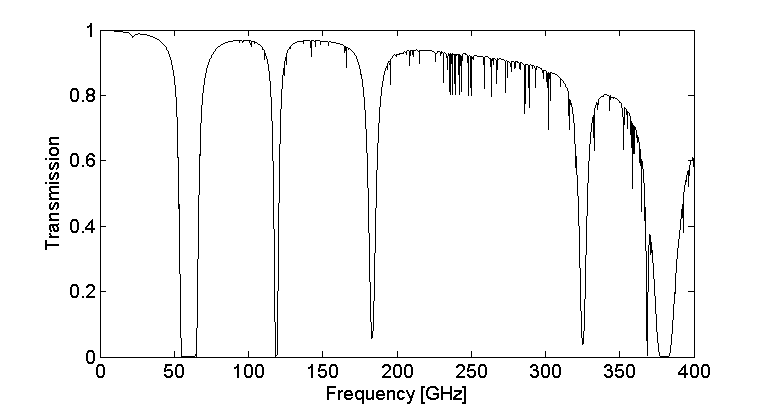}
\caption{(Left) Root mean square (RMS) brightness temperature for polarization as a function of frequency and astrophysical component~\cite{Adam:2015wua}. (Right) Atmospheric transmission coefficient for a precipitable water vapor level of 1\,mm and an observing elevation angle of 60 degrees~\cite{am}.}
\label{fig:foreground}
\end{figure}

For a ground-based microwave telescope, the atmospheric transmission profile defines four discrete frequency windows that are useful for observation: a low-frequency band that extends from $\sim$30--50\,GHz, mid-frequency bands from $\sim$75--110\,GHz and $\sim$130--170\,GHz, and a high-frequency band above $\sim$190\,GHz as shown in Figure~\ref{fig:foreground} \cite{obsast_book}.
These windows are separated by molecular oxygen lines at 60 and 120\,GHz and a water line at 183\,GHz.
Above 200\,GHz, atmospheric transmission and sky noise get steadily worse.  There may however be useful bandwidths for mapping dust foregrounds up to the 325\,GHz water line if the level of dust foregrounds increases faster than atmospheric noise.
While mapping speed considerations would favor designing instruments that cover as much of this bandwidth as possible, the problem of separating the CMB signal from astrophysical foregrounds will require CMB-S4 to feature a number of somewhat narrower frequency bands.

The two dominant polarized astrophysical foregrounds for CMB observations are synchrotron emission from free electrons and thermal emission from microscopic dust grains.
Foreground emission can be distinguished from the CMB by its spectrum. Relative to the $2.73\,K$ blackbody of the CMB, synchrotron emission grows brighter at low frequencies while dust is brighter at high frequencies as shown in Figure~\ref{fig:foreground}.
Multi-frequency data allows us to identify and remove foreground signals. Indeed, the science goals of CMB-S4 lead to stringent requirements on the accuracy and precision of foreground separation.
Even over a small region of clean sky, the power spectrum of polarized dust at 95\,GHz exceeds the $r=0.001$ tensor spectrum by more than an order of magnitude, highlighting the difficulty of this problem (see Figure 7 in Reference~\cite{cmbs4_sciencebook}).

With current data, we are just beginning to be able to measure the properties of polarized foregrounds at high Galactic latitude \cite{pip_xxx}.
As the signal-to-noise ratio on foregrounds improves, we will likely find that the simple parametrizations in use today are inadequate, for instance due to spatial variation of the spectral index or frequency-dependent variation of the polarization angle \cite{pip_l}.
Failure to account for the full complexity of the foreground signals could lead to bias on cosmological parameters. The job of detecting and constraining these foregrounds requires better frequency resolution.
To account for this yet unknown complexity, the projections for inflation science from tensor modes with CMB-S4 make a baseline assumption of eight frequency bands, splitting each atmospheric window into two sub-bands \cite[Section 2.3]{cmbs4_sciencebook}.
Our understanding of this problem will improve with data from Stage-III experiments, but CMB-S4 sensitivity will remain at the leading edge of our ability to separate components.

%% file: detector_rf/Background_Bandwidth.tex
\subsection{Total bandwidth and spectral resolution}
\label{sec:bandwidthresolution}

Current bolometric detector technology has reached the noise limit set by CMB photon noise. 
Once individual detectors are limited by photon noise, the mapping speed for a fixed field of view can be increased by use of multichroic detectors.
Designing an array of diffraction-limited, multi-chroic pixels in a limited field of view introduces a sensitivity optimization challenge. 
Optimizing pixel size given a fixed focal plane area must balance two competing effects: small pixel diameter allows for more detectors but degrades aperture illumination efficiency while large pixel diameter improves aperture illumination efficiency but reduces the detector count. 
The product of these opposing effects gives a mapping speed peak at some optimal pixel diameter. 

%Figure \ref{fig:bb} shows mapping speed \cite{griffin, tokiThesis} as a function of detector pixel diameter assuming 
%a small-aperture, cryogenically cooled reflective optics telescope (i.e. 50 K filter, 50 K half-wave plate, and 4 K mirrors amounting to a $\sim$ 10 K telescope temperature \cite{abs}) Large-aperture, ambient temperature mirror telescope with a receiver with cryogenically cooled optics (i.e. 300 K mirrors,  100 K HWP, 50 K filter, and 4 K reimaging optics amounting to a $\sim$ 18 K telescope temperature \cite{pb2}), multi-chroic receiver with a 100 mK focal plane of fixed area.  
%This calculation shows that given a single observation wavelength $\lambda$ and a telescope F/\# ``F'', one ought to set the pixel diameter at $\sim 0.65 $F$ \lambda$. 
%However, given multiple observation bands on a single multi-chroic detector pixel, the mapping speed at some frequencies will be greater than others. 

\begin{figure}[h]
\centering
\includegraphics[height = 2.2in]{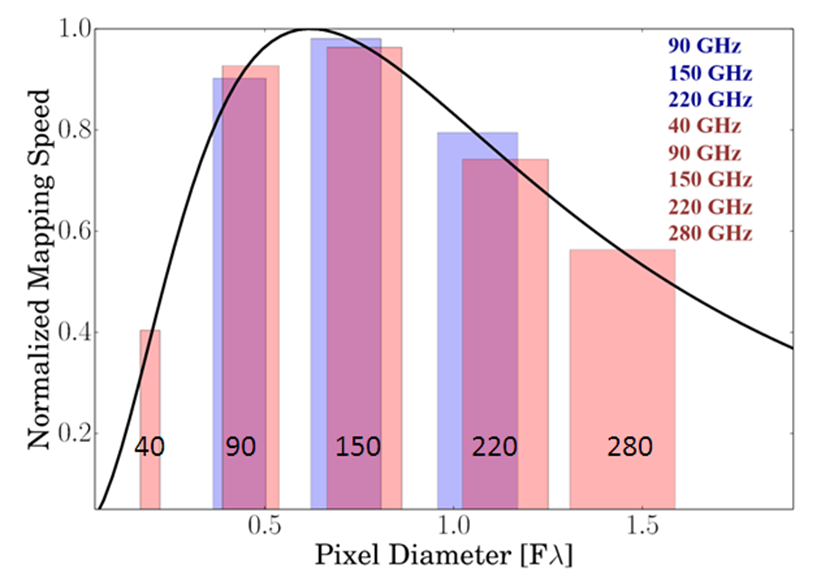}
\caption{
Mapping speed (CMB spectrum) versus pixel size in units of F$\lambda$, where F is the final F/\# at the focal plane and $\lambda$ is the observation wavelength. We assume a fixed focal plane area filled with diffraction limited multi-chroic pixels. Plotted mapping speed is from a 10\,K telescope with a 100\,mK focal plane. We show example band locations for a three (90/150/220\,GHz) and five (40/90/150/220/280 GHz) -band receiver. Band positions for given a pixel diameter are chosen to optimize integrated mapping speed with more weight on CMB bands. The optimal pixel size for a three band detector and five band detector is shifted due to this optimization.}
\label{fig:bb}
\end{figure}

Figure \ref{fig:bb} shows mapping speed (referenced to a CMB) as a function of detector pixel diameter assuming multi-chroic receiver with a $\sim$ 10\,K telescope temperature and a 100\,mK focal plane of fixed area\cite{griffin}.  
This calculation shows that given a single observation wavelength $\lambda$ and a F/\# ``F'' at a focal plane, one ought to set the pixel diameter at small F$ \lambda$, near $\sim 0.6$. 
However, given multiple observation bands on a single multi-chroic detector pixel, the optimal pixel size at some frequencies will be different from others.

Also, example band locations for a three/five-band receiver given a pixel diameter that optimizes integrated mapping speed across the experiment's bandwidth are shown in Figure \ref{fig:bb}.
As an experiment adds more total bandwidth, there is a decrease in mapping speed in channels away from the optimal frequency. 
Therefore, even though building multi-chroic detectors can improve sensitivity by enhancing foreground removal and total optical throughput, the relative sensitivity in each frequency channel should be considered carefully when choosing the total bandwidth of a pixel.

As discussed in Section~\ref{sec:foreground}, dividing the atmospheric windows into sub-bands helps resolve foregrounds using their spectral dependencies.
On-chip multi-chroic band pass filter techniques that are used to divide broadband signals into sub-bands are described in Section~\ref{sec:onchipfilter}.
There are challenges associated with packing spectral bands close together %, including increased sensitivity to filter dimensions as overlaps increase, and dielectric loss in the filters that increases as the roll off becomes sharper. Additionally, 
, such as increasing the number of bands introduces readout challenges, including the possibility of a greater multiplexing factor, more complicated wiring schemes, and higher interconnect density. 
Therefore, the experiment's design should be optimized to find a balance between the capability of a focal plane and the complexity of a design.

%% file: detector_rf/Feed_Intro.tex
The choice of microwave antenna influences the angular response,
polarization properties, bandwidth and efficiency of a detector. An
ideal antenna has a polarization-symmetric beam pattern across its
entire spectral bandwidth.

%Size of beam waist should be close to physical pixel size to maximize efficiency of a focal plane. 
%Also coupling between between a feed and a transmission line should be efficient.

Multiple antenna technologies have been used for CMB experiments:
horns, lenslet-coupled antennas, antenna arrays, and direct absorber
coupling.  Broadband horn antennas have been observing the CMB on the
ACTPol and Advanced ACTPol experiments.  Broadband lenslet coupled
antennas were deployed on SPT-3G in early 2017, will be deployed
on \Pb/Simons Array in 2017, and cover two to three atmospheric
windows with one pixel.  Single-band antenna arrays are being used
on \bicepII/Keck Array/\bicepIII\ experiments, and development aimed
at increasing the bandwidth for antenna arrays is ongoing.  Direct
absorber detectors are being developed for future balloon and
satellite CMB experiments.

This section will review the basic properties and current state of
these detector along with supporting technologies.  Demonstrated performance
and future prospects are given for each topic.

%% file: detector_rf/Feed_Horn.tex
%\documentclass[a4paper,12pt]{article}
%\usepackage{setspace}
%%\usepackage[english]{babel}
%\usepackage{graphicx}
%\usepackage{fullpage}
%\usepackage{parskip}
%\usepackage{multicol}
%\usepackage{subfigure}
%\usepackage{multirow}
%\usepackage{tabularx}
%\usepackage[font=small,labelfont=bf]{caption}
%%\bibliographystyle{apj}
%%\RequirePackage{natbib}
%
%\usepackage{float}
%\floatstyle{boxed}
%
%%\doublespace
%\begin{document}
%\noindent {\bf \Large Feedhorns}
%
%\noindent{Sara M. Simon, Jeff McMahon, Hannes Hubmayr}

\subsection{Feedhorn coupling}
Feedhorns have been widely used in radio astronomy. 
Horn antenna defines angular response of a detector.
Planar ortho-mode transducer (OMT) probe can be used to couple RF power to microwave circuit on a chip. 
RF sensitive detectors, can be placed directly at one end of a horn for direct detection or behind microwave circuit after on-chip signal processing.
In this subsection, technologies for feedhorn-coupled detectors designed for CMB polarimetry experiment is described.

\subsubsection{Feedhorns}
\label{sec:horn}

\begin{figure}[h!]
\centering
\includegraphics[height=2 in]{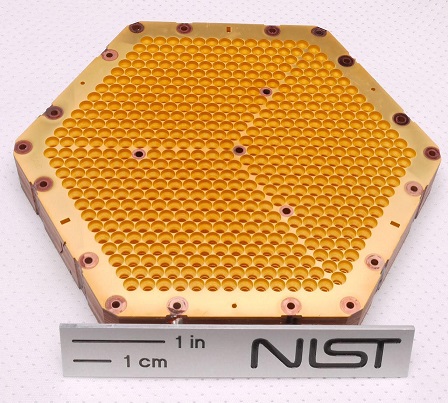}
\includegraphics[height=2 in]{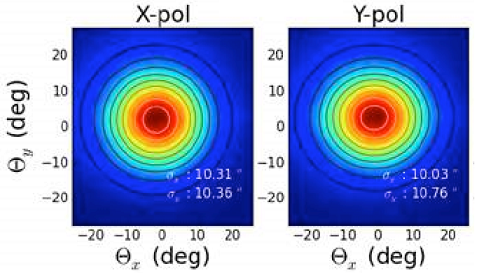}
\caption{(Left) A photograph of the fully assembled and gold coated 150/230\,GHz AdvACT feedhorn array. The array consists of 503 spline-profiled feeds that were optimized for low beam systematic errors and high coupling efficiency with a small aperture. (Right) 2D angular response measurements of an ACTPol single-pixel detector consisting of a single corrugated feedhorn and a single 90/150 GHz dichroic pixel~\cite{HubmayerLTD16}}.
\label{fig:HF_array}
\end{figure}

\paragraph{Description of the technology} 
Feedhorns have been a work horse of radio astronomy for generations as they offer the ability to minimize polarization systematic errors and adjust the detector beam size with no need for AR coatings. The leading approach for control of beam systematics has been the corrugated feed which produces a nearly Gaussian beam shape with small polarization leakage over a wide band~\cite{olver,5398898,Simon_2016} (see Figur~\ref{fig:HF_array}). Recently, advances in computer driven optimization have facilitated new feed designs based on a smooth spline-profiled taper~\cite{Granet_Horn}. These spline-profiled designs can achieve beam properties comparable to what has been demonstrated with corrugated feeds, while providing opportunities to optimize for a combination of beam systematic errors and increased array packing densities. Both spline-profiled and corrugated feeds have been demonstrated with more than an octave of bandwidth. Other feedhorn design approaches, including dielectric-loaded feeds, offer paths to extend this technology to achieve broader bandwidth while maintaining attractive beam shapes and low beam systematic errors.

\paragraph{Demonstrated performance} 
Feedhorns have been used widely in observatories for the CMB including COBE, WMAP, PLANCK, SPTpol, ACTPol, and many other experiments. The ACT collaboration has recently deployed two dichroic arrays using feedhorns to define the detector coupling over more than an octave of bandwidth. The first array of 256 horns was deployed in early 2015 and covered (75-165)\,GHz using ring-loaded corrugated feeds \cite{Datta_2014}. The second array was comprised of 503 spline-profiled horns that covered the (120-280)\,GHz observation band and deployed in mid-2016 \cite{Simon_2016}. These horn arrays were fabricated out of stacked silicon wafers that are each etched with a pattern of holes and plated in gold after assembly. The use of silicon waferseliminates the need to account for differential thermal contraction between the horn array and the silicon detector wafers and has lower mass than metal horn arrays. Further, the use of photolithography allows for tight tolerances of 1-2\,$\mu$m ovre 150-mm wafer. The spline-profiled feeds were optimized to maximize the packing density of the feed array while controlling beam systematic errors to the level required for the AdvACT experiment. The 90/150\,GHz spline-profiled feedhorn designed for AdvACT improves the mapping speed of the array by a factor of $\sim$1.8 over the 90/150 GHz corrugated ACTPol array and has a cross-polarization lower than -18\,dB. The analysis of the data from these arrays is ongoing, but simulations of estimated polarization leakages show that the feeds are not expected to limit the measurements. 

\paragraph{Prospects and R\&D path for CMB-S4}
The technology status level of the feed horn system is 5 for single frequency operation and 4 for dichroic design. Feed horn detector array has deployed in SPT-pol and ACT-pol. Data from dichroic horn system from ACT-pol is currently being actively analyzed. 
The production status level for the current feed horn system is 3 due to challenges associated with scalability of silicon platelette fabrication. 

Several technical aspects of producing feedhorns will need to be addressed for CMB-S4. Current methods of fabricating platelets (a stack of micro-machined silicon wafers) can be time consuming, and production is currently limited to 150\,mm wafers. Mass producing platelets on wafers up to 305\,mm is achievable, but needs to be demonstrated. The deep reactive ion etch (DRIE) rate dictates that a typical feedhorn array requires $\sim$20\,hours of etching to produce all platelets. Additional time is also required for etch preparation and post-etch wafer cleaning. Such work can be outsourced to an industrial micro-electromechanical (MEMs) facility. Laser etching could further expedite the processing time for wafers and can be considered as an alternative. In addition, improved methods of platelet metalization are likely required. On the design side, if broader bandwidth is desired, it is possible to further optimize the spline-profiled design or develop new approaches including a dielectric-loaded feed based on silicon metamaterials. The current OMT design limits the bandwidth ratio (ratio of the highest frequency and the lowest frequency) to $\sim$2.3:1, but the use of a quadridge architecture (a horn with four internal ``fins'') in combination with dielectrically-loaded feeds could open the possibility of 6:1 bandwidth coupling. Finally, there are trade-offs between beam systematic errors and coupling efficiency, especially at small aperture sizes, that must be evaluated based on a system level optimization that includes the telescope and detector array design.

%To best prepare for S4 we should: (1) explore more efficient approaches to manufacturing feed arrays like laser machining, (2) explore broadband designs based on existing and novel approaches to feeds, and (3) continue to develop tools to handle the complex optimization of feeds in light of the full system requirements.

%\noindent\fbox{
%\parbox{\textwidth}{
%\noindent\textbf{Lab demonstration:} 2.3:1 bandwidth, round beam, $<$-18dB cross-pol.  90/150/220/350 GHz\\
%\noindent\textbf{Sky demonstration:} Deployed 90/150 GHz and 150/230 GHz multi-chroic horn with ACT-Pol\\
%\noindent\textbf{Path to CMB-S4:} Speed up production rate with industry fab. Increase total bandwidth (6:1)
%}
%}
\begin{table}[h]
\begin{center}
\begin{tabular} {|l l|}
\hline  
\textbf{Lab Demonstration:} & 2.3:1 bandwidth, round beam, $<$-18dB cross-pol. 90/150/220/350 GHz\\
\textbf{Sky Demonstration:} & Multichroic horns 90/150 GHz (ACTPol) 150/230 GHz (AdvACT)\\
\textbf{Path to CMB-S4:} & Speed up production rate, advance spline-profile design\\
\hline 
\end{tabular}
\end{center}
\end{table}

%
%
%
%\bibliographystyle{unsrt}
%\bibliography{feedhorn.bib}
%
%\end{document}

%% file: detector_rf/Feed_OMT.tex
%\documentclass[11pt, oneside]{article}   	% use "amsart" instead of "article" for AMSLaTeX format
%\usepackage{geometry}                		% See geometry.pdf to learn the layout options. There are lots.
%\geometry{letterpaper}                   		% ... or a4paper or a5paper or ... 
%%\geometry{landscape}                		% Activate for rotated page geometry
%%\usepackage[parfill]{parskip}    		% Activate to begin paragraphs with an empty line rather than an indent
%\usepackage{graphicx}				% Use pdf, png, jpg, or eps§ with pdflatex; use eps in DVI mode
%								% TeX will automatically convert eps --> pdf in pdflatex		
%\usepackage{amssymb}
%\usepackage{sidecap}
%%SetFonts
%
%%SetFonts
%
%
%
%\begin{document}
%%\maketitle
%%\section{}
%%\subsection{}
%
%\noindent {\bf \large Planar OMT coupling} \\
%%\date{}							% Activate to display a given date or no date

\subsubsection{Planar OMT coupling}
\label{sec:hornomt}

\begin{figure}[h]
   \centering
       \includegraphics[width=0.87\textwidth]{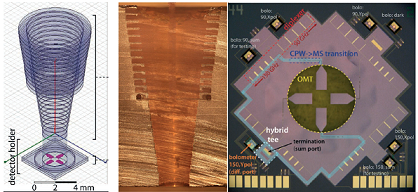} % requires the graphicx package
    \caption{The dual-band, dual-polarization sensitive pixel consists of a silicon-platelet corrugated-feedhorn (cross-section of an actual horn shown) coupled to a polarimeter chip containing Nb probes, diplexers, hybrid tees and TES bolometers.} 
   \label{fig:hornomt}
\end{figure}

\paragraph{Description of the technology}
A feedhorn couples to a planar circuit on a silicon wafer by use of a broad-band, planar OMT comprised of four niobium probes fabricated on a low-stress, silicon nitride membrane as shown in Figure~\ref{fig:hornomt}. 
These fins separate the two orthogonal polarizations and launch radiation onto superconducting coplanar waveguide (CPW) transmission lines (TLs).  
A wide bandwidth stepped impedance transformer is used to switch from CPW to low impedance micro-strip lines that travel to diplexers comprised of resonant stub filters that separate the two frequency bands.  
Light from each pair of opposite OMT probes within a given frequency band are symmetrically fed into a hybrid tee \cite{knochel1990broadband,1742-6596-155-1-012006,4408512} that differences the two signals and leads to single-moded (TE11) output over 2.3:1 ratio bandwidth. 
Unwanted higher order modes are dissipated on the substrate and relative power changes from the lowest order waveguide mode are sensed with TES bolometers. 
%Full details are given in \cite{mcmahon2012multi}. 
Details of planar OMT coupling can be found in \cite{mcmahon2012multi,1742-6596-155-1-012006}, and in \cite{doi:10.1117/12.927056,doi:10.1117/12.2057266,doi:10.1117/12.2234308,Chuss:2016} for implementation on single-crystal silicon for the CLASS focal planes.

\paragraph{Demonstrated performance}
Multiple experiments have been deployed with single color planar OMT coupled feed horns~\cite{abs,actpol,Essinger-Hileman14,sptpol}.
A multichroic polarimeter array covering the 90 and 150\,GHz bands was deployed in 2015 as part of the ACTPol experiment and has logged 2 seasons of observations \cite{actpol,henderson/etal:2016}.
In mid 2016, a second 150/230 GHz array was deployed on the ACT telescopes as the first installment of the Advanced ACTPol instrument.   
%Preliminary analysis shows that the 90/150 multichroic array is the most sensitive of the  three ACTPol arrays, with a noise below $10 \mu$K $\sqrt{s}$ and 85\% end-to-end yield, limited by readout integration \cite{datta2015design} and the data quality from the 150/230 GHz array is on going.
The beams are defined by feedhorns, which offers flexibility to optimize sensitivity and control of systematic errors, and the OMT defines a frequency independent polarization angle offering an advantage for control of polarization mixing effects. 
Multiple deployments of experiments with single colored OMT coupled horns and multichroic OMT coupled horn arrays represent full system demonstrations of this technology and all the ancillary systems, paving the way for their use on even more ambitious future experiments.

\paragraph{Prospects and R\&D path for CMB-S4}
The technology status level and production status level for OMT coupling carries over from the horn coupled detector system. 
The technology status level of the feed horn system is 5 for single frequency operation and 4 for dichroic design. Feed horn detector array has deployed in multiple CMB polarimetry experiments. Data from dichroic focal plane is currently being actively analyzed. 
The production status level for the current feed horn system is 3 due to challenges associated with scalability of silicon platelet fabrication. 

CMB-S4 requires frequency coverage from roughly 30-300\,GHz with potentially finer spectral resolution than what has been deployed to date. 
Frequency scaling above 300\,GHz has been demonstrated with OMT coupled feedhorns.
Work must be done to fully demonstrate detectors with (i) improved spectral resolution, (ii) a low enough frequency limit, and (iii) improved optical coupling efficiency through design and control of dielectrics.  

%Due to the limitations of a single diffracting aperture, octave bandwidth provides much, if not most of the sensitivity gains possible with multichroic polarimeters so we don't view the 2:3:1 ratio bandwidth as a tremendous limitation for this technology.   (This was discussed in bandwdith section)
%However, in some cases there can be benefits from even wider bandwidth. 
OMTs based on a quadruple ridge wave guide are under development to increase the bandwidth of this technology. 
Quad ridge wave guide has been demonstrated in systems with single moded performance in excess of 6:1 ratio bandwidth into Vivaldi style feeds. 
The current design is based on the eVLA 1-2\,GHz receiver and achieves 3.3:1 bandwidth, in line with the current bandwidth limits of our feed horns. 
%We expect to prototype this approach to wide bandwidth in the coming years, but this development could be accelerated if it were deemed crucial to S4. 
%Resources should be allocated to improving dielectrics, optimizing filters for the S4 bands and for low dielectric losses as described above. [This will be discussed in dielectric section]
%Work should be done on quad-ridge wave guide designs with the priority set by with the overall system optimization.

%\noindent\fbox{
%\parbox{\textwidth}{
%\noindent\textbf{Lab demonstration:} SiN dielectric feed achieved 70\% efficiency\\
%\noindent\textbf{Sky demonstration:} Deployed 90/150 GHz and 150/230 GHz multichroic horn with ACT-Pol\\
%\noindent\textbf{Path to CMB-S4:} Low frequency, extend total bandwidth. Quad-ridge 6:1, eVLA like 3.3:1
%}
%}

\begin{table}[h]
\begin{center}
\begin{tabular} {|l l|}
\hline  
\textbf{Lab Demonstration:} & SiN dielectric feed achieved 70\% efficiency\\
\textbf{Sky Demonstration:} & 90/150 GHz (ACTPol) 150/230 GHz (AdvACT)\\
\textbf{Path to CMB-S4:} & Improved spectral resolution, low frequency, coupling efficiency\\
\hline 
\end{tabular}
\end{center}
\end{table}

%\end{document}  

%% file: detector_rf/Feed_MKIDHorn2.tex
\subsubsection{Direct kinetic inductance detector coupling}
\label{sec:mkid_coupling}
%%%%%%%%%%%%%%%%%%%%%%%%%%%%%%%%%%%%%%%%%%%%%%%%%%%%%%%%%%%%%%%%%%%%%%

\begin{figure}[h]
\centering
\includegraphics[width=\textwidth]{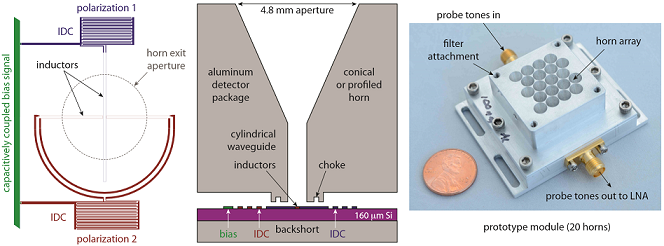}
\caption{
(Left) Schematic of LEKID that is sensitive to one spectral
band centered on 150\,GHz~\cite{mccarrick_2016}.
The $LC$ resonator sensitive to the horizontal polarization is colored
red, while the resonator sensitive to the orthogonal polarization is
colored blue.
The inductor in the resonator is the photon absorber.
The dotted circle represents the waveguide exit aperture at the back
of the horn.
The resonators are driven by a probe tone capacitively coupled to a
TL for read out, which is colored green.
(Center) A cross-sectional view of a single array element.
%
%The LEKIDs are fabricated on silicon and directly illuminated.
%%
%The horn aperture tapers to a cylindrical waveguide which also acts as
%a high-pass filter.
%%
%A choke matches the impedance between the waveguide and the LEKID
%absorber, while also controlling lateral radiation loss along the
%array inside the detector module.
%%
%The aluminum bottom of the module acts as the backshort, and the
%backshort distance is set by the silicon wafer thickness.
%
(Right) A photograph of a 20-element dual-polarization LEKID
module.
}
\label{fig:lekid_figure1}
\end{figure}
%%%%%%%%%%%%%%%%%%%%%%%%%%%%%%%%%%%%%%%%%%%%%%%%%%%%%%%%%%%%%%%%%%%%%%

\paragraph{Description of the technology}
Two RF coupling strategies are currently being developed for MKIDs:
(i) Horn-coupled, multichroic MKIDs and (ii)
dual-polarization lumped-element KID (LEKIDs), which are shown in Figure~\ref{fig:lekid_figure1}.
For the multichroic MKIDs, RF coupling is achieved with an OMT-coupled feedhorn design as described in Sections~\ref{sec:horn} and~\ref{sec:hornomt}~\cite{zeng_2010}.
Multichroic MKID detectors share common RF circuit elements with multichroic TES detectors between the OMT coupling and the RF termination.
RF termination design for MKID detectors is described in Section~\ref{sec:termination}. 
% which are shown in Figure~\ref{fig:mkid_coupling1}~\cite{johnson_2016,mccarrick_2016}.
%
LEKIDs are placed directly at output of a horn. It makes direct polarization sensitive detection of RF power without on-chip RF signal processing.
For the dual-polarization LEKIDs, the planar resonators are made from
a thin aluminum film deposited on a silicon substrate, and they
consist of two orthogonal inductors connected to inter-digitated capacitors (IDCs).
Each resonator is capacitively coupled to a TL, which
carries a GHz probe tone that drives each resonator at its resonant
frequency.
The inductor in the resonator acts as the absorber, which is fed by a
horn that is perpendicular to the silicon substrate.
Each inductor is naturally polarization sensitive, preferentially
absorbing radiation with the E-field aligned to the thin inductor
traces.
The dimensions of the inductor are optimized so the wave impedance is
well matched to the incoming radiation~\cite{bryan_2015a}.
Millimeter-wave photons from the sky absorbed in the inductor break
Cooper pairs, which changes the quasiparticle density.
The quasiparticle density affects the kinetic inductance and the
dissipation of the superconducting film, so a changing optical signal
will cause the resonant frequency and internal quality factor of the
resonator to shift.
These changes in the properties of the resonator can be detected as
changes in the amplitude and phase of the probe tone.

\paragraph{Demonstrated performance}
LEKID detectors that are sensitive to polarized millimeter wave signal were deployed to study galactic magnetic field at 140 GHz and 260 GHz by NIKA \cite{ritacco_2016}.

The 150\,GHz LEKID technology for CMB polarimetry observation has been extensively studied in the
laboratory~\cite{mccarrick_2016,flanigan_2016,mccarrick_2014}, but not
yet demonstrated on the sky.
Similar 1.2\,THz devices are being developed for
BLAST-TNG~\cite{galitzki2014,dober2014}.
Prototype arrays of the multichroic MKIDs will be fabricated starting
in the summer of 2016.
Laboratory studies of these prototype arrays will follow.
%
%Since the multichroic MKID devices are based on the polarimeters that
%were developed for the Advanced ACTPol experiment, the on-sky
%demonstration of the TES version of this technology provides some
%indication of how well the feed and couplign scheme could perform in the
%future~\cite{henderson2016,MCdetectorsDatta}.

%%%%%%%%%%%%%%%%%%%%%%%%%%%%%%%%%%%%%%%%%%%%%%%%%%%%%%%%%%%%%%%%%%%%%%

\paragraph{Prospects and R\&D path for CMB-S4}
%Both RF coupling strategies are being developed with CMB-S4 in mind.
%
The multiplexing factor, which is one of the key advantages of MKIDs,
is largely determined by the quality factor of the resonator and the
bandwidth of the readout.
%
%ROACH-based readout systems currently have approximately 500~MHz of
%bandwidth (see section on MKID readout of the readout instrumentation white paper), and resonator quality
%factors greater than 10,000 are achievable with GHz resonant
%frequencies, so the state-of-the-art multiplexing factor is
%approximately 500.
%%
%Multiplexing factors of 1,000 or more could be possible with some
%research and development.
%
%[Toki Comment: This is readout topic]
In terms of scalability, it should be possible to make the required
multi-kilo-pixel arrays of the dual-polarization LEKIDs now given the
manufacturability of the design.

On sky demonstration of 1.2 THz LEKID detetctors for BLAST-TNG, and further demonstration of 150GHz LEKID detector will reveal competitiveness of this technology.

The technology status level of the LEKID detector system is 2. LEKID system has been demonstrated in laboratory, but deployable focal plane with LEKID detector has not been fabricated and demonstrated. 
The production status level for the current feed horn system is 3. Demonstration of simplicity in microfabrication shows that LEKID is promising technology bring scalability to CMB-S4. PSL for LEKID will advance once the fabrication of large pixel-count LEKID arrays has been demonstrated.

%
%The multi-chroic MKID technology is designed to be scaleable to arrays
%of 10,000 single-polearization detectors or more.
%
%Laboratory demonstration of the multichroic MKIDs in 2016/2017 will reveal how scalable the technology is.

%Horn coupled LEKID\\
%\noindent\fbox{
%\parbox{\textwidth}{
%\noindent\textbf{Lab demonstration:} 150 GHz LEKID studied in the laboratory \\
%\noindent\textbf{Sky demonstration:} 1.2 THz devices are being developed for BLAST-TNG, 140/230 GHz NIKA\\
%\noindent\textbf{Path to CMB-S4:} Scalability is KID detector's strength. Lab demonstration of scalability.
%}
%}
\begin{table}[h!]
\begin{center}
\begin{tabular} {|l l|}
\hline 
Horn coupled MKID & \\ 
\textbf{Lab Demonstration:} & Design is complete. Fabrication will start soon \\
\textbf{Sky Demonstration:} & - \\
\textbf{Path to CMB-S4:} & Scalability is KID detector's strength. Lab demonstration of scalability\\
\hline 
\end{tabular}
\end{center}
\end{table}

\begin{table}[h!]
\begin{center}
\begin{tabular} {|l l|}
\hline 
Horn coupled LEKID & \\ 
\textbf{Lab Demonstration:} & 150 GHz LEKID studied in the laboratory \\
\textbf{Sky Demonstration:} & 1.2 THz devices are being developed for BLAST-TNG, 140/230 GHz NIKA\\
\textbf{Path to CMB-S4:} & Scalability is KID detector's strength. Lab demonstration of scalability\\
\hline 
\end{tabular}
\end{center}
\end{table}

%%  LocalWords:  feedhorn MKID TES inductors IDCs quasiparticle THz
%%  LocalWords:  TNG scalability manufacturability detetctors NIKA
%%  LocalWords:  CMB

%% file: detector_rf/Feed_LensletAntenna.tex
%\documentclass[a4paper,12pt]{article}
%\usepackage{setspace}
%\usepackage{graphicx}
%\usepackage{fullpage}
%\usepackage{parskip}
%\usepackage{multicol}
%\usepackage{subfigure}
%\usepackage{multirow}
%\usepackage{tabularx}
%\usepackage[font=small,labelfont=bf]{caption}
%
%\usepackage{float}
%\floatstyle{boxed}
%
%\begin{document}
%\noindent {\bf \Large Lenslet Coupled Broadband Antenna}
%
%\noindent{Aritoki Suzuki}

\subsection{Lenslet coupled antennas}
Lenslet coupled planar antenna is used in millimeter and sub-millimeter frequencies. 
In this subsection, performance of lenslet coupled broadband antenna and manufacture challenge for anti-reflection coated lenslet array will be discussed. 

\subsubsection{Lenslet coupled broadband antennas}
\label{sec:lensletantenna}

\begin{figure}[h!]
\centering
\includegraphics[width=1.00\textwidth]{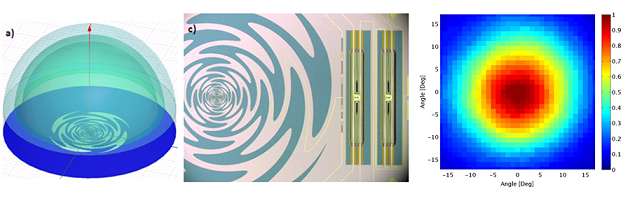}
\caption{(Left) A CAD drawing of a lenslet coupled sinuous antenna designed to cover 90\,GHz and 150\,GHz. The lenslet in this drawing is AR coated with two layers of dielectric. The lenslet is 5.68\,mm in diameter, and the sinuous antenna is 3\,mm in diameter. (Center) A microscope photograph of a fabricated sinuous antenna detector. (Right) A measurement of the beam of the 150\,GHz channel with a two layer anti-reflection coated silicon lenslet.}
\label{fig:LensSinuous_LA}
\end{figure}

\paragraph{Description of the technology} 
Using a contacting lens to increase a planar antenna's gain has been a common technique over a wide range of frequencies, 
including millimeter and sub-mm wavelengths~\cite{Rebeiz}. 
In the ray-optics limit, an elliptical lens collimates rays from a point source placed at the far focus of an ellipse.
%The elliptical lens should have a minor axis to major axis ratio equal to $\sqrt{1-1/n^2}$ where $n$ is the optical index of refraction of the lens.
%In the physical optics limit, a diffraction aperture forms at the tip of an elliptical lens. 
However, a true elliptical lens is difficult and expensive to fabricate. 
To approximate an elliptical lens, it is common to synthesize a lens that is a combination of a hemisphere and an extension. 
%Ratio between radius of a hemisphere and length of a cylinder is a function of optical index of lens material. 
%Although exact value is a function of integrated antenna, this ratio is typically around 0.4 for silicon ($\epsilon_r = 11.7$). 
Coupling this style of lenslet to a planar antenna has multiple benefits. 
First, it increases the antenna's forward gain, efficiently coupling the antenna to the telescope optics. 
%A planar antenna has beam that is as wide as $\pm45^\circ$ half-angle in dielectric. 
%A synthezied elliptical lens collimates beam to form diffraction limited beam. 
Second, a planar antenna placed in a space filled with half air and half dielectric favorably radiates into the dielectric,
and the fraction of energy that radiates into the dielectric increases with increasing dielectric constant. 
Third, the required antenna size for a given frequency also becomes smaller with a higher surrounding material dielectric constant,
freeing up wafer real estate for RF filters, detectors, and inter-pixel wiring.
%Exact fraction depends on integrated antenna, but silicon lenslet {\it pulls} $\sim 95\%$ of radiation towards sky side. 
%Since backlobe of silicon lenslet-coupled antenna is small fraction of a total energy, it is safe to terminate with a cold absorber with minimal efficiency loss \cite{Filipovic, Edwards}. 
%The curved geometry of the dielectric lenslet also suppresses a substrate mode, preventing cross-talk between adjacent pixels. 
%The antenna is smaller than the effective aperture size of a pixel, which is defined by the diameter of the pixel. 
%This gives space on detector wafer to place RF components and bolometers. 
A high dielectric constant lens needs an anti-reflection (AR) coating to suppress reflections, and 
broadband AR coatings for high dielectric constant lenslets have been developed \cite{Siritanasak2016}.
AR coated lenslet arrays are discussed in Section~\ref{sec:lensletarray}.

%A planar antenna interacts with a lenslet to set frequency bandwidth, beam shape, and polarization characteristics. 
A slot antenna is preferred over a wire antenna for the on-chip integrated circuit design because the 
slot antenna and TL can share a ground plane.
This ground plane also provides continuous RF shielding for cryogenic readout electronics.
A lenslet-coupled double-slot dipole antenna was deployed for single-frequency
observations that cover a $\sim 30\%$ fractional bandwidth \cite{pb, Filipovic}. 
%Double slot dipole antena is sensitive to linear polarization. 
%Spacing between two opposite slots can be tuned to create round beam \cite{Zmuidzinas, Filipovic}. 
For broadband applications, a lenslet-coupled sinuous antenna was developed \cite{Edwards, o'brient:063506}.
The sinuous antenna is in a class of antennas called log-periodic antennas, for which the antenna's characteristics repeat every log-frequency cycle.
%For a dual-linear polarization application, the sinuous antenna can be designed with a self-complementary design where the metal and the  slot have an identical shape as shown in Figure~\ref{fig:LensSinuous_LA}. 
%The self-complementary design further stabilizes the antenna's impedance over a wide frequency band \cite{Booker,Deschamp}. 
%A sinuous antenna on silicon-air half-space has impedance of $\sim 100 \Omega$. 
The sinuous antenna's lowest and highest operation frequencies are set by the largest and smallest radius of the antenna respectively, so 
there are no theoretical limits on the operable frequency range of a sinuous antenna. 
Practical limits, such as finite lithography resolution and finite bandwidth of the AR coating on the lenslet, restrict the frequency range \cite{Suzuki2014}. 
%A microstripline is still capable of matching to this impedance. 
%Transmission feed line uses a metal arm of a sinuous as a ground plane to extract signal. 
%Two opposing arms of antenna radiates to create round beam. Antenna is also sensitive to linear polarization. 
Log-periodic antennas are known to have a linear polarization axis that oscillates as a function of frequency, but the amplitude of
this oscillation for a sinuous antenna is relatively small ($\sim\pm 5^\circ$) \cite{Edwards}, 
%Broadband HWPs have a similar polarization angle rotation as a function of frequency.
and data analysis technique has been developed to deal with this effect \cite{Bao1,Bao2}. 
Hardware mitigation has also been implemented in the \Pb-2 and SPT-3G detector array design; 
two types of pixels, each with the opposite handedness of sinuous antenna, were used in the 
detector array, canceling the polarization rotation effect.

\paragraph{Demonstrated performance} 
The \Pb-1 experiment has been observing in the 150\,GHz atmospheric window with a focal plane filled with double-slot dipole antennas with silicon lenslets \cite{pb,arnoldpb1}. 
%The silicon lenslets were AR coated with thermoformed Ultem-1000 plastic \cite{Quealy}. 
The ellipticity of the feed is $< 1\%$, and the cross-polar response is better than -20\,dB in a plane 45 deg rotated from polarization angle where cross-polar response is expected to be highest (D-plane).
The \Pb-2/Simons Array and SPT-3G are deploying with lenslet coupled sinuous antennas \cite{Suzuki2014}. 
%The \Pb-2/Simons Array will use silicon lenslet; SPT-3G will use alumina lenslet. 
%Alumina lenslet is cheaper option. 
%Alumina has slightly different dielectric constant ($\epsilon_r \sim 10$) than a silicon detector wafer, although reflection from such interface is negligible. 
%Silicon lenslet for the \Pb-2 is anti-reflection coated with two-layer epoxy anti-reflection coating. Stycast 2850 FT ($\epsilon_r = 4.95$) and Stycast 1090 ($\epsilon_r = 2.05$) were applied to lenslets with precision machined mold \cite{Rosen, Siritanasak}. Alumina lenslet for SPT-3G is anti-reflection coated with three-layer plastic coating \cite{Joaquin}. Expanded Teflon, RO3035 and RO3006 were pressed formed onto lenslet array. Sinuous antenna for both experiments have inner radius of 20 micron and outer radius of 1500 micron. This sinuous antenna is sensitive from around 50 GHz to around 300 GHz. 
Each \Pb-2 pixel covers the 90\,GHz and 150\,GHz bands; each SPT-3G pixel covers the 90\,GHz, 150\,GHz, and 220\,GHz bands.
The lenslet-coupled sinuous antenna has been demonstrated from the 40\,GHz band to the 350\,GHz band \cite{Westbrook2016},
and has been shown to have a round beam with an ellipticity of $\sim 1\%$ \cite{Siritanasak2016} and a measured cross-polar response of $\sim 1\%$ \cite{SuzukiThesis}.

\paragraph{Prospects and R\&D path for CMB-S4}
The technology status level of the single frequency band lenslet coupled antenna detector system is 5. Single band lenset-coupled antenna detector was deployed in POLARBEAR-1. Systematic errors were studied in detail using on-sky data for the polarization result. The technology status level for multichroic version of lenslet coupled antenna detector with sinuous antenna is 3. Multichroic version has been deployed for SPT-3G, and multichroic focal plane for POLARBEAR-2 is in final stage of integration for deployment. 
The production status level for the current lenslet coupled antenna system is 3. Large quantity ($\mathcal{O}$ 100) of lenslet coupled detector wafers were fabricated for the stage-III experiments. Production rate shows promise to meet high demand ($\mathcal{O}$ 1000) from CMB-S4. 

As described in Section~\ref{sec:bandwidthresolution}, small pixel size (in unit of $F\lambda$) is preferred for a ground based experiment. 
A lenslet-coupled antenna's sensitivity to beam and polarization systematic errors as a function of radius of a lenslet and wavelength should be studied in detail with 3D EM simulators such as HFSS.
A scale model test at lower frequency ($\sim10$ GHz) should also be performed~\cite{Edwards}.

The sinuous antenna's polarization oscillation amplitude can be reduced by decreasing period of fractal repetition as a function of frequency \cite{Edwards}, but 
micro-fabrication becomes more challenging as the expansion factor becomes smaller.
Fabrication of a sinuous antenna at a smaller expansion factor is possible for low frequency ($< 100\,$GHz) with current fabrication methods,
and sub-micron lithography can be explored for sinuous antennas with smaller expansion factors for higher frequencies.

An on-sky demonstration of lenslet-coupled sinuous pixels is happening in 2017 with the \Pb-2 and the SPT-3G experiments. 
A detailed study of systematic errors with actual on-sky data will be important for the development.

%\noindent\fbox{
%\parbox{\textwidth}{
%\noindent\textbf{Lab demonstration:} 5:1 bandwidth. 40 GHz to 350 Ghz band. 2,3, and 7 band pixels. \\
%\noindent\textbf{Sky demonstration:} 90 GHz /150 GHz will deploy with \Pb-2. 90/150/220 GHz will deploy with SPT-3G.\\
%\noindent\textbf{Path to CMB-S4:} On sky demonstration of sinuous antenna detector. Study of systematics with actual data. 
%}
%}

\begin{table}[h]
\begin{center}
\begin{tabular} {|l l|}
\hline  
\textbf{Lab Demonstration:} & 5:1 bandwidth. 40 GHz to 350 GHz band. 2,3, and 7 band pixels. \\
\textbf{Sky Demonstration:} & 90 GHz /150 GHz PB-2 (2017). 90/150/220 GHz SPT-3G (2017).\\
\textbf{Path to CMB-S4:} & On sky demonstration and systematic error study\\
\hline 
\end{tabular}
\end{center}
\end{table}

%\end{document}

%% file: detector_rf/Feed_LensletArray.tex
\subsubsection{Lenslet arrays}
\label{sec:lensletarray}

\paragraph{Description of the technology}
An array of lenslets coupled to planar antennas increases the gain of the antenna array as described in Section~\ref{sec:lensletantenna}.
In this section manufacturing challenge for anti-reflection coated lenslet array will be discussed. 
\Pb/Simons Array and SPT-3G use silicon ($\epsilon_r = 11.7$) and alumina ($\epsilon_r = 9.6$) lenslets respectively,
and a broadband AR coating is applied to the lenslets to suppress reflections.
The details of the AR technologies are given in the AR coating section of the optics chapter (Section~\ref{sec:ARepoxy} and~\ref{sec:ARplastic}).
The fabrication and assembly processes for a monolithic lenslet array will be described in this section.

\begin{figure}[h]
\centering    
\includegraphics[width=6in]{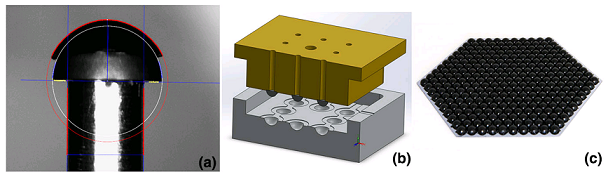}
\caption{(Left) Photograph of a single layer AR coating on a silicon lenslet under a microscope. Photographs were used to inspect the shape of the AR coating. The solid red line indicates a fit to the AR coating and the solid white line indicates a fit to the hemispherical lens. (Middle) The mold for one layer of coating is shown in cross section. The hemispherical lens, which is placed on the seat, is also shown in the drawing. (Right) A photograph of the fully populated broadband AR-coated hemispherical lenslets on the 150\,mm wafer.}
\label{fig:epoxylenslet}
\end{figure}

\begin{figure}[h]
\centering    
\includegraphics[width=6in]{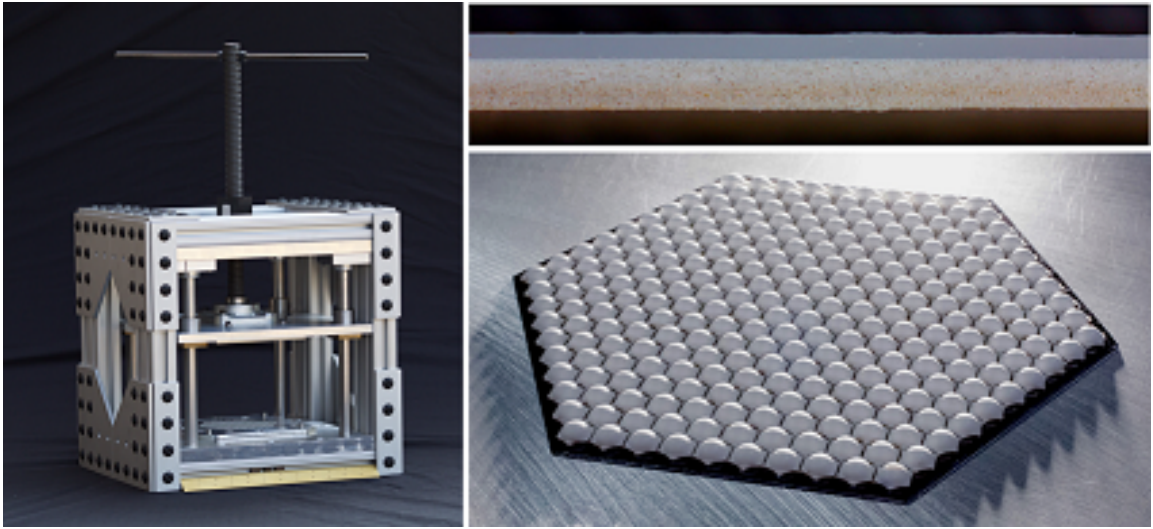}
\caption{(Left) Photograph of the press used for the molding process (Top Right) Cross-sectional image of laminated loaded Teflon sheets. Laminates are Teflon, RO3035, and RO3006. RO3035 and RO3006 are dielectric loaded plastic sheet from Roger Corporation. (Bottom Right) Photograph of a lenslet array for the SPT-3G experiment with 271 lenslets. The Teflon laminates are laser ablated to physically separate lenslets from each other.}
\label{fig:plasticlenslet}
\end{figure}

\paragraph{Demonstrated performance}
Figure~\ref{fig:epoxylenslet} and Figure~\ref{fig:plasticlenslet} show lenslet arrays and assembly jigs for the \Pb{} and SPT-3G experiments respectively~\cite{Siritanasak2016}. 
%Both methods use hexagonal silicon wafers with $\sim100\,\mu$m deep circular pockets to align lenslets. 
%The pockets were patterned with a photo-lithography process and etched with DRIE process. 
%The pockets are larger than the lenslet diameter by $10\,\mu$m to absorb the lenslet manufacturing tolerance.
Both methods populate silicon hemispheres in $\sim100\,\mu$m deep circular pockets etched by micro-fabrication process. 
%The epoxy method coats individual lenslets prior to array assembly.
%Two layers of AR with Stycast 2850 FT and Stycast 1090 were used to make a broadband AR coating \cite{Rosen:13}.
%No extra adhesion layer is required for the process as the epoxy is also the coating material. 
%Metal molds with precisely machined cavities are used to control the thickness of the AR coating to $10 \mu$m, as shown in Figure~\ref{fig:epoxylenslet}.
Two layers of AR coating made of Stycast 2850 FT and Stycast 1090 are applied with metal molds \cite{Rosen:13,Siritanasak2016}, and 
the application of the AR coating and the assembly of the lenslet array are done manually.

%Hemispherical lenslets are placed in each pocket by hand.
%AR coated lenslets are secured to silicon pockets with six small drops Stycast 2850 FT, and drops are deposited with pneumatically controlled dropper.

%For the plastic sheet method, the alumina lenslets were fixed to the silicon wafer before application of AR coating.
%A hemispherical lenslet is placed in each pocket and seated securely by hand.
%The lenslet is then secured to the seating wafer by means of a Stycast 1266 fillet around the lenslet-seating wafer joint.
%The fillet is deposited by hand using a stepper motor driven syringe to ensure a consistent, calibrated volume of epoxy is dispensed about each lenslet.
%The populated array is then allowed to cure to full hardness before further processing.
For the plastic sheet method, three types of loaded PTFE sheets are laminated together with a thermal cycling process, and the 
laminated sheets are molded to conform to the populated lenslet array with a screw-driven die press and system of molds.
%One half of the mold is attached to a moving plate, the other is attached to a stationary plate.
%The two halves are held parallel to to one another by a system of four steel rods and linear bearings; their relative rotation is fixed by dowel pins. 
%The moving plate is driven by a case-hardened ball screw, which is turned by hand.
%The force of the screw is transmitted through the moving plate, through an assembly of high-load die springs, through the upper mold half, and onto the AR coating.
%The known properties of the springs allow us to exert a calibrated, repeatable amount of force on the coating to ensure consistency between molding operations. 
Once the coating has been molded, it is attached to the populated lenslet array using a calibrated volume of Stycast 1266 and allowed to cure.
%The process of molding the AR coating decreases the thickness by $\sim10$\%.
The final 3\,mm-thick molded AR coating is repeatable to $\sim 20\,$$\mu$m.
%Without some form of relief, at cryogenic temperatures the differential thermal contraction between the silicon wafer substrate and the PTFE AR coating causes the lenslet array to break.
A 30W CO$_{2}$ laser was used to cut relief cuts between lenslets to mitigate delamination from differential thermal contraction between the AR coating and the lenslet.
%catastrophe by minimizing the total contractile area of any one PTFE element.
%The laser ablation is fast, accurate, repeatable, and---unlike traditional dicing and cutting methods---exerts no tool pressure on the fragile silicon substrate.
%Each lenslet is physically decoupled from its neighbors; after dicing, the lenslet array consists of 271 individual, AR coated optical elements.

\paragraph{Prospects and R\&D path for CMB-S4}
The technology status level of the lenslet array for lenslet coupled antenna detector system is 5 for single frequency band system and 3 for multichroic detector. As mentioned in previous section, lenset coupled antenna detector was deployed in POLARBEAR-1. Systematic errors were studied in detail using on-sky data for the polarization result. Multichroic version has been deployed for SPT-3G, and multichroic focal plane for POLARBEAR-2 is in final stage of integration for deployment. 
The production status level for the current lenslet coupled antenna system is 3. Large quantity ($\mathcal{O}$ 30) of lenslet coupled detector wafers were fabricated for the stage-III experiments at production rate of approximately 1 array per week. Current production method still involves manual labor, which may limit throughput for mass production in future.

Several improvements can be made to the lenslet array fabrication process in order to increase throughput and repeatability.

At present, populating the silicon lenslet array is performed by hand.
While by-hand assembly is feasible now---for experiments with $\mathcal{O}(10,000)$ pixels---it will not be feasible for future experiments with $\mathcal{O}(100,000)$ pixels.
Many, if not all, of the epoxy dispensing steps in the fabrication process can be adapted for computer numerical controlled devices, increasing precision and repeatability, and decreasing the time spent in fabrication.
%The requirements for CNC adaptation are modest; a two-axis gantry, rotating stage, a series of stepper motors, and control software are the primary components.
%The CNC components require no development; the CNC industry is very mature, and there are countless resources and vendors to choose from.

The plasma spray technique described in Section~\ref{sec:ARthermalspray} can also be used to AR-coat lenslet arrays.
The process is fast and fully automated, and is a scalable technology for CMB-S4.
R\&D is required to verify coating thickness uniformity across a lenslet array.

%\noindent\fbox{
%\parbox{\textwidth}{
%\noindent\textbf{Lab demonstration:} Silicon and alumina lenslet array. Two layer epoxy AR coating and three layer plastic AR coating.\\
%\noindent\textbf{Sky demonstration:} 271 pixel array: 90 GHz /150 GHz will deploy with \Pb-2. 90/150/220 GHz will deploy with SPT-3G.\\
%\noindent\textbf{Path to CMB-S4:} Increase production rate with automation, new AR coating method such as plasma spray
%}
%}

\begin{table}[h]
\begin{center}
\begin{tabular} {|l l|}
\hline  
\textbf{Lab Demonstration:} & Silicon and alumina lenslet array. Epoxy AR coating and plastic AR coating\\
\textbf{Sky Demonstration:} & 90/150 GHz PB-2, 2017. 90/150/220\,GHz SPT-3G, 2017\\
\textbf{Path to CMB-S4:} & Increase production rate with automation and faster AR coating method\\
\hline 
\end{tabular}
\end{center}
\end{table}

%% file: detector_rf/Feed_GRIN.tex
%\documentclass[12pt,letterpaper]{article}
%\usepackage{times}
%\usepackage{amssymb,amsmath}
%\usepackage{dcolumn}
%\usepackage{array}
%\usepackage[pdftex]{graphicx}
%\usepackage{sidecap}
%\usepackage{wallpaper}
%\DeclareGraphicsExtensions{.jpg,.pdf,.png}
%\graphicspath{{figures/}}
%\usepackage{color}
%\usepackage{multirow}
%\usepackage{sidecap}
%
%\topmargin=0.0in \headheight=0.0in \headsep=0pt \textheight=9.in
%\textwidth=6.5in \oddsidemargin=0.in
%\setlength{\parskip}{5pt}
%
%\begin{document}
%
%
%\onecolumn
%
%\onecolumn
%
%\setlength{\baselineskip}{0.95\baselineskip}
%%use below for double space
%%\setlength{\baselineskip}{1.95\baselineskip}
%%use below for single space
%\setlength{\parskip}{1.\parskip}
%
%\subsubsection{Metamaterial Lenslet Arrays}

\subsubsection{Metamaterial lenslet arrays}
\label{sec:grin}

\paragraph{Description of the technology}
As an alternative to hyper-hemispherical lenslet arrays,
planar lenslet arrays using metamaterials can be fabricated using
silicon wafers. Instead of curved optical surfaces, the lenslets
consist of a stack of silicon wafers each patterned with a periodic
array of subwavelength features. Two approaches can be used,
gradient index (GRIN) lenslets produced by etching radially varying
holes in the wafers, and metal-mesh lenslets produced by depositing a
radially varying metal mesh grid that acts as a series of TL lumped element filters to control the wavefront phase delay
across the lenslet. Metamaterial lenslets can
be fabricated using standard lithographic techniques on silicon wafers
in only a few steps, they are precise, repeatable, and scalable to mass
production, and the flat optical surface lends itself to a variety of
broadband AR coating techniques, including impedance
matching to free space using the metamaterial itself. Also, since both the
detector arrays and lenslet arrays are patterned on silicon, there is
no differential thermal contraction between the two, and this allows
them to be designed together in close proximity, accounting for
electromagnetic interactions.

\begin{figure}
\vspace{-0.25in}
        \begin{center}
        \includegraphics[height = 1.6in]{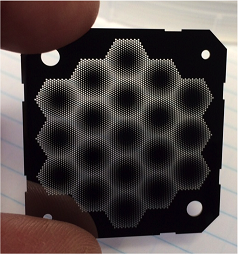}
        \includegraphics[height = 1.6in]{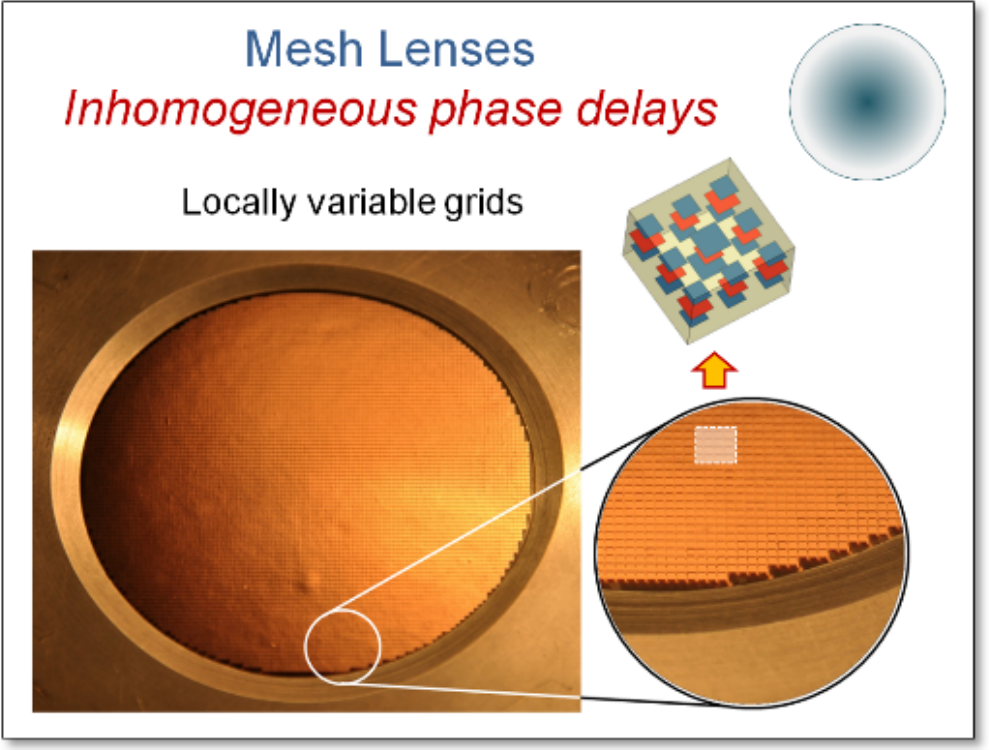}
        \includegraphics[height = 1.6in]{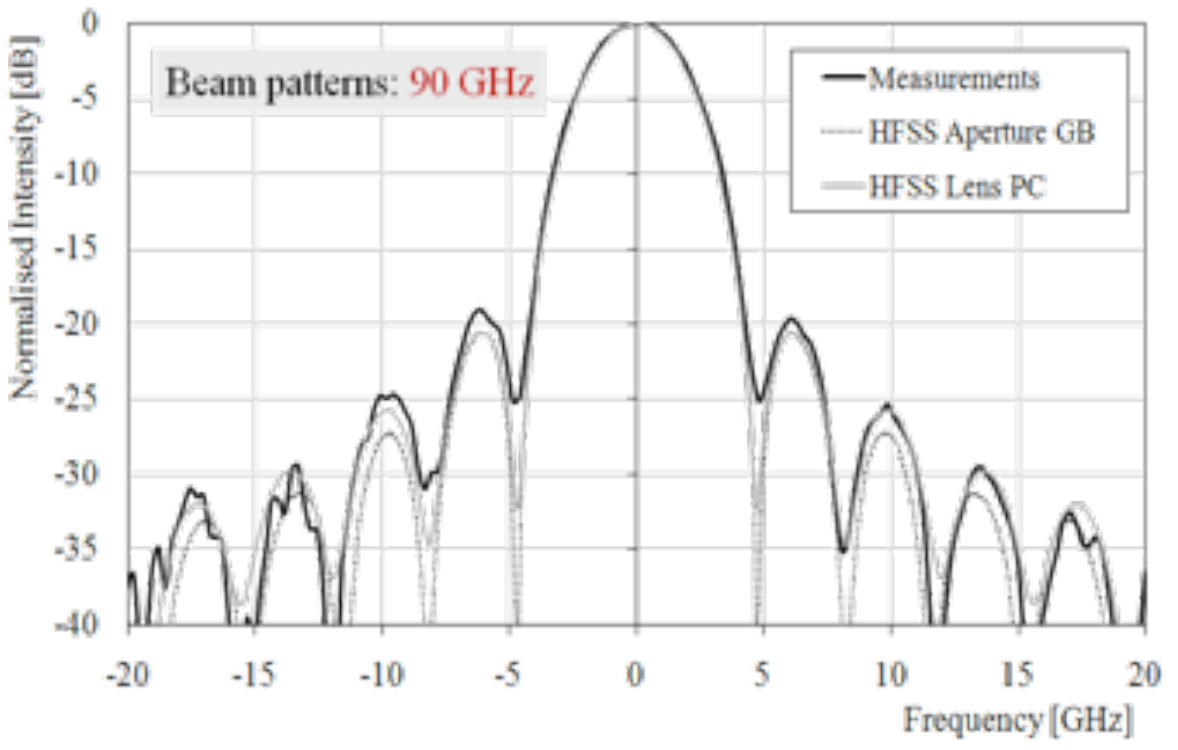}
        \end{center}

        \caption{(Left) Photograph of a single GRIN
          lenslet array layer showing the etched hole pattern. Eight
          wafers are stacked together to form the prototype 19-element
          GRIN lenslet array. Each lenslet is 6.8-mm
          diameter. (Center) 54-mm diameter metamaterial
          metal-mesh lens photograph and concept \cite{pisano13}. (Right) Experimental
          measurements of the beam created by the mesh lens. Notice
          the agreement between models and data down to the fourth
          sidelobes \cite{pisano13}.
       \label{fig:metamateriallenslets}}
\end{figure}     

\paragraph{Demonstrated performance}

Only recently have grooved or perforated dielectrics been studied to
produce GRIN lenses at submillimeter wavelengths. For example, a
single-layer etched GRIN was tested as a candidate lenslet array at
350\,$\mu$m wavelength with the MAKO \cite{swenson12} instrument (Chris
McKenney, private communication), and a single wafer GRIN lenslet
array using a $120\,\mu$m hole pitch on a $100\,\mu$m thick silicon wafer has
been demonstrated with broadband operation from 0.3--1.2\,THz
\cite{park14}. 
%Recently, the University of Colorado Boulder (CU) and
%the National Institiute of Standards and Technology (NIST) have
Recently, a 19-element prototype GRIN lenslet subarray was designed and fabricated for mm-wave application
(Fig.~\ref{fig:metamateriallenslets}, left panel). The
prototype array is being optically tested using a single-pixel
prototype \Pb-2 sinuous-antenna coupled dual-polarization
90/150~GHz TES detector. Preliminary measurements show that the
optical efficiency is similar to that of the same detector mounted to
a conventional AR-coated hemispherical lenslet.

Single meshes or combinations of different grids have long been used
to form low-pass, high-pass, band-pass, and dichroic spectral filters
(see, e.g., \cite{ade:meshfilters}), and the same technology has been further
developed to realize phase retarders such as mesh half-wave plates and
mesh quarter-wave plates \cite{pisano12}. Recently, a metal-mesh metamaterial lens was developed\cite{pisano13}. 
A 54-mm diameter W-band (70--115\,GHz) metal-mesh lens was demonstrated by using stacks of spatially varying inhomogeneous grids
(Figure~\ref{fig:metamateriallenslets}, center panel). The lens does not
need an AR coating since all the cells of the surface are
optimized to be impedance matched to free space. Experimental measurements of the
mesh lens beam pattern agree well with HFSS simulations
(Figure~\ref{fig:metamateriallenslets}, right panel).

\paragraph{Prospects and R\&D path for CMB-S4}
%Collaborators at CU, NIST, Cardiff, and UC Berkeley have proposed to develop metamaterial lenslet arrays for CMB and submillimeter applications. 
In principle, the technology is scalable to mass production.
For both the etched-hole and metal-mesh lenslets, stacking and
alignment will be performed using alignment features and notches
fabricated on the individual wafers to align the layers, and the
layers will then be glued together using Stycast in vertical channels on
the edge of the stack, similar to the method developed for the silicon corrugated feed arrays. 

The technology status level for the technology is 2. Laboratory test has been done to show that technology works at small array level. 
Demonstration of deployable size array with study of beam and polarization systematics will push the technology status level to next level.
The production status level for the technology is 1. Fabrication of proto-type 19 pixel model for demonstration of technology status level 1 and 2 has been done. 

%\noindent\fbox{
%\parbox{\textwidth}{
%\noindent\textbf{Lab demonstration:} Designed and fabricated 19-element lenslet array. Efficienct is similar to hemispherical lenslet.\\
%\noindent\textbf{Sky demonstration:} - \\
%\noindent\textbf{Path to CMB-S4:} Fabricate and demonstrate matamaterial lenslet arrays for mm and sub-mm array \\
%}
%}

\begin{table}
\begin{center}
\begin{tabular} {|l l|}
\hline  
\textbf{Lab Demonstration:} & Fabricated 19-element lenslet array. Efficiency is similar to hemispherical lenslet \\
\textbf{Sky Demonstration:} & - \\
\textbf{Path to CMB-S4:} & Demonstrate metamaterial lenslet arrays with antenna-detector array\\
\hline 
\end{tabular}
\end{center}
\end{table}

%
%\renewcommand{\refname}{\section{References and Citations}}
%
%\newpage
%
%\bibliographystyle{plain}{}
%\bibliography{mmlenslet}%, bib,refs}
%
%
%
%\end{document}

%% file: detector_rf/Feed_AntennaArray.tex
\subsection{Antenna array coupling}
\label{sec:antennarray}

\begin{figure}[!htbp]
\centering
\includegraphics[height=2 in]{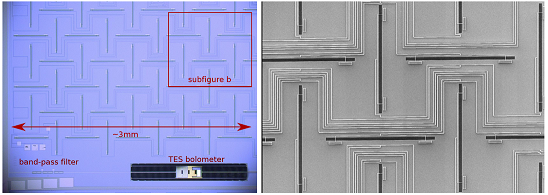}
\caption{(Left) Pixel design of the antenna array. (Right) a zoom in of the left-hand panel.  Dark lines are slots in Nb ground plane. \label{fig:phasearraydesign}}
\end{figure}

\paragraph{Description of the technology}
To facilitate rapidly deploying over 10,000 detectors in the \bicepI, Keck Array, and \spider\ experiments, planar antenna-array coupled detectors have been developed. This design eschews large bulk coupling optics such as horns or contacting lenses and instead synthesizes a beam from coherently fed sub-antennas \cite{Kuo_SPIE}, all fabricated entirely through photolithographic means.  Figure~\ref{fig:phasearraydesign} shows the design of the antenna array from one pixel. The sub-antennas are slots carved into a superconducting niobium film and their waves are captured and summed in an integrated niobium microstrip circuit that uses the metal around the slots as a ground plane. 

Each pixel contains two interleaved co-centered antenna arrays that receive the two orthogonal linear polarizations and couples them to two independent microstrip feeds. The feed combines waves in microstrip T-junctions. It is possible to control the optical mode to which the detectors couple by choosing the impedance of the lines at the junctions as well as the length of line to the adjacent junctions.  This design avoids microstrip cross overs, simplifying the fabrication by reducing the number of required depositions and etches as well as obviating superconducting vias between layers. 

\paragraph{Demonstrated performance}
The antenna-array coupled design is mature, thanks to deployment of 88~tiles into scientific experiments where they were subjected to exhaustive analysis. These measurements have demonstrated an antenna band that is nearly 50\% wide, but limited to~30\% by integrated band-defining microstrip filters centered at 90, 150, and 220\,GHz. The end-to-end optical efficiencies are~40\% in the deployed cameras. Detectors have been developed that cover the 40\,GHz and 270\,GHz bands for the \bicepArray\ that will deploy in 2018. 
%\begin{figure}[!htbp]
%\centering
%\includegraphics[height=2 in]{figure/det_spectra.eps}
%\includegraphics[height=2 in]{figure/OE_B2.eps}
%\caption{Optical Performance of Detectors.  \label{fig:phasearraybandwidth}}
%\end{figure}

\begin{figure}[!htbp]
\centering
\includegraphics[height=2 in]{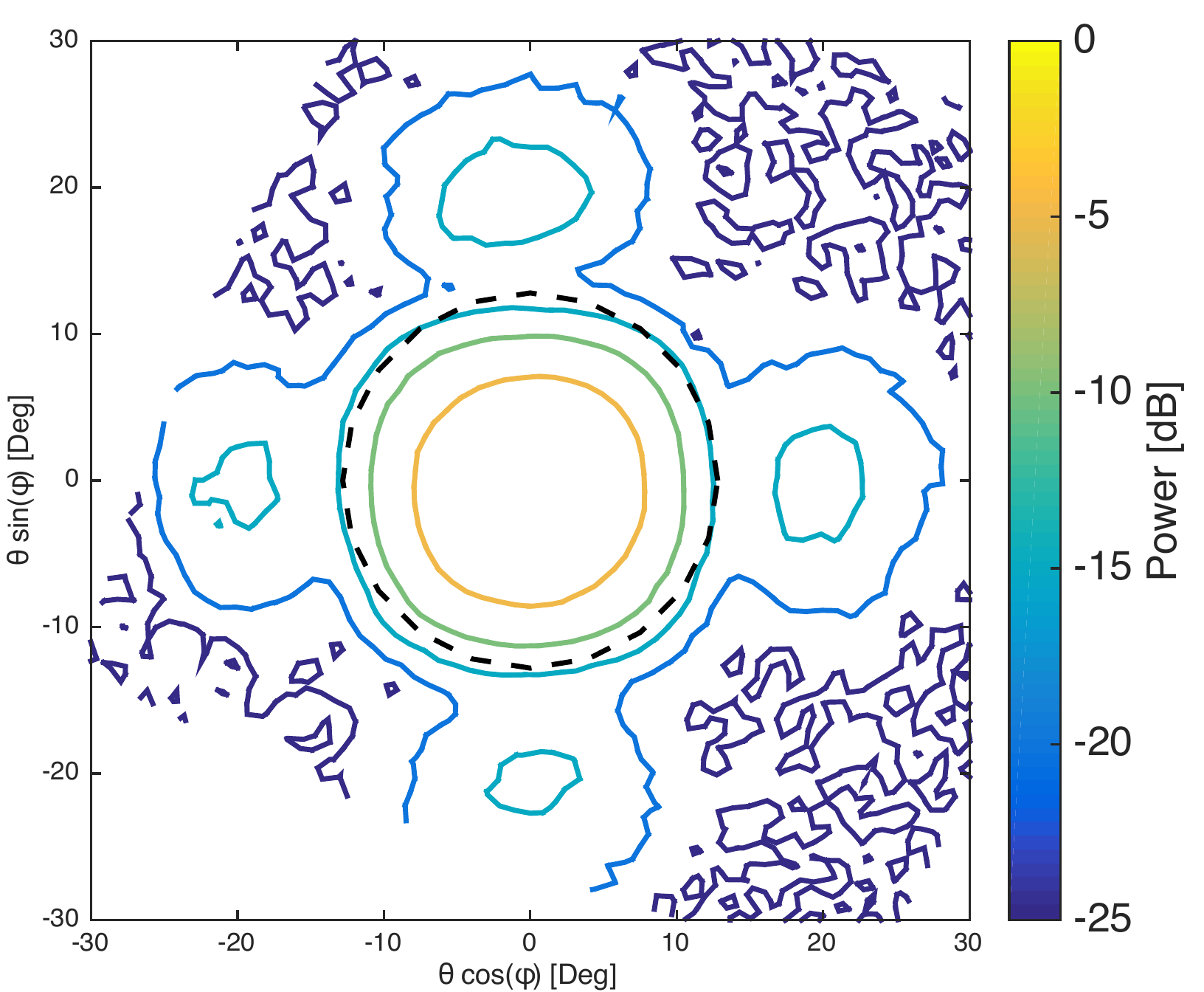}
\includegraphics[height=2 in]{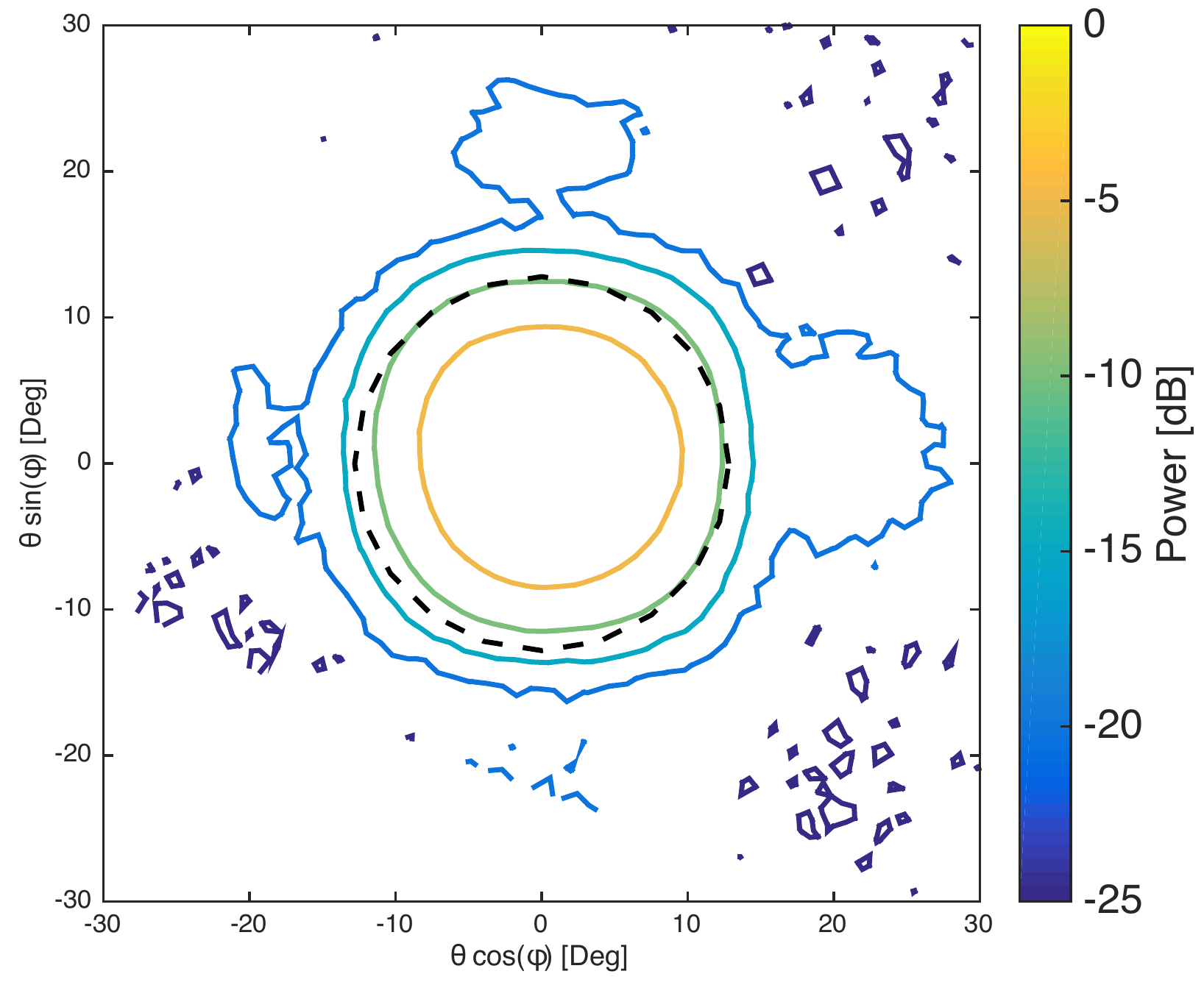}
\caption{Measured beam patterns for pixels formed from antenna arrays, without refracting optics. (Left) Antenna beam pattern from an antenna array with top hat illumination pattern. (Right) Antenna beam pattern with Gaussian illumination pattern. \label{fig:phasearraybeam}}
\end{figure}

Early designs used a top-hat illumination of the antenna, which couples to sinc-patterned modes in the detectors' far field. While these are acceptable for \bicepI-style refracting telescopes with a well-controlled cold 4-K aperture, other optical design require lower-sidelobe levels to limit detector loading from warmer surfaces. Detectors in the \bicepIII\ telescope have a Gaussian illumination, controlled through the impedance of the transmission lines at the T-junctions. These receive more power in the center than edge, dropping side-lobe levels by nearly 10\,dB~\cite{OBrient2012}. Figure~\ref{fig:phasearraybeam} shows a comparison of the feeds' performance. In principle, it is possible to match to more exotic illuminations, such as sinc-patterns that overlap between pixels in the illumination tails and synthesize top-hats in the telescope aperture, providing very high optical throughput. However, implementing such a design would require multiple ground planes and myriad microstrip cross-overs.

\paragraph{Prospects and R\&D path for CMB-S4}
The technology status level of the antenna array system is 5. The antenna array system has has deployed in multiple CMB polarimetry experiments (series of BICEP experiments and SPIDER). Detailed systematic error studies have been done for published results. 
The production status level for the current feed horn system is 4. Large quantity of antenna arrays have been fabricated and deployed to multiple experiments. Lithographed beam forming elements that is unique to this technology provides scalability to the technology. 

Mode coupling can be further customized by altering the relative phase between sub-antennas. For example, by increasing the length of the lines leading to the sub-antennas in a way that linearly increases across a pixel's array, it is possible to couple to modes whose boresight is angled away from the focal-plane's normal vector. In this way, the detector naturally accommodates non-telecentric optical designs. The phase could be varied quadratically across the pixel, which would allow pixels to couple to waves that have a waist off the physical detector tile locations, accommodating optics with curved focal surfaces as well. Detectors with linear phase shifts have been fabricated, but we have yet to fabricate higher order phase profiles. 

%\begin{figure}[!htbp]
%\centering
%\includegraphics[height=2 in]{figure/steer_histo.png}
%\caption{Histogram of steered centroids, measured relative to other un-steered polarization.  No refracting optics used.  Beam width is 14$^o$, as in Figure ~\ref{fig:phasearraybeam} \label{fig:phasearraysteer}}
%\end{figure}

Multiple antenna-array designs have been explored to extend detector bandwidth. Arrays of ``figure-eight'' antennas, reminiscent of bow-ties, can provide in excess of an octave bandwidth-- more than enough for multi-color pixels. These are currently under development.  Another far more ambitious possibility is building focal planes where slot array from two orthogonal polarizations and frequency bands are interleaved. If the detectors are all at the edge of such an array, then the beams from different frequency bands could be independently tuned to match the optics, providing a highly efficient use of focal plane real estate. Implementation of this concept presents similar engineering challenges as for the sinc-illuminations described above.

\begin{figure}[!htbp]
\centering
\includegraphics[height=2 in]{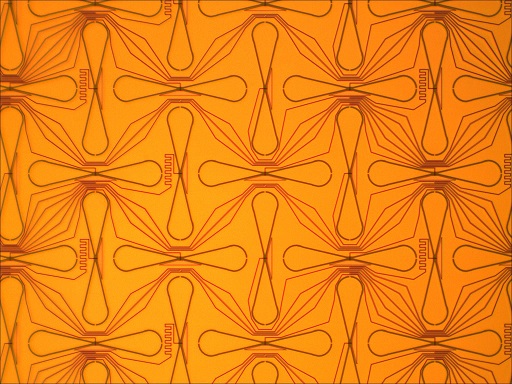}
\includegraphics[height=2 in]{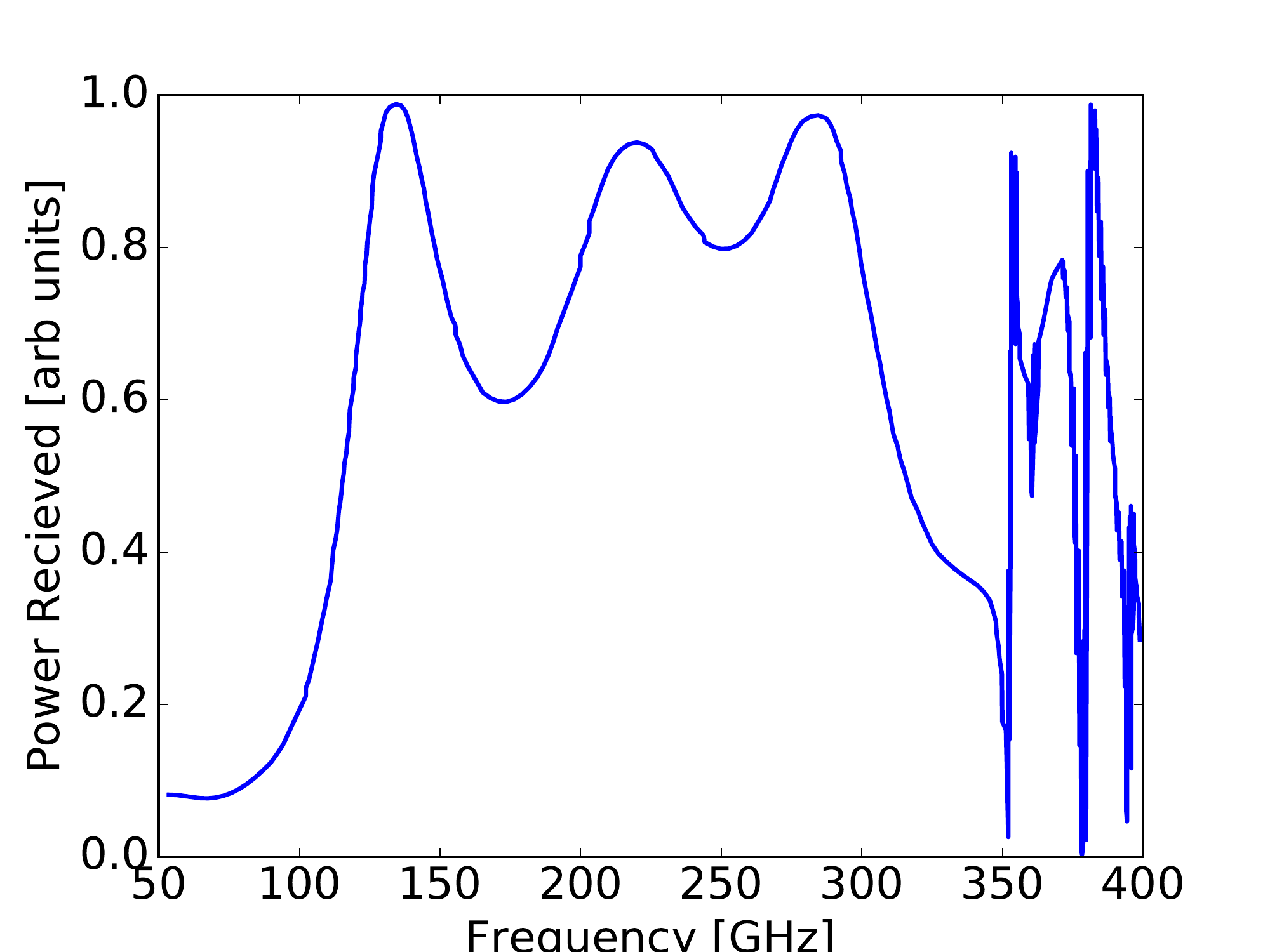}
\caption{(Left) Photograph of a broadband antenna array. (Right) Simulation of received power by the broadband antenna array as a function of frequency. \label{fig:phasearrayrd}}
\end{figure}

%\noindent\fbox{
%\parbox{\textwidth}{
%\noindent\textbf{Lab demonstration:} 270 GHz detector arrays. Design and fabricated broadband (2:1) antenna array \\
%\noindent\textbf{Sky demonstration:} Deployed 90, 150, 220 GHz in the \bicepI, Keck Array, and \spider\ \\
%\noindent\textbf{Path to CMB-S4:} Beam steering with customized phasing for non-telecentric coupling
%}
%}

\begin{table}[h]
\begin{center}
\begin{tabular} {|l l|}
\hline  
\textbf{Lab Demonstration:} & Design and fabricated broadband (2:1) antenna array \\
\textbf{Sky Demonstration:} & Deployed 90, 150, 220, 270\,GHz in the \bicepI, Keck Array, and \spider\ \\
\textbf{Path to CMB-S4:} & Beam steering with customized phasing for non-telecentric coupling \\
\hline 
\end{tabular}
\end{center}
\end{table}

%% file: detector_rf/Feed_Multimode.tex
\subsection{Direct coupling to single and multimoded resistive absorber bolometers}
\label{sec:multimode}

\paragraph{Description of the technology}
To meet the evolving demands of CMB science, technological advancements have focused on improving array sensitivity. 
Rather than increase the number of sensing elements in the focal plane, multimode devices use fewer, larger absorbing structures to collect photons in more than one spatial mode. 
The simplest such absorber is a resistive sheet. 
Depending on the optical coupling between the absorber and the sky, such a sheet can be operated in single-moded or multimoded configurations. 

\paragraph{Demonstrated performance} 
As an example of a single-moded implementation, the Millimeter Bolometric Array Camera (MBAC) on the Atacama Cosmology Telescope used pop-up bolometers with impedance-matched solid silicon sheet absorbers \cite{swetz:2011}. 
More generally, planar absorbers can be a realized on a thin membrane, i.e. SiN or SiN$_x$, or solid dielectric substrate.
This technology was first used by the SHARC-II instrument for the Caltech Submillimeter Observatory \cite{dowell:2003}
A further evolution of the pop-up design, the Backshort-Under-Grid (BUG) detector array with individual absorber backshorts has found use in sub-mm and IR polarimetric experiments when combined with a polarizing wire-grid analyzer. 
Examples are the HERTZ polarimeter \cite{schleuning:1997}, the polarimeter for SCUBA-2 \cite{bastien:2005}, the SHARC polarimetric instrument SHARP \cite{li:2006}, and the CMB instrument PIPER \cite{benford:2010}.

In contrast, a multimoded polarimeter can be formed from inherently polarization-sensitive sheet absorbers. 
For the PIXIE mission \cite{kogut:2011}, a freestanding grid of doped silicon wires forms the detecting element of a space-based FTS. 
This ``harpstring'' absorber enables broadband detection of polarized optical power between 30\,GHz and 6\,THz. 
The detectors would achieve low levels of crosspolar response, measured as individual detector response to incoming orthgonal polarization, when two orthogonally-sensitive harpstring detectors are mounted together \cite{kusaka:2012}.

\paragraph{Prospects and R\&D path for CMB-S4}
The technology status level of the direct coupling to single and multimoded resistive absorber bolometer system is 3. Experimental capable version has been fabricated and tested in laboratory for CMB instrument PIPER.
The production status level for the current feed horn system is 3. Fabrication for resistive absorber bolometer is done using micro-fabrication technique. Such technique is suitable for scalable fabricatio. The production status level will advance once mass fabrication for this type of detector has been demonstrated.

Polarization-sensitive multi-moded detector pairs can also be formed into arrays. Using many fewer detectors, such arrays would achieve equivalent sensitivity to current or future kilopixel arrays \cite{kusaka:2014}.
The reduced angular resolution of a multimode detector would not affect a target of measuring B-modes at large angular scales sourced by primordial gravitational waves.
Current detector development is focused on verifying thermal transport and optical response of the harpstring absorber. Initial results for prototype bolometers reported on performance of ion-implanted semiconductor thermistors in the harpstring frame \cite{nagler:2016}.

%\noindent\fbox{
%\parbox{\textwidth}{
%\noindent\textbf{Lab demonstration:} Direct absorber with polarizing wiregrid analyzer, BUG Detector for PIPER, PIXIE \\
%\noindent\textbf{Sky demonstration:} MBAC, SHARC-II \\
%\noindent\textbf{Path to CMB-S4:} Fabrication of polarization-sensitive multi-moded detector array.  Verifying thermal transport and optical response of the harpstring absorber. 
%}
%}

\begin{table}[h]
\begin{center}
\begin{tabular} {|l l|}
\hline  
\textbf{Lab Demonstration:} & Direct absorber with wiregrid polarizer, BUG Detector for PIPER, PIXIE \\
\textbf{Sky Demonstration:} & MBAC, SHARC-II \\
\textbf{Path to CMB-S4:} & Fabricate polarization-sensitive multi-moded detector array \\
& Verify thermal transport and optical response of the harpstring absorber\\
\hline 
\end{tabular}
\end{center}
\end{table}

%% file: detector_rf/RF_Intro.tex
%Coupling optical signal onto a lithographed superconducting circuit enabled a possibility of operating on a signal before detection. 
%RF engineering techniques were applied to perform multiple useful tasks such as multiplexed band pass filtering and mode rejection. 
%RF engineering techniques for the CMB application matured through stage-2 and stage-3 CMB experiments.
%For a optimal RF performance RF design, it is desireble to have low loss dielectric, stable material properties and fabrication capability to fabricate designed layout repeatedly. 
%Low loss dielectric decreases absorption loss in a trasmission line, and it also allows for more flexible band-pass (and band-stop) filter design without incurring extra loss.
%Stable material property allows EM simulation to predict how detector would perform accurately.
%Fabrication capability and uniformity ensures that designed RF circuit will be realized uniformily across detector array.
%This section will survey different RF design approaches employed by multiple CMB experiments.

Contemporary CMB detectors typically employ low-loss superconducting TL to convey the optical signal from the RF feed to the detector where the signal is thermalized and measured. The use of planar TL enables implementation of traditional RF circuit elements for signal processing prior to detection. Realized applications include beam synthesis as part of phased antenna arrays, mode rejection and passband definition with the latter including channelizing the signal into multiple passbands. Applying these RF engineering techniques to CMB applications is now a mature technology having been successfully implemented in Stage-II and upcoming Stage-III experiments. The RF circuit design needs to occur within the broader context of detector fabrication and testing in order to yield structures that can be reliably and uniformly fabricated without repercussions to other detector components. In this section, we survey different RF circuit components employed across multiple CMB experiments.

%% file: detector_rf/RF_Tline.tex
%%%%%%%%%%%%%%%%%%%%%%%%%%%%%%%%%%%%%%%%%%%%%
%\documentclass[12pt, letter]{article}
%\begin{document}

\subsection{Superconducting RF transmission line}
\label{sec:rf_tline}

\paragraph{Description of the technology}
Microfabrication of CMB detectors on silicon wafers with lithography technique enabled on-chip RF signal processing. 
Typical RF circuitry used in CMB detectors utilizes both CPW and microstrip TLs where the conducting metal is a superconducting film, typically $\sim$300\,nm of $\mathrm{Nb}$. To maximize detector performance, impedance, absorptive loss, and radiative loss of TLs should be carefully thought through. CPW structures have higher impedance and typically lower loss compared to microstrip. Radiative losses are important for CPW structures and need to be minimized through design considerations. Microstrip TLs have substantially less radiation compared to CPW, but suffer from losses in the dielectric material separating the conductor strip from the ground plane. A review of superconducting planar TL technology detection of the CMB is given by U-yen, Chuss and Wollack \cite{U-yen:2008}.

\paragraph{Demonstrated performance}
Low-loss TL is essential for providing flexibility in the detector RF circuit design.
The dielectric loss can be parameterized by the loss tangent, defined as $\tan~\delta = \epsilon''/\epsilon'$, where $\epsilon=\epsilon'+i\epsilon''$ is the complex dielectric constant. 
Fielded systems have typical loss tangents of $\tan \delta <5\times10^{-3}$. 
Current CMB detectors have explored a number of dielectric materials including: silicon oxide, silicon nitride and single crystal silicon. 
Silicon oxide is the most common dielectric and has a dielectric constant of $\sim 3.8$ and loss tangent of $\sim 5\times10^{-3}$.
Increasing the silane-to-oxygen ratio during plasma-enhanced chemical-vapor deposition of the silicon oxide improves the dielectric loss-tangent from $6\times10^{-3}$ for stoichiometric silicon dioxide to $2\times10^{-3}$ for a more silicon-rich silicon oxide \cite{DaleNISTOxide}.
Similar to silicon oxide, silicon nitride, which has a dielectric constant of $\sim 7.0$, % and loss tangent of $\sim 1\times 10^{-3}$.
can also be made silicon-rich thus reducing the dielectric loss tangent from $1.2 \times10^{-3}$ to $2.5 \times10^{-5}$ \cite{Paik}.
For example, the 150/230\,GHz Advanced ACTPol array with low-loss SiN dielectrics and appears to have a dielectric efficiency in line with the predictions of $\sim$70\%. 
Single crystal silicon has a dielectric constant of $\sim 11.7$ and loss tangent of $\sim 1 \times 10^{-5}$ or better.
Microstrip TL can be fabricated with single crystal silicon dielectric by using Silicon-on-Insulator (SOI) wafer \cite{CLASSFab,Denis16} processing.
%Detector fabricated for CLASS experiment has feed-to-detector efficiency of greater than $\sim$90\%. 
 Detectors fabricated for the 38 GHz channel of the CLASS experiment have been demonstrated to achieve feed-to-detector efficiency of $\sim$90\% \cite{doi:10.1117/12.2057266, doi:10.1117/12.2234308, Chuss:2016}. 
 %These detectors uniquely employ an integrated silicon backshort for the OMT probes [190, 217]. The metalized silicon enclosure that defines the backshort also serves to reduce the coupling of high-frequency stray radiation to the detectors, mitigating the so-called "blue-leak" effect [190, 217, C, D, E, 194]
%Single crystal silicon has a predictable EM performance due to its controlled dielectric properties and thickness in the SOI process \cite{Cataldo}.
Single-crystal silicon has a predictable EM performance [191] and provides uniformity in microwave properties and substrate thickness over wafer batches\cite{Cataldo}.

The typical conducting material for the RF TL is niobium, a superconductor with superconducting transition temperature $\sim 9$ Kelvin. 
%Niobium can support signals with frequencies up to $\sim 700$\,GHz without breaking Cooper pairs.
%The practical London penetration depth for sputtered niobium is $\sim 100\,\mu$m. 
%Changes in this penetration depth from various effects such as film stress, contamination and temperature can change the kinetic inductance of the superconducting niobium film. 
%For this reason, typical superconducting TLs are a few penetration depths thick ($\sim$300\,nm), providing a stable wave speed, which is important for RF filter performance. % as a function of conductor thickness.
%Thus $\sim$300~nm is used for CMB detectors where stable wave speed is important for RF filter performance.
Changes in quality of Nb from from various effects such as film stress, contamination and temperature can change loss and the kinetic inductance of the superconducting niobium TL. 
%Also, it is reported that detectors fabricated with tensile niobium have poor detector efficiency \cite{BICEPFab}.
Most CMB detectors use compressive Nb films where the stress is tuned to be around $0 \sim 500$\,MPa.
%Both the dielectric and metal films need to be near zero stress because the TES bolometer is a suspended structure on a thin low stress silicon nitride membrane.

For a microstrip line, as the dielectric constant increases, the impedance of the TL drops as approximately $1/\sqrt{\epsilon_r}$. 
To compensate for this effect, either the thickness of the dielectric needs to be increased or the stripline needs to be made thinner. A balance is required between the two because a thicker dielectric increases the amount of field fringing whereas micro-fabrication capabilities limit the width of the stripline.

\paragraph{Prospects and R\&D path for CMB-S4}
The technology status level of the superconducting RF transmission line is 5. 
Every CMB experiment uses micro-lithographed superconducting RF transmission line as part of detector system. 
Multiple scientific results were published with systematics effect that came about due to transmission line coupling problem.
This problem was understood, simulated and solution to mitigate that problem was found and demonstrated. 

The production status level for the superconducting RF transmission line is 5. Hundreds of detectors were fabricated for stage-III experiments. During fabrication for stage-III experiments, we learned various ways performance of superconducting RF transmission line could degrade. Prevention methods were found for these problems, and large quantity of stage-III detector arrays were fabricated without problem with superconducting transmission line.

Low-loss TL is essential for providing flexibility in the circuit design and improving detector efficiency.
A reliable dielectric constant is important for predictable RF circuit performance and 
dielectric films with loss tangent lower than $1\times10^{-3}$ are needed for dielectric loss to be negligible. 
%Thus, a micro-fabrication process that reliably produces low-loss dielectric films with a stable dielectric constant needs to be established.
Silicon nitride and single crystal silicon have desirable properties, and multiple Stage-III experiments are going to deploy detectors with these dielectrics.
Demonstration of detector fabrication with these films will pave the way for the CMB-S4 detector fabrication.

Reliable fabrication of high quality niobium films requires a dedicated Niobium sputter machine that is under tight control.
Multiple CMB detector fabrication facilities already have dedicated niobium sputtering systems for superconducting film process.
A similar degree of control needs to be implemented for CMB-S4 detector fabrication to realize predictable detector performance.

%\noindent\fbox{
%\parbox{\textwidth}{
%\noindent\textbf{Lab demonstration:} Improved efficiency with silicon nitride dielectric \\
%\noindent\textbf{Sky demonstration:} Multiple CMB experiments deployed with SiO2 microstrip line \\
%\noindent\textbf{Path to CMB-S4:} Improve efficiency with low-loss dielectric such as silicon nitride and single crystal Si
%}
%}

\begin{table}[h]
\begin{center}
\begin{tabular} {|l l|}
\hline  
\textbf{Lab Demonstration:} & Improved efficiency with silicon nitride dielectric \\
\textbf{Sky Demonstration:} & Multiple CMB experiments deployed with SiO$_2$ microstrip line \\
\textbf{Path to CMB-S4:} & Improve efficiency with low-loss dielectric such as SiN and single crystal Si\\
\hline 
\end{tabular}
\end{center}
\end{table}

%\bibliographystyle{unsrt}
%\bibliography{bibtex/feedhorn,bibtex/OnChipFilter_Bib,bibtex/CMBS4_BandwidthReview,bibtex/Termination_Bib,bibtex/foreground,bibtex/lensletantenna,bibtex/mmlenslet.bib,bibtex/bjohnson.bib,bibtex/array_pix.bib,bibtex/spt.bib,bibtex/bibcover.bib,bibtex/multimode_v1.bib} 
%%%%%%%%%%%%%%%%%%%%%%%%%%%%%%%%%%%%%%%%%%%%%
%
%\end{document}

%% file: detector_rf/RF_Filter.tex
%\documentclass{article}

%\usepackage[width = 8.5in, height = 11in, margin = 1in]{geometry}
%\usepackage{graphicx}
%\usepackage{subfig}
%\usepackage{verbatim}

%\newcommand{\bicepII}{{\sc bicep}2}

%\newcommand{Figure~\ref}[1]{Fig.~\ref{#1}}

%\begin{document}

\subsection{On-chip microwave filters}
\label{sec:onchipfilter}

\begin{figure}
\begin{center}
%\subfloat[ABS~$150$-$\mathrm{GHz}$ passband~\cite{Simon:2014}.]{\includegraphics[height = 0.22\textwidth]{figure/ABS_passband}
\includegraphics[height = 0.22\textwidth]{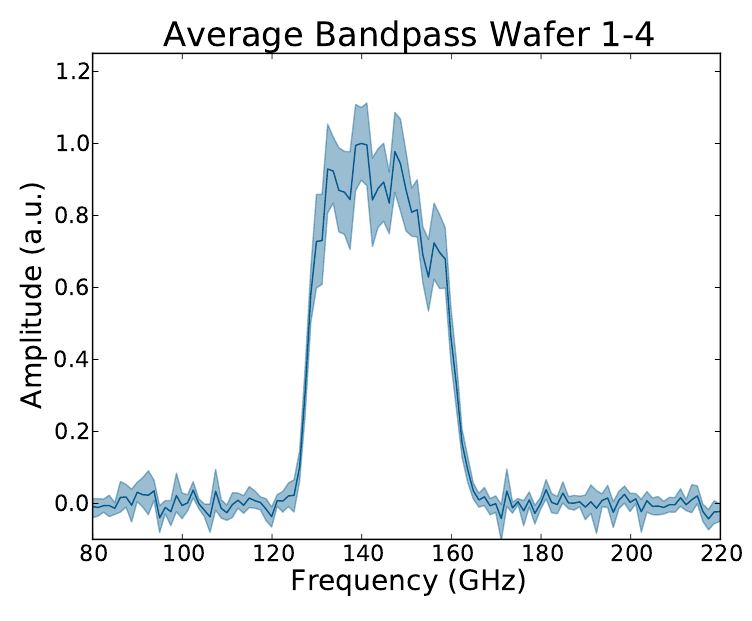} \label{fig:ABSpassband}
\hspace{0.01\textwidth}
%\subfloat[Passbands for the SPT-3G $90$/$150$/$220$-GHz triplexer~\cite{Posada:2015}.]{\includegraphics[height = 0.22\textwidth]{figure/SPT3G_bands_measured} \label{fig:SPT3Gbands}}
\includegraphics[height = 0.22\textwidth]{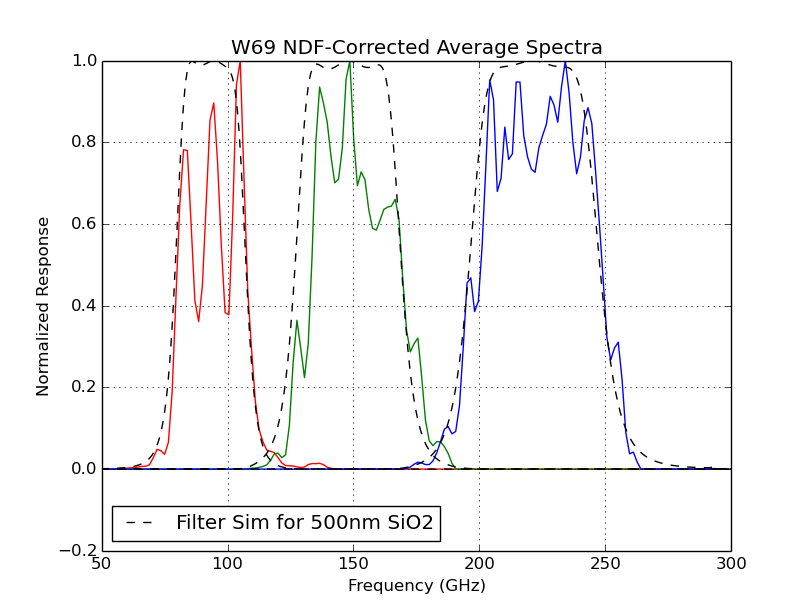} \label{fig:SPT3Gbands}
\hspace{0.01\textwidth}
%\subfloat[Channelizer bands for the pixel shown in Figure~\ref{fig:channelizerPixel}~\cite{OBrient:2013}.]{\includegraphics[height = 0.22\textwidth]{figure/channelizer_bands} \label{fig:channelizerBands} }
\includegraphics[height = 0.22\textwidth]{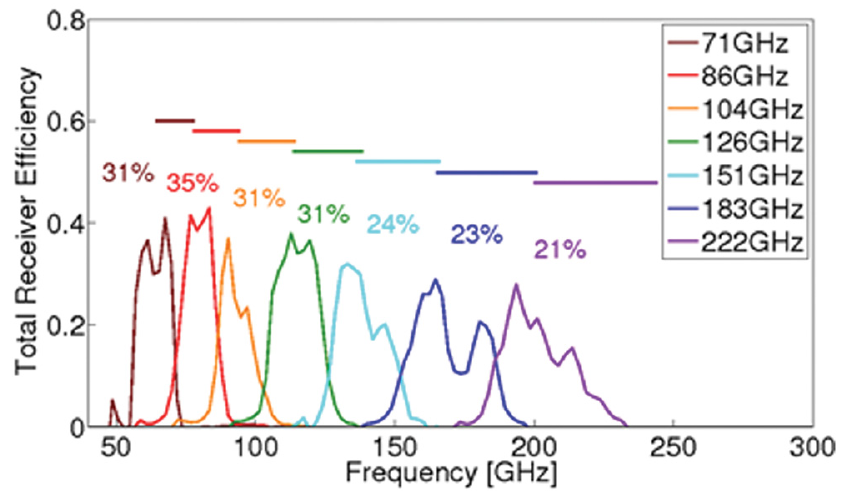} \label{fig:channelizerBands}
\end{center}
\caption{Example passbands (Left) Horn coupled band pass filter from ABS~$150$-$\mathrm{GHz}$ passband~\cite{Simon:2014}. (Center) Passbands for the SPT-3G $90$/$150$/$220$-GHz triplexer~\cite{Posada:2015}. (Right) Channelizer bands for the pixel shown in Figure~\ref{fig:filters}~\cite{OBrient:2013}.}
\label{fig:filtersmeas}
\end{figure}

\paragraph{Description of the technology}
Many experiments employ band-defining filters on the detector wafer \cite{arnoldpb1,BICEP2_II:2014,BK_SPIDER:2015,Chuss:2016,Datta:2016,Posada:2015,Simon:2014,Suzuki:2014}. These are planar microwave structures that lie between the antennas and the detectors are typically composed of sections of microstrip lines and coplanar waveguides. Most current ground-based experiments design for bandwidths of~$\sim30\%$. An example passband is shown in Figure~\ref{fig:filtersmeas}.

\begin{figure}
\begin{center}
%\subfloat[ACT-Pol~pixel with $5$-pole single-band $150$-$\mathrm{GHz}$ stub filters~\cite{arnoldpb1}. The filters are labeled (v).]{\includegraphics[height = 0.2\textwidth]{figure/ACT_pixel} \label{fig:ACT_pixel}}
\includegraphics[height = 0.2\textwidth]{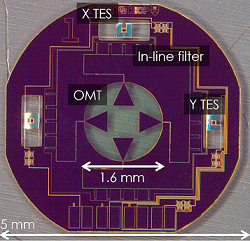} \label{fig:ACT_pixel}
\hspace{0.01\textwidth}
%\subfloat[\bicepII~single-band $150$-$\mathrm{GHz}$ lumped-element filter with corresponding circuit diagram~\cite{BICEP2_II:2014}.]{\includegraphics[height = 0.2\textwidth]{figure/BICEP2_filter} \label{fig:bicep2Filter} }
\includegraphics[height = 0.2\textwidth]{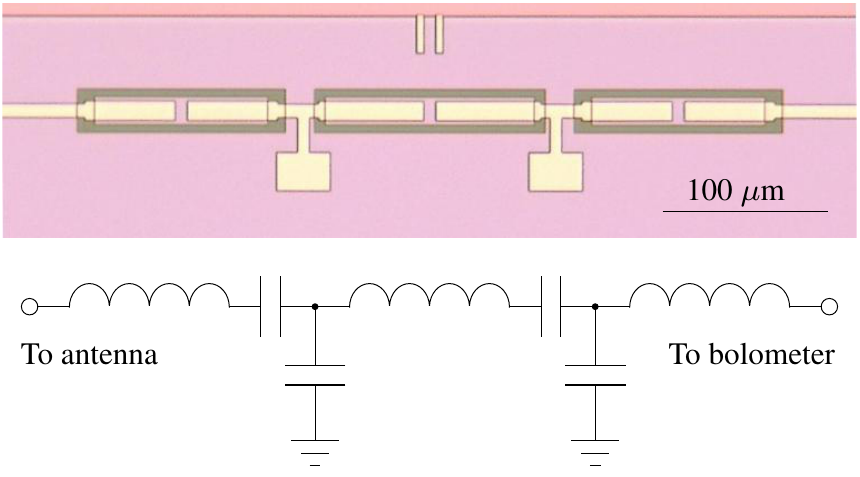} \label{fig:bicep2Filter}
\hspace{0.01\textwidth}
%\subfloat[Log-periodic lumped-element channelizer with $7$ contiguous bands between $50$~and $230~\mathrm{GHz}$~\cite{OBrient:2013}.]{\includegraphics[height = 0.2\textwidth]{figure/channelizer} \label{fig:channelizerPixel}}
\includegraphics[height = 0.2\textwidth]{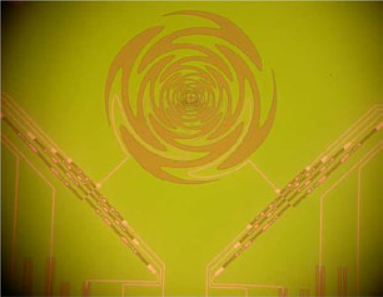} \label{fig:channelizerPixel}

\end{center}
\caption{Photographs of RF filters from CMB detectors. (Left) ACT-Pol~pixel with $5$-pole single-band $150$-$\mathrm{GHz}$ stub filters~\cite{arnoldpb1}.  (Center) \bicepII~single-band $150$-$\mathrm{GHz}$ lumped-element filter with corresponding circuit diagram~\cite{BICEP2_II:2014}. (Right) Log-periodic lumped-element channelizer with $7$ contiguous bands between $50$~and $230~\mathrm{GHz}$~\cite{OBrient:2013}.}
\label{fig:filters}
\end{figure}

%Typically, a circuit model is used as an ideal to be strived for; an example is shown in the lower of Figure~\ref{fig:bicep2Filter}. 
Typically, the filters are modeled by an ideal circuit composed of exclusively reactive elements (e.g. see Figure~\ref{fig:filters}).
%A bandpass filter is usually modelled using exclusively reactive circuit elements. 
The number of degrees of freedom in the filter design is often referred to as the number of \emph{poles} and is related to the order of the polynomial that describes the passband. A higher-pole filter has, by definition, more degrees of freedom and is, therefore, able to achieve a steeper roll-off in the passband. The disadvantage of a higher-pole filter is that dielectric loss is more severe due, heuristically, to multiple reflections within the filter, so that the loss is much greater than would be incurred by an equivalent length of transmission line. For a microstrip filter with a dielectric loss tangent of $3 \times 10^{-3}$, the expected loss is $\sim 5\%$ for a $3$-pole filter and $\sim 10\%$ for a $5$-pole filter. In designing a microwave filter, then, a balance must be struck between efficiency and band shape.

There are two main strategies for implementing a given filter circuit. One is to use quarter-wavelength short-circuited TL stubs. These types of microwave structures are sometimes called distributed filters, because they do not consist of discrete circuit elements but instead exploit the similarity in behavior of $LC$ resonators and quarter-wavelength TL stubs. The degrees of freedom translate to the impedances of the stub sections, which are usually controlled by microstrip width. The other main approach is to use lumped circuit elements. In this paradigm, each geometric structure corresponds to a specific circuit element, e.g., an inductor or a capacitor. 

\begin{figure}
\begin{center}
%\subfloat[A bandstop filter can be using to notch out unwanted frequencies within a band, e.g., $\mathrm{CO}$ lines. Shown is a three-pole bandpass filter in series with a three-pole bandstop filter.]{\includegraphics[height = 0.23\textwidth]{figure/Bandpass_220_Notch_225_3pol_compareWithoutNotch_noGrid.pdf} \label{fig:bandstopCukierman} }
\includegraphics[height = 0.23\textwidth]{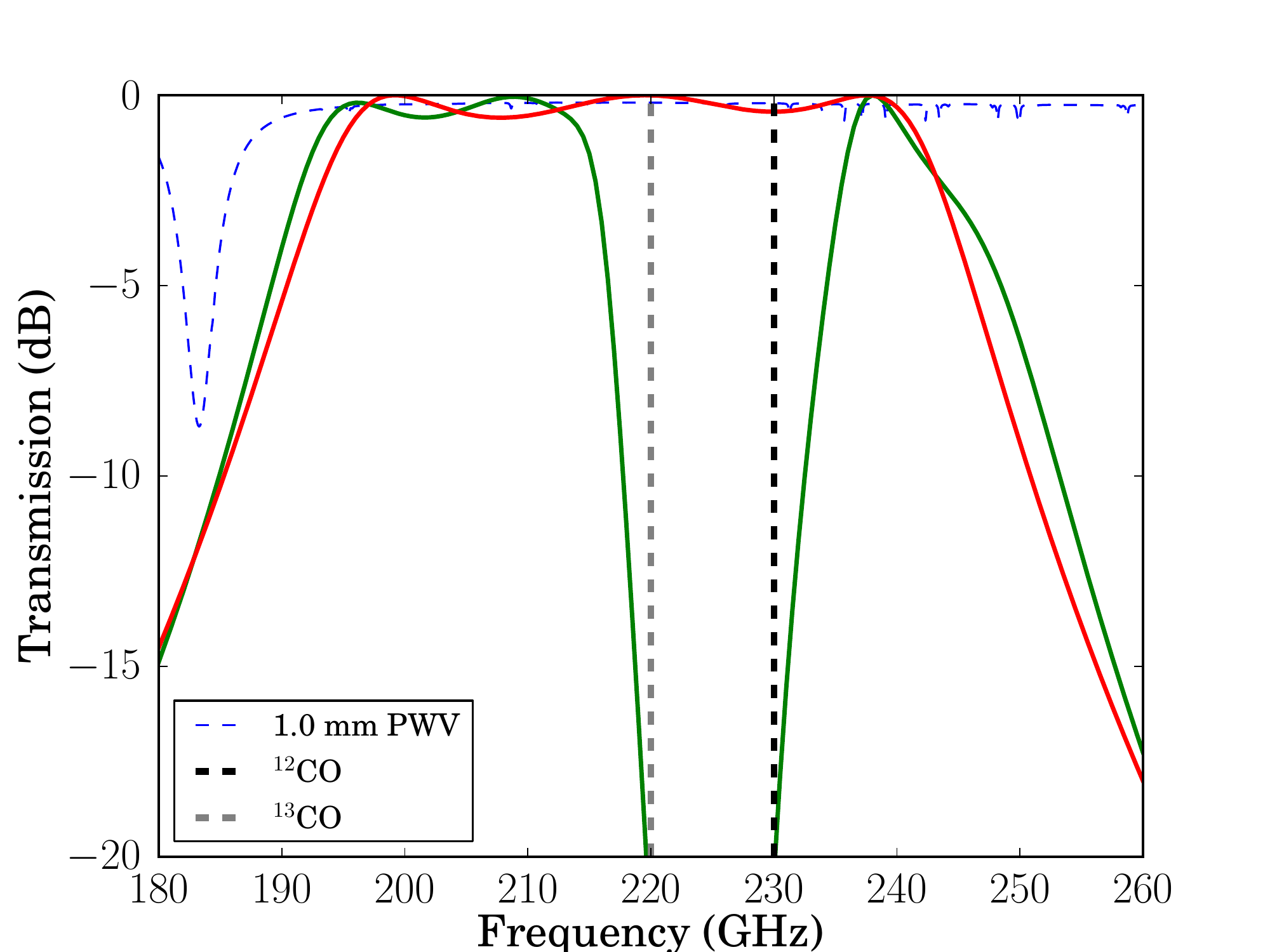} \label{fig:bandstopCukierman}
\hspace{0.01\textwidth}
%\subfloat[A lower-loss dielectric allows for higher-pole filters. Shown is a simulated comparison of a three-pole filter (red) and a seven-pole filter (green) on dielectric with $\tan\delta = 3\times10^{-3}$. The higher-pole filter has a more rapid roll-off of the passband, but loss increases.]{\includegraphics[height = 0.23\textwidth]{figure/Compare_3poleSiO2_7poleSiO2_Linear.pdf} \label{fig:3pole_vs_7pole_Cukierman} }
\includegraphics[height = 0.23\textwidth]{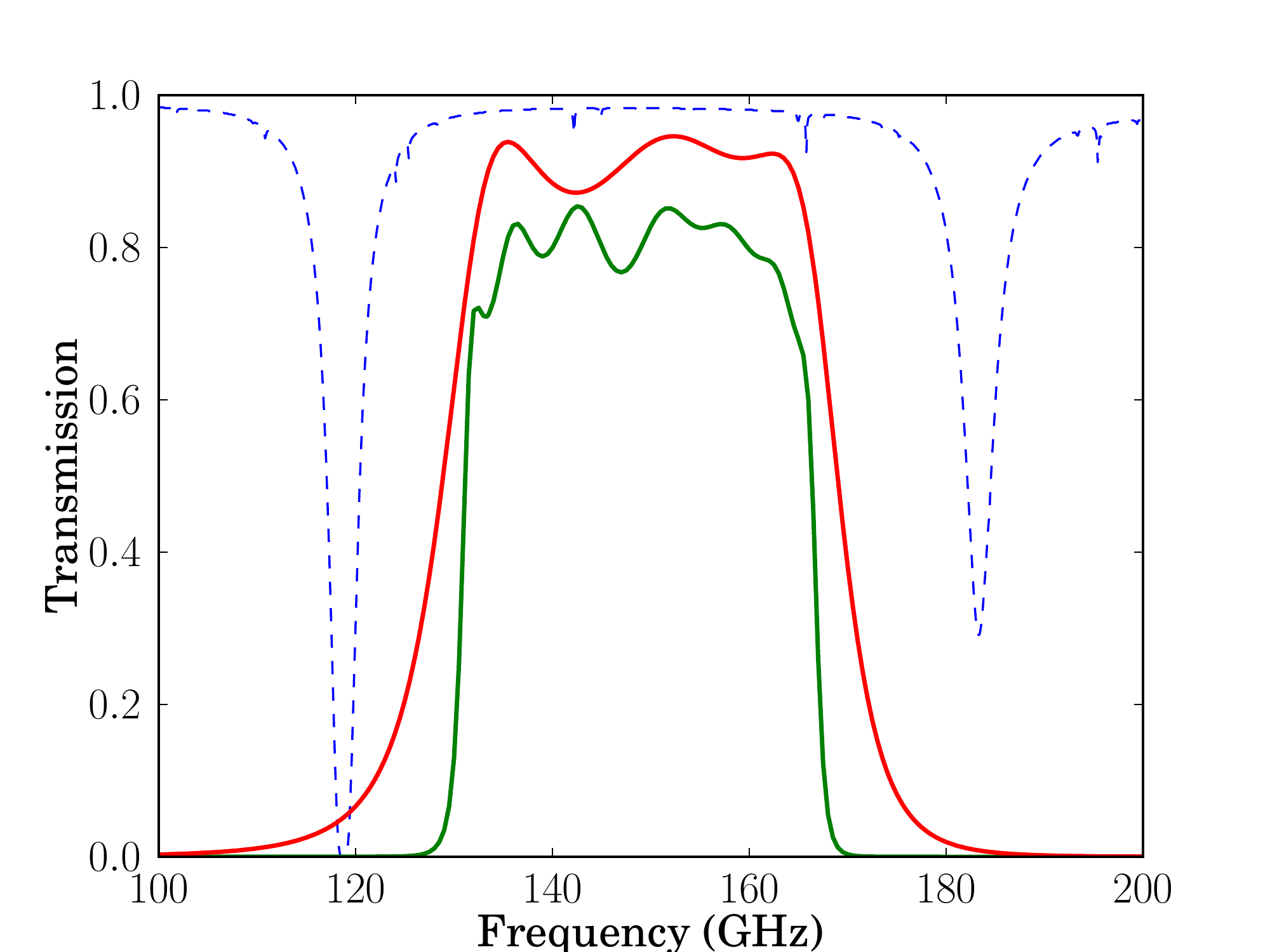} \label{fig:3pole_vs_7pole_Cukierman} 
\hspace{0.01\textwidth}
%\subfloat[The effect of dielectric loss on a seven-pole filter. The red curve is from a simulation with $\tan\delta = 3\times 10^{-4}$; the green is with $\tan\delta = 3 \times 10^{-3}$.]{\includegraphics[height = 0.23\textwidth]{figure/Compare_7poleSiO2_7poleSiN_Linear_noGrid.pdf} \label{Compare_7poleSiO2_7poleSiN_Linear.pdf} }
\includegraphics[height = 0.23\textwidth]{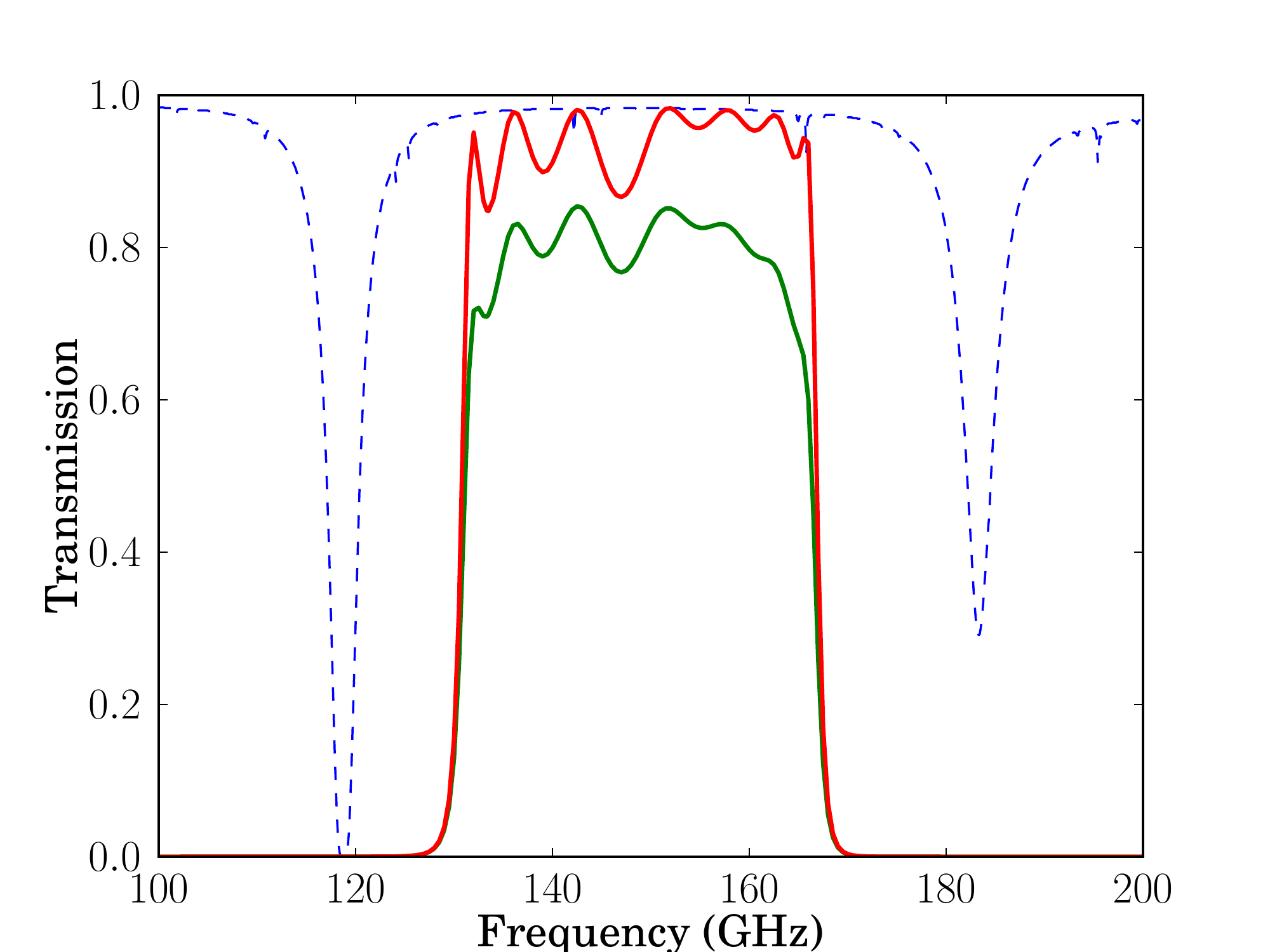} \label{Compare_7poleSiO2_7poleSiN_Linear.pdf}
\end{center}
\caption{(Left) Simulated performance of bandstop filter. A bandstop filter can be using to notch out unwanted frequencies within a band, e.g., $\mathrm{CO}$ lines. Shown is a three-pole bandpass filter in series with a three-pole bandstop filter. (Center) Comparison of roll off speed for different number of poles. A lower-loss dielectric allows for higher-pole filters. Shown is a simulated comparison of a three-pole filter (red) and a seven-pole filter (green) on dielectric with $\tan\delta = 3\times10^{-3}$. The higher-pole filter has a more rapid roll-off of the passband, but loss increases. (Right) Simulated band with different dielectric loss tangent. The effect of dielectric loss on a seven-pole filter. The red curve is from a simulation with $\tan\delta = 3\times 10^{-4}$; the green is with $\tan\delta = 3 \times 10^{-3}$.}
\label{fig:multpolefilters}
\end{figure}

In implementing an on-chip filter, it is vital to have good knowledge and control of the material properties. Most current experiments use a microstrip paradigm, which involves a metal-dielectric-metal tri-layer stack. The metals are superconducting, e.g., $\mathrm{Nb}$, which eliminates resistive losses for high quality films. The kinetic inductance of the superconductor, however, can affect the impedance of the TL. The dielectric constant of the middle layer controls the impedance as well, and the loss tangent of this dielectric is often the limiting factor in transmission efficiency. Some typical microstrip dielectrics include $\mathrm{SiO}_2$ with $\tan \delta \sim 10^{-3}$, $\mathrm{Si}_3\mathrm{N}_4$ with $\tan\delta \sim 10^{-4}$ and single-crystal~$\mathrm{Si}$ with $\tan \delta < 10^{-5}$. Lower-loss dielectrics allow for higher-pole filters. A file-pole filter with $\tan\delta = 3 \times 10^{-3}$ dissipates $\sim 10\%$ of the incident power, whereas the same filter with $\tan\delta = 3 \times 10^{-4}$ would dissipate $\sim 1\%$. A comparison of filters is shown in Figure~\ref{fig:multpolefilters}.

\paragraph{Demonstrated performance}

The \Pb, \Pb-2, \bicepII~and SPT-3G experiments use $3$-pole filters~\cite{arnoldpb1,Suzuki:2014,BICEP2_II:2014,Posada:2015}; the ACTPol experiment uses a $5$-pole filter~\cite{Datta:2016}. ACTPol, CLASS, and \Pb are using distributed filters; %an example filter is shown in
an example passband is shown in Figure~\ref{fig:filtersmeas}. The CLASS detectors also employ filters to reject out-of-band radiation coupling to the microstrip circuit \cite{doi:10.1117/12.927056,doi:10.1117/12.2057266,doi:10.1117/12.2234308,Chuss:2016}. The \bicepII~experiment and the upcoming \Pb-2 and SPT-3G experiments are using lumped-element filters; an example is also shown in Figure~\ref{fig:filtersmeas}. 
%The CLASS experiment uses a waveguide cutoff to define the lower edge of the band and an on-chip low-pass filter composed of both distributed and lumped elements to define the upper edge of the band \cite{U-yen:2008}.

Recently, the ACTPol experiment deployed a diplexing $90$/$150$-$\mathrm{GHz}$ distributed filter, which is essentially a T-junction with a different bandpass filter on each branch. The \Pb-2 experiment will deploy a $90$/$150$-$\mathrm{GHz}$ lumped-element diplexer, and the SPT-3G experiment will deploy a $90$/$150$/$220$-$\mathrm{GHz}$ lumped-element triplexer (see Figure~\ref{fig:filtersmeas}). 

Two separate arrays of multichroic detectors have been deployed and additionally several experiments are near deployment with multichroic detector focal planes~\cite{Datta:2016,Posada:2015,Suzuki:2014}. Also laboratory demonstration was made on filter designs that increase the bandwidth even further~\cite{Westbrook:2016,OBrient:2013}. 
%All of these devices employ broadband antennas connected to on-chip microwave filter to partition signals into multiple bands. 
%An advantage of the lumped-element paradigm over the distributed stubs is that the filters take up less physical space, which allows for multiple bandpass filters to branch off of an incoming transmission line without colliding. 

\paragraph{Prospects and R\&D path for CMB-S4}
In~\cite{Westbrook:2016}, some $3$- and $4$-band filters are achieved by sprouting several bandpass filters from a common node. If the passbands are not overlapping, a given frequency is admitted by at most one of the branches, so that there is very little interaction among the filters. This is the paradigm used in the SPT-3G triplexer, whose passbands are shown in Figure~\ref{fig:filtersmeas}.
Another method, which is shown in Figure~\ref{fig:filters}, is to construct a ``channelizer,'' in which the bandpass filters branch off log-periodically from a TL trunk. 
%For the lower frequencies, the series inductance of the main trunk and the shunt capacitance of the higher-frequency filters combine to form an effective TL that allows propagation of the signal until a resonant filter is reached. 
The channelizer produces an arbitrary number of contiguous bands; the design shown in Figure~\ref{fig:filters} has $7$~filters, and the passbands are shown in Figure~\ref{fig:filtersmeas}.

An extreme version of the channelizing filter is a filter bank that subdivides the telluric windows into many channels, either to provide additional spectral information for foreground characterization and removal or to pursue ancillary science opportunities. Today several groups are designing compact, on-chip spectrometers that use either superconducting transmission line resonators as filter elements or phased delay lines to create a grating-waveguide analogue \cite{Hailey-Dunsheath:2016, Endo:2012, OBrient:2014, Cataldo:2014}. 
%These spectrometers employ either broad-band lithographic antennas or feed horns and couple the filtered radiation to an array of either TES or MKID broad-band detectors. 
Laboratory demonstrations have shown excellent rejection of out-of-band direct pickup and NEPs suitable for background limited performance at $R = \nu / \Delta \nu \lesssim100$ for ground-based operation at mm-wavelengths, and on-sky demonstrations are planned within the coming year. 
%A straightforward adaption of these technologies could be applied to design low-resolution filter banks optimized for ground-based or orbital CMB observation.

Bandstop filters can be implemented to reject certain frequencies, e.g., atmospheric or $\mathrm{CO}$ lines. The design approach is similar to that for bandpass filters with the main difference being in the ideal circuit model. Bandstop filters can be implemented in series with bandpass filters to notch out unwanted frequencies. An example is shown in Figure~\ref{fig:multpolefilters}, in which a $3$-pole bandstop filter notches out the $220$- and $230$-$\mathrm{GHz}$ $\mathrm{CO}$ lines while leaving the rest of the $220$-$\mathrm{GHz}$ band mostly intact.

The technology status level of the on-chip microwave filter is 5. 
Multiple CMB polarization resutls were published using detector system that uses on-chip microwave filters to define its spectral response. 
The technology status level of the multi-chroic on-chip filter is 4. Technology is deployed on ACT-pol, Adv-ACT and SPT-3G. 

The production status level for the on-chip microwave filter is 5. 
On-chip microwave filter is part of lithographed detector architecture. 
Large quantity of detector array was fabricated for stage-III experiments with on-chip microwave filter.
It was demonstrated that process step to define on-chip microwave filter is not limiting throughput of detector fabrication. 

\begin{figure}
\begin{center}
%\subfloat[$3$-pole Chebyshev bandpass filter with $0.5$-$\mathrm{dB}$ ripples. The combined integrated bandwidth is~$47~\mathrm{GHz}$.]{\includegraphics[width = 0.48\textwidth]{figure/subdivide150_3pole.pdf}}
\includegraphics[width = 0.48\textwidth]{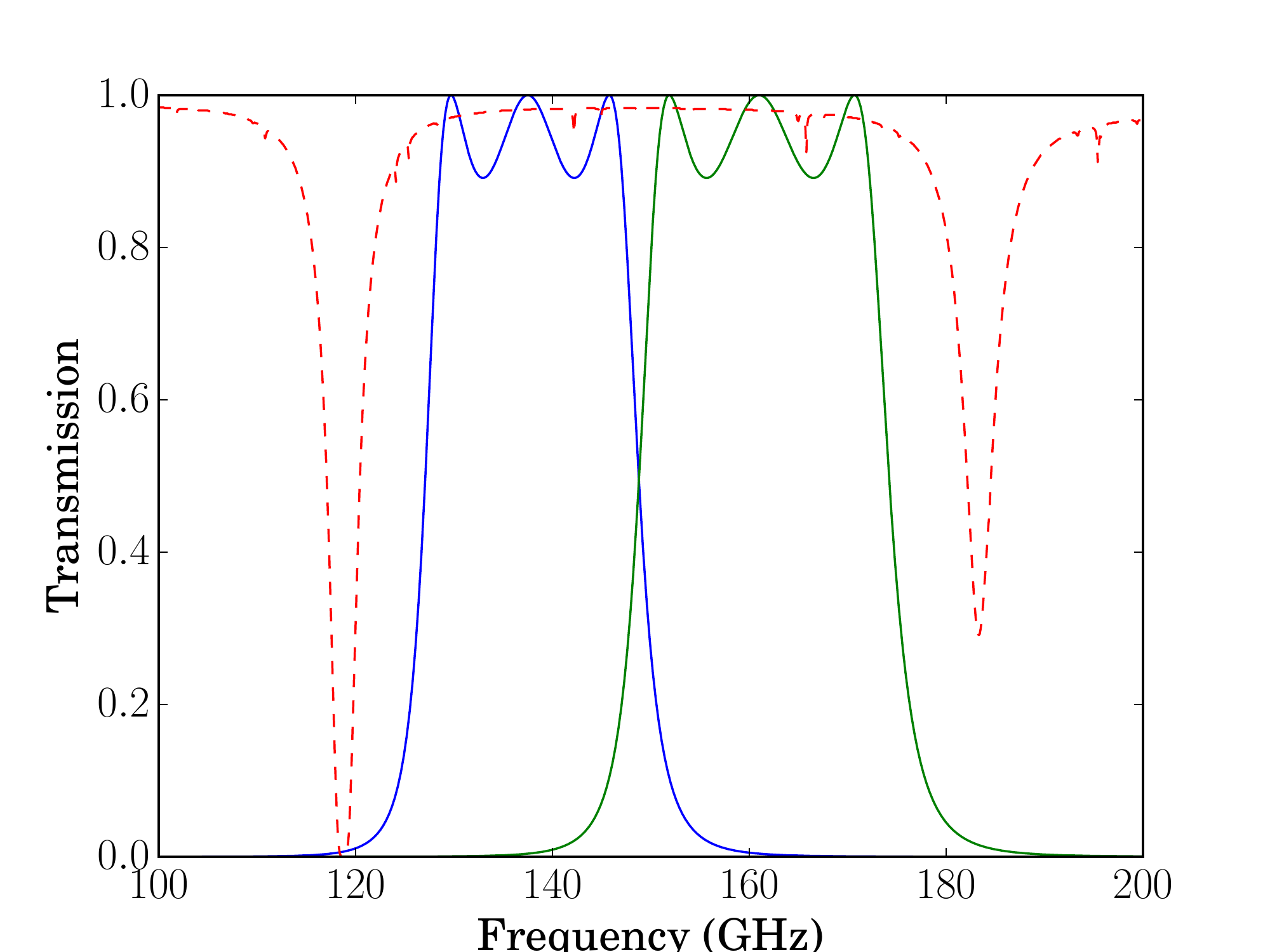}
\hspace{0.01\textwidth}
%\subfloat[$7$-pole Chebyshev bandpass filter with $0.5$-$\mathrm{dB}$ ripples. The combined integrated bandwidth is~$50~\mathrm{GHz}$.]{\includegraphics[width = 0.48\textwidth]{figure/subdivide150_7pole.pdf}}
\includegraphics[width = 0.48\textwidth]{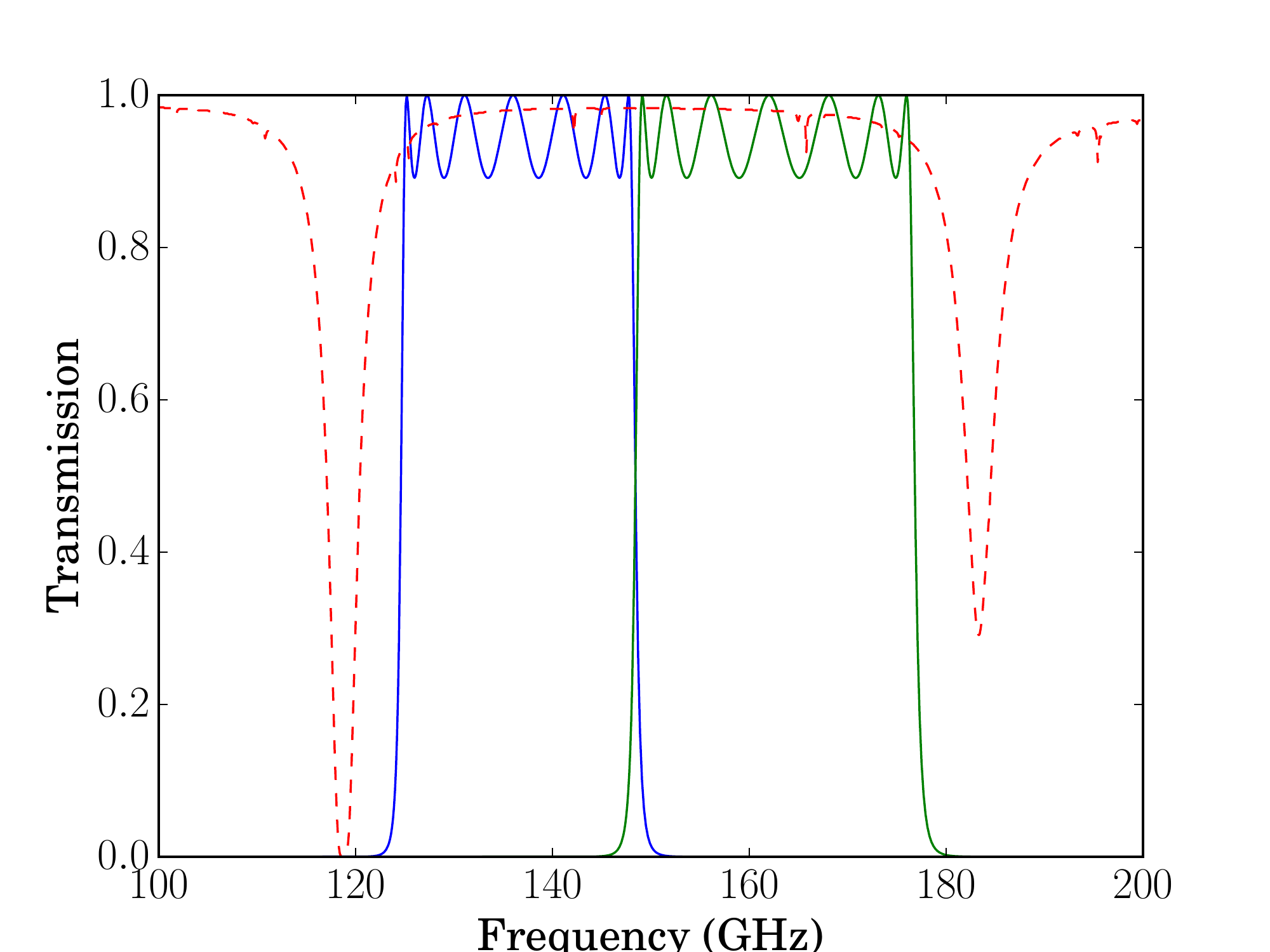}
\end{center}
\caption{Subdividing the $150$-$\mathrm{GHz}$ for better spectral resolution. In this case, higher-pole filters do not provide a substantial increase in transmitted power, but they do help to isolate the bands and, thereby, avoid mutual coupling. A proper treatment would involve a simulation of the microwave circuit as, e.g., a diplexer where the power can be shared between the two filters, but the raw Chebyshev transmission curves give a good approximation of what to expect from these simulations. (Left) $3$-pole Chebyshev bandpass filter with $0.5$-$\mathrm{dB}$ ripples. The combined integrated bandwidth is~$47~\mathrm{GHz}$. (Right) $7$-pole Chebyshev bandpass filter with $0.5$-$\mathrm{dB}$ ripples. The combined integrated bandwidth is~$50~\mathrm{GHz}$.}
\label{fig:subdivideAtmWindow}
\end{figure}

For better spectral resolution, the atmospheric windows can be subdivided into multiple bands. This is illustrated in Figure~\ref{fig:subdivideAtmWindow} for both three- and seven-pole Chebyshev filters.
Subdividing the atmospheric window does not reduce overall transmission and there is only a small gain from increasing the number of poles. % does help but not significantly. 
Integrated bandwidths are given in the captions of Figure~\ref{fig:subdivideAtmWindow}. Realizing these narrower-band filters presents some challenges. For stub filters, the impedance of each stub is proportional to the fractional bandwidth. For a microstrip implementation, this requires wider stubs for narrower bands. When the stub width is comparable to $\lambda/4$, the stubs can no longer be treated as quarter-wave resonators and will not produce the designed passband. For lumped-element filters, the required inductances are roughly inversely proportional to the fractional bandwidth. It is difficult to realize a large inductance while keeping the effective length much smaller than a wavelength, i.e., maintaining the lumped-element approximation. These challenges are not insurmountable but will undermine the naive application of current techniques.

%\noindent\fbox{
%\parbox{\textwidth}{
%\noindent\textbf{Lab demonstration:} triplexer, tetraplexer and 7-band channelizer filter \\
%\noindent\textbf{Sky demonstration:} Single and dual band filter deployed with multiple CMB detectors \\
%\noindent\textbf{Path to CMB-S4:} High-Q band to sub-divide atmospheric window
%}
%}

\begin{table} [h]
\begin{center}
\begin{tabular} {|l l|}
\hline  
\textbf{Lab Demonstration:} & Triplexer, tetraplexer and 7-band channelizer filter \\
\textbf{Sky Demonstration:} & Single and dual band filter deployed with multiple CMB detectors \\
\textbf{Path to CMB-S4:} & High-Q band to sub-divide atmospheric window\\
\hline 
\end{tabular}
\end{center}
\end{table}

%\bibliographystyle{unsrt}
%\bibliography{OnChipFilter_Bib}

%\end{document}

%% file: detector_rf/RF_CrossOver.tex
\subsection{Microwave cross-over}\label{sec:crossover}
%Dan Becker, Shannon Duff, Hannes Hubmayr
\paragraph{Description of the technology}
Microwave cross-overs allow two transmission lines to cross on a wafer.
It provides flexibility for detector design. 
It is important to design cross-over with high transmission efficiency and without coupling between orthogonal lines.
Cross-overs are used on most dual polarization-sensitive CMB detector arrays \cite{arnoldpb1,Chuss:2016,Datta:2016,Posada:2015,Simon:2014,Suzuki:2014,CLASSFab,Duff2016,VialessCrossOver}, except for the antenna array detector 
design (used in the \bicepI\ series
of experiments), which avoids the cross-over process.  
%In a typical design, microstrip lines carrying the vertical polarization must come together at one detector, while microstrip lines carrying the horizontal polarization must come together at a different detector. 
%The geometry of the probes requires at crossings of the microstrip lines.
%Antenna array detector design for the BICEP series experiments avoids cross-over process.
Planar RF cross-over designs are well established, with 
multiple experiments implementing %adapted similar design, and they designed a 
cross-over designs that are compatible with the rest of the detector fabrication steps. 

\begin{figure}[h]
\begin{center}
\includegraphics[height = 2in]{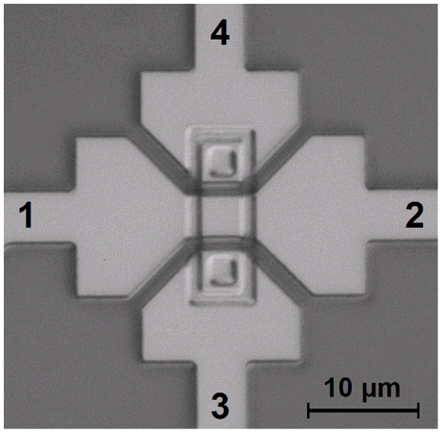}
\includegraphics[height = 2in]{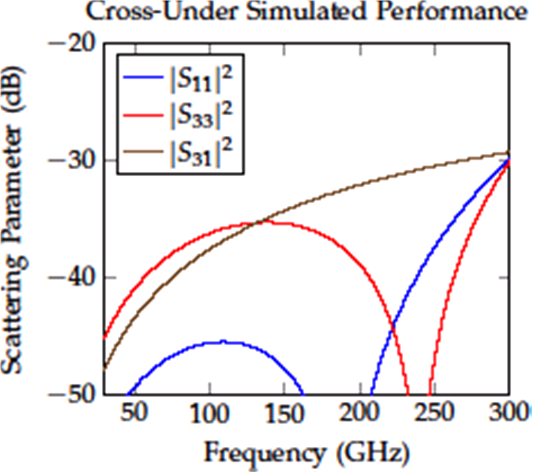}
\end{center}
\caption{(Left) Optical microscope image of a microstrip cross-under. (Right) Simulated Performance of a microstrip cross-under.}
\label{fig:crossover}
\end{figure}

A typical microstrip cross-over uses two metal wiring layers separated by an insulator; a 
microscope photograph and the simulated performance of a cross-over design is shown in Figure~\ref{fig:crossover}.
In this approach, the lower layer is a common ground plane with a section cut out in the area of the ``cross-over.'' 
The top wiring layer carries the primary runs of microstrip, and the line which ``crosses under'' connects down to the lower layer through vias. 
%A short section of metal within the cut-out ground plane connects this line underneath the other line. 
In the area of the ground plane cut-out, neither line truly has a ground plane, introducing a deficit of capacitance to ground, which is compensated for by adding ``wings'', small sections of widened trapezoidal transmission lines, to both lines.
Simulations of this ``cross-under'' predict cross-talk and reflection below -30\,dB over nearly all of the 30-300\,GHz range.

%Other experiments use similar designs.
\Pb-2 and SPT-3G use a design that has additional insulator and metal layers that form a cross-over \cite{Posada:2015,Suzuki:2014}.
In this design, the conductor at the cross-over is narrowed to minimize capacitive coupling between the two orthogonal channels. 
The extra inductance introduced by the short narrow section is compensated for by widening the transmission line section via wings, similar to the cross-over shown in Figure~\ref{fig:crossover}.

It is also possible to design a cross-over without using a via, a vertical short to connect two conductors at different layers, demonstrated for narrow-band applications \cite{VialessCrossOver}. The CLASS detectors employ a via-less crossover design. A broadband version of the via-less design has been recently reported \cite{doi:10.1117/12.2234308}. In addition, an air-bridge crossover, extending the bandwidth to $\sim$500 GHz has been fabricated \cite{doi:10.1117/12.2234308, 7802647}. 
A via-less cross-over has the benefit of simplifying fabrication.

%Signal from one polarization (Pol A) couples to slot lines via inductively.
%Microstrip line from orthogonal polarization (Pol B) crosses over the slot lines, but impedance of both slot and microstrip is chosen to minimize cross-polarization coupling. 
%Signal from the slot lines (Pol A) are coupled to microstrip line with identical inductive coupling scheme to complete the cross-over circuit. 

%[Toki comment: include a figure of CLASS via-less cross over?]

\paragraph{Demonstrated performance}
Multiple Stage-II CMB experiments successfully deployed detector arrays with cross-overs \cite{pb,sptpol,actpol}. 
There is no measurable difference in efficiency between the two orthogonal polarizations and the differential spectra between two orthogonal polarization channels are small indicating that cross-overs work well.
There are multiple Stage-III CMB experiments that have designed and demonstrated cross-overs spanning multiple frequency bands. 
The cross-over for the \Pb-2 experiment was designed to cover the 90\,GHz and 150\,GHz band \cite{SuzukiThesis}.
Microwave cross-unders targeting the frequency range 60-300\,GHz were deployed for the AdvACT experiment in 2016. 
Cross-unders targeting 240-340\,GHz have also been designed and fabricated, and will be deployed in the \spider\ experiment.

\paragraph{Prospects and R\&D path for CMB-S4}
The technology status level for microwave cross-overs is 5. Multiple Stage-II CMB experiments successfully deployed detector arrays with cross-overs. There is no measurable difference in efficiency between the two orthogonal polarizations and the differential spectra between two orthogonal polarization channels are small indicating that cross-overs work well.
The production status level for microwave cross-over is also 5.  Microwave cross-overs are a mature technology that is compatible with other detector fabrication steps. There are well established designs that achieve reflection and cross-talk below -30\,dB and there are no scaling issues for CMB-S4.

For a receiver configuration that does not use polarization modulation, fabricating symmetric detectors for the two orthogonal polarizations will be important. 
The dominant issue for implementing cross-overs is quality control during fabrication, an issue which would benefit from developing a simple method for validating cross-over performance without necessitating a full optical test.
%We will need to develop a method to test cross-over performance reliably to give accurate feed back to detector fabrication design. 

%\noindent\fbox{
%\parbox{\textwidth}{
%\noindent\textbf{Lab demonstration:} Multiple designs robustly designed \\
%\noindent\textbf{Sky demonstration:} With via, via-less cross-filters demonstrated in multiple CMB experiments \\
%\noindent\textbf{Path to CMB-S4:} Technology is mature. Detail systematics study for asymmetric cross-over
%}
%}

\begin{table}[h]
\begin{center}
\begin{tabular} {|l l|}
\hline  
\textbf{Lab Demonstration:} & Multiple designs robustly designed \\
\textbf{Sky Demonstration:} & With via, via-less cross-filters demonstrated in multiple CMB experiments \\
\textbf{Path to CMB-S4:} & Technology is mature. Detail systematic error study for asymmetric cross-over\\
\hline 
\end{tabular}
\end{center}
\end{table}

%% file: detector_rf/RF_Termination2.tex
%\documentclass{article}

%\usepackage[width = 8.5in, height = 11in, margin = 1in]{geometry}
%\usepackage{graphicx}
%\usepackage{subfig}
%\usepackage{verbatim}

%\newcommand{\bicepII}{{\sc bicep}2}

%\newcommand{Figure~\ref}[1]{Fig.~\ref{#1}}

%\begin{document}

\subsection{Microstrip termination}\label{sec:termination}

\paragraph{Description of the technology}
In most current-generation experiments, the incident radiation couples to an antenna which is fed by a microstrip line. The signal then passes through bandpass filters and/or mode rejectors on its way to the bolometer island, where the power is dissipated as heat for the bolometer to detect. A resistive element is used to dissipate the heat. There are two common techniques in the field: a lossy meandering microstrip line and a lumped resistor. 

Most experiments use superconducting metals and low-loss dielectrics for their microstrip lines in order to minimize attenuation of the microwave signal. At the bolometer island, however, it is necessary to dissipate the power. This can be achieved by transitioning to a non-superconducting, purposefully resistive metal. The signal will be attenuated along the length of this non-superconducting microstrip line, and the power will be dissipated as heat. The microstrip can be designed to meander so that the path length is large while occupying a relatively small area on the bolometer island. The resistivity of this non-superconducting metal must be relatively low in order to prevent an impedance mismatch between the incoming superconducting microstrip line and the lossy meandering microstrip line. Since the resistivity is low, the attenuation per unit length is relatively small; therefore, the meander must have a large path length in order to dissipate most of the power. This tends to make the lossy meanders large, which also increases the size of the bolometer island. A desirable property of this termination is that it requires only a signal unbalanced microstrip line coming in to the bolometer island. The end of the meander can be left open-circuited, since the reflected power is heavily attenuated by the lossy metal. Since different frequencies pass through a different number of wavelengths in the meander, the absorption efficiency is frequency dependent.

The other main type of termination is a lumped resistor. The incoming microstrip line is terminated by an impedance-matched resistor, and the power is then dissipated as heat on the bolometer island. An advantage of this paradigm is that a lumped resistor tends to be relatively small and represents a minor contribution to the size of the bolometer island. The lumped resistor typically consists of a short section of high-resistivity metal, where the particular geometry is important in determining the lumped resistance. For a single unbalanced microstrip line, the lumped resistor should be shorted to ground; the disadvantage here is that a via is required. For two balanced microstrip lines, the resistor can be differentially fed and, if its resistance is chosen to be twice the microstrip impedance, will dissipate all of the power without a via. Another advantage of the differentially fed termination is that it accepts odd modes but rejects even modes. The lumped-resistor paradigm is relatively insensitive to the termination resistance, because the reflected power goes as
\begin{equation}
| \Gamma |^2 = \left | \frac{ R_0 - R_L }{R_0 + R_L} \right |^2 , 
\end{equation}
where~$\Gamma$ is the reflection amplitude, $R_0$ is the characteristic impedance of the microstrip line and $R_L$ is the termination (load) resistance.
%
%\begin{figure}
%\begin{center}
%\includegraphics[height = 2in]{figure/reflection.pdf}
%\includegraphics[height = 2in]{figure/Pd_Straight_30Ohm.pdf}
%\end{center}
%\caption{(Left) Fraction of power reflected as a function of the termination (load) resistance. The amount of reflection is relatively insensitive to the load resistance~$R_L$: even a factor-of-$2$ discrepancy produces only~$\sim 10\%$ reflection.
%(Right) The simulated reflected power from a 1.5-mm long gold meander. The absorption increases with frequency. The ripples are due to the impedance mismatch between the superconducting microstrip and the lossy meander, which creates an extra reflection in addition to the one generated at the open-circuited end of the lossy meander. The two reflections interfere with each other.}
%\label{fig:reflection.pdf}
%\end{figure}
%
%This relation is plotted in Figure~\ref{fig:reflection.pdf}, where both impedances are assumed to be pure real as is the case in most experiments. We see that 
The reflection increases relatively slowly as~$R_L$ deviates from~$R_0$. Even when the termination resistance differs from the microstrip impedance by a factor of~$2$, the reflection is only~$\sim10\%$.

%%%%%%%%%%%%%%%%%%%%%%%%%%%%%%%%%%%%%%%%%%%%%%%%%%%%%%%%%%%%%%
%A broadband CPW-to-microstrip transition composed of alternating sections of CPW and microstrip is used to
%transition the radiation onto microstrip lines.
%%
%Next, diplexers composed of two separate band-pass filters separate the radiation into 125\,to\,170\,GHz and
%190\,to\,280\,GHz pass bands.
%%
%The signals from opposite probes within a single sub-band are then
%combined using a hybrid tee.
%%
%Signals at the sum output of the hybrid are routed to a termination
%resistor and discarded, while the difference port is evenly divided
%in-phase onto two microstrips each with twice the impedance of the
%incoming microstrip (see the right panel of
%Figure~\ref{fig:mkid_coupling2} and~\cite{surdi2016}).
%

\begin{figure}[h]
\centering
\includegraphics[width=0.8\textwidth]{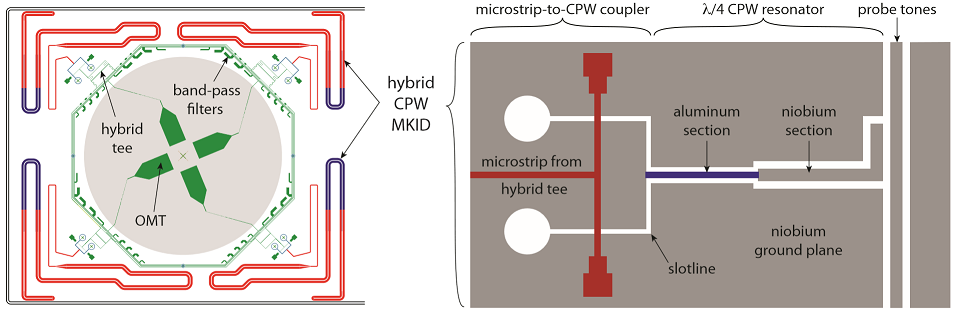}
\caption{
%
%(Left)
%%
%One polarization sensitive multichroic MKID array element.
%%
%Each array element is sensitive to two polarizations and two
%polarizations, so there are four MKIDs per element.
%%
%(Right)
%%
A schematic of the microstrip-to-CPW coupling schematic for MKID detectors.
The millimeter-wave power is coupled from the microstrip output of the
hybrid tee to the CPW of the MKID using a novel, broadband
circuit~\cite{surdi2016}.
}
\label{fig:mkid_coupling2}
\end{figure}

RF termination for MKID detectors also uses lossy metal to generate quasi-particles, but its implementation is slightly different from terminations used for TES bolometers. The coupling scheme for RF termination is shown in Figure~\ref{fig:mkid_coupling2}.
A microstrip line feeds a standard broadband microstrip-to-slotline transition, where the slotline is formed in the niobium ground plane that is common to the microstrip and the MKID CPW.
The two slotlines are then brought together and become the gaps of the
CPW transmission line, efficiently coupling the radiation into the
aluminum CPW center line, where it dissipates by exciting
quasiparticles and thereby changes the resonant frequency of the
device.
The slotline is electrically short at the resonant frequency of the
MKID, and thus it does not impact the microwave characteristics of the
resonators.
Each CPW resonator is capacitively coupled to a
transmission line and driven by a probe tone; sky signals are detected
as changes in the amplitude and phase of this probe tone.
HFSS/Sonnet simulations show the expected absorption efficiency of the
detector is approximately 90\%.

\paragraph{Demonstrated performance}
Experiments that use lossy meanders include \bicepII, ABS, ACTPol and SPTpol~\cite{BICEP2_II:2014,Simon:2014,Grace:2014,Henning:2012}. Gold is a popular low-resistivity metal for this purpose.
\begin{figure}
\begin{center}
%\subfloat[\bicepII~bolometer island with a lossy gold meander (left) and TES (right)~\cite{BICEP2_II:2014}.]{\includegraphics[height = 0.16\textwidth]{figure/boloIsland_BICEP2} \label{fig:boloIsland_BICEP2} }
\includegraphics[height = 0.16\textwidth]{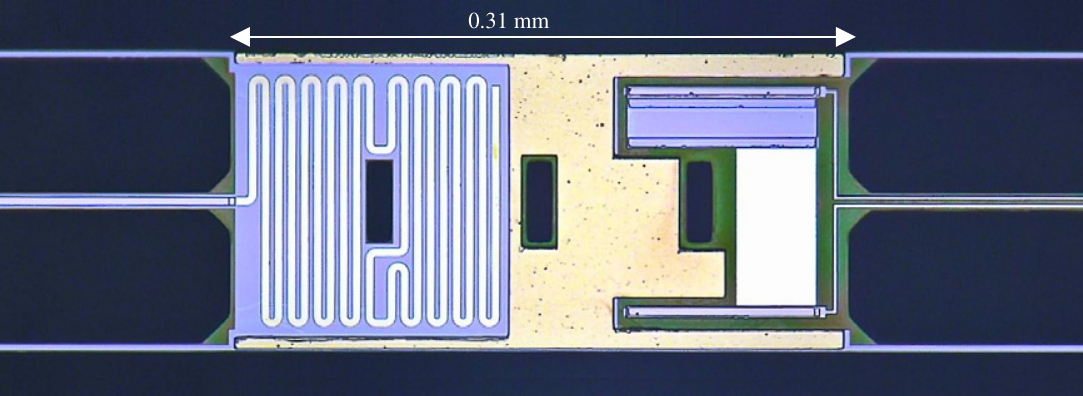} \label{fig:boloIsland_BICEP2} 
\hspace{0.008\textwidth}
%\subfloat[\Pb-2-style bolometer island with a lumped titanium resistor (right) and TES (left)~\cite{Westbrook:2016}.]{\includegraphics[height = 0.27\textwidth]{figure/boloIsland_Westbrook2016} \label{fig:boloIsland_Westbrook2016}}
\includegraphics[height = 0.27\textwidth]{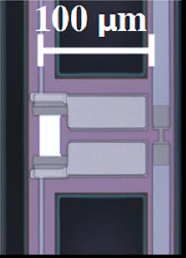} \label{fig:boloIsland_Westbrook2016}
\end{center}
\caption{Example microstrip terminations for TES bolometers. (Left) \bicepII~bolometer island with a lossy gold meander and TES~\cite{BICEP2_II:2014}. (Right) \Pb-2-style bolometer island with a lumped titanium resistor (right) and TES (left)~\cite{Westbrook:2016}.}
\label{fig:ustripTerminations}
\end{figure}

Experiments that use lumped resistors include \Pb, \Pb-2, SPT-3G and CLASS~\cite{pb,Suzuki2014,Posada:2015,Chuss:2016}. Titanium is a popular high-resistivity metal for this purpose. The critical temperature of titanium, which is $\sim500\,\mathrm{mK}$, is low enough for frequencies above $\sim 40\,\mathrm{GHz}$ to break Cooper pairs and see titanium as an effectively normal metal. A bolometer islands are shown in Figure~\ref{fig:ustripTerminations}, where the lumped resistor can be seen. Notice that the lumped resistor is substantially smaller than the lossy gold meander.

\paragraph{Prospects and R\&D path for CMB-S4}
The technology status level of the RF termination for TES bolometer is 5. 
Multiple CMB polarization results were published using microwave coupled TES bolometers. 

The production status level for the RF termination for TES bolometer is also 5.
RF termination is micro-fabricated as part of the standard detector fabrication process.
Large quantity of detector array was fabricated for stage-III experiments to demonstrate scalability. 

The technology status level of the RF termination for CPW coupled MKID detector system is 1. 
A laboratory demonstration of the RF termination scheme described above with multichroic MKIDs will happen in 2017

The production status level for the RF termination for CPW coupled MKID detector system is 1. 
MKID technology is designed to provide high throughput necessary for CMB-S4. Demonstration in 2017 will inform the scalability of the technology.

%
%\noindent\fbox{
%\parbox{\textwidth}{
%\noindent\textbf{Lab demonstration:} -  \\
%\noindent\textbf{Sky demonstration:} Lumped and distributed termination deployed successfully with high efficiency \\
%\noindent\textbf{Path to CMB-S4:} Technology is mature
%}
%}

\begin{table}[h]
\begin{center}
\begin{tabular} {|l l|}
\hline  
\textbf{Lab Demonstration:} & RF termination for MKID detectors\\
\textbf{Sky Demonstration:} & Lumped and distributed termination deployed successfully with high efficiency \\
\textbf{Path to CMB-S4:} & Technology is mature for TES coupling\\
 & MKID termination could provide additional path for scalability\\
\hline 
\end{tabular}
\end{center}
\end{table}

%\bibliographystyle{unsrt}
%\bibliography{Termination_Bib}

%\end{document}

%% file: detector_rf/Array_PixelSize.tex
%\subsection{Pixel Size and Wiring Consideration}

\paragraph{Description of the technology}
Finding an optimal pixel size is a complex problem that requires balancing the sensitivity of a detector array, the finite size of RF components, and the density of interconnects. 
Sensitivity of the detector array as a function of pixel size can be determined by calculating the sensitivity of each pixel and the number of pixels in a focal plane.
For ground-based experiments that have to observe through the atmosphere, an optimal pixel size for a fixed field of view typically is in the range 0.5 to 1.5 $f\lambda$. 
Knowing the $F$/\# of the optics at the focal plane and the wavelength of interest translates pixel size in $f\lambda$ units to a physical pixel size.
%Ground based experiment generally favors smaller pixel.
The challenge is then to find a balance between the desire to have smaller pixels and the need to fit RF components within a limited space.
%Cost and complexity of interconnects and read-out drives design to favor large numbers of detectors.

\begin{figure}[h]
\centering
\includegraphics[height = 2in]{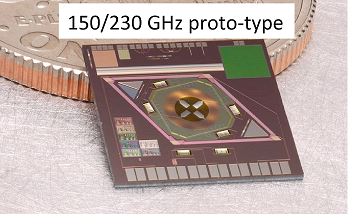}
\includegraphics[height = 2in]{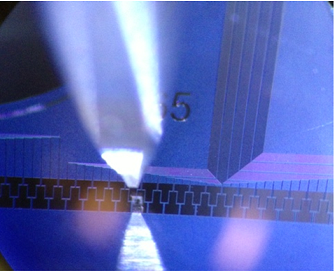}
\caption{(Left) Advanced ACTPol pixel for 150\,GHz and 230\,GHz dual-color dual-polarization pixel. Pixel spacing is 4.75\,mm. (Right) Array of wire bond pads for the \Pb-2 detector array with a wire bond head. Pads are 90 micron wide with 10 micron gap between pads.}
\label{fig:pixelsize}
\end{figure}

For example, a mutichroic pixel from AdvACT covers the 150\,GHz and 230\,GHz bands. 
The $f$/\# of the AdvACT optics at the focal plane is 1.35, so $f\lambda$ for the center frequency (185\,GHz) is 2.2\,mm. 
If the optimal pixel size is between 0.5 to 1.5 $f\lambda$ (see Figure~\ref{fig:bb}), the pixel spacing should be around 1.1\,mm to 3.3\,mm.
Figure~\ref{fig:pixelsize} shows a proto-type pixel for the AdvACT experiment that has one side of the rhombus at 4.75\,mm. 
Inside the pixel, the OMT feed is the largest element in the pixel, and 
there is not much freedom to tune antenna size, as it is constrained by wavelength. 
Lenslet-coupled antennas use the fact that the wavelength is shorter inside a dielectric to shrink the antenna size for a given frequency.
As the number of bolometers increases, the filters and bolometers start to take up significant space. It is possible to ease the pixel size challenge in isolation by designing optics with a larger $f$/\#; however, increasing the $f$/\# has multiple consequences, such as an increase in the number of detector wafers that need to be fabricated and a need for larger optical components.

Interconnects between detector wafer and readout electronics become challenging as the number of bolometers on a wafer increases.
CMB detectors make the connection between a wafer and the readout cable at the perimeters of the wafer. A larger wafer is more challenging because the pixel count increases as length-squared, whereas the length of perimeter grows as length. 
A multi-chroic detector multiplies the number of required interconnects by the multiplexing factor.
Stage-III experiments use automatic wire bonders to make wire-bond connections at $\sim 100\,\mu$ pitch.
As shown in Figure~\ref{fig:pixelsize}, the current bond pad size is approaching the size of the wire bonding tip. 

\paragraph{Demonstrated performance}
\Pb-2/Simons Array (90 and 150\,GHz) and the SPT-3G (90, 150, and 220\,GHz) detector array have 6.789\,mm spacing fabricated on a 150\,mm diameter wafer.
Each hexagonal detector array has 271 hexagonal-shaped pixels with a total of 1,626 bolometers. 
The largest RF element in the pixel is a sinuous antenna that takes up 3\,mm in diameter.
Both experiments use a lumped filter design since it takes up less space compared to the equivalent distributed stub filter design. 
There are six TES bolometers per pixel that are each roughly 100 micron x 1000 micron. 
Readout wiring is routed between hexagonal pixels and consists of a 5 micron wide line with a 5 micron spacing between lines. 
An automatic wire bonder makes wire bonds between Nb pads on the wafers to a flexible cable at 100~micron pitch.

The Advanced ACTPol detector array (150 and 230\,GHz) was also fabricated on a 150\,mm wafer \cite{Duff2016}.
A pixel is shown in Figure~\ref{fig:pixelsize}; the array has tiles of rhombus-shaped pixels. 
Wiring from each pixel travels between rhombuses with a 5 micron wide line with 1 micron spacing. 
There are 503 pixels on a wafer with 2,012 bolometers per array \cite{henderson/etal:2016}.
The bond-pad size is 120 microns wide with a 140 micron pitch, and bond-pads are staggered, so the separation between two wire bonds is 70 microns.
Wire bonding with an automatic wire bonding process \cite{ho/etal:2016}.

%\begin{figure}[h]
%\centering
%\includegraphics[width = 6in]{figure/OnChipLCs.png}
%\caption{Photograph of superconducting resonators fabricated on same wafer as multi-chroic detector.}
%\label{fig:LConChip}
%\end{figure}

\paragraph{Prospects and R\&D path for CMB-S4}
Pixel size optimization is a global optimization problem.
It is important to understand the pros and cons to come up with a good design for CMB-S4.

For example, there is a trade-off between smaller pixel size and beam performance of a pixel.
Also as the density of pixels and wiring increases, cross-talk between detectors needs to be studied carefully. 
EM simulations would be helpful to study these effects for various pixel sizes.

As the number of pixels on a wafer increases, physical space that inter-pixel wiring takes up becomes a problem.
One idea to increase readout wiring density is to run readout lines on top of each other,
requiring a pin-hole free dielectric layer to prevent shorts between two lines.
Increased detector count also makes the detector to readout cable interconnect challenging.
An alternative to wirebonding for high-density interconnections is bump bonding, which was used on SCUBA-2 and will be used on PIPER. 
It is generally less reversible than wirebonding; however, it warrants further study because wirebonding becomes difficult to reverse when thousands of wirebonds have been installed on large format detector arrays.

Dead space between detector modules due to mounting hardware could hurt coupling efficiency. 
A modular optics tube that physically separates adjacent detector modules allows mounting hardware to be present between detector modules. 
This approach is used in Advanced ACTPol.  
The same approach could be improved by matching the shape of the first refractive optic to the hexagonal shape of the detector array (e.g \cite{niemack2016}). 
This approach seems to reduce the detector array constraints in exchange for increasing the refractive optics constraints. 
It facilitates deploying additional frequencies on the same telescope though, because each optics tube can easily be used for different frequencies.

Integrating a multiplexing circuit with a detector wafer will greatly reduce the number of wire bonds required, as shown in Figure~\ref{fig:ReadoutLConChip}. 
%Number of required bond pad will be reduced by multiplexing factor.
%Lawrence Berkeley National Lab collaborated with a commercial microfabrication foundary to integrate superconducting resonators on detector wafers as shown in Figure~\ref{fig:LConChip}
Such resonators can be coupled to \umux{} or high frequency DfMux readout as outlined in Chapter~\ref{chp:readout}.
Currently 50-100\,MHz resonators are being developed for DfMux readout. 
The same method can be used to fabricate $\sim$1~GHz resonators for \umux{} readout, which would be significantly smaller and take less space on a wafer.

Because array packing density problem is global optimization problem, it is hard to come up with the technology status level for the topic.
We assigned TSL level 4, since multiple experiments successfully deployed with densely packed focal plane. However, this does not guarantee focal plane density won't be a problem for CMB-S4. Small f-number at focal plane, physically small pixels for high frequency channels, and multichroic pixel could pose new level of packing density challenge for CMB-S4. 
We assigned slightly lower status level for the production status level. The production status level is 3. 
There are new ideas such as bump bonding to solve interconnect challenge, but these new approaches have not been demonstrated ability to be mass produced at scale of CMB-S4.

%A benefit of fabricating large single detector arrays is decreasing dead space due to detector holders and interfaces.  
%It is possible to eliminate dead space between wafers by designing a for seamless wafer module.
%If there are no dead space, then detector can be diced into smaller array with more perimeter to wire bond to. 
%CCD detectors already successfully implemented such seamless detector array tiling. 

%\noindent\fbox{
%\parbox{\textwidth}{
%\noindent\textbf{Lab demonstration:} 90/150/220 GHz lenslet detector array packed 1600 bolometers on a wafer \\
%\noindent\textbf{Sky demonstration:} 150/230 GHz horn array packed 2000 bolometers on a wafer \\
%\noindent\textbf{Path to CMB-S4:} High density interconnect, integrate multiplex readout on chip, new telescope design
%}
%}

\begin{table}[h]
\begin{center}
\begin{tabular} {|l l|}
\hline  
\textbf{Lab Demonstration:} & 90/150/220 GHz lenslet detector array packed 1,600 bolometers on a wafer \\
\textbf{Sky Demonstration:} & 150/230 GHz horn array packed 2,000 bolometers on a wafer \\
\textbf{Path to CMB-S4:} & High density interconnect, integrate multiplexer on chip, new telescope design\\
\hline 
\end{tabular}
\end{center}
\end{table}

%% file: detector_rf/DetectorCharacterization.tex
%\subsection{Detector characterization}
\paragraph{Introduction}
%Detector characterization is key part of detector fabrication.
%With accurate feed back, it is possible to fine tune detector design to meet various requirements that a detector array need to satisfy. 
%Detector characterization need to provide feed backs that allow design team to identify which RF element need to be adjusted to achieve the goal.

Detector testing is an essential part of detector fabrication, a common need for all the technologies discussed in this paper. 
Detector testing is challenging for two fundamental reasons: (i) CMB detectors utilize superconducting technology and can only be fully characterized using sub-Kelvin testbeds, and (ii) complete optical characterization of CMB detectors requires broadband incoherent light sources spanning $\sim$1--3\,mm wavelengths, a spectrum where there is little or no commercial instrumentation. As such, the testing and feedback associated with developing detector RF architectures can only be fulfilled through research groups at universities and national labs. In this section, we review common methods and challenges associated with characterizing RF performance.

\paragraph{``Room temperature ($>$1~K)'' inspection}
Room temperature measurements are typically used as a first pass assessment of fabrication quality. These measurements include visual inspection and electrical resistance measurements, where the latter provides some information regarding electrical connectivity (or isolation) and materials properties. In general, these measurements primarily help with preparing devices for cryogenic testing. The critical limitation to these measurements arises from the fact that the CMB detectors need to be superconducting in order to operate. Similarly, measurements at $\sim$70\,K are limited in their utility. There is some benefit to measurements at 4\,K, as at this temperature, the detector microstrip structures are functional. Though it isn't possible to characterize the integrated performance of a detector at 4\,K, it is possible to understand generic microstrip properties using dedicated test structures. For example, it is possible to measure a microstrip test device that couples radiation from one polarization, transmits that signal through an RF test circuit (including filters and calibration structures), and then re-radiates the signal into the orthogonal polarization. This test structure can be cooled to 4\,K and analyzed using more conventional room temperature network analyzers.

%the RF circuit can be probed by injecting a linearly polarized signal and receiving radiated signal in orthogonal polarization.
%For this setup, RF circuit is setup such that signal from one polarization is connected to orthogonal polarization on a detector wafer. 
%Polarized signal is injected through one polarization, then it goes through series of RF circuit to orthogonal polarization.
%Signal is radiated out through orthogonal polarization.
%Efficiency can be calculated by comparing setup with RF circuit with same setup with shorted transmission line.

\begin{figure}[h]
\begin{center}
\includegraphics[height = 1.8in]{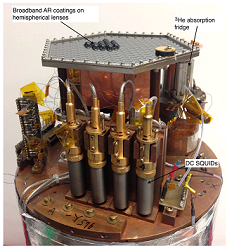}
\includegraphics[height = 1.6in]{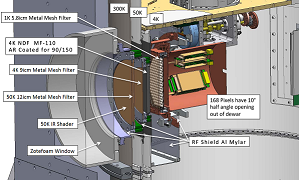}
\includegraphics[height = 1.8in]{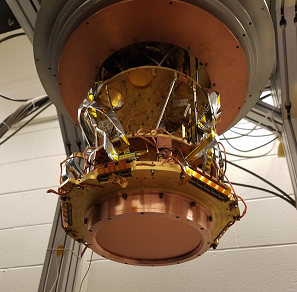}
\end{center}
\caption{(Left) Photograph of a 8-inch wet dewar with a He-3 adsorption refrigerator. TES bolometers are read out by commercially available DC SQUID. (Middle) Cross-section of PTC cooled detector test cryostat with ADR. (Right) Photograph of a PTC cooled dilution refrigerator with an Advanced ACTPol array installed prior to deployment.}
\label{fig:testbeds}
\end{figure}

\paragraph{Sub-Kelvin testbeds}
%Full characterization of CMB detectors operate at cryogenic temperature.
The necessary measurements for developing the detector RF design require operating devices at temperatures below the detector critical temperature with base temperatures ranging from $\sim$50\,mK-300\,mK. Current test beds (see Figure~\ref{fig:testbeds}) include smaller cryostats, often using liquid cryogens, and larger cryostats typically cooled using cryogen-free pulse tube coolers (PTC)s. The advantage of the smaller cryostats is that they can typically reach base temperature in less than 12 hours allowing for rapid turnaround. If cooled using liquid cryogens, the small size efficiently utilizes the liquid cryogens, though regular servicing and monitoring is required to keep the system cold. PTC-cooled cryostats are now commercially available, though they have higher startup costs and require careful design to minimize electrical and microphonic pickup. The advantage of PTC systems is in their low operating overhead, which makes them efficient for tests requiring large cryogenic volumes.

%A wet dewar with liquid cryogens and a dry dewar with a pulse tube cooler are commonly used to cool devices.
%A wet dewar is simple to operate, and it is very useful for quick lab test.
%Typical wet dewar can reach its base temperature in less than 12 hours. 
%Cost of liquid helium becomes non-negligible for many tests.
%Also a wet dewar requires an operator to be around to refill the dewar to prevent the dewar from warming up.

%Pulse-Tube cooler (PTC) is an alternate method to cool a dewar.
%Example of a test dewars with PTC is shown in Figure~\ref{fig:testbeds}.
%Setting up PTC system is simple as multiple ``turn-key'' commercial pulse-tube systems exist.
%Initial cost of PTC system can be high, but after initial purchase only recurring cost is electric power.
%PTC system has a benefit that it can stay cold without an operator being present.
%Care must be taken to set up the PTC system to prevent electrical and microphonic noise from contaminating measurements.

\begin{table}[th]
\begin{center}
\begin{tabular}{l c c c}
\hline
\hline
 & \textbf{He-3 Adsorption} & \textbf{ADR} & \textbf{Dilution}\\
\hline
\hline
Operation  & One-shot & One-shot or Continuous & Continuous\\
Stages [Kelvin] & 2, 0.35, 0.25 & 1, 0.5, $<$ 0.1 & 1, $<$ 0.1\\
Cooling power [$\mu$W] & $\sim 5$ & $\sim 5$ & $\sim 100$\\
\hline
\hline
\end{tabular}
\end{center}
\caption{Comparison of sub-Kelvin cooler}
\label{tab:subfridge}
\end{table}

%CMB the detectors are cooled to 
CMB detector test beds achieve sub-Kelvin operating temperatures using either a Helium-3 adsorption refrigerator, an ADR, or a dilution refrigerator. Comparison of refrigerator characteristics are tabulated in Table~\ref{tab:subfridge}. 

The cryogenic testing technology for CMB detector development is mature and well understood. The primary challenge for CMB-S4 detector development is in the sparsity of this critical resource. Investment into building up sub-Kelvin testing capabilities at universities and national labs is a high priority for CMB-S4 R\&D.
%ADR or dilution refrigerators should be used for testing detector that requires 100 milli-Kelvin base temperature.

\paragraph{Detector loading}
%CMB detectors are designed to observe the CMB through dry atmoshpere and transparent optics that has effective temperature of approximately 10 to 30 Kelvin.
%Laboratory optical tests use black body sources that are around 77 Kelvin to 1300 Kelvin. 
The difference in optical power loading between actual observation and the laboratory environment places requirements on the detector characterization procedure in the lab.
To prevent optical power from the calibration source from saturating CMB detectors, a cold attenuating filter is often installed inside the dewar. 
A commonly used attenuating filter is MF-110, a castable mm-wave absorber.
%It is mounted on a 4-Kelvin stage to reduce optical loading.
There is literature on emissivity of MF-110, but the exact details of the filter performance depend on its temperature and AR coating.
The attenuator has a steep attenuation profile versus frequency, so a filter optimized for one frequency band is not suitable for testing other bands.

Another way to characterize the RF performance is to couple the RF circuit to a detector that is designed for high optical loading. 
A TES bolometer can be fabricated to accept a higher optical load by increasing the transition temperature of the thermistor.
The \bicepII/Keck Array/\bicepIII\ experiments and \Pb-1 had a high Tc superconducting metal (Aluminum) in series with a superconducting metal with transition temperature tuned for observations. 
The detector is biased onto the high Tc superconducting metal for laboratory testing, and the 
detector is biased to the low Tc superconducting metal during actual observation.
This method has the benefit of not requiring a special attenuating filter at the cost of having to fabricate an additional TES for each detector for lab testing.
In-situ testing of the final optical system, including beam and bandpass calibrations, also benefit from having a second, high-loading detector.

\begin{figure}[h]
\begin{center}
\includegraphics[height = 1.8in]{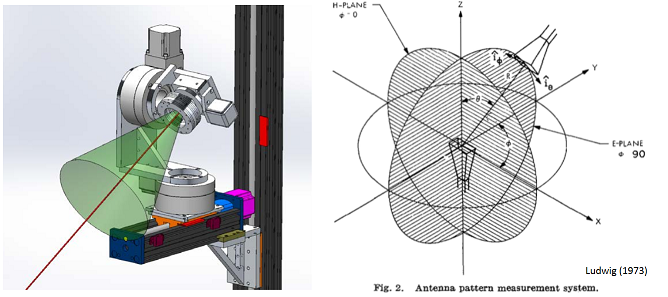}
\includegraphics[height = 1.8in]{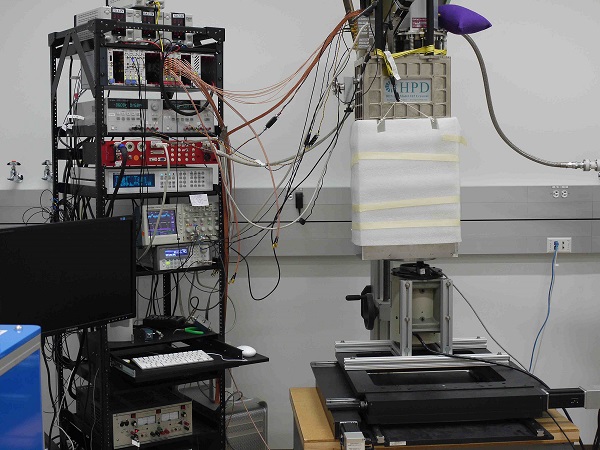}
\end{center}
\caption{(Left) CAD drawing of multi-axis beam map setup. (Middle) Graphical representation of antenna pattern measurement setup. (Right) Photograph of a 2D beam mapping system at NIST. The detectors look down through the bottom of the dewar, while the chopped source points upward and is mounted on a two axis stage.}
\label{fig:beammapmultiaxis}
\end{figure}

\paragraph{Beam}
Angular response of feed is characterized by sweeping a source in front of a detector. 
%Power received by a feed is product of gains of a source antenna and a feed under test.
Simple beam mapping approach is to sweep unpolarized temperature modulated incoherence source with circular aperture in flat 2-dimensional linear stage.
%This setup have simple $\cos(\theta)/r^2$ dependancy with no polarization direction to deal with.
More elaborate setup involves linear translation stages with a source antenna attached to multi-axis rotation head. 
A CAD drawing of a multi-axis system is shown in Figure~\ref{fig:beammapmultiaxis}.

\paragraph{Polarization}
Co-polar (E-plane) beam and cross-polar (H-plane) beam of a feed are characterized by attaching a source with a well-defined polarization to beam mapping setup. 
%Characterizing off-axis polarized beam is challenging as Ludwig's third definition of co-pol and cross-pol requires source polarization to rotate as a function of off-axis angle as shown in Figure~\ref{fig:beammap}
The multi-axis setup described previously allows accurate mapping of the co-polar and cross-polar response of a feed.

\paragraph{Spectrum}
An FTS is used to characterize the frequency response of each detector. 
A plastic sheet is often used as a beam splitter for a Michelson FTS, and a wire grid is used as a beam splitter for a Martin-Puplett FTS.
A dielectric beam splitter has frequency response that needs to be taken into account in data analysis.
Optical coupling between a FTS and a detector needs to be optimized for an accurate spectrum measurement.
Inserting an integrating sphere could mitigate the coupling problem at the cost of degradation of signal strength.
\begin{figure}[h]
\begin{center}
\includegraphics[height = 2in]{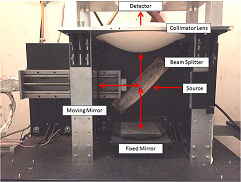}
\includegraphics[height = 2in]{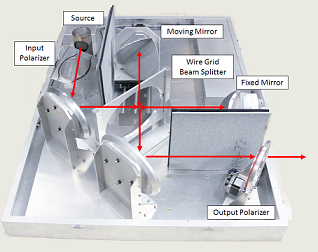}
\end{center}
\caption{(Left) Photograph of a Michelson FTS with a Mylar beam splitter. UHMWPE lens is placed at output of the FTS to collimate output to a detector. (Right) Photograph of a Martin-Puplett FTS that uses wire grid as a beam splitter.}
\label{fig:beammapfts}
\end{figure}

\paragraph{Efficiency}
Accurate characterization of detector (feed to detector) efficiency is necessary to optimize the detector design, and consistency in measured detector efficiencies is a good indicator of fabrication repeatability.
Efficiency is measured by changing the input temperature of a black body source to known temperatures, and comparing the change in power received versus the total optical power from the source at different
input temperatures.
For a single-moded detector in the Rayleigh-Jeans limit, the change in optical power from beam-filling black body source is simply $\Delta P = k_B \Delta T \Delta \nu$ where $k_B$ is Boltzmann's constant, and $\Delta \nu$ is the detector's integrated bandwidth. 
In a case where the temperature modulated source is outside of a dewar, it is important to know the in-band efficiencies of IR filters, attenuator and windows of the dewar.
It is also possible to insert a temperature modulated blackbody inside of cryostat, or to
%This approach also requires infrared filter to prevent focal plane from heating up.
inject signal into one polarization channel, route the signal through on-chip RF circuits, and then re-emit directly through the same antenna on the orthogonal channel. 

\paragraph{Prospects and R\&D path for CMB-S4}
CMB-S4 will deploy an order of magnitude more detectors than Stage-III CMB experiments. 
%To keep up with throughput need, CMB-S4 detector test setup need to shorten detector testing time.
Detector characterization throughput needs to keep up with detector fabrication. 
A significant amount of time for detector testing is taken up by cool down time for the test cryostat, so a 
robust method to shorten cool down time should be demonstrated. 
Automation of testing procedures allow detector characterization to be done in parallel at multiple places, and standardizing the test setup is important to be able to distribute testing to multiple institutions and still be able to compare test results.
As the experiment's sensitivity increases, the requirements on detector systematic errors becomes tighter, and it becomes important to understand details such as characteristics of the attenuating filter used for testing and any reflections that happen between various optical elements in the test setup. 

Development of specific RF circuit components would benefit greatly from new measurement techniques. 
Current practice requires end-to-end measurements, typically with free-space coupling, where the measurement includes only the integrated detector response plus the response from optics external to the device under test. 
Isolation of a specific RF component in the circuit is challenging. 
Cryogenic mm-wave VNAs would allow designers to isolate and develop specific circuit elements of the detector design. 

We assigned technology status level 4 for characterization.
Multiple CMB experiments successfully characterized focal plane system at a level that is good enough for Stage-III experiments.
Expected high sensitivity of CMB-S4 will require tighter control of detector systematic errors. 

The production status level for characterization is 2. 
Stage-III experiments characterized large quantity of focal plane units, but characterization needs much higher throughput if we are to test all detectors. 

%\noindent\fbox{
%\parbox{\textwidth}{
%\noindent\textbf{Lab demonstration:} 250 mK and 100 mK test bed for TDM FDM and MKID system at multiple institutions \\
%\noindent\textbf{Sky demonstration:}  - \\
%\noindent\textbf{Path to CMB-S4:} Standarization. Improve throughput
%}
%}

\begin{table}[h]
\begin{center}
\begin{tabular} {|l l|}
\hline  
\textbf{Lab Demonstration:} & 250/100\,mK test bed for TES and MKID at multiple institutions \\
\textbf{Sky Demonstration:} & N/A \\
\textbf{Path to CMB-S4:} & Standardization. Improve throughput. Dedicated cryo RF test setup\\
\hline 
\end{tabular}
\end{center}
\end{table}

%% file: detector_rf/ConclusionShort.tex
%Given the simultaneous need for high sensitivity and multiple bands to discriminate foregrounds multichroic detector designs are an attractive solution to total bandwidth and spectral resolution requirements of CMB-S4.  Broadband antennas have demonstrated promising performance, and new developments are in place to extend their capabilities.
%RF design for CMB applications is becoming mature and designs have been successfully implemented in Stage-2 and upcoming Stage-3 experiments.
%Efficient coupling of millimeter waves to both TES bolometers and MKIDs have been demonstrated. 
%Developing dielectric insulators with low loss is essential for boosting detector efficiency and providing flexibility in  circuit design.
%Reduced loss dielectric films make it feasible to build higher order narrow band filter designs that subdivide single atmospheric windows.
%Stable material properties are important for predictable RF performance.
%Fabrication facilities with a stable environment are necessary for mass production of CMB-S4 detectors.

There are a wealth of promising RF designs for CMB detectors.  
Given the simultaneous need for high sensitivity and multiple bands to discriminate foregrounds, recent work on implementing multichroic detector designs suggests having two or more frequency bands for each pixel could be a viable path to achieving the total bandwidth and spectral resolution requirements of CMB-S4. 
Two-channel designs have been successfully implemented in Stage-II and Stage-III experiments, and will soon be deployed in more.  
Broadband antennas have demonstrated promising performance, and new developments are in place to extend their capabilities. 
Efficient coupling of millimeter waves to both TES bolometers and MKIDs have been demonstrated. 

Scaling up detector fabrication for CMB-S4 requires an increase in production throughput.
National labs and universities are exploring ways to expand their production and testing capability.
Fabrication facilities with a stable environment are necessary for mass production of CMB-S4 detectors.
Groups are studying the feasibility of automating repetitive tasks, simplifying designs by integrating parts, and outsourcing to commercial fabrication foundries.
Emerging RF techniques, such as metamaterial lenslet arrays, may facilitate mass production.

Developing dielectric insulators with low loss is essential for boosting detector efficiency and providing flexibility in circuit design. 
Reduced loss dielectric films make it feasible to build higher order narrow band filter designs that subdivide single atmospheric windows. 

Technology to improve detector packing density should be developed, and assembly should be simplified for mass production. 
Several new developments such as using stepper lithography to shrink wiring real estate, integrating resonators for multiplexing readout on a detector wafer, and new telescope designs that give more design flexibility for detector array are on going.

Detector characterization is an essential part of detector fabrication.
Timely feedback with accurate information is necessary to fabricate high performance detectors. 
The CMB community has many years of experience characterizing detector arrays. 
Development for high throughput testing is required to meet the demands of CMB-S4.
New test cryostat designs, automation of testing and standardization of detector characterization will be necessary to increase detector testing throughput.
%Standard test setups and methods should be established to characterize detector arrays at different institutions.
Systematic error requirements on RF performance will be tighter for a more sensitive future CMB experiment and higher accuracy detector. 
Detector characterization will be needed to meet these requirements. 

Multichroic detectors are deploying in the field, and new ideas are being tested in laboratories.
Different designs have unique strengths and short comings. 
Feedback from up coming Stage-III experiments will allow us to make informed decisions for CMB-S4 development prioritization. 
Systematic errors that arise from non-ideality in detector performance and cost evaluation were not addressed in this chapter, but they will be addressed in a future iteration of this document.
Detector RF design will be decided based on a global optimization that maximizes scientific return.

%% file: detector_rf/DetSumTable.tex
%TSL and PSL definition.
%
%\begin{table} [!h]
%\begin{center}
%\begin{tabular} {cl}
%\hline
%\textbf{TSL} & \textbf{Description} \\
%\hline
%1 & Lab test of technology to show principle\\
%2 & Lab test of technology but with full feature set and performance suitable for ground test\\
%3 & Experiment capable version built and tested in the lab\\
%4 & Deployed in a CMB experiment and data taken\\
%5 & Data fully analyzed, systematic errors understood\\
%\hline
%\hline
%\textbf{PSL} & \textbf{Description} \\
%\hline
%1 & Fabrication of a TS1/TS2 prototype demonstrated\\
%2 & Fabrication of a one or more experimental capable units\\
%3 & Conceptual plan of methods for production at scale\\
%4 & Demonstrated the critical steps for production at scale\\
%5 & Capability for production at scale exists and is demonstrated\\
%\hline
%
%\end{tabular}
%\end{center}
%\end{table}

\begin{landscape}
\begin{table}
\begin{center}
\begin{tabular} {|l|c|c|c|c|c|}
\hline  
 & \textbf{Lab Demonstration} &  \textbf{Sky Demonstration} & \textbf{Path to CMB-S4} & Section & T/PSL\\
\hline 
\hline
\textbf{Antenna and Feed}  &  &  &  &  & \\
\textbf{Single band}  &  &  &  &  & \\
Feed horn/Planar OMT  & $\surd$ &  90,150\,GHz ACTPol/SPTpol & Mass prod of horn & \ref{sec:horn}, \ref{sec:hornomt} & 5/3 \\
LEKID + Horn & 150\,GHz dual pol &  1.2\,THz BLAST-TNG (2017) & On sky demo, fast prod & \ref{sec:mkid_coupling} & 2/3 \\
Lenslet coupled antenna & $\surd$ &  150 GHz PB-1 & Mass prod & \ref{sec:lensletantenna}, \ref{sec:lensletarray} & 5/3 \\
Antenna Array & $\surd$ & 90/150/220/270\,GHz {\bicepI}s & Steerable beam & \ref{sec:antennarray} & 5/4 \\
Multimode Absorber & BUG for PIPER & MBAC, SHARC-II & Pol sensitive det array & \ref{sec:multimode} & 3/3 \\
\textbf{Multi-chroic}  &  &  &  &  & \\
Feed horn/Planar OMT  & $\surd$  &  90/150, 150/230\,GHz ACTPol & Mass prod, coupling & \ref{sec:horn}, \ref{sec:hornomt} & 4/3 \\
MKID + Horn & Design completed &  - & RF coupling, scalability & \ref{sec:mkid_coupling} & 1/1 \\
Lenslet coupled antenna & 5:1, 40 to 350\,GHz &  90/150, 220/280\,GHz SA(2017) & Mass prod, on sky syst & \ref{sec:lensletantenna}, \ref{sec:lensletarray} & 3/3 \\
 &  &  90/150/220\,GHz SPT-3G &  & & \\
Meta material lenslet & Coupled with antenna & - & Lab study, mass prod & \ref{sec:grin} & 1/2 \\
Antenna Array & 2:1 bandwidth & - & Total bandwidth & \ref{sec:antennarray} & 5/4 \\
Multimode Absorber & PSB for PIXIE & - & Pol sensitive det array & \ref{sec:multimode} & 3/2 \\
\hline 
\textbf{RF Technology}  &  &  &  &  & \\
T-Line & SiNx & SiOx \& Si & Low loss dielectric & \ref{sec:rf_tline} & 5/5 \\
Filter & 3,4,7-channels & 2,3 channel lumped \& stub &  Split atm. window & \ref{sec:onchipfilter} & 5/5 \\
Cross-over & via \& via-less & via \& via-less & $\surd$ & \ref{sec:crossover} & 5/5 \\
Termination & meander \& lumped & meander \& lumped & $\surd$ & \ref{sec:termination} & 5/5 \\
\hline
\textbf{Array Layout}  &  &  &  &  & \\
Array Layout & 90/150/220\,GHz  & 150/23\,GHz0, 2,000 ch/wafer & Integ mux, int connect & \ref{sec:array} & 4/3 \\
 &  1,600 ch/wafer &  &  &  & \\
\hline 
\textbf{Characterization}  &  &  &  &  & \\
Characterization & $\surd$ & mult exp tested \& deployed & Std, throughput, syst & \ref{sec:detchar} &  4/2 \\
\hline
\end{tabular}
\end{center}
\end{table}
\end{landscape}

%% file: readout/detectorreadout_cmbs4.tex
\chapter{ Focal plane sensors and readout} 
\label{chp:readout}
\vspace*{\baselineskip} % title is multiline, and without this no space between title & text
\vspace{1cm}
%\include{readout/exec_summary}

%\tableofcontents

%\include{readout/introduction}
\input{readout/introduction}

\input{readout/tes}

\input{readout/mkids}
%\input{readout/tdm}
\input{readout/tdm}
%\include{readout/fdm}
\input{readout/fdm}
%\include{readout/umux}
\input{readout/umux}

%\include{readout/microwave_readout}
\input{readout/microwave_readout}

\input{readout/conclusion}
\input{readout/readoutSumTable}

% moved this into acronym_readout ; otherwise there's a blank page w/
% ``List of acronyms'' section title followed by another page with the
% list ...
%\section*{List of acronyms}
%\input{readout/acronym_readout}

%
%%%%%%%%%%%%%%%%%%%%%%%%%%%%
%\clearpage
%\bibliographystyle{unsrt}
%\bibliography{bibtex/refs,bibtex/refsTDM,bibtex/refsMKID,bibtex/refsumux}
%
%\end{document}

%% file: readout/introduction.tex
\section{Introduction}
\vspace{0.25in}
%\ref{section:TES} -> 5.2
%\ref{section:mkids} -> 5.3
%\ref{section:tdm} -> 5.4
%\ref{section:fdm} -> 5.5
%\ref{section:umux} -> 5.6
%\ref{sec:mkid_readout}  -> 5.7
%\ref{sec:readout_conclusion} -> 5.8
%\ref{sec:redsum} -> 5.9 

In this section, we briefly review the state of low-noise sensors and
signal readout suitable for CMB polarimetry, focusing on scalable
technologies that hold promise for CMB-S4. For each technology
described here, we provide (1) an overview and references for further
study, (2) a summary of current performance as demonstrated on-sky for
technologies with established heritage, and laboratory performance for
technologies with promising initial results, and (3) challenges and
the requisite R\&D path to scale and/or refine the technology for
CMB-S4 requirements.

We describe low-noise sensors for detecting the CMB in
Sections~\ref{section:TES} and~\ref{section:mkids}, which address
transition edge sensor (TES) bolometers and microwave kinetic
inductance detectors (MKIDs) respectively. Highly multiplexed readout
is crucial for operating large arrays of sensors at sub-Kelvin
temperatures.  MKIDs were conceived to be read out in a frequency-division
 multiplexing (FDM) scheme at approximately GHz frequencies,
as described in Section~\ref{section:mkids}.  Several different
readout techniques exist for TESs.  Sections~\ref{section:tdm}
describes time-division multiplexing (TDM), while Sections~\ref{section:fdm}
and~\ref{section:umux} overview two different FDM schemes. All the TES
readout methods rely on cold-stage signal amplification using
SQUIDs. The FDM techniques using GHz interrogation frequencies use a
cold-stage low-noise amplifier such as a high-electron mobility
transistor (HEMT) amplifier to read out groups of 500-1000 detectors.
Section~\ref{sec:mkid_readout} reports on the room-temperature
electronics for FDM in its various forms. Finally,
Section~\ref{sec:readout_conclusion} gives conclusions from this
sensor and readout review, and Section~\ref{sec:redsum} provides
summary tables.

The large number of detectors required for CMB-S4 puts a premium on developing
robust methods for assembly of the sensors with the cryogenic readout
(often called ``packaging'') and/or techniques for integrated
fabrication of sensors with readout elements. The present work
provides context for these assembly and integration issues, but
further elaboration is delayed for future work, as are direct
performance comparisons and detailed cost discussions.

%% file: readout/tes.tex
\section{Transition edge sensor (TES) bolometers}
\label{section:TES}

\subsection*{Description of the technology}
\begin{figure}[t]
\centering
\includegraphics[trim=2.5in 0.25in 2.0in .25in,clip,scale=0.4576]{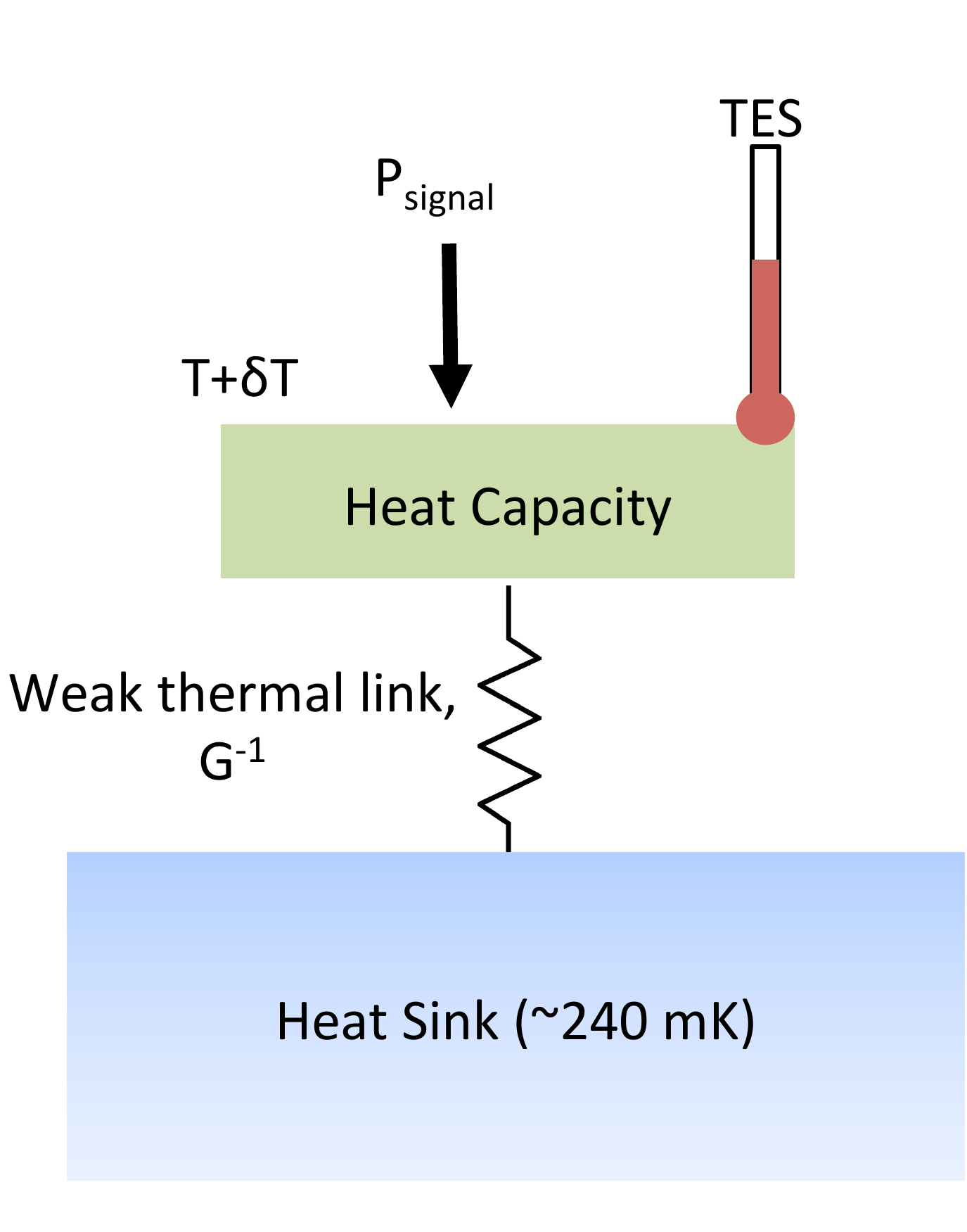}
\includegraphics[trim=2.25in 1.0in 2.5in 1.75in,clip,scale=0.385]{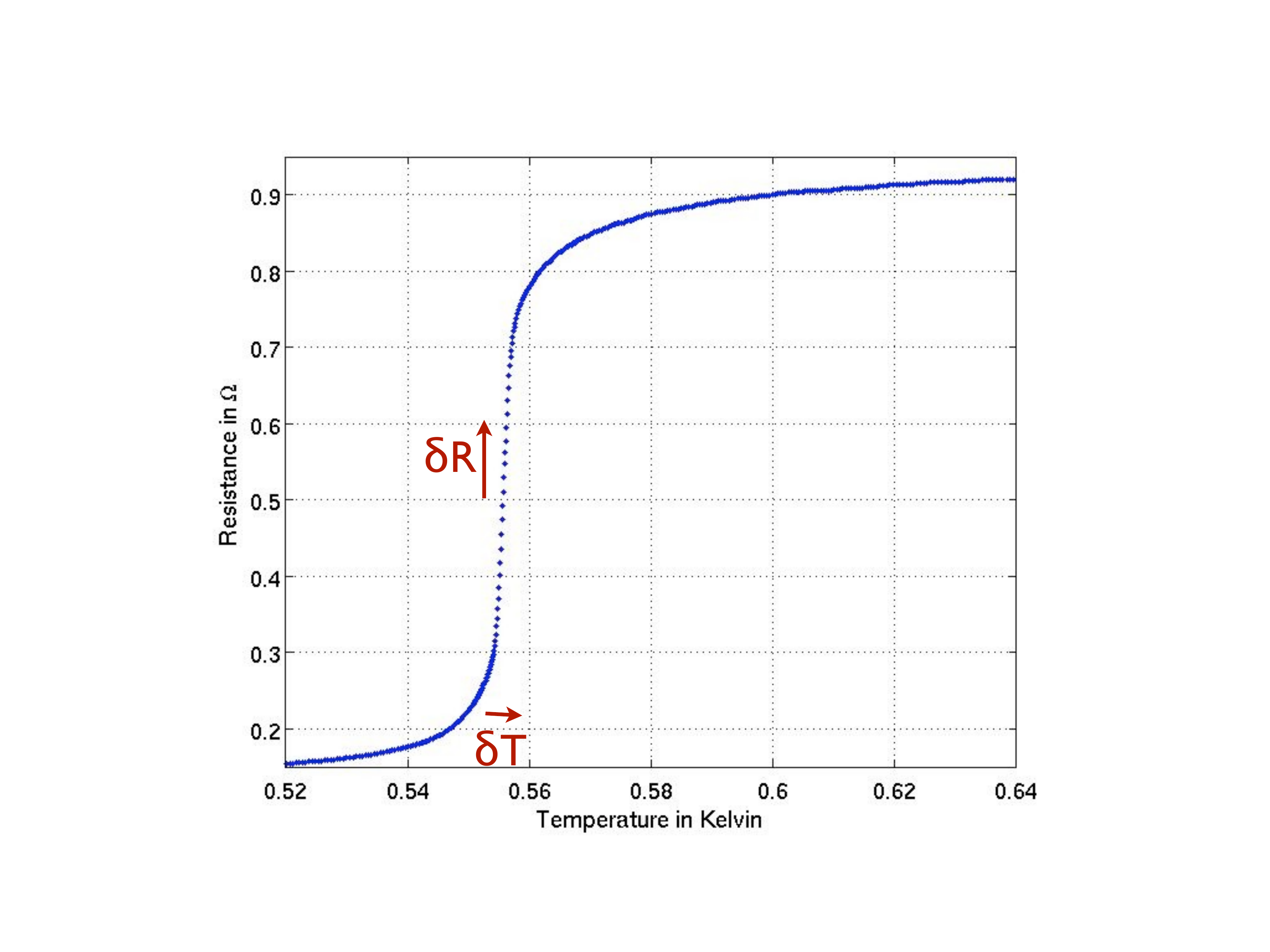}
\vskip 0pt
\caption{(Left) Illustration of a thermal circuit for a typical TES detector highlighting the principles of signal detection. A weakly thermally sunk heat capacity absorbs power, P$_{\rm signal}$, which is to be measured. Variations in the absorbed power change the heat capacity's temperature, which is measured by a TES operating under strong electrothermal feedback. (Right) Plot of resistance versus temperature for a typical TES illustrating the principles of negative electrothermal feedback~\cite{Irwin95}. The TES is voltage biased onto its superconducting-to-normal transition. Small changes in the TES temperature produce large changes in the TES resistance. Since the TES is voltage biased, an increase (or decrease) in the temperature produces an increase (or decrease) in the resistance leading to a decrease (or increase) in the Joule heating power supplied by the bias. This canceling effect corresponds to a strong negative electrothermal feedback making the current through the TES nearly proportional to P$_{\rm signal}$.}
\label{fig:TEScartoon}
\vskip -12pt
\end{figure}

A TES bolometer is a highly sensitive thermometer consisting of a superconducting thin film weakly heat-sunk to a bath at much lower temperature than the superconductor $T_{c}$ (see Fig.~\ref{fig:TEScartoon}). 
Arrays of these devices are fabricated via micro-machining of thin films deposited on silicon wafers.
%By supplying electrical power to the TES, we can raise the temperature of the sensor so that the film is in the middle of its superconducting-to-normal transition (see Fig.~\ref{fig:TEScartoon}, right).
When supplied with a voltage bias, a TES sensor can operate in its superconducting-to-normal transition such that small changes to the TES temperature, arising from changes in the absorbed power, lead to large changes in the TES electrical resistance. The combination of voltage bias and sharp transition (large $dR/dT$) lead the TES to experience strong electrothermal feedback~\cite{Irwin95}: the TES Joule power dissipation, $V^{2}/R$, opposes changes in the incident power, maintaining the TES at a nearly constant temperature. This negative feedback linearizes the detector's response, expands its bandwidth, and ensures a simple relationship (``self-calibration'') between observed TES current and incident power.

%Due to the shape of the TES transition, the applied voltage bias establishes a negative feedback loop~\cite{irwin:1998} where the change in TES resistance cancels external changes in absorbed power. This negative electro-thermal feedback is very strong because the transition is very sharp linearizing the detector response and expanding the detector bandwidth.

Operationally, a TES is voltage biased using either an AC or DC signal and is read out using a SQUID. SQUIDs have much lower noise than TES or photon noise, enabling multiplexed detector readout schemes 
%(see TDM, FDM and uMUX \textbf{$\mu$MUX ?})
(see Sections~\ref{section:tdm},~\ref{section:fdm}, and~\ref{section:umux}). Multiplexed readout is important for operating large arrays of detectors at sub-Kelvin temperatures. An important consideration in TES detector design is operational stability of the electrothermal circuit. The detector's operational time constant needs to be fast relative to the sky signal, but slow relative to the per-channel readout bandwidth. 
%Fielded TES detectors satisfy these constraints by engineering the detector transition shape, internal heat capacities and conductivities to realize operational time constants of $\sim$1\,ms.
Fielded TES detectors have satisfied these constraints with tuned bolometer heat capacities and thermal conductivities, which in combination with the detector transition shapes $R(T, I)$ have yielded time constants of  order 1\,ms.

The theoretical foundations for use of TESs in detectors are well developed~\cite{Irwin2005} providing good descriptions of the noise and response for real devices. In CMB applications, the irreducible noise for a TES detector arises from statistical fluctuations in the absorbed photons~\cite{Zmuidzinas2003}. For ground-based experiments, this noise is typically ${\cal O}$(10) aW/$\sqrt{\mathrm{Hz}}$, though values vary depending on platform/site, observation frequency/bandwidth, and the instrumental throughput/efficiency. The second source of fundamental noise for TES bolometers comes from fluctuations in the thermal carriers of the TES's weak thermal link to the bath~\cite{Mather82}. With appropriate thermal isolation structures and $T_{c}$ ranging from 100--500\,mK, TES detectors can achieve thermal conductivities of $\sim$50-200\,pW/K, where the thermal fluctuation noise becomes comparable or subdominant to the photon noise. Together with sufficiently low-noise readout electronics, TES bolometers have achieved nearly background limited sensitivities.

\subsection*{Demonstrated performance}

TES bolometric detectors have been applied across a diverse set of CMB
experimental platforms. Current detector architectures utilize
a low-loss superconducting microstrip transmission line coupled to planar structures to
realize optical bandpass definition, polarization separation, beam
synthesis and radiation coupling (see
Chapter~\ref{ch:detetcorrf}). Examples of implemented TES
architectures include in-phase combined antenna arrays~\cite{Kuo_SPIE} used by the
\spider, \bicepII, \bicepIII, and the Keck Array experiments, lenslet coupled
antennas~\cite{arnoldpb1} used by the \Pb{} and SPT-3G experiments, absorber
coupled devices used by the EBEX~\cite{Reichborn2010,EBEXPaper2} and SPTpol (90
GHz)~\cite{Sayre12} experiments, and feedhorn coupled devices with
planar orthomode transducers used by the ABS, CLASS~\cite{Denis16},
ACTPol~\cite{Thornton16}, and SPTpol (150\,GHz)~\cite{Henning:2012}
experiments. For these detector architectures, the RF performance can
be modeled and simulated with results in good agreement with measured
performance (see Chapter~\ref{ch:detetcorrf}).  

CMB detector performance is typically reported as a noise equivalent power (NEP) in
units of W/$\sqrt{\textrm{Hz}}$, defined as the amount of detected
signal power required to obtain a signal-to-noise ratio of unity in a
1~Hz bandwidth.  
Detector NEP is often converted into a noise equivalent temperature (NET), 
which refers the noise to the equivalent sky signal in units of K$\sqrt{\textrm{s}}$. This conversion requires knowledge of (or assumptions about) the end-to-end optical efficiency of the completed receiver system, as well as the frequency spectrum of the observed power.
CMB experiments have deployed TES detectors sensitive to frequencies
spanning the entire range envisioned for CMB-S4: 40--300\,GHz, with
detectors achieving noise equivalent power (NEP) of (30-50)\, aW/$\sqrt{\textrm{Hz}}$ (nearly
background limited at CMB frequencies).
%TES detectors have been deployed across experiments spanning
%(40-300)\,GHz, the entire optical frequency range envisioned for
%CMB-S4, with detectors achieving NEPs of (30-50)\,
%aW/$\sqrt{\textrm{Hz}}$ (nearly background limited at CMB
%frequencies).
Detectors deployed at low optical frequencies ($\sim$40\,GHz) and balloon-borne payloads should realize even lower NEPs of $\sim$10\,aW/$\sqrt{\textrm{Hz}}$. In multiple deployed experiments, the TES noise is consistent with what is predicted from theoretical modeling with realized ground-based experimental sensitivities (array NET) in the range of $\sim$10-20~$\mu$K$\sqrt{\textrm{s}}$~\cite{OBrient15, Thornton16, pb, George12}.%Kermish12 is pb

\subsection*{Prospects and R\&D path for CMB-S4 for TES bolometers}
%Given the maturity, diversity and demonstrated performance of TES-based CMB detectors, the TES bolometer technology is at a high technical readiness level. R\&D to scale up TES array production is the most critical element in advancing TES technology for CMB-S4.
%%The R\&D focus for using TES detectors in CMB-S4 is on scaled production.
%TES detectors are fabricated via micro-machining of thin films deposited on silicon wafer substrates.
%%The materials and contamination requirements for superconducting device fabrication (including TES detectors) is incompatible with microfabrication foundries used in the integrated circuit industry. R\&D is required for TES technology to meet the needs of CMB-S4. Three lines of research are as follows:
%%[Toki: I took this out because this sentence is only true for selected MOS fabrication line in a foundary, even at UCB MOS fab and MEMS (TES fab falls into this category) fab are co-existing in same fab space]
% However, unlike semiconductor-based detectors, the absence of a developed superconducting integrated circuit industry means commercial production options having the requisite economies of scale are not readily available. Here we identify three lines of research for preparing TES detector technology for CMB-S4.

Given the maturity, diversity and demonstrated performance of
TES-based CMB detectors, TES detectors are at a high technology status
level (see Section~\ref{sec:redsum}). Multiplexed readouts are
required for implementing TES focal planes with thousands of
detectors.  Current multiplexer technologies
(Sections~\ref{section:tdm} and~\ref{section:fdm}) already enable
arrays of ${\cal O}$(10,000) detectors, though readout R\&D described
in Sections~~\ref{section:fdm}--\ref{sec:mkid_readout} could improve
scalability and lower cost.  The most critical R\&D elements needed to
advance TES technology for CMB-S4 is that associated with scaling up
TES array production for large sensor counts and sufficient quality
assurance.

\begin{itemize}
\item {\bf Increase production throughput:} Current TES bolometer array fabrication typically involves processing $\sim$10 layers of materials on substrates that are 100-150\,mm in diameter. A 150-mm wafer supports $\sim$1,000 detectors at 150\,GHz, a density which varies strongly with observing frequency, and which can be multiplied with multichroic optical coupling designs. Arrays are typically fabricated by a team of 2-3 experts producing 5-10 arrays in approximately 3-6 weeks. Improvements in fabrication throughput will come from parallelizing fabrication resources, both person-power and equipment, and by developing modest changes to fabrication techniques and logistics.
%Modest and low risk improvements to fabrication techniques and logistics (e.g. interleaved fabrication batches) are expected to increase this throughput.
The primary requirement for increasing TES production is access to micro-fabrication resources with a particular need for dedicated thin film deposition systems to guarantee cleanliness and control of exotic materials. % This exclusive access to microfabrication tooling can be addressed with national lab resources.

\item{\bf Optimize materials and establish a quality assurance
  program:} Subsets of TES bolometer arrays for CMB-S4 will by design possess small
  variations in device parameters to accommodate
  different operating conditions associated with different observing frequencies, sites and instrument throughput. It is also possible that
  different RF coupling schemes will be employed to optimize use of
  different platforms. An important R\&D goal is to identify the best
  materials and processing to accommodate these minor variations in
  TES designs such as optimal operation temperatures (100 vs 300\,mK)
  and different RF couplings (see Section~\ref{sec:rf}). This R\&D
  should proceed in parallel with a program focused on understanding
  the connection between variations in fabrication processing and
  superconducting RF circuit performance and thermal properties. In
  addition to materials and process optimization, it is important to
  establish test facilities and a quality assurance program among the
  universities, national labs and fabrication facilities that is
  commensurate with the increased fabrication throughput. The ultimate
  goal of this R\&D would be an end-to-end production line yielding
  TES bolometer arrays with uniform properties across each wafer and consistent
  performance from wafer-to-wafer for a given set of device parameters.

%\item {\bf Improved production reliability} %The primary challenge
%for scaling A critical challenge for scaling TES detector arrays for
%CMB-S4 is improving the fabrication consistency. Addressing this
%challenge requires a short term R\&D program focused on understanding
%the connection between variations in fabrication processing and
%superconducting RF circuit performance and mechanical thermal
%properties. A parallel goal of this work will be to Improving
%production reliability also requires

%\item {\bf Multiplexed TES readout:} Multiplexed TES readouts are
%required for implementing focal planes with more than 1,000 detector
%elements and will continue to be an active component for
%R\&D. Current multiplexer technologies already enable operation of
%arrays of $O$(10,000) detectors. Continued improvement of these
%multiplexing schemes will further extend these capabilities. Recent
%developments of new readout techniques may lead to new multiplexer
%technologies with broader applicability and lower cost.

\end{itemize}

%% file: readout/mkids.tex
\section{Microwave Kinetic Inductance Detectors}
\label{section:mkids}

\subsection*{Description of the technology}

MKIDs are superconducting
thin-film GHz resonators that are also designed to be photon
absorbers~\cite{zmuidzinas2012a}.
Absorbed photons with energies greater than the superconducting gap
of the film
($\nu > 2 \Delta/h \cong 74~\mbox{GHz} \times (T_c/1~\mbox{K})$) break
Cooper pairs, changing the density of quasiparticles in the device.
The quasiparticle density affects the dissipation of the
superconducting film and the inductance from Cooper pair inertia
(kinetic inductance), so that a changing optical signal will cause the
resonant frequency and internal quality factor of the resonator to
shift.
These changes in resonator properties can be detected as
changes in the amplitude and phase of a probe tone that drives the
resonator at its resonant frequency.
This detector technology is particularly well-suited for sub-Kelvin,
kilo-pixel detector arrays because each detector element can be
dimensioned to have a unique resonant frequency, and the probe tones
for hundreds to thousands of detectors can be carried into and out of
the cryostat on a single pair of coaxial cables (see
Section~\ref{sec:mkid_readout}).

The total instrument noise is the quadrature sum of the detector noise
and the photon noise, and the fundamental performance goal is to
achieve a sensitivity that is dominated by the random arrival of
background photons.
For an MKID, the detector noise includes contributions from three sources:
generation-recombination (g-r) noise, two-level system (TLS) noise,
and amplifier noise~\cite{zmuidzinas2012a}.
In general, g-r noise comes from the generation and recombination of
quasiparticles.
Under typical operating conditions for ground-based CMB observations,
any thermal g-r noise is negligible, so the two main noise sources are
quasiparticle generation noise from photons (photon noise) and the
associated random quasiparticle recombination noise.
TLS noise is produced by dielectric fluctuations due to quantum two
level systems in amorphous dielectric surface layers surrounding the
MKID.
The scaling of TLS noise with operating temperature, resonator
geometry, and readout tone power and frequency has been extensively
studied experimentally and is well described by a semi-empirical
model~\cite{Gao2008b}.
Finally, the amplifier noise is the electronic noise of the readout
system, which is dominated by the cryogenic microwave low-noise
amplifier.

%%%%%%%%%%%%%%%%%%%%%%%%%%%%%%%%%%%%%%%%%%%%%%%%%%%%%%%%%%%%%%%%%%%%%%

\subsection*{Demonstrated performance}
A range of MKID-based instruments have already shown that MKIDs work
at millimeter and sub-millimeter wavelengths.
Early MKIDs used antenna coupling~\cite{Day2006}, and these
antenna-coupled MKIDs were demonstrated at the Caltech Submillimeter
Observatory (CSO) in 2007~\cite{Schlaerth2008} leading to the
development of MUSIC, a multichroic antenna-coupled MKID
camera~\cite{golwala+12}.
A simpler device design that uses the inductor in a single-layer $LC$
resonator to directly absorb the millimeter and sub-millimeter-wave
radiation was published in 2008~\cite{doyle}.
This style of MKID, called LEKID, was first demonstrated in 2011 in the 224-pixel NIKA
dual-band millimeter-wave camera on the 30\,m IRAM telescope in
Spain~\cite{monfardini}.
This pathfinder NIKA instrument led to an upgraded
polarization-sensitive NIKA2 receiver with approximately 3,300
detectors~\cite{calvo_2016,ritacco_2016}.
A large format sub-millimeter wavelength camera, called A-MKID, with
more than 20,000 pixels and a readout multiplexing factor greater than
1,000 has been built and is currently being commissioned at the APEX
telescope in the Atacama Desert in Chile~\cite{baryshev+2014}.

Photon-noise-limited horn-coupled LEKIDs sensitive to 1.2\,THz were
recently demonstrated~\cite{Hubmayr_2015} for
use in the balloon-borne experiment
BLAST-TNG~\cite{galitzki2014,dober2014}.
Laboratory studies have shown that state-of-the-art MKID and LEKID
designs can achieve photon noise limited
performance~\cite{flanigan_2016,mccarrick_2014,mauskopf_2014,mckenney_2012}.
Finally, MKID-based, on-chip spectrometers for sub-millimeter
wavelengths (SuperSpec and Micro-Spec) are currently being
developed~\cite{superspec2012,microspec2013}.

Two scalable varieties of MKID -- using two completely different RF
coupling strategies -- are currently being developed for CMB
polarization studies with CMB-S4 in mind: (i) dual-polarization
LEKIDs and (ii) multichroic MKIDs.
%, which are shown in Figure~\ref{fig:mkid_coupling}~\cite{johnson_2016,mccarrick_2016}.
The details of the RF coupling designs are discussed in
Section~\ref{sec:mkid_coupling} and Section~\ref{sec:termination}.
The horn-coupled, multichroic devices are based on the polarimeters
that were developed for the Advanced ACTPol
experiment~\cite{henderson/etal:2016,Datta_2014}.  However in this new
MKID-based version, the TES bolometers are replaced with hybrid
coplanar-waveguide
(CPW) MKIDs, and the millimeter-wave circuit is
fully re-optimized for SOI wafers.
The multichroic MKIDs are still in the development stage, and a
laboratory performance demonstration is planned for
early 2017.
The NET, NEP,
in-band spectral response, pulse response (time constant),
low-frequency noise performance, and multiplexing performance of
LEKIDs have all been studied extensively in the
laboratory~\cite{mccarrick_2016,flanigan_2016,mccarrick_2014}.
These studies have revealed that the performance of LEKIDs can be 
comparable to that of state-of-the-art TES bolometers -- especially
for ground-based experiments when the optical loading is greater than
approximately 1\,pW.

Development work is underway to make the sensing element in various
MKID architectures out of materials with a tunable transition
temperature, such as aluminum manganese (AlMn), titanium nitride
(TiN), TiN trilayers, and aluminum-titanium
bi-layers~\cite{deiker_2004,lowitz_2016,catalano_2016}.
With these materials it is possible to decrease the transition
temperature below that of thin-film aluminum in a controllable way,
which does two critical things.
First and foremost, near 150\,GHz photons are energetic enough to break
multiple Cooper pairs in the sensing element, so that the detector noise
will be further suppressed below the photon noise improving the
sensitivity.
Second, a lower $T_c$ makes the detector technology sensitive to lower
frequencies ($\sim$30\,GHz), so that one MKID architecture with a tunable
transition temperature could be used for all of the spectral bands in
CMB-S4.

\subsection*{Prospects and R\&D Path for CMB-S4 for MKIDs}
MKIDs are a new detector option for CMB studies, and they may have
appreciable advantages worth considering for CMB-S4.
For example, the technology was invented with high multiplexing
factors in mind, the readout uses low-power commercially available
hardware (see Section~\ref{sec:mkid_readout}), some device
architectures can be made from a single superconducting film, and
high-performance prototype LEKIDs have been fabricated in small
commercial foundries.
Therefore, although MKIDs lack the heritage of TES bolometers in the
CMB community, it is reasonable to anticipate that the technology
could flourish in a large-scale program like CMB-S4.
To make MKIDs a viable candidate for CMB-S4 instruments, research and
development work must be done in the following areas:

%The readout system for MKIDs requires hardware to generate and
%demodulate hundreds of individual microwave tones, a cryogenic low
%noise amplifier (LNA), and low-loss cryogenic microwave transmission
%lines.
%%
%A well-established example uses the open-source ROACH FPGA hardware
%with associated ADC/DAC boards, a SiGe amplifier, and superconducting
%coaxial cables (see Figure~\ref{fig:mkid_readout}).
%%
%%For this example construction, 
%All of these elements are already
%commercially available, so the technical readiness level is high.
%%
%For CMB-S4 any readout development work would likely focus on FPGA
%programming.
%%
%Open-source software packages are already available as starting points
%for this work.

\begin{itemize}

\item \textbf{Build deployment-quality arrays:}
To date, in the spectral bands for CMB-S4 (30--300\,GHz), only comparatively small arrays
and scalable prototype arrays of MKIDs have been built, mostly at frequencies of 150\,GHz and above.
These existing technologies will need to be scaled up and optimized
for performance, yield and manufacturability.

\item \textbf{Demonstrate MKIDs on the sky:}
An on-sky test demonstrating that MKIDs can be used for high-precision
CMB polarimetry is the critical next step.
NIKA2 is starting to make polarization measurements now, and this work
will be informative.
LEKID-based CMB polarimeter concepts have been considered but not yet
funded or built~\cite{johnson_2014,araujo_2014,lowitz_2016}.
Dual-polarization LEKID arrays~\cite{mccarrick_2016} with
approximately 500 single-polarization detectors are currently being
fabricated, and a demonstration using this array could take place in
the next year or two.

\item \textbf{Increase production throughput:}
MKIDs can be fabricated using the tools and techniques currently
available in the foundries in national laboratories.
However, the number of detectors required for CMB-S4 is unprecedented,
so improvements in fabrication throughput will be required.
A coordinated effort among existing foundries will likely be needed.

%\item \textbf{Develop Warm Electronics Firmware:}
%%
%Well-established examples exist of warm readout for MKID arrays using
%open-source ROACH FPGA hardware and commercially available components,
%so the technical readiness level is high.  For CMB-S4 any readout
%development work would likely focus on FPGA programming.

%% STS suggested moving this into text at start of R\&D section
%\item \textbf{Develop warm readout Readout Software:}
%%
%The readout system for MKIDs requires hardware to generate and
%demodulate hundreds of individual microwave tones, a cryogenic low
%noise amplifier (LNA), and low-loss cryogenic microwave transmission
%lines.
%%
%A well-established example uses the open-source ROACH FPGA hardware
%with associated ADC/DAC boards, a SiGe amplifier, and superconducting
%coaxial cables (see Figure~\ref{fig:mkid_readout}).
%%
%%For this example construction, 
%All of these elements are already
%commercially available, so the technical readiness level is high.
%%
%For CMB-S4 any readout development work would likely focus on FPGA
%programming.
%%
%Open-source software packages are already available as starting points
%for this work.

\end{itemize}

%%%%%%%%%%%%%%%%%%%%%%%%%%%%%%%%%%%%%%%%%%%%%%%%%%%%%%%%%%%%%%%%%%%%%%

%% file: readout/tdm.tex
\section{Time-division multiplexing (TDM) using DC SQUIDs}
\label{section:tdm}

\subsection*{Description of the technology}
\label{sec:overview}
%In time-division multiplexing (TDM), a group of detectors share a common readout amplifier chain by addressing each detector one at a time in a sequence.  The detectors are arranged into a two-dimensional logical array, with a dedicated readout chain for each column.  Only one row of the array is routed to the amplifiers at any given time.  The various rows are cycled through in rapid succession to record the entire array.
In TDM, a group of detectors is arranged into a two-dimensional logical array. Each column of detectors shares a dedicated readout amplifier chain, and only one row of the array is routed to the amplifiers at any given time. The various rows are addressed cyclically in rapid succession to record the entire array.
%We focus primarily on the system architecture developed at NIST~\cite{Irwin:TDM2002,deKorte:2003mux} and readout electronics developed at UBC~\cite{Battistelli:2008} as applied to the readout of large arrays of transition-edge sensors (TESs).
%\subsubsection{Cold hardware}

\begin{figure}[htb]
   \centering
   \includegraphics[width=4in]{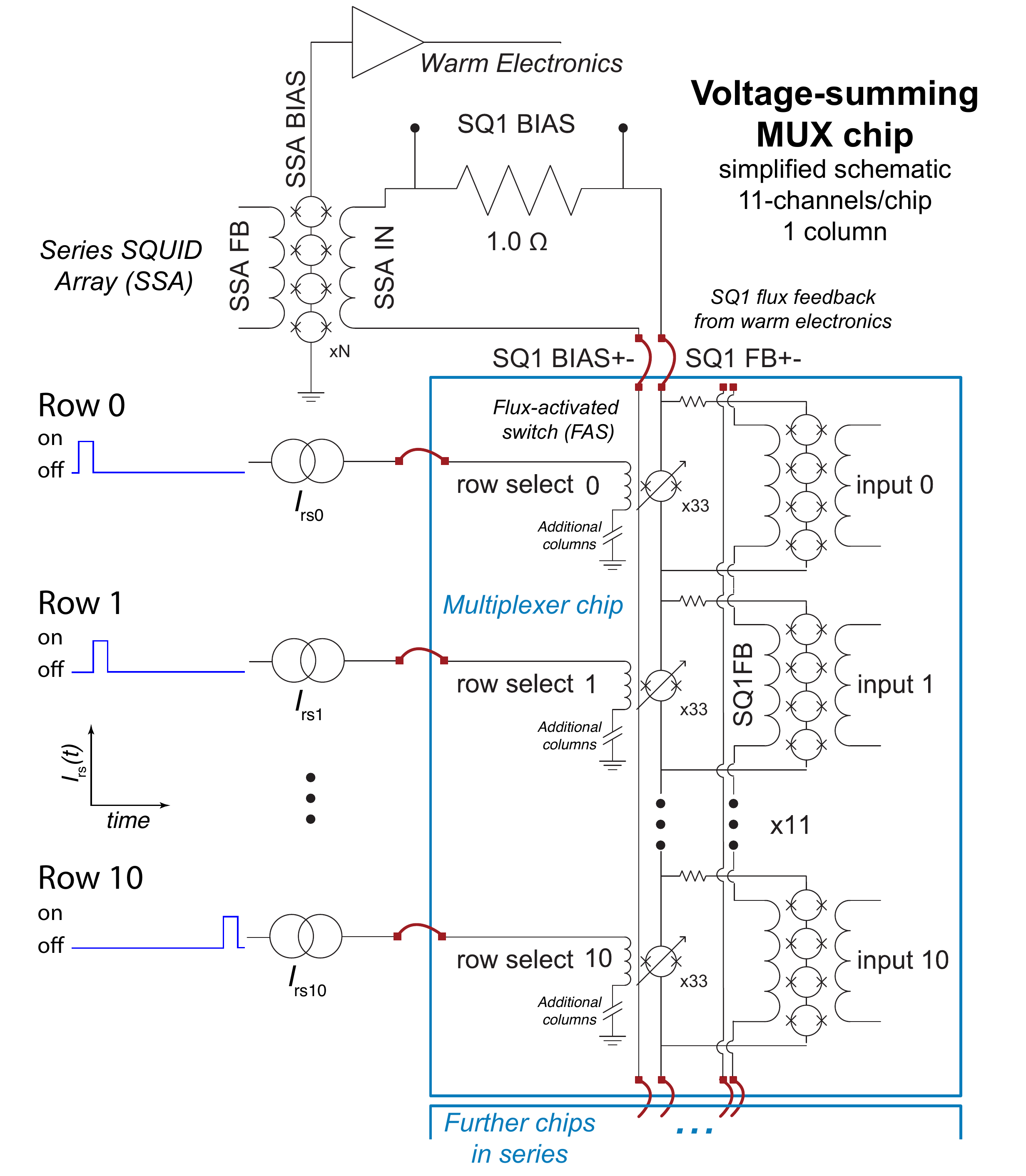} % requires the graphicx package
   \caption{Schematic illustration of a single column of the voltage-summing NIST SQUID multiplexer system. Each TES is coupled inductively to a SQ1. All SQ1s in a column are wired in series to the input of a SSA, but at any given time all but one row of SQ1s is bypassed by a flux-activated switch. The various row-select lines are biased in sequence with low-duty-cycle square waves, as shown at left.
   %Figure adapted by Jeff Filippini from one by Kent Irwin.
   }
   \label{fig:ReadoutChainVS}
\end{figure}

In the latest generation of the system architecture developed at NIST~\cite{Irwin:TDM2002,deKorte:2003mux}, the current signal from each TES is amplified by a dedicated first-stage SQUID (SQ1)\footnote{Each first-stage SQUID in the current system is actually itself a small series SQUID array, but we ignore that in this discussion to avoid confusion with each column's SQUID array}. Each first-stage SQUID is wired in parallel with a Josephson junction switch, and the series voltage sum of all such units in the column is amplified by a series SQUID array (SSA) for transmission to the warm electronics. During multiplexing all but one of the switches are closed to short out the inactive SQUIDs, so that only a single first-stage SQUID feeds the SSA at any given time. This arrangement is shown schematically in Figure~\ref{fig:ReadoutChainVS}. This two-stage voltage-summing TDM architecture was first deployed by \bicepIII\ in 2015~\cite{Hui:2016pfm}; previous instruments such as SCUBA-2, \bicepII\, and ACT employed a three-stage flux-summing architecture~\cite{deKorte:2003mux}.

The first-stage SQUIDs and flux-activated switches for 11 rows of a single readout column are patterned on a single ``multiplexer chip''.  Each multiplexer chip is mated to a corresponding `interface chip', which contains the parallel (shunt) resistors to voltage-bias each TES and series inductors to define the TES bandwidth. The lines connecting the multiplexer and Nyquist chips to the TESs must have low parasitic resistance (typically superconducting), so these chips are typically operated at the detector temperature (0.1-0.3)\,K.

%\subsubsection{Warm electronics}
%The complete open-loop response function of this system is a sinusoidal composition of the modulation curves of the two SQUID stages.
The Multi-Channel Electronics (MCE), developed 
%at UBC 
for SCUBA-2~\cite{Battistelli:2008}, 
provide bias currents and flux offsets for all SQUID stages and switches, thus controlling the shapes and relative alignments of the various modulation curves. The warm electronics linearize this complex response function through flux feedback to the first-stage SQUIDs, keeping each locked at an appropriate point along its modulation curve. This feedback constitutes the recorded signal for each detector. A single MCE crate contains all of the low-noise DACs, ADCs, and digital processing necessary to operate a full TDM array of up to 32 columns and 64 rows. Each circuit board in the crate is controlled by a field-programmable gate array (FPGA), allowing for feature additions and bug fixes through firmware updates.  The entire crate communicates with a control computer through a single fiber optic pair, ensuring electrical isolation. Multiple MCEs can be synchronized with one another (and with external hardware) using a shared ``sync box'', which distributes trigger signals and time stamps from a crystal oscillator.

%The MCE is controlled and read out by a mature and versatile software code base, developed by the UBC group and the larger user community.  Robust automation of biasing, data acquisition, and error recovery with minimal human intervention has been demonstrated for kilopixel arrays, both for terrestrial receivers and for the balloon-borne \spider\ instrument.  Both software and firmware development are ongoing, regularly incorporating feature requests from the various instrument teams.

%\subsubsection{Interconnects}
Like all systems in which the multiplexing operation takes place outside the detector wafer, the NIST TDM system makes heavy hybridization demands. The connections between TES and SQ1 must have parasitic resistances that are small compared to the TES shunt resistor (typically a few m$\Omega$), which precludes most connector types. We thus must typically make at least eight superconducting wire bonds per TES: two each from detector to circuit board, from circuit board to multiplexer chip, and between multiplexer and interface chips, plus two for the row select lines. The present division between multiplexer and interface chips is largely to allow interchangeability among instruments with multiple TES architectures; in mass production these chips could easily be integrated, saving one wire bond pair per channel. Connections between the multiplexer chips and SSAs can be made with superconducting Nb wiring for low parasitic resistance and acceptable thermal isolation. The requirements on parasitic resistance are much weaker here, so connectors may be used.

In the TDM system the number of wires to ambient temperature scales roughly as the perimeter of the 2D readout array, while the pixel count scales as the area. It requires one pair per row (row select) and four pairs per column (bias and feedback for the first-stage SQUIDs and SSA). These connections are typically twisted pairs with few-MHz bandwidth.

\subsection*{Demonstrated performance}
\label{sec:parameters}

The TDM architecture described above is now very mature and has extensive field heritage on a variety of CMB instruments, including ABS~\cite{EssingerHileman:2010hh}, ACT~\cite{Swetz:2010fy}, ACTPol~\cite{actpol}, \bicepII~\cite{BICEP2_II:2014}, \bicepIII~\cite{ahmed2014}, CLASS~\cite{Essinger-Hileman14}, Keck Array~\cite{Ade:2015fwj}, and \spider~\cite{Fraisse2013}. As with the other multiplexer systems we describe, the primary figures or merit to consider are (1) wire count per detector, both cold (hybridization effort) and warm (thermal load); (2) thermal loading, both from wiring and amplifier dissipation; and (3) noise, which should be sub-dominant to expected detector and photon noise.

%\subsubsection{Achieved multiplexing factor}
The achievable multiplexing factor is constrained by the ratio of readout bandwidth to TES bandwidth. For a science signal bandwidth of $\lesssim$100\,Hz, considerations of stability typically demand a TES bandwidth of order a few kHz~\cite{Irwin2005}. This bandwidth is defined by the TES resistance (typically $<$1\,$\Omega$) and the inductor on the interface chip (typically 0.1-2\,$\mu$H).
%, so higher multiplexing factors require higher readout bandwidth.
Readout chain bandwidth is typically defined by the SQUID amplifier and interconnects, notably by the $L/R$ time constant of the first-stage SQUIDs driving the SSA input coil and (in some cases) by the $RC$ time constant of the cables to ambient temperature. AdvACT is currently deploying the highest achieved multiplexing factor of 64 TES channels per readout column using the NIST TDM chips and the 
%University of British Columbia (UBC) 
MCE electronics~\cite{henderson/etal:2016}. There is no intrinsic limit on the number of columns, given sufficient warm readout electronics.

%\subsubsection{Noise}
Since the readout chain's bandwidth must be much higher than the sampling rate of any given TES, noise from the SQUIDs and warm amplifiers is heavily aliased. The aliasing penalty for RMS noise is proportional to the square root of the multiplexing factor. There is some freedom to limit the aliasing impact by reducing detector resistance or adding turns to the SQUID input coil, so in practice the impact from the SQUID/amplifier alone has been small:
%If the detector bandwidth is not sufficiently below the multiplexing rate, the detector noise can be aliased as well:
\bicepII\ with a 25\,kHz TDM revisit frequency experienced $\sim$14\% aliased noise penalty to its total (photon-noise-dominated) NET, mostly from aliased detector noise~\cite{OBrient15}. %Ade:2015pez

%\subsubsection{Thermal budget}
Current instruments dissipate $\sim$1.8~nW per readout column at the detector temperature (100-300~mK)~\cite{Doriese:2016,Doriese:2016err}. This should not scale strongly with multiplexing factor, since it is dominated by the single first-stage SQUID that is operational at any given time.
The SSAs dissipate substantially more power: $\sim$20~nW per readout column. This power may be dissipated at a somewhat higher temperature (typically 1--4\,K), and so is typically not a limiting factor.

%The cryogenic cabling transports power to the cold stages, which must be accounted for in cooler specifications.  As an example, a 1.5~m run of 38~AWG Manganin wire, intercepted at 50~K halfway down, is expected to dump $\sim$40~$\mu$W per 32-channel readout column on the 4~K stage.

%\subsubsection{Crosstalk}
TDM has several known crosstalk mechanisms, generally of modest amplitude ~\cite{deKorte:2003mux,BICEP2_II:2014}. The largest form of crosstalk is inductive: each SQ1 detects current from neighboring input coils (adjacent rows in the same readout column) inductively at the $\sim$0.3~\% level, and at a yet smaller level to more distant rows. In a well-designed system all other forms of crosstalk are subdominant.

%\subsubsection{Warm electronics power and space requirements}
A typical full-sized (72-HP) MCE crate serving a $\sim$2000~pixel (32 column by 64 row) array consumes 85 watts, supplied by custom linear or switched DC supplies. The crate dimensions are approximately $40 \times 43 \times 34$\,cm (depth / width / height) and it weighs approximately 13\,kg, not including separate DC supplies.  %There has been some effort to modestly reduce power consumption for specific experiments, \eg the \spider\ balloon-borne instrument.

\subsection*{Prospects and R\&D path for CMB-S4 for TDM}
\label{sec:discussion}

TDM benefits from almost a decade of field experience in CMB
instruments, which has yielded dozens of publications involving more
than 10,000 detectors. The hardware and software are
well-characterized and well-supported. Systematic errors are controlled and
understood for arrays with as many as 64 rows. The interconnect
technologies are also relatively simple: twisted-pair cryogenic cables
and aluminum wire bonds.

\bicepIII\ and Advanced ACTPol have successfully deployed CMB receivers using TDM at the
$\sim$2,000-detector scale, comparable to the channel counts targeted
for CMB-S4's lower frequency (\eg\ 30 -- 40~GHz) channels.  
Despite
these successes, there are significant development challenges to
scaling this technology to the high pixel counts envisioned for
CMB-S4's higher-frequency receivers.
%TDM's challenges are more daunting at the $>10^4$-detector
%scale envisioned for the higher frequency receivers. 
TDM is nonetheless a natural back-up alternative to more ambitious
multiplexing schemes.

Among other schemes, modified versions of the TDM system known as
``code-division multiplexing'' (CDM), now under development, may prove
to be more viable for larger multiplexing
factors~\cite{Fowler:2011vr,Irwin:2011yt,Stiehl:2012rt}. Rather than
switching among individual detectors, a CDM system switches among
measurements of various Walsh code combinations (alternating-sign
sums) of the various TES signals. In this configuration all TES
signals are sampled at all times, eliminating the $\sim\sqrt{N_{mux}}$
amplifier noise aliasing penalty. This allows for much more efficient
use of readout bandwidth and thus higher multiplexing factors.

%TDM brings excellent performance and extensive field experience to
%CMB-S4, but the challenges above dampen its prospects as the sole
%solution for the program. \bicepIII\ and Advanced ACTPol have
%successfully deployed TDM at the $\sim$2,000-detector scale,
%comparable to the channel counts targeted for CMB-S4's lower
%frequency (\eg\ 30 -- 40~GHz) channels. TDM's challenges are more
%daunting at the $>10^4$-detector scale envisioned for the higher
%frequency receivers. TDM is nonetheless a natural back-up alternative
%to more ambitious multiplexing schemes.

R\&D items for CMB-S4 include:
\begin{itemize}
\item{\bf Decrease assembly complexity:} Since TDM row-switching is
  carried out at ambient temperature, wires to room temperature are
  required for each row as well as each column. That leads to a
  relatively high wire count per pixel: roughly 264 wire pairs to
  sub-Kelvin for a 32$\times$200 array. This may be ameliorated
  somewhat through the development of a custom cold switching
  system~\cite{Prele:2016}.  The standard TDM system also has no
  provision for individually-tuned TES bias values down a common
  line. Larger multiplexing factors thus make heavier demands on TES
  fabrication uniformity (in order to use a common bias), or demand
  additional TES bias lines. Cold hybridization requirements are also
  substantial: at least eight bonds per TES, plus four per column for
  SQUID and TES biasing. This hybridization effort may be reduced with
  fully-automated wire bonding systems or development toward indium
  bump-bonded systems (\eg~\cite{Jhabvala:2014tia}). The number of
  interconnects could be drastically reduced by fabricating the SQUIDs
  alongside the TESs on the same wafer, though this would require
  development effort to ensure adequate uniformity and yield.

\item{\bf Increase production throughput for cold components} 
  Extrapolating from current technology, a 32$\times$200 (6,400 TES)
  readout array would incorporate more than 70,000 Josephson junctions,
  50,000 wire bonds, and $\sim$60\,nW of power dissipation at detector
  temperature.  The manufacture of large quantities of high-quality
  Josephson junctions is relatively complex, demanding careful control
  of superconducting film deposition. Such arrays are now manufactured
  routinely at \eg\ NIST, but are rare in industrial fabrication.
%The sheer quantity of junctions required in this system (currently
%several per TES) would thus require some development effort to reach
%the full CMB-S4 scale at a competitive price.

\item{\bf Increase multiplexing factor} The large number of detectors
  per telescope envisioned for CMB-S4, particularly for the
  higher-frequency instruments, will demand a higher multiplexing
  factor than has been demonstrated thus far.
%Starting from the Advanced ACTPol 64-way multiplexer,
  Careful tuning of TES and SQUID properties could potentially double
  readout bandwidth over Advanced ACTPol while halving TES bandwidth,
  for a total multiplexing factor of order $\sim$200. Larger factors
  seem difficult to reconcile with current interconnect bandwidth and
  TES stability.
%concerns.
  Significant increases in multiplexing factor may also be achieved
  in the exploration of hybrid FDM/TDM or modified TDM architectures,
  including CDM-based systems presently under study.

\end{itemize}
%Extrapolating from current technology, each 32$\times$200 (6,400 TES) readout array would incorporate more than 70,000 Josephson junctions, 50,000 wire bonds, and $\sim$60\,nW of power dissipation at base temperature.

%TDM brings excellent performance and extensive field experience to CMB-S4, but the challenges above dampen its prospects as the sole solution for the program. \bicepIII\ and Advanced ACTPol have successfully deployed TDM at the $\sim$2,000-detector scale, comparable to the channel counts targeted for CMB-S4's lower frequency (\eg\ 30 -- 40~GHz) channels. TDM's challenges are more daunting at the $>10^4$-detector scale envisioned for the higher frequency receivers. TDM is nonetheless a natural back-up alternative to more ambitious multiplexing schemes.

%% file: readout/fdm.tex
\section{Frequency-division multiplexing using MHz LC resonators (DfMux)}
\label{section:fdm}

%Frequency-division multiplexed readout (FDM) has been demonstrated on the sky with multiplexing factors of $8\times$ - $16\times$ in several stage-2 CMB polarization experiments, including \Pb-1, \cite{Kermish2012, PB_ClBB_2014}, SPTpol\cite{SPTPolAustermannSPIE, Hanson2013}, and EBEX\cite{Reichborn10}.
%The current generation of FDM readout has been developed for deployment on stage-3 experiments starting at the end of 2016\cite{Rotermund2016, Bender_SPIE2014}. \Pb-2 will have a multiplexing factor of 40\cite{Barron_SPIE2014, Hattori2013}, and SPT-3G will have a multiplexing factor of 68\cite{Bender_SPIE2016}.

\subsection*{Description of the technology}

Frequency-division multiplexing (FDM) takes advantage of the relatively large bandwidth of the SQUID amplifier (1--100\,MHz) compared to the small bandwidth of CMB signal incident on a TES bolometer.
For FDM using in-series MHz LC resonators, each detector is assigned a channel in frequency space, defined by a resonant series RLC circuit, with the bolometer $R_{\mathrm{TES}}$ acting as a variable resistor.
Each detector is AC biased with a unique sinusoidal carrier at its resonant frequency.
Sky signals modulate $R_{\mathrm{TES}}$, which causes amplitude modulation in the carrier current, encoding the signals as sidebands of the carrier frequency.
A key feature of this strategy is that the bias power provided to each detector can be chosen independently, allowing the readout system to compensate somewhat for non-uniformities amongst detector parameters.

\begin{figure*}
    \centering
    \includegraphics[width = 450pt]{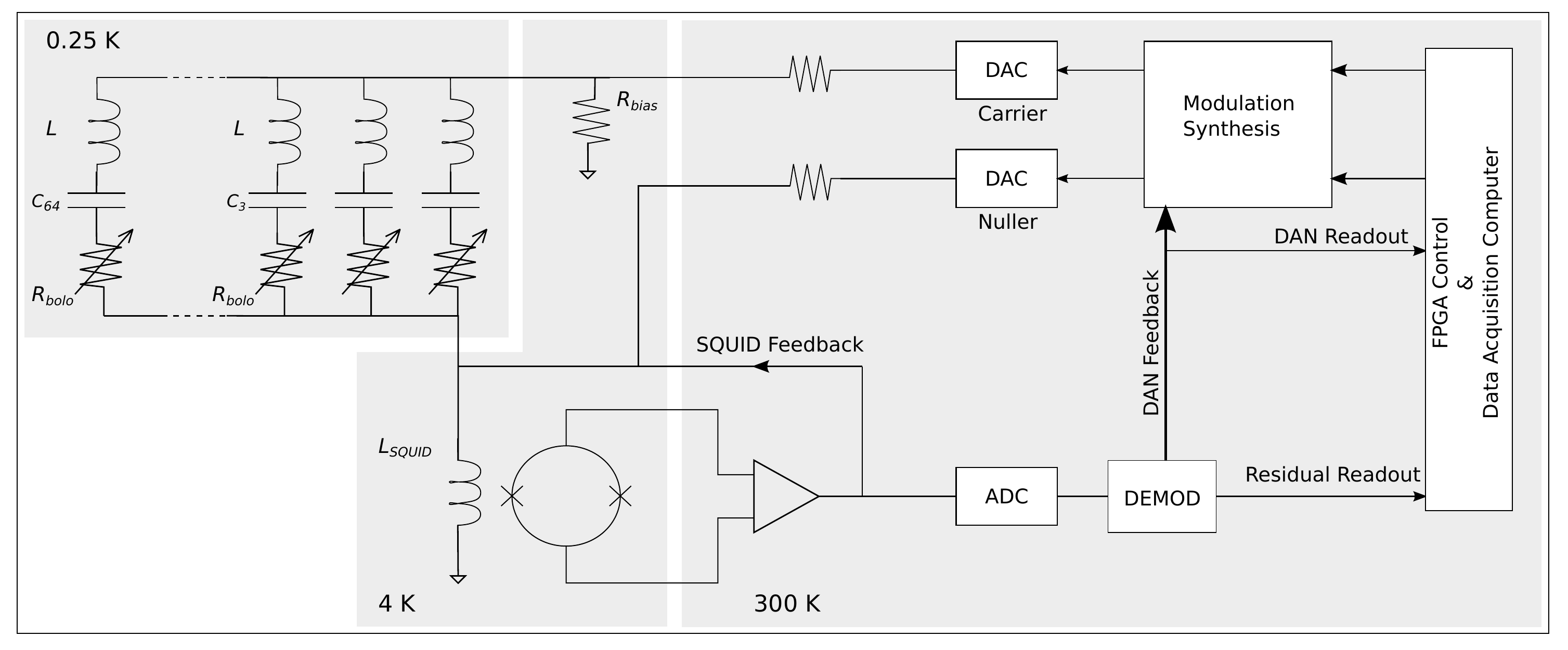}
    \caption[Circuit diagram of DfMux readout system]{A circuit diagram of the DfMux readout system is shown, with the cryogenic portion at the left, and the room temperature electronics at right. Figure from \cite{Bender_SPIE2016}.}
    \label{fig:FDMcircuit}
  \end{figure*}

The current system of FDM using in-series MHz LC resonators which has been used in CMB experiments is known as DfMux and is described in Reference \cite{Dobbs2012}. 
A circuit diagram of the DfMux readout system is shown in Figure \ref{fig:FDMcircuit}.
A bias resistor is wired in parallel with the bolometer LCR circuit, with $R_{\mathrm{bias}} << R_{\mathrm{TES}}$, creating a voltage bias $V_\mathrm{bias}$ on the bolometers.
In the current system, this bias resistor is located at 4~K so that the voltage bias can be supplied by a single pair of wires to the sub-Kelvin focal plane for each comb of bolometers.
A current-biased series array of DC SQUIDs\cite{Huber2001} (referred to here as a ``SQUID'') is used to read out a comb of multiple channels.
The current from the bolometers is summed at the SQUID, whose output is modulated by the bolometer currents.
To maintain linearity the voltage bias input must be nulled at the amplifier input and the magnitude of the remaining sky signal must also stay within the linear regime of the SQUID.
%Negative feedback is required to linearize the SQUID response and provide a large dynamic range.
%The voltage bias input must be nulled by sending in its inverse, to cancel out its contribution to current through the SQUID.
%The magnitude of sky signals must also stay within the linear regime of the SQUID.
The current generation of DfMux uses a form of baseband feedback known as Digital Active Nulling (DAN) \cite{deHaan2012}, where feedback is applied only around the bolometer carrier frequencies.
With DAN the sky signal is also nulled, so that the SQUID acts as an error sensor, and the nulling current is the sky signal.

The readout bandwidth is set by the inductance $L$ and the resistance $R_{\mathrm{TES}}$.
The inductance of each channel is constant, and the capacitance is varied to set the resonant frequency.
In current implementations of the DfMux system, there is a comfortable margin between the necessary optical time constant, the detector time constant, and the readout time constant ($L = 60\,\mu$H, $\tau_{e}\sim 0.5\,$ms)\cite{Rotermund2016}.
The spacing of the channels in a frequency comb must be large enough so that the off-resonance current from neighboring channels does not interfere with the voltage bias on-resonance, and so that crosstalk between neighboring channels is small. 
%In the current system we enforce a minimum spacing of 40 kHz between resonant frequencies to keep crosstalk below 0.5\%.
To keep crosstalk below 0.5\%, minimum spacing between channels  are greater than 40 kHz. 
%In the current system, the resonant frequencies range from 1--6~MHz, with a minimum spacing of 40\,kHz to keep crosstalk below 0.5\% .
Crosstalk here is defined as the ratio of power fluctuations from other channels to the expected signal power fluctuation.

The operation of the resonant RLC circuit depends on there being negligible impedance in series with the well-defined components of the circuit.
The bolometer resistance $R_{TES}$ must be the dominant resistance, and there also must be minimal stray inductance from wiring and circuit boards.
These components and wiring are all at sub-Kelvin temperatures, which helps to achieve these specifications.
The wiring from the SQUID and bias resistor at 4\,K to the sub-Kelvin focal plane must be low inductance while acting as a thermal break; this is achieved with broadside-coupled NbTi striplines with lengths $\sim 50$--$100\,$cm.
For systems where the bias resistor and SQUID input sit at different cryogenic temperatures,
the practical lengths and inductances of the current sub-Kelvin wiring and components requires $R_{TES} \approx 1\,\Omega$, to keep the bolometer impedance large compared to other impedances.
FDM for much lower $R_{TES}$ has been implemented by placing the bias resistor and a first stage SQUID at low temperature, such that the wire lengths are short and wiring inductance is negligible \cite{Bruijn2012}.

There are two pairs of wirebonds per detector: from detector wafer to cable, and across the LC resonator.
There is no power dissipation at the sub-Kelvin stages from readout components.
The cryogenic wiring is simple: there is just one pair of wires running to the sub-Kelvin stages for each multiplexed comb of bolometers.
There are only two connectors used in the readout chain so as to minimize stray inductance and resistance: after the wafer readout cable, and at the 4\,K SQUID.
The thermal load on the focal plane from these wires is about $\sim 1\,$nW per comb,
depending on wire length and the number of thermal interfaces.
There is also minimal power dissipated at the other temperature stages from readout wiring.
The readout system dissipates power at the 4~K stage in the current system, from SQUIDs and bias resistors, $ \lesssim  1\,\mu$W per comb.
If the bias resistor is moved to a sub-Kelvin stage with limited cooling power, it could be replaced with an inductive or capacitive divider to reduce power dissipation to zero.
The standard twisted-pair cryogenic wiring which runs between the cold components and the room temperature electronics requires three pairs for each multiplexed comb. 
The length of this wiring is practically limited by its total resistance ($\sim20\Omega$), since it acts as a voltage divider for the voltage output of the SQUID. 
Exact total length could be adjusted by varying gauge of low thermal conductivity wire. 

The DfMux readout system uses custom warm electronics
%designed at McGill University
%\cite{Dobbs2012,deHaan2012,Bender_SPIE2014}. These
to synthesize the bolometer voltage biases (labeled ``Carrier Bias
Comb''), the nulling signal that is applied to the SQUID to increase
its dynamic range (labeled ``Nulling Comb''), and the
demodulators~\cite{Dobbs2012,deHaan2012,Bender_SPIE2014}.  The SQUID
has a transimpedance that is high enough to convert small current
through the bolometers into a voltage that can be read out with a
room-temperature amplifier.  The power and space requirements for the
warm electronics are relatively small.  The ``ICE'' room temperature
readout electronics~\cite{Bandura:2016dpm} being deployed for SPT-3G
with a 68x multiplexing factor will operate 8,700 detector channels
per 9U crate (40\,cm tall, 25\,cm deep, 50\,cm wide), with less than
1\,kW of power draw.

\subsection*{Demonstrated performance}

Stage-II CMB polarization experiments such as \Pb-1~\cite{pb,
  PB_ClBB_2014}, SPTpol~\cite{sptpol, Hanson2013}, and
EBEX~\cite{Reichborn2010} have demonstrated
frequency-division multiplexing factors of $8\times$--$16\times$ on a
single pair of cryogenic wires.  The Stage-III experiments
SPT-3G~\cite{Bender_SPIE2016} and \Pb-2~\cite{Rotermund2016,
  Barron_SPIE2014, Hattori2013} are deploying in 2016 and 2017 with
multiplexing factors of $68\times$ and $40\times$,
respectively. %Kermish2012 is pb

The introduction of DAN feedback (demonstrated on-sky with SPTpol) has extended the usable bandwidth with stable SQUID feedback, allowing channels to be distributed anywhere within the 120\,MHz bandwidth of the current SQUID series arrays.

%The usable bandwidth with stable SQUID feedback was extended by changing to a form of baseband feedback known as DAN \cite{deHaan2012}, where feedback is applied only around the bolometer carrier frequencies.
%Channels can be placed anywhere in the SQUID bandwidth, which for the series arrays presently used is 120\,MHz.
%DAN feedback has been demonstrated on the sky with the SPTpol experiment.
To improve the precision of channel placement and to reduce loss at higher frequencies, superconducting resonator components were developed \cite{Hattori2013, Rotermund2016}, with an interdigitated capacitor along with a spiral inductor in a single layer of superconducting traces.  These developments increased the potential multiplexing factor by a factor of five for the DfMux system used in Stage-III experiments.

Two of the dominant noise sources are related to the SQUID: the current noise of the SQUID itself, and the voltage noise of the SQUID's first-stage amplifier.
Both of these noise sources are far from fundamental limits and could be further reduced.
There is also noise associated with the generation of the carrier and nuller signals, dominated by the output current noise of the DAC,
presently limited by available off-the-shelf DAC technology.
The system can be designed so that the expected noise equivalent current of the dominant readout noise sources is sub-dominant to the bolometer power noise terms ($\sim 20$\,--\,$30 \mathrm{pA}/\sqrt{\mathrm{Hz}})$, and a readout noise equivalent current of $\sim 7$\,--\,$10\mathrm{pA}/\sqrt{\mathrm{Hz}}$ has been demonstrated on several experiments \cite{pb,AubinSPIE2010,Ruhl2004}.
This results in the readout noise being negligible compared to the photon noise for detectors with appropriate parameters.

In the DfMux system, signal crosstalk onto a detector can only occur if
the crosstalk signal lies within its frequency bandwidth. Other
crosstalk can introduce excess loading on the SQUID and noise, but
does not introduce false sky signals. Since the fraction of total
SQUID bandwidth that is occupied by detector signals is very small,
there is a small amount of crosstalk from its nearest neighbors in
frequency space and physical space in the comb.  For \Pb-1, the highest
level of signal crosstalk came from neighbors in frequency space, with a
maximum level of about $1\%$~\cite{PB_ClBB_2014}.  Stage-III
experiments expect similar low levels of crosstalk, but this remains
to be demonstrated on the sky for the larger multiplexing factors
used.

\subsection*{Prospects and R\&D path for CMB-S4 for DfMux}

Expanding the channel capacity of the current DfMux readout systems,
without any improvements to the cold or warm readout electronics and
maintaining the present multiplexing factor of 68 per wire-pair, would
mean increasing the number of readout modules needed for a single
telescope.  The number of readout modules that would be required to
read out 50,000 detectors is a factor of $\sim 4$ larger than the
number currently in use by Stage-III experiments (SPT-3G has 15,234
detectors at $68\times$ multiplexing).  The ``ICE'' readout boards
that are used for DfMux are also used in radio astronomy
correlators~\cite{Bandura:2016dpm}.  For example, in 2017, the CHIME telescope
 started using 128 readout boards (4.5 times more than SPT-3G), demonstrating 
that systems with this number of boards is tractable. 

\begin{figure}[h]
\centering
\includegraphics[width = 6in]{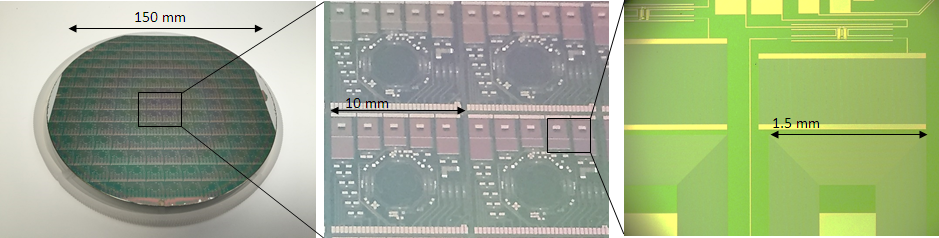}
\caption{Demonstration of superconducting resonators integrated onto a multichroic detector wafer.
Left: Photograph of the device wafer. 
Center: Close-up view of four multichroic pixels with integrated superconducting
resonators at the top of each pixel. 
Right: Close-up view of a superconducting resonator, with interdigitated capacitor at the top and spiral inductor at bottom.}
\label{fig:ReadoutLConChip}
\end{figure}

To increase the multiplexing factor per readout module, and simplify the complexity of cold component integration,
an alternate approach is under development, referred to here as 50 MHz fMUX.
%With the currently fielded DfMux readout systems, resonators have frequencies from $\sim$ 1 -- 6~MHz, 
%with the usable bandwidth limited by stray impedance from wiring.
Integration of the frequency multiplexing circuit onto the detector wafer
simplifies interconnects by reducing the number of required connections by the multiplexing factor.
This requires shrinking the physical size of the resonators to a scale smaller than the detector pixel size,
which increases the resonant frequencies to $\sim$ 50 -- 100~MHz.
With this approach, parasitic impedance can be accurately simulated and controlled with standard
micro-fabrication techniques used in detector RF circuit design.
As a test of this approach, superconducting resonators were integrated on detector wafers as shown in
Figure~\ref{fig:ReadoutLConChip}. 
Currently 50 -- 100~MHz resonators are being developed for 50 MHz DfMux, but
the same method can be used to fabricate $\sim$1\,GHz resonators for microwave SQUID readout (see Section~\ref{section:umux}).

R\&D items for CMB-S4 include:
\begin{itemize}
\item{\bf Demonstrate electrothermal design for 5~MHz DfMux for 50,000 detectors}

Scaling to cryogenic FDM circuits up is more challenging.
The technology status level of the 5 MHz DfMux system is currently 4/5.
Data has been fully analyzed for lower multiplexing factors used in \Pb-1 and SPTpol,
and $68\times$ multiplexing deployed with SPT-3G at the end of 2016, which
should advance the TSL to 5 in near future.
The overall production status level for the current DfMux system is 4, 
and could be increased to 5 with R\&D effort.
Scaling cryogenic DfMux circuits up to the channel densities required by
CMB-S4 presents a challenge.  One potential issue is that it could require
long lengths of wiring to the bolometers (compared to the current
lengths of $\sim 50$--$100\,$cm).  The thermal loads on the sub-Kelvin
stages from $\sim 1,500$ readout wires would require a significant
portion of the cooling capacity of a three-stage helium sorption
fridge.  In a dilution refrigerator cooled cryostat, the SQUID and
bias resistor could be moved from the 4~K stage to the 1~K buffer stage if it could
accommodate the $\sim 1\,$mW of dissipated power, along with the
thermal load from three pairs of wires per comb.  This would greatly
reduce the physical distance and necessary wiring lengths to the
bolometers.

\item{\bf Increase multiplexing factor}

The issues associated with the cryogenic wiring complexity would be
addressed by increasing the multiplexing factor up to
128--256$\times$.  This can either be achieved by packing the channels
closer together by means of narrower inductor-capacitor resonances, or
by extending the readout bandwidth to accommodate additional frequency channels.
Both strategies are being actively explored.  Packing channels more
closely together requires high uniformity in the channel spacing, and
excellent control of stray impedances in the wiring and interconnects.
Control over stray impedances in particular may be improved by keeping
all components of the cold-multiplexer at sub-Kelvin temperature,
ensuring short wire lengths.  Significant increases in system
bandwidth may also be achieved with 50 MHz fMUX, as described above.
The technology and production status levels (TSL/PSL) for the 50 MHz fMUX system
are currently only at 1, since so far only prototypes have been fabricated. There is significant unknown physics involved in operating TES devices at high AC modulation frequencies including increased SQUID backaction noise, high TES kinetic inductance, pair breaking near $T_{c}$ at modulation frequencies, and the potential for other new sources of bias instability and noise. If these issues are investigated and overcome, readout R\&D efforts can be implemented that would advance to higher TSL levels.
Modest R\&D efforts then would be expected to advance this technology to TSL/PSL 3, based on their similarity to existing technologies and production methods. If successful, further R\&D efforts leading up to CMB-S4 could advance the 50 MHz fMUX system to TSL/PSL 5.

%\item{\bf Warm readout electronics capabilities}
\item{\bf Further develop warm electronics}
Increasing the backend electronics multiplexing factor is not an issue.
Firmware for the ``ICE'' backend electronics already supports a multiplexing factor of 128x and uses about half the FPGA resources.
Exploiting full FPGA resources and optimizing firmware should allow an increase to 256x without warm hardware changes ($\sim$32,000 detector channels for a single 9U crate).
The ICE backend electronics used with TES detectors with 5 MHz DfMux could also be specialized for higher frequency (100\,MHz or 1\,GHz) readout of MKIDs or \umux{} by using higher frequency digitizer daughter-boards that are available commercially, or by developing custom daughter boards.
For telescopes that plan to support deployment of both TES and KID focal planes, a system that can support fMUX at high and low frequencies would allow the same core electronics to be used for the readout of both detector systems.
The technology status level of the warm readout electronics is currently 5, and the production status level is 4. 

\end{itemize}

%\begin{itemize}
%\item Path to achieve MUX factors of ~125, 250 or 500, 1000 (as appropriate for technology)

%\item Path to achieve other CMB-S4 specifications (We'll work on this after first round drafts)
%\end{itemize}

%% file: readout/umux.tex
% moved to top doc
%\def\umux{{\sc $\mu$mux}}
\newcommand{\figref}[1]{Fig.~\ref{#1}}
\newcommand{\tabref}[1]{Tab.~\ref{#1}}
\newcommand{\secref}[1]{$\S$~\ref{#1}}

%\title{Microwave SQUID Multiplexers}
%\author{J. Hubmayr}
%\date{}                                           % Activate to display a given date or no date

%\begin{document}
%\maketitle
%\begin{abstract}
%This document describes the current status of the microwave SQUID multiplexer.
%This text will be used in a larger document to inform the readout technology for CMB-S4.
%This technical development is the result of many peoples' work.
%Hannes is just writing about it.
%\end{abstract}

%%%%%%%%%%%%%%%%%%%%%%%%%%%%%%%%%%%%%%%%%%%%
%%%%%%%%%%%%%%%%%%%%%%%%%%%%%%%%%%%%%%%%%%%%
%%%%%%%%%%%%%%%%%%%%%%%%%%%%%%%%%%%%%%%%%%%%
\section{Frequency-division multiplexing using SQUID-coupled GHz resonators (\umux)} % title stays. pick a fight with zeesh.
%\section{Frequency Division Multiplexing (\umux)}
\label{section:umux}
\subsection*{Description of the technology}

\begin{figure}[t]
        \begin{center}
        \includegraphics[width = .90\textwidth]{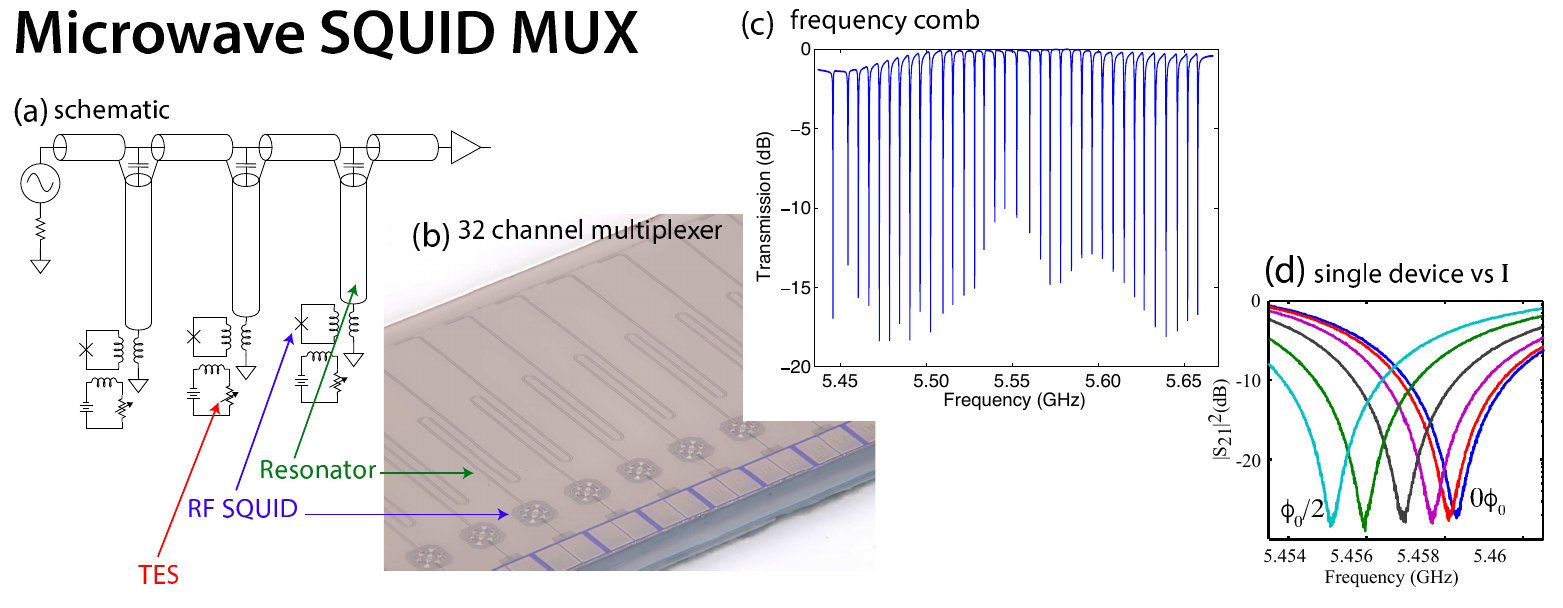}
        \end{center}
        \caption{ \footnotesize Overview of the the microwave SQUID multiplexer.
        {\bf (a)} Schematic of the circuit.
        {\bf (b)} Photograph of a 32-channel \umux\ chip.
        {\bf (c)} $S_{21}$ transmission measurement of the \umux\ with 32 active channels.
        {\bf (d)} Variation of single readout channel transmission curves to applied input magnetic flux (or equivalently applied current when inductively coupled).
        \label{fig:umux}}
\end{figure}

The microwave SQUID multiplexer (\umux) \cite{irwin2004microwave, mates2011thesis} is a readout scheme intended to greatly increase the focal plane pixel count of TES bolometer arrays, using inspiration from the GHz FDM approach of MKIDs \cite{zmuidzinas2012a,day2003}.
\figref{fig:umux} illustrates the concept.

%The TES bias circuit is identical to that of time-division SQUID multiplexing.
%The TES is DC biased by means of applying a constant current to a shunt resistor that is wired in parallel with the TES.
%Many detectors may be biased using a single DC current source and two wires.
%The readout mechanism, however, is very different than TDM.
%Each TES is inductively coupled to it's own resonator at a unique resonance frequency.
%
The TES bias circuit is identical to that of time-division SQUID multiplexing: each TES is voltage-biased with a DC current and parallel shunt resistor, with multiple TESs sharing a single pair of bias wires.  The readout mechanism is very different from TDM, however: each TES is inductively coupled to its own resonator and addressed by its unique resonant frequency.
Current sourced from a TES produces a frequency shift in a microwave resonator by means of a flux-coupled RF-SQUID.
In this manner, many readout channels can be densely packed onto a single superconducting transmission line with a total readout bandwidth of several GHz.
Similar to MKID readout, signals are determined from the transmission properties of microwave resonators, monitored by use of homodyne readout techniques.
The only difference in \umux\ readout, with respect to MKID readout, is the addition of ``flux-ramp demodulation'' \cite{mates2012flux}, which linearizes the response and substantially decreases 1/f readout noise.
Cabling per module consists of a pair of DC wires and two coaxial cables.
%Additional detector bias lines are required.

%%%%%%%%%%%%%%%%%%%%%%%%%%%%%%%%%%%%%%%%%%%%
%%%%%%%%%%%%%%%%%%%%%%%%%%%%%%%%%%%%%%%%%%%%
%%%%%%%%%%%%%%%%%%%%%%%%%%%%%%%%%%%%%%%%%%%%
\subsection*{Demonstrated performance}

Performance has been demonstrated through extensive lab-based measurements and with on-sky observations in the MUSTANG2 receiver.
Readout noise levels relevant for CMB-S4 have already been demonstrated in the lab.
The architecture was used to read out a 3$\times 10^{-17}$ W/$\sqrt{\mathrm{Hz}}$ NEP TES bolometer, which was optimized for CMB polarization measurements \cite{mates2011thesis}.
In this demonstration, the readout noise was negligible compared to the system noise, to modulation frequencies as low as 1\,Hz.
Lower modulation frequencies were not investigated.
By altering the input coil coupling, the current generation of \umux\ chips have achieved an input-referred white current noise level of 17~pA/$\sqrt{\mathrm{Hz}}$ \cite{bennett2015integration}.
This noise level is nearly a factor of 10 below the expected photon noise level in the types of cryogenic receivers envisioned for CMB-S4.
(Here we assume 3\,pWs of optical load at 150\,GHz, and an optimized TES bolometer with $R_{TES}~\sim~5$\,m$\Omega$.)

On-sky observations have been made in two engineering runs of the MUSTANG2 receiver on the Green Bank Telescope (GBT).
In 2015, MUSTANG2 used the architecture in a 32-channel per module configuration to make first-light images \cite{stanchfield2016development}.
In 2016, on-sky, background-limited sensitivity has been demonstrated in pixels coupled to 64-channel multiplexers.

In addition to bolometric applications, the \umux\ is under development for several TES microcalorimeter-based instruments.
A DOE-funded, 512-pixel gamma-ray spectrometer demonstration, called SLEDGEHAMMER, is underway \cite{bennett2015integration}, and the readout approach is baselined for a first-light instrument at the Linac Coherent Light Source II (LCLS-II).
The current technical state-of-the-art for calorimetric applications is a successful demonstration of undegraded energy resolution in a 4-pixel array that was read out using the scalable ROACH-II warm electronics.

The total readout bandwidth and resonator frequency spacing set the number of detectors multiplexed on a single coaxial cable.
To date, the \umux\ development has focused on the (4--8)\,GHz band since this matches the bandwidth of existing cryogenic HEMT amplifiers.
Resonator spacing of 6\,MHz has been demonstrated with few resonator collisions.
Hence in the implemented approach and using the full HEMT bandwidth, 660 detectors can in principle be multiplexed on a single line.

Multiple generations of 33-channel multiplexers using a standard 3\,mm~$\times$~19\,mm form factor have been fabricated and tested.
Different frequency band 33-channel chips have been wired in series to increase the number of readout channels per multiplexer module.
This demonstrates both successful frequency scaling of the devices and the ability to daisy chain chips together, as a means to increase the multiplexing factor.
The silicon footprint is currently identical to that of time-division SQUID multiplexing.

The total power dissipated on the cold stage is $\sim$~10\,pW/channel.
When resonators are spaced in frequency at least ten times their bandwidth, nearest neighbor crosstalk is measured to be $<$~0.1~\%.
Linearity has been measured to 1 part in 1,000.

%%%%%%%%%%%%%%%%%%%%%%%%%%%%%%%%%%%%%%%%%%%%
%%%%%%%%%%%%%%%%%%%%%%%%%%%%%%%%%%%%%%%%%%%%
\subsection*{Prospects and R\&D path for CMB-S4 for microwave SQUIDs}

The \umux\ is less mature than time-division SQUID multiplexing (TDM) or MHz frequency-division SQUID multiplexing (FDM), which together have been used to read out $\sim$30,000 deployed TES detectors that observe in the mm/sub-mm/FIR.
However, the technology is rapidly gaining maturity through its use in several instruments.
The demonstrated multiplexing factor is equal to that of contemporary TDM instruments (x64, AdvACT \cite{henderson/etal:2016}) and FDM instruments (x68, SPT-3G \cite{benson2014}).
The envisioned multiplexing factor is at least an order of magnitude higher than this.

R\&D items for CMB-S4 include:
\begin{itemize}
\item{\bf Increase multiplexing factor} With current technology, the cold multiplexing density achieves a multiplexing factor of 660.
Recent developments in fabrication have reduced the frequency scatter by several factors, and thus the multiplexing density may be increased by this same factor.
Near-term efforts to demonstrate $\sim$~1~MHz frequency spacing would be beneficial, as the quantity of warm readout electronics boards and cryogenic HEMT amplifiers reduces by this same factor.

\item{\bf Demonstrate array performance on the sky}
The majority of experimental data on the \umux\ is at the few pixel demonstration level.
On-sky results from MUSTANG2 are promising (which uses a $\times$64 multiplexing factor), but a detailed study of performance, including low frequency noise properties, cross-talk and linearity is needed.
MUSTANG2 offers a nice platform for this study, but other instruments or lab-based work will be essential since MUSTANG2 will not probe the target $\sim$~1\,MHz frequency spacing.

\item{\bf Shrink cryogenic circuit elements}
Smaller circuit elements will reduce the cost of any cryogenic readout technology, since fewer wafer need to be processed for a fixed number of readout channels.
The footprint of the circuitry is currently comparable or several factors smaller than the leading TES multiplexing approaches.
By moving to a lumped element design, the readout footprint could shrink substantially.

\item{\bf Demonstrate integrated sensor and \umux\ fabrication}
The current \umux\ implementation decreases the number of wires running between temperature stages.
However, four to six wirebonds per channel are still required at the cold (isothermal) focal plane stage.
Placing the readout onto the detector wafer solves this issue, drastically reduces the complexity of focal plane assembly, and eliminates the need for separate SQUID multiplexer chip fabrication.
Beam forming elements, such as lenslets or feedhorns, which create space on the wafer for the readout components make integrated fabrication a possibility.
In the near term, preliminary designs should be pursued, and steps to demonstrate an integrated fabrication process flow should be taken.

\item{\bf Further develop warm electronics}
Warm readout for the \umux\ has heavily benefitted from the developments in MKID readout.
But future R\&D is required and discussed in the following section.
\end{itemize}

Lastly, we note that the \umux\ may be developed as a stand-alone multiplexing technique for CMB-S4, or it may find use in a hybrid multiplexing scheme.
Hybrid multiplexing is a common way to use available bandwidth more efficiently, e.g. in 3G mobile phone technology.
For readout of TES bolometers, a lower bandwidth multiplexing scheme, such as TDM or code-division multiplexing, is embedded within a wider bandwidth GHz resonator.
Therefore each GHz tone carries the signals from $N$ bolometers, where $N$ may be 32 or 64.
A proof-of-concept demonstration was shown in 2008 \cite{reintsema2008tdma}.
In the near term, a design study for hybrid multiplexing should be undertaken to inform CMB-S4, and if determined viable, an R\&D path laid out.

%\begin {itemize}
%\item Comments on applying technology as is.
%\item Path to achieve MUX factors of 1000+
%\item Path to achieve other CMB-S4 specifications (We'll work on this after first round drafts)
%\end{itemize}

%%%%%%%%%%%%%%%%%%%%%%%%%%%%%%%%%%%%%%%%%%%%
%%%%%%%%%%%%%%%%%%%%%%%%%%%%%%%%%%%%%%%%%%%%
%%%%%%%%%%%%%%%%%%%%%%%%%%%%%%%%%%%%%%%%%%%%
%\bibliographystyle{hunsrt}
%\bibliography{ref}%, bib,refs}

%\end{document}

%% file: readout/microwave_readout.tex
\section{Room-temperature electronics for frequency-division-multiplexed readout}
\label{sec:mkid_readout}

\subsection*{Description of the technology}

%Two of the three readout schemes for TES detectors discussed in
%this paper as well as MKID readout operate in the frequency domain.
%The response to a tone played at the fundamental resonant
%frequency of an MKID or a resonator coupled to a TES is measured
%for amplitude and/or phase shift.
In the various FDM readout schemes described above for TESs and MKIDs, the response to a tone played at a specified resonant frequency is measured for amplitude and/or phase shift after modulation by a cryogenic detector.
A signal in DfMux is an amplitude modulation,
in \umux\ is a phase shift in the resonance,
and in MKIDs is a shift in both phase and amplitude.
In this section, we focus on the warm readout electronics for \umux\
and MKIDs, which operate in GHz frequencies and enjoy extensive
commonality in the architecture of their readout electronics.
MHz FDM shares the same overall strategy, albeit at a lower
frequency. We will also discuss the possibility of the same backend
electronics supporting all three of these techniques.

\begin{figure}[h!]
\centering
\includegraphics[width=0.90\textwidth]{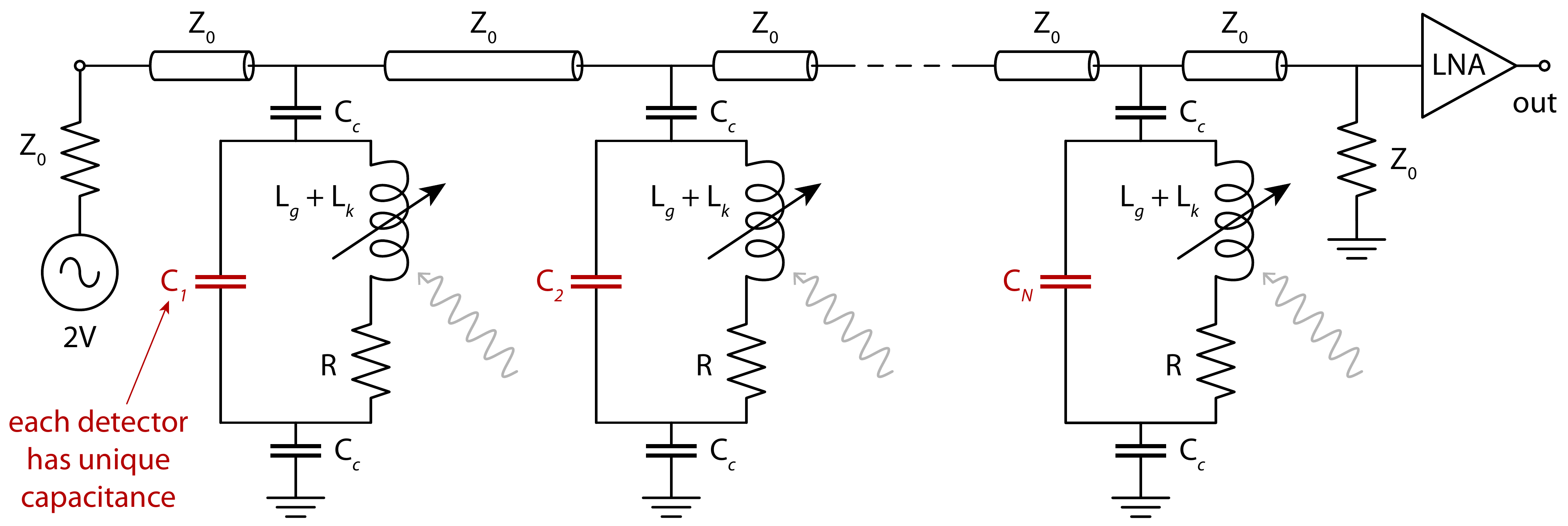}
\vspace{1cm}
\hrule
\vspace{1cm}
\includegraphics[width=0.90\textwidth]{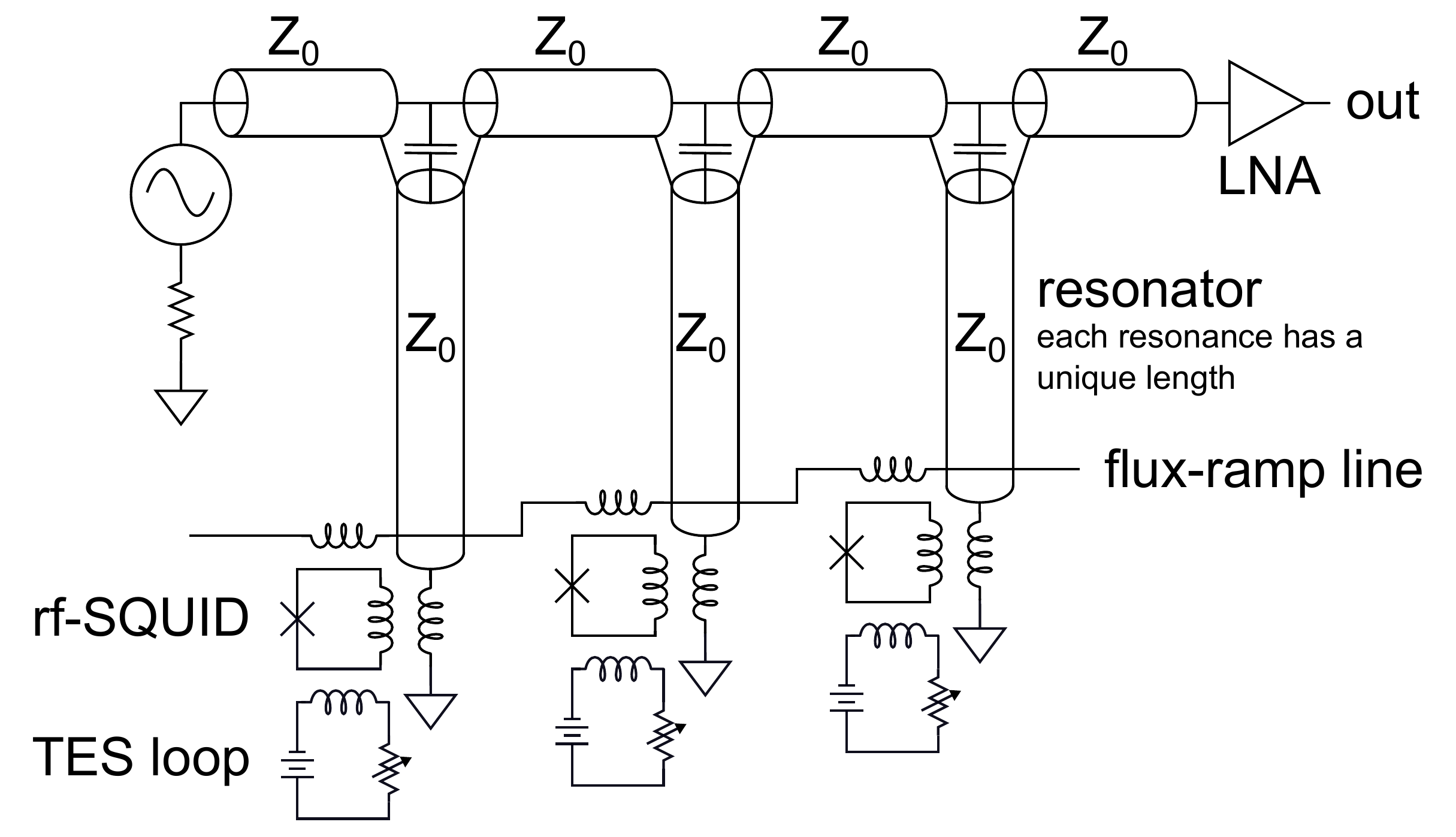}
\caption{
Circuit schematics showing how MKIDs and \umux\ TESs are multiplexed.
(Top) Each MKID has a unique resonance frequency, which is set by the
capacitor coupled to the resonator, for example.
A comb of probe tones is routed to the MKID array using a single
transmission line, and single cryogenic LNA is
used to read out all of the detectors.
(Bottom) Each \umux\ readout channel has a unique resonance frequency
set by the length of the quarter-wavelength resonator.
Like MKID readout, \umux\ also uses a comb of probe tones and a single
LNA to read out many detectors.  
}
\label{fig:mkid_multiplexing}
\end{figure}

Figure~\ref{fig:mkid_multiplexing} shows the cryogenic readout schematic for MKID arrays
and the \umux\ multiplexer.
These systems operate at the range of the resonance frequencies of
the detectors, which is typically 100 -- 8,000\,MHz.
They are designed to support a sufficiently large bandwidth (500 -- 2000\,~MHz) to read out
hundreds or thousands of detectors at once, depending on the resonator quality factors and frequency
spacing. The readout noise is much less than the intrinsic detector
noise (below $\sim -90$\,dBc/Hz) with frequency resolution to probe resonators
with very high quality factors (Q $\sim
100,000)$~\cite{SRON_4kreadout,mkid_readout,mazin_1decade,highspeed_readout,
duan_readout,McHugh2012,Bourrion2012,nikel2016,litebird_readout,yates_fftreadout}.

A common readout design implemented by various MKID experiments including
AMKID~\cite{Schlaerth2008}, BLAST-TNG~\cite{dober2014},
MAKO~\cite{mako2012}, MUSIC~\cite{MUSIC_2012}, NIKA~\cite{nika2014}, and
NIKA-2~\cite{calvo_2016} makes use of a homodyne readout technique.
%An FPGA and DAC generate complex IQ signals.
%
%The signals are mixed with a local oscillator to the required frequency of the resonators.
%
%The tones are passed into the cryostat on a single coax line and excite the detectors resonances.
%
%The resulting signal is then demodulated by an IQ mixer and digitized by the ADC.
%
%The detector signal channels are selected and read out using a
% polyphase filterbank on the FPGA and sent to a computer for processing
% and storage.
%
%The resulting amplitude and phase shifts of the tones are then
%analyzed and calibrated into intensity
%units~\cite{mazin_readout,SRON_4kreadout,nikel2016,Bourrion2012,mccarrick_2014,yates_fftreadout,McHugh2012,highspeed_readout}.
%
A digital tone generator, such as an FPGA, is
connected to a DAC to produce the probe tones.
The waveforms are generated on the FPGA by taking a length N inverse fast Fourier transform (IFFT) of a
delta function comb.
The length of the IFFT sets the frequency resolution of the tones.
For example, an FPGA with 500\,MHz of bandwidth divided into $2^{18}$
bins gives a frequency resolution of about 1.9\,kHz.
The initial waveform amplitude should be maximal within the range of
the DAC and the waveform crest factor (the ratio of peak to r.m.s. amplitudes) should be minimized.
This is achieved by randomly generating the probe tone phases (more
advanced techniques are unnecessary because the MKID devices
themselves will quasi-randomly shift the tone phases)~\cite{SRON_4kreadout}.
A mixing circuit is used to bring the signals to the required
frequency.
For example, to read out devices with resonances between 1000 to
1500\,MHz, one would use the FPGA to generate complex tones from -250 to
250\,MHz and mix them with a 1250\,MHz local oscillator (LO) and IQ modulator
to the required frequency.
%The same LO is used to downmix the tones after passing through the
%MKID array.
The tones are then fed into the cryostat via coaxial cables and vacuum
feedthroughs.
The coax is then wired through to the required cold stages and
attenuated before interacting with the detectors.
The signal is then passed into a cold low noise amplifier and then
back out of the cryostat.
The signal is again amplified and mixed down before going into an ADC
and back into the digital readout.
Signals are then demodulated into amplitude and phase shifts, which
can be calibrated to intensity variations on the detectors.

%An alternative readout design uses direct digital synthesis and demodulation to
%eliminate the tuning parameters necessary for the IQ scheme.

%Notes: Need more here about the cold components for the coax,
%ADC/DAC, and LNA specs. table of systems with fpga and adc/dac and
%reference

The warm electronics currently used for \umux\ borrows largely from MKID readout
developments. One additional requirement for the \umux\ system is a flux
modulation applied to the RF SQUID to linearize its response. The demodulation of this signal is typically done in the same FPGA that is used for digital tone generation and readout. 

%MUSTANG2 and NIST have enhanced the homodyne readout infrastructure described above to include this capability for \umux.% \comred{(cite)}.

%%%%%%%%%%%%%%%%%%%%%%%%%%%%%%%%%%%%%%%%%%%%%%%%%%%%%%%%%%%%%%%%%%%%%%

\subsection*{Demonstrated performance}

A combination of the digital readout bandwidth and resonator quality
factors determine the maximum number of detectors that can be read out
on a single coaxial line.
State of the art microwave readout systems can support thousands of
resonators~\cite{mazin_readout,SRON_4kreadout,McHugh2012,yates_fftreadout,litebird_readout,araujo_2014,dober2014,duan_readout}.
In lab systems have demonstrated multiplexing factors up to 1000 while
maintaining the required noise performance~\cite{SRON_4kreadout,farir_mkid2016}.
NIKA2 has demonstrated the highest on-sky multiplexing factor of
400~\cite{calvo_2016,nikel2016}. %\comred{blast tng is dober2014?}%dober2014 \comre
% corrected blast_tng with dober2014
%For a typical system with 500 MHz of bandwidth centered on 1 GHz and
%resonant frequencies spaced by 10 times the width, the MKIDs must have
%a quality factor greater than 10,000 while optically loaded.
%Based on these results, multiplexing factors up to 4000 appear plausible for MKIDs.

The readout heat loading in the cryostat is due to the RF signal
and LNA power usage. The RF power dissipated at the cold stage depends
on the design of the resonators and the input power. For the \umux\ system,
this turns out to be 10~pW per channel. The LNA dissipates 5--10 milliwatts of
power, but it is thermalized at a warmer stage ($\sim$4\,K) which has
substantially more cooling power. The total power consumption of the
system is dominated by the warm electronics and totals around
50--100\,W~\cite{SRON_4kreadout,farir_mkid2016,nikel2016}.

The readout noise should be sub-dominant to the detector noise at all
frequencies, and this has been demonstrated in a variety of MKID
instruments
at sub-millimeter and millimeter frequencies~\cite{calvo_2016,nikel2016,SRON_4kreadout,farir_mkid2016,ritacco_2016}.

%Notes: modern hardware can do 1k resonators in 500 MHz. what limits
%the muxing factor, check 4k paper. tone power at adc and dac,
%quantization noise, and lna power limit / noise floor.

%%%%%%%%%%%%%%%%%%%%%%%%%%%%%%%%%%%%%%%%%%%%%%%%%%%%%%%%%%%%%%%%%%%%%%

%%%%%%%%%%%%%%%%%%%%%%%%%%%%%%%%%%%%%%%%%%%%%%%%%%%%%%%%%%%%%%%%%%%%%%

\begin{figure}[h!]
\centering
\includegraphics[width=\textwidth]{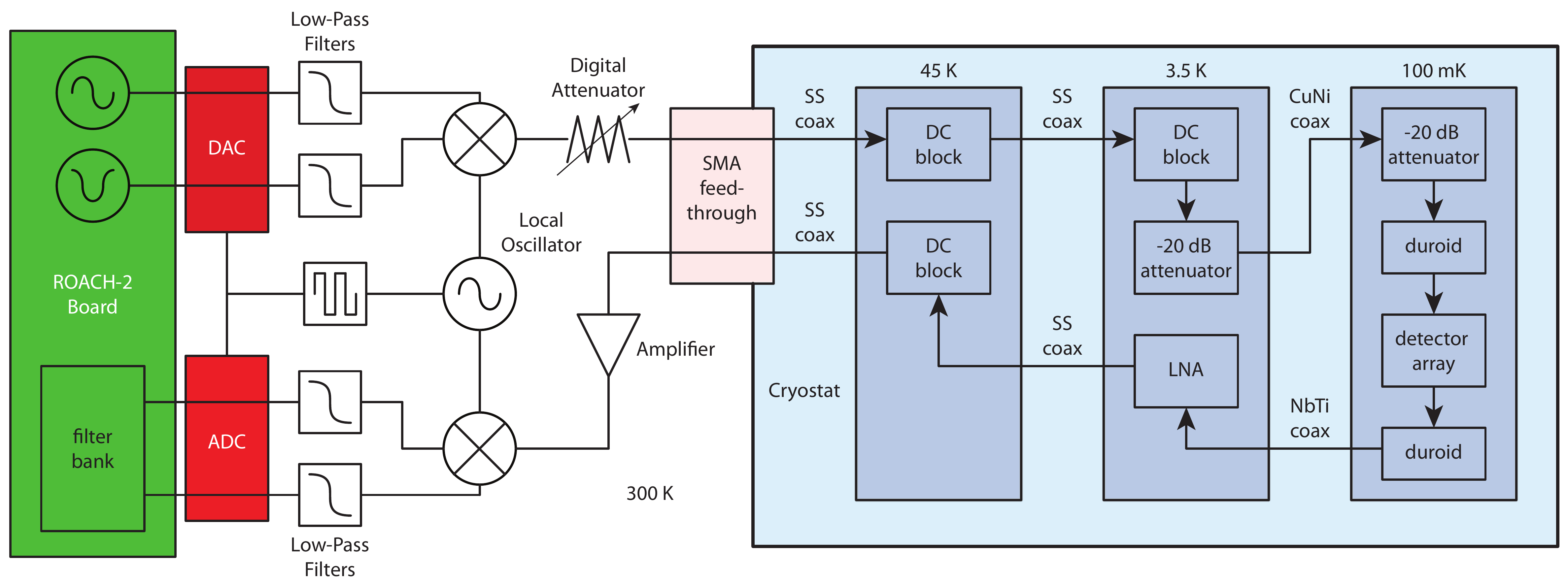}
\includegraphics[width=\textwidth]{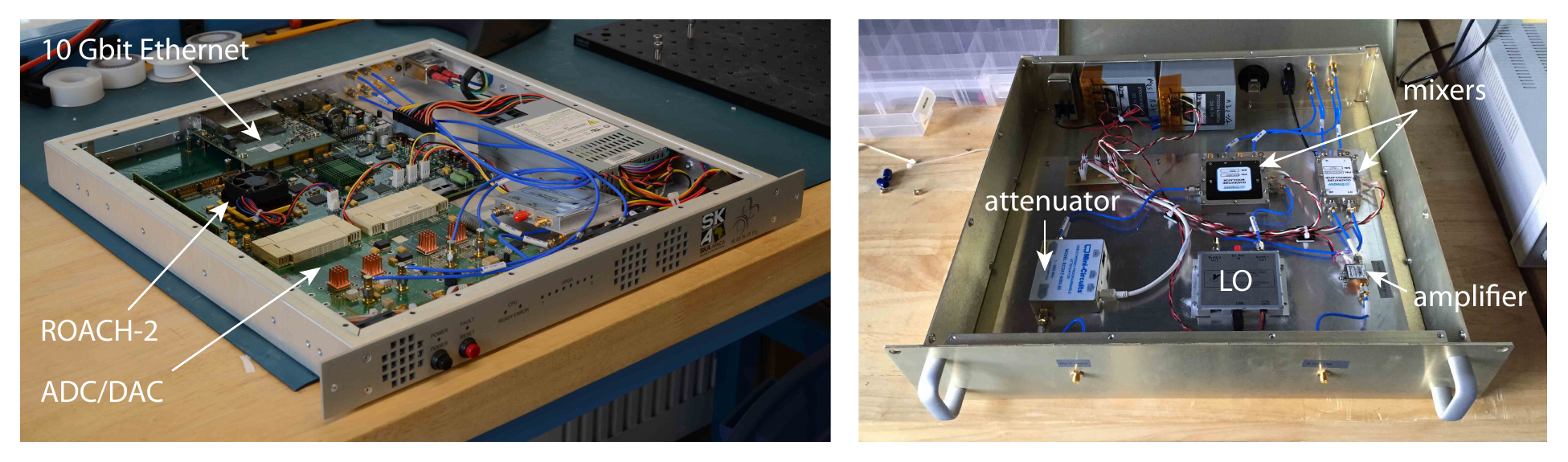}
\caption{
\textbf{(Top)} Readout schematic for the ROACH system showing the
probe tone path in an MKID readout system.
The top left shows the signal generation, digital-to-analog
conversion, and IQ mixing.
The blue portion shows the cryogenic part of the system.
The bottom left shows the demodulation and filtering scheme.
\textbf{Bottom Left:} The ROACH-2 with the DAC/ADC.
\textbf{Bottom Right:} The analog signal conditioning hardware. This
chassis houses the filters, room-temperature mixers, attenuator, warm
amplifier, and the local oscillator shown in the schematic above.
Existing hardware has a multiplexing factor of approximately 500 and
each readout consumes only 20\,W.
Figure from \cite{johnson16mkids}
}
\label{fig:mkid_readout}
\end{figure}

%%%%%%%%%%%%%%%%%%%%%%%%%%%%%%%%%%%%%%%%%%%%%%%%%%%%%%%%%%%%%%%%%%%%%%

\subsection*{Prospects and R\&D path for CMB-S4 for microwave readout}

%MKID readout is typically limited by the detector devices rather than
%the electronic readout.
%
Current tone generation and multiplexing schemes are capable of reading out
thousands of detectors on a single pair of coaxial cables.
%, while meeting the stability, noise, and power requirements for CMB detectors. .
%
%Multiplexing limitations arise from resonator quality factors and resonance frequency collisions within a set readout bandwidth.
%
%Current microwave readout efforts are focused on the implementation of robust systems that capable of multiplexing thousands of resonators on a pair of coax cables.
%
In order to meet the stability, noise, and power requirements for CMB-S4 detectors,
the readout systems would benefit from development several following areas. For some of these technology needs (e.g. faster ADCs and more powerful FPGAs) we can expect to benefit from existing massive development efforts for the communications and computation industries. New developments in microwave readout are exploring direct digital synthesis and demodulation to
eliminate the tuning parameters necessary for the IQ scheme, oversampled polyphase filter banks as an alternative to long IFFTs, and resonator tone tracking and feedback for improved system linearity.  
\begin{itemize}
\item {\bf Implement systems with faster ADC/DACs}
    Increasing the bandwidth of the digital electronics directly increases the
    multiplexing factor of the system.

\item {\bf Validate and/or improve low-frequency noise performance}
    Minimizing the low frequency noise of the readout system is critical for CMB-S4. Instability, drift, and jitter in reference clocks and reference voltages are common sources for such noise.
    To mitigate the low-frequency noise, readout systems commonly place tones off
    resonance to characterize correlated electronic noise,
    which is subsequently subtracted from the detector data~\cite{SRON_4kreadout, farir_mkid2016}. 
    Work should be done to assess if this is sufficient and improve the
    LNA, ADC, and DAC stability. 

\item {\bf Increase linearity}
    The total drive power on a line must be sufficiently low to avoid cross talk between channels caused by intermodulation distortion in the LNA, ADCs and DACs.
    The drive power for MKIDs is -90\,dBm per resonator, allowing for nearly 10,000 detectors on a
    single line before affecting LNA linearity for instance. However, the \umux\ system is driven at a higher power and runs into
    this limit at $\sim$1,000 detectors. Development of drive systems that track and feedback on a resonator's fundamental frequency (tone following) will greatly relax such limits and are being actively pursued.  
    
%\item {\bf Improve clock distribution stability}
%    Instability, drift, and jitter in the clock distribution will introduce
%    1/$f$ noise into the system.

\item {\bf Investigate universal backend electronics}
    The readout electronics for DfMux, \umux\ and MKIDs rely on similar core electronics but operate at different frequencies. Universal backend electronics with swappable RF daughter boards should be investigated. This will be especially useful if CMB-S4 will support deployment of both TES and MKID focal planes. A system that can support FDM at high and low frequencies would allow the same core electronics to be used for the readout of both detector systems.

%\item {\bf Develop FPGA firmware for warm electronics} For MKIDs in
%  particular, FPGA-based readout hardware implementations suitable for
% CMB-S4 using open-source and commercially available components
% exist.  However for CMB-S4 every microwave readout system considered
% here will require FPGA firmware development.  Open-source software
% packages are already available as starting points for this work.

%\item {\bf Superconducting coaxial cables}
%    Improvements in the robustness and flexibility of the superconducting cables
%    would help for instruments with 100s of coax pairs.

\end{itemize}

%% file: readout/conclusion.tex
\section{Conclusion}
\label{sec:readout_conclusion}

%\ref{section:TES} -> 5.2
%\ref{section:mkids} -> 5.3
%\ref{section:tdm} -> 5.4
%\ref{section:fdm} -> 5.5
%\ref{section:umux} -> 5.6
%\ref{sec:mkid_readout}  -> 5.7
%\ref{sec:readout_conclusion} -> 5.8
%\ref{sec:redsum} -> 5.9 

We are in the fortunate situation of having low-noise sensors and
associated readout with sufficient performance for CMB-S4's likely technical
specifications already in hand. We also have sufficient time to pursue R\&D in areas
that have high probability of lowering the overall CMB-S4 detector
budget, including the costs of R\&D as well as of production, assembly
and quality assurance.  The existing solution comprises TES bolometers
with either of the existing readout methods outlined in Sections~\ref{section:tdm}
(TDM) or \ref{section:fdm} (DfMux with interrogation frequencies $\lesssim$~10\,MHz).

TES bolometers possess a long record of CMB science results from
well-characterized kilopixel arrays. In particular their noise
properties have been demonstrated to be sufficient for CMB-S4
needs. TES bolometers are ready for R\&D investment in scaling up
their production rate and expanding quality assurance testing
facilities in universities and national labs.  All the TES readout
techniques would benefit from R\&D investment in scaling up the
production rate for SQUIDs.  Several promising avenues exist for R\&D
to reduce the assembly complexity and thus the cost and schedule for
the readout for TESs, including continued development of \umux{}
readout at GHz interrogation frequencies (Section~\ref{section:umux}) and exploration
of FDM RLC with frequencies $\gtrsim$~50~MHz (Section~\ref{section:fdm}).

Another approach would be to use MKIDs, which were designed for highly
multiplexed readout at GHz interrogation frequencies with minimal
assembly complexity since the readout is integrated on the detector
wafer. Readout of arrays with large multiplexing factors (O(1000)) through a single coaxial cable has been demonstrated.
 Lab tests with MKIDs show nearly comparable noise performance
to that from TES bolometers, though adequate low frequency noise
performance has been demonstrated only in a lumped element MKID
design. However, MKIDs have not yet been deployed in any CMB
instrument.  An on-sky CMB mapping demonstration is essential to
validate MKIDs in the field before considering continuing their
development for CMB-S4 production.

The GHz-interrogation readout techniques would all benefit from R\&D
investment into scaling up production of low-noise,
low-dissipation-power 4K IF amplifiers.  In all cases, the warm
readout electronics appear scalable with R\&D. FDM schemes for both TESs
and MKIDs use similar room-temperature biasing and readout electronics,
enabling common development. In particular, schemes to ensure
linearity of these systems at high multiplexing count should be
validated.

%% file: readout/readoutSumTable.tex
\section{Summary of sensor and readout technologies}\label{sec:redsum}
%TSL and PSL definition
%
%\begin{table} [!h]
%\begin{center}
%\begin{tabular} {cl}
%\hline
%\textbf{TSL} & \textbf{Description} \\
%\hline
%1 & Lab test of technology to show principle\\
%2 & Lab test of technology but with full feature set and performance suitable for ground test\\
%3 & Experiment capable version built and tested in the lab\\
%4 & Deployed in a CMB experiment and data taken\\
%5 & Data fully analyzed so systematic errors understood\\
%\hline
%\hline
%\textbf{PSL} & \textbf{Description} \\
%\hline
%1 & Fabrication of a TS1/TS2 prototype demonstrated\\
%2 & Fabrication of a one or more experimental capable units\\
%3 & Conceptual plan of methods for production at scale\\
%4 & Demonstrated the critical steps for production at scale\\
%5 & Capability for production at scale exists and is demonstrated\\
%\hline
%
%\end{tabular}
%\end{center}
%\end{table}

\begin{landscape}
\begin{table}
\begin{center}
\begin{tabular} {|l|c|c|c|c|c|}
\hline
 & \textbf{Lab Demonstration} &  \textbf{Sky Demonstration} & \textbf{Path to CMB-S4} & Section & T/PSL\\
\hline
\hline
\textbf{Sensors}  &  &  &  &  & \\
\hline
Transition edge sensors & - & ACTPol/SPTpol/{\bicepI}s/PB & mass prod\ &  \ref{section:TES} & 5/4 \\
                        &   &                         & material optimization & &\\
\hline
MKID & 150 GHz LEKID & 1.2THz BLAST-TNG (2017) & On sky pol demo, &  \ref{section:mkids} & 2/3 \\
& & NIKA2 & fab uniformity, & &\\
& & & mass prod &  &\\
\hline
\hline
\textbf{Cold Multiplexers}  &  &  &  &  & \\
\hline
Time-division  & 64x mux & ACTPol/{\bicepI}s & interconnects, & \ref{section:tdm} & 5/4 \\
& & & mux factor & & \\
\hline
Frequency-division: 5MHz DfMux & 68x mux & SPT/PB & higher freq resonances, & \ref{section:fdm} & 5/4 \\
50 MHz fMux & & & integrated fab & & 1/1 \\
\hline
Microwave-multiplexed  & 66x mux & MUSTANG2 & Resonator freq density, & \ref{section:umux} & 3/3 \\
SQUIDs (\umux{}) & & & resonator size, & & \\
& & & integrated det fab & & \\
\hline
MKID & 1000x & 400x NIKA2 & resonance spacing & \ref{section:mkids} & 2/3 \\
\hline
\hline
\textbf{Room temperature readout}  &  &  &  &  & \\
\hline
TDM & 64x mux & ACTPol/{\bicepI}s & mux factor & \ref{section:tdm} & 5/4 \\
\hline
DfMUX & 256x mux & SPTpol/PB & mux factor & \ref{section:fdm} & 5/4 \\
\hline
Microwave & 1000x mux SRON & - & increased linearity, & \ref{sec:mkid_readout} & 3/3 \\
& & & universal back-end & & \\
\hline
\end{tabular}
\end{center}
\end{table}
\end{landscape}

%% file: conclusion_cmbs4.tex
\chapter{Conclusion and Future Work} \label{ch:conclusion}
\vspace*{\baselineskip} 
\vspace{1cm} 

The design of the experimental configuration of CMB-S4 will be dictated by technical requirements determined by the science objectives.  The top-level requirements will be 
on 1) the instrumental sensitivity $\cal{S}$  as a function of angular scale (or multipole $\ell$) and frequency $\nu$,   $\cal{S}(\ell, \nu)$; 2) the suppression of systematic errors that could overwhelm $\cal{S}(\ell, \nu)$;  and 3) the area of sky and amount of overlap with existing and planned surveys in other wavelength bands.  
The recent explosion of progress in measuring the CMB, in interpreting data from the CMB, and in designing and deploying  new technologies  for even better measurement of the CMB has made the ambitious scope of the  CMB-S4 science objectives possible.  This Technology Book represents an essential first step in planning for the design of CMB-S4:  assessing the status and potential of the myriad technical options.  

We have  presented the progress to date in developing technologies relevant for achieving the science goals of CMB-S4.  The book focused on the technical challenges in four broad areas, demonstrated that multiple pathways exist in each area,  quantified the readiness of each of the pathways in terms of TSL and PSL,  presented pros and cons of competing methods, and indicated the breadth of  R\&D required to advance each alternative sufficiently for use in the CMB-S4 construction project.   A partial list of the elements not yet addressed  includes: cryogenics for cooling massive focal planes, instrumental control and monitoring, tools and techniques for in situ calibration and validation during integration and test, data storage and management, and power management for remote sites.  These elements  do not drive the overall instrument concept.

Armed with the facts tallied here, the CMB-S4 community will now proceed to the next steps required  to plan the experimental configuration. These include:

\begin{enumerate} 
 \item Development of good estimates of cost, schedule, performance and risk or the technological alternatives, with the aim of prioritizing R\&D needs; 
 \item Definition of performance and risk metrics for evaluating systematic errors; 
 \item End-to-end propagation of the performance properties of each subsystem to estimates of $\cal{S}(\ell, \nu)$; and 
 \item Assessment of and planning for the instrumental elements not included in this version of the Technology Book.  
 \end{enumerate}

This book is the most comprehensive compendium of instrumentation for the CMB ever compiled.  It benefited dramatically from the cooperation among instrumentalists from more than a half dozen CMB experiments.

%% file: acronym_total.tex
%\documentclass[a4paper]{article}
%\begin{document}
\section*{List of acronyms}

%total
\begin{flushleft}
\begin{tabular} {ll}
%\textbf{3-D} & tridimenSsional\\
%\textbf{ABS} & Atacama B-Mode Search \\
\textbf{ADC} & Analog to Digital Converter\\
%\textbf{AdvACT} & Advanced Atacama Cosmology Telescope\\
\textbf{ADM} & Artificial Dielectric Material \\
\textbf{ADR} & Adiabatic Demagnetization Refrigerator \\
\textbf{AHWP} & Achromatic Half Wave Plate \\
%\textbf{ANSYS} & \\
\textbf{AR} & Anti Reflection \\
\textbf{ARC} & Anti Reflection Coating \\
\textbf{AZ} & Azimuth \\
\textbf{BUG} &Backshort-Under-Grid \\
\textbf{CAD} & Computer Assisted Design \\
%\textbf{CCD} &  \\
\textbf{CD} & Critical Decision \\
\textbf{CDM} &code-division multiplexing \\
\textbf{CFRP} & Carbon Fiber Reinforced Polymer  \\
\textbf{CMB} & Cosmic Microwave Background \\
\textbf{CHES} & Controlled Heat Extraction System \\
\textbf{CLASS} & Clear Large Aperture Sapphire Sheets \\
%\textbf{CLASS} & Cosmology Large Angular Scale Surveyor \\
\textbf{CMM} &  Coordinate Measuring Machine\\
\textbf{CNC} & Computer Numerical Control  \\
\textbf{CPW} & Co-Planar Wave guide\\
%\textbf{CSO} &Caltech Submillimeter Observatory  \\
\textbf{CTE} & Coefficient of Thermal Expansion \\
\textbf{DAC} & Digital to Analog Converter \\
\textbf{DAN} & Digital Active Nulling \\
%\textbf{DfMUX} & Digital frequency Multiplexing\\
%\textbf{DOE} & Department of Energy \\
\textbf{DRIE} & Deep Reactive Ion Etching \\
\textbf{EFG} &  Edge-defined Film-fed Growth \\
\textbf{EL} & Elevation \\
%\textbf{EM} &electromagnetic \\
\textbf{FDM} & Frequency Division Multiplexing\\
\textbf{FIR} & Far Infrared Radiation \\
\textbf{FOV} & Field Of View \\
%\textbf{FNL} & \comred{???} \\
\textbf{FP} & Fabry-Perot \\
\textbf{FTS} & Fourier Transform Spectrometer \\
\textbf{FWHM} & Full Width at Half-Maximum \\
\textbf{FPGA} &  field-programmable gate array\\
\textbf{GRIN} & GRradient-INdex\\
\textbf{GBT} &Green Bank Telescope \\
\textbf{HDPE} & High Density PolyEthylene \\
\textbf{HEM} & Heat Exchanger Method \\
\textbf{HEMT} & High-electron-mobility transistor\\
\textbf{HF} & High-Frequency \\
%\textbf{HTT} & Huan Tran Telescope \\
\textbf{HWP} & Half Wave Plate \\
\end{tabular}
\end{flushleft}

\begin{flushleft}
\begin{tabular} {ll}
\textbf{IDC} & InterDigitated Capacitor \\
\textbf{IGW} &Inflationary Gravitational Wave\\
\textbf{IR} & InfraRed \\
\textbf{JFET} & Junction gate Field-Effect Transistor \\
\textbf{LAIS} &  Laser Ablated Infrared Shaders\\
\textbf{LCLS-II} & Linac Coherent Light Source II \\
\textbf{LDPE} & Low-Density PolyEthylene \\
\textbf{LED} & Light-Emitting Diode \\
\textbf{LEKID} & Lumped-Element Kinetic Inductance Detector \\
\textbf{LF} & Low Frequency \\
%\textbf{LHE} & Liquid Helium \\
%\textbf{LN2} & Liquid Nitrogen \\
\textbf{LNA} & Low Noise Amplifier \\
%\textbf{MBAC} & Millimeter Bolometric Array Camera \\
\textbf{MCE} & Multi-Channel Electronics\\
\textbf{MEM} & Micro-ElectroMechanical \\
\textbf{MF} & Medium Frequency \\
\textbf{MKID} & Microwave Kinetic Inductance Detector \\
\textbf{MMF} &  Metal Mesh Filter \\
\textbf{MMARC} & Meta-Material Anti-Reflection Coating\\
\textbf{MML} & Meta-Material Lens \\
\textbf{NET} & Noise Equivalent Temperature\\
\textbf{NEP} & Noise Equivalent Power\\
\textbf{NIR} & Near InfraRed \\
\textbf{NTD} & Neutron Transmutation Doped \\
\textbf{OMT} & OrthoMode Transducer \\
%\textbf{PB-2} & \Pb-2 \\
%\textbf{PIPER} & Primordial Inflation Polarization Explorer \\
\textbf{PPTFE} & Porous PolyTetraFluoroEthylene \\
\textbf{PSL} & Production Status Level \\
\textbf{PTC} & Pulse-Tube Cooler \\
\textbf{PTFE} & PolyTetraFluoroEthylene \\
\textbf{QWP} & Quarter Wave Plate \\
\textbf{R\&D} & Research and Development \\
\textbf{RF} & Radio Frequency \\
%\textbf{RMS} & Root mean square \\
\textbf{RT-MLI} & Radio-Transparent Multi-Layer Insulation \\
\textbf{SMB} & Superconducting Magnetic Bearings \\
\textbf{SQUID} & Superconducting Quantum Interference Device \\
\textbf{SOI} &Silicon-on-Insulator \\
\textbf{SQ1} & first-stage SQUID \\
\textbf{SSA}& series SQUID array \\
\textbf{SWS} & SubWavelength Structures \\
\textbf{Tc} & critical Temperature \\
\textbf{TES} & Transition Edge Sensor  \\
\end{tabular}
\end{flushleft}

\begin{flushleft}
\begin{tabular} {ll}
\textbf{TL} & Transmission Line\\
\textbf{TLS} & Two-Level System \\
\textbf{TDM} & Time Division Multiplexing\\
%\textbf{TMM} & \comred{??} \\
\textbf{TOD} & Time Ordered Data  \\
\textbf{TSL} & Technology Status Level\\
%\textbf{UBC} & University of British Columbia\\
\textbf{UHMWPE} & Ultra-High Molecular Weight PolyEthylene \\
%\textbf{\umux{}} &microwave SQUID multiplexer\\
\textbf{VNA} & Vector Network Analyzer \\
\textbf{VPM} & Variable Polarization Modulator\\
\end{tabular}
\end{flushleft}
%
%\end{document} 